%% file: main.tex
\documentclass{book}
\usepackage[T1]{fontenc}
\usepackage[hidelinks]{hyperref}
\usepackage{graphicx}
\usepackage{xcolor}
\usepackage{tcolorbox}
\usepackage[numbers,sort&compress]{natbib}
\usepackage{bm,bbold}
\usepackage{bbm}
\usepackage{amsfonts, amsmath, amsthm, amssymb}
\usepackage{mathtools}
\usepackage{tikz}
\usepackage{tikzsymbols}

\usepackage{marvosym}
\usetikzlibrary{decorations.pathmorphing}
\numberwithin{equation}{chapter}
\usepackage[normalem]{ulem}
 
\usepackage[utf8]{inputenc}

\usepackage{subcaption}
\usepackage{color}
\usepackage{comment}
\usepackage{algorithm} 
\usepackage{algpseudocode}
\usepackage{braket}
\usepackage{svg}       
\usepackage{tikz-3dplot}
\usepackage{pdfpages}
\usepackage{fancyhdr}
\renewcommand{\vec}[1]{\bm{#1}}

\usepackage{tcolorbox}
\newtcolorbox{mybox}[1][]{
    colback=gray!20, 
    colframe=gray!80, 
    fonttitle=\bfseries, 
    coltitle=black, 
    boxrule=0.5pt,
    title=#1 
}
 
\begin{document}

\begin{titlepage}
    \centering
    \vspace*{4cm} 

    {\huge Deep Learning in Classical and Quantum Physics}
    \vspace{1.5cm}
 
   {\Large Lecture Notes \par}
    \vspace{1cm}

        {\large 
        Timothy Heightman\textsuperscript{1, 2}, 
        Marcin P{\l}odzie\'n\textsuperscript{1,3} \par
        \vspace{1cm}
        \textsuperscript{1}ICFO – Institut de Ciències Fotòniques, \\
        The Barcelona Institute of Science and Technology, \\
        08860 Castelldefels (Barcelona), Spain \par
        \vspace{0.3cm}
        \textsuperscript{2}Quside Technologies SL, Carrer d’Esteve Terradas, 1, 08860 Castelldefels, Barcelona, Spain\par
        \vspace{0.3cm}
        \textsuperscript{3}Qilimanjaro Quantum Tech, Carrer de Veneçuela 74, 08019 Barcelona, Spain\par
        \vspace{1cm}
    }

    \vspace{1.5cm}

    {\large \today \par}
    {\textcopyright \quad Timothy Heightman and Marcin P{\l}odzie\'n (2025)}    
    \vfill
\end{titlepage}

 \include{chapters/chapter_0_preface_short}

\tableofcontents

    \include{chapters/chapter_1_introduction}

\include{chapters/chapter_2_NN_fundamentals}

  \include{chapters/chapter_3_unsupervised_learning}

   \include{chapters/chapter_4_quantum_basics}

  \include{chapters/chapter_5_deep_learning_quantum}

  \include{chapters/chapter_6_conclusion}

\bibliographystyle{unsrt}  
\bibliography{references} 


\end{document}

%% file: chapters/chapter_0_preface_short.tex
\section*{Preface}

Scientific progress has always gone hand in hand with the emergence of new tools. From the telescope and microscope in centuries past, to the high-energy particle accelerators and supercomputers of the modern age, each technological leap has opened the door to novel discoveries and deeper understanding. Today, the rise of machine learning (ML)—and in particular its deep learning (DL) subfield—represents the next step in this tradition of transformative research tools.

In quantum science and technology, where the systems under study are inherently complex, deep learning offers unprecedented opportunities. When combined with quantum-based methods, ML can help navigate vast parameter spaces, identify patterns in experimental data, and suggest new directions for research. Quantum physicists, chemists, engineers, and computer scientists can already see how modern AI methods assist with tasks that range from refining quantum control protocols to discovering novel materials with quantum properties. It has become increasingly clear that an understanding of ML and DL concepts is not just a side skill but an essential part of the toolkit for the next generation of quantum scientists.

However, with any powerful method, it is just as important to recognize the boundaries and potential pitfalls as it is to celebrate the successes. Deep learning excels at finding correlations in large datasets, but it can also overfit to noisy data or produce results that lack direct physical insight. Understanding these limitations—and devising strategies to mitigate them—will help ensure that ML-driven methods remain scientifically rigorous.

These lecture notes aim to bridge the gap for graduate-level students coming from physics, mathematics, engineering, or computer science. Through a combination of conceptual explanations and hands-on examples, the notes emphasize both the power and the practical constraints of deep learning as applied to quantum technologies. The overarching goal is to equip readers with the foundational knowledge to recognize when and how to use deep learning effectively, while remaining aware of the subtleties involved in adapting AI methods to quantum science. They are intended to be a complete, detailed description of deep learning, with some of its applications in the quantum sciences. The material builds progressively, presenting ideas in a sequence that mirrors how they are best understood and applied to physics problems. For this reason, \textit{it is recommended to read the notes from beginning to end—each chapter sets the stage for the next, and the storyline is continuous}. By working through the chapters in order, readers will develop a more cohesive understanding of both the theoretical and practical aspects of deep learning in quantum science.

As you work through these notes, keep in mind that the greatest breakthroughs often arise when bold ideas and new tools intersect. We hope you will be inspired to explore how deep learning can propel your own research within quantum sciences, and, in doing so, contribute to the next generation of scientific and technological innovation.

\section*{Acknowledgements}

We would like to thank Ruth Mora-Soto, Carlos Pasqual, Marie-Ange, and Jose Ramon-Martinez, who helped with proof-reading. We would also like to thank 
Edward Jiang for Fig.~5.15 in Chapter 5, and Luke Mortimer for Figs.~5.19, 5.20 and 5.21.

Please reach out to \texttt{timothyheightman@gmail.com}, \texttt{mplodzien@gmail.com} if you see any typos or errata in formulae so we can correct them for future revisions. 

TH acknowledges support from the Government of Spain (Severo Ochoa CEX2019-000910-S, Quantum in 
Spain, FUNQIP and European Union NextGenerationEU PRTR-
C17.I1), the European Union (PASQuanS2.1, 1011 13690  
and Quantera Veriqtas), Fundació Cellex, Fundació Mir-
Puig, Generalitat de Catalunya (CERCA program), the 
ERC AdG CERQUTE and the AXA Chair in Quantum 
Information Science. MP acknowledges support from the European Research Council (ERC) under the Advanced Grant \emph{NOQIA}. This work has received funding from the Spanish Ministerio de Ciencia, Innovación y Universidades (MCIN) and Agencia Estatal de Investigación (AEI) (MCIN/AEI/10.13039 
\\ /501100011033) through projects PGC2018-091013-B-I00, CEX2019-000910-S, Plan Nacional FIDEUA PID2019-106901GB-I00, Plan Nacional STAMEENA PID2022-139099NB-I00, and the FPI Programme. Additional support was provided by the European Union NextGenerationEU/PRTR (PRTR-C17.I1).
Funding has been received from the QuantERA programme, project \emph{DYNAMITE} (PCI2022-132919), under the QuantERA II Programme co-funded by the European Union’s Horizon 2020 research and innovation programme (Grant Agreement No. 101017733). Further support comes from the Ministry for Digital Transformation and Public Service of the Spanish Government through the \emph{QUANTUMENIA} call Quantum Spain project—and by the European Union through the Recovery, Transformation and Resilience Plan—NextGenerationEU within the framework of the \emph{Digital Spain 2026} Agenda.
We acknowledge the contributions of Fundació Cellex, Fundació Mir-Puig, and the Generalitat de Catalunya (European Social Fund—FEDER and the CERCA Programme), as well as computing resources provided by the Barcelona Supercomputing Center—MareNostrum (FI-2023-3-0024).
This project has also received funding under the European Union programmes HORIZON-CL4-2022-QUANTUM-02-SGA \emph{PASQuanS2.1} (Grant Agreement No. 101113690), EU Horizon 2020 FET-OPEN \emph{OPTOlogic} (Grant No. 899794), \emph{QU-ATTO} (Grant Agreement No. 101168628), and EU Horizon Europe \emph{NeQST} (Grant Agreement No. 101080086). Internal support from ICFO through the \emph{QuantumGaudí} project is also acknowledged.
The views and opinions expressed are those of the author(s) only and do not necessarily reflect those of the European Union, the European Commission, the European Climate, Infrastructure and Environment Executive Agency (CINEA), or any other granting authority. Neither the European Union nor any granting authority can be held responsible for them.

%% file: chapters/chapter_1_introduction.tex
\chapter{Introduction}
\label{CH:INTRODUCTION}

How then can we understand Deep Learning and Neural Networks, to later apply them problems in quantum and classical physics? Nowadays AI has become somewhat of a popular science. Deep learning is often stated to be Artificial Intelligence, however this is not the case. The latter is in fact a much bigger field. It is therefore important to have a broader view on the AI research over the last few decades to understand how deep learning fits into the bigger picture. 

This chapter briefly covers the key elements of AI, and how they came about to become an active and collective research field. This will allow you to get a feel for the landscape of this field, and see where deep learning is playing a major role.

\section{Artificial Life}

Research on Artificial Life explores the principles of life through simulations, synthetic biology, and computational models. This field intersects with complex systems modelling, computational sciences, and biology, providing insights into emergent behaviour, self-organization, and adaptation. By studying simplified models such as cellular automata and Langton’s ant, researchers uncover fundamental properties of life and evolution while advancing tools for solving real-world problems.

Cellular automata (CA) are mathematical models composed of grids of cells, each of which evolves over discrete time steps according to a set of rules based on the states of neighbouring cells. Despite their simplicity, cellular automata can exhibit extraordinarily complex behaviour, making them valuable tools for exploring emergent phenomena and simulating processes such as growth, diffusion, and pattern formation.

One of the most famous examples of cellular automata is John Conway’s \textit{Game of Life}. Introduced in the 1970s, the \textit{Game of Life} operates on a two-dimensional square grid with simple rules for cell survival, birth, and death based on its eight nearest neighbours (above, below, and diagonal). The rules are as follows:
\begin{itemize}
    \item Any live cell with two or three live neighbours survives to the next generation.
    \item Any dead cell with exactly three live neighbours becomes a live cell (is "born") in the next generation.
    \item All other live cells die in the next generation, either due to isolation (fewer than two live neighbours) or overcrowding (more than three live neighbours).
\end{itemize}

\begin{figure}
    \centering
    \includegraphics[width=\linewidth]{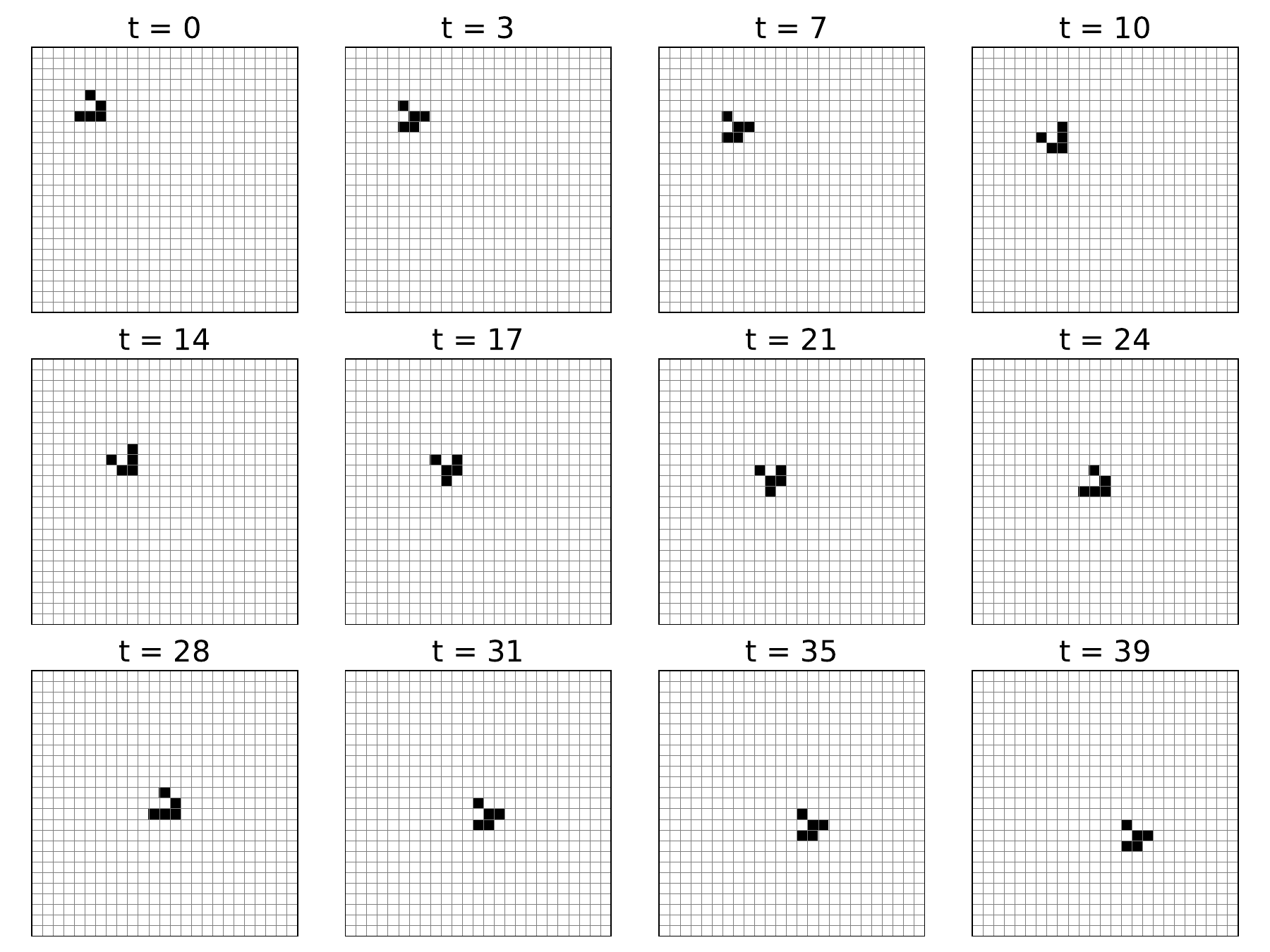}
    \caption{
    Time-lapse of Conway’s \textit{Game of Life} “glider",  first discovered by Richard K. Guy in 1969.  
The animation begins with the canonical five-cell seed ( $t=0$)  and advances one update per frame. The pattern repeats every four steps while translating one cell down and one cell right, illustrating both the glider’s period-4 oscillator nature and its ability to propagate information diagonally across the lattice.  
    The sequence showcases how simple birth-and-survival rules can produce a self-propelling, information-carrying structure on an otherwise empty grid.
    }
    \label{fig:conway_gol_glider}
\end{figure}

These simple rules lead to stunningly diverse outcomes, including static structures ("still lifes"), periodic patterns ("oscillators"), and even patterns that move across the grid ("spaceships"), see Fig.~\ref{fig:conway_gol_glider}. The \textit{Game of Life} also demonstrates the potential for Turing completeness \cite{Rendell2002,Myreen2025}, meaning that it can simulate any computational process given the right initial configuration.

The Conway model has profound implications for understanding complexity and emergence. It shows how intricate behaviors can arise from basic interactions, a principle that resonates across natural and artificial systems. Researchers have used the \textit{Game of Life} to study self-replication, computation, and even artificial evolution, cementing its place as a cornerstone of artificial life studies.

\begin{figure}
    \centering
    \includegraphics[width=\linewidth]{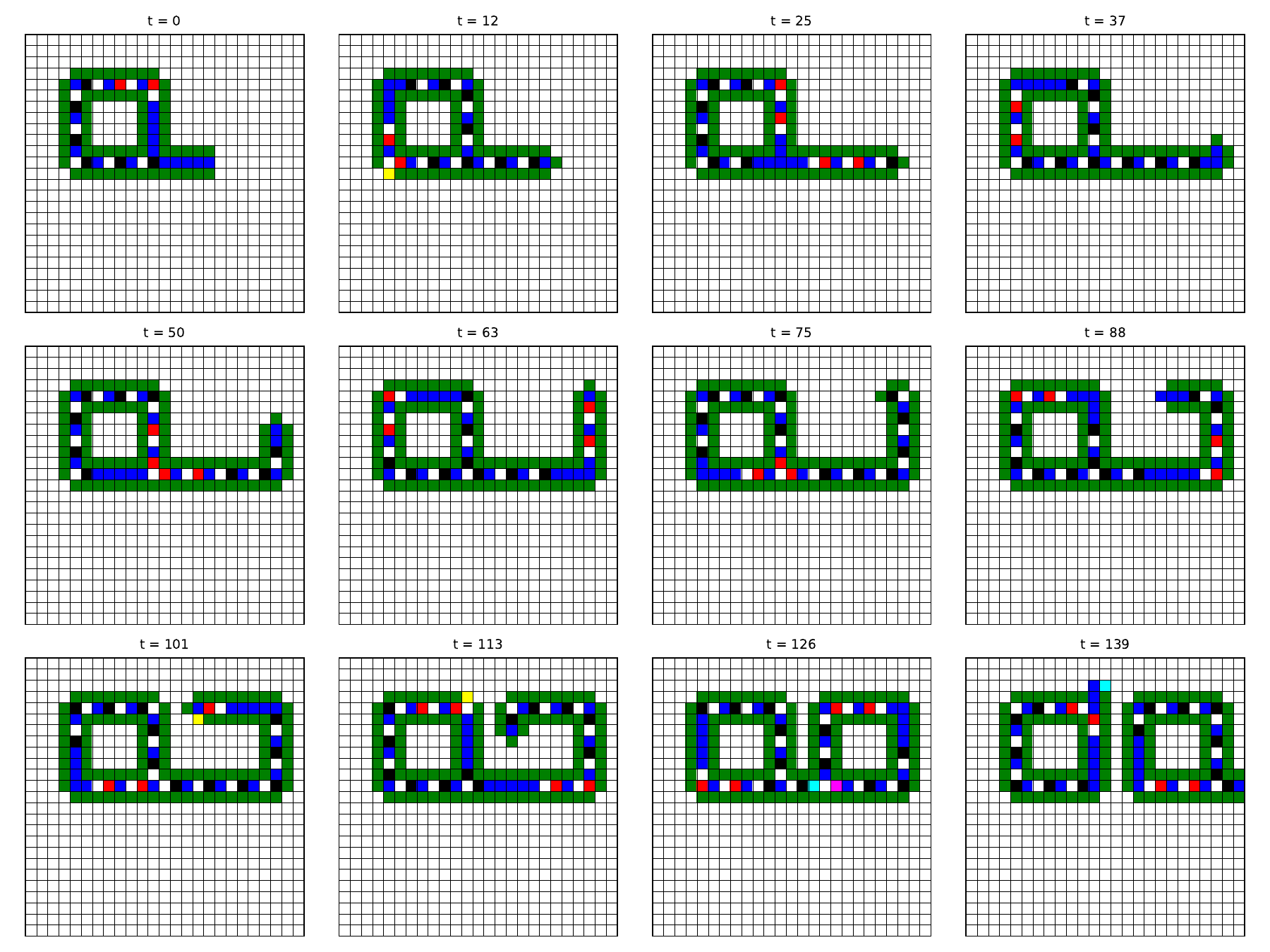}
    \caption{
    Langton’s self-replicating loop: an initial loop follows the finite set of rules resulting in copying itself, producing an identical “offspring” after 130 generations—illustrating how simple local rules in a cellular automaton can give rise to autonomous reproduction and the hallmarks of artificial life.
    }
    \label{fig:langton_loop}
\end{figure}

Langton's loop is a landmark illustration of self-replication in cellular automata and, by extension, artificial life. Working on an eight-state square lattice, the loop begins as a compact 'parent' ring whose interior wiring encodes the rules of its own construction. During each update cycle, it extends a narrow arm that writes a duplicate ring, then neatly seals the offspring before withdrawing. The result is a proliferating garden of identical loops, each able to copy itself, demonstrating that purely local, deterministic rules can generate the lifelike hallmarks of growth, heredity, and colonization across the grid; see Fig.~\ref{fig:langton_loop}.

Langton’s Loop, similarly to Conway's \textit{\textit{Game of Life}}, highlights how simple rules can lead to the appearance of order and complexity, a principle applicable to fields such as robotics, autonomous systems, and artificial intelligence. It serves as a metaphor for understanding self-organization in biological systems and demonstrates how localized interactions can give rise to macroscopic patterns without centralized control.

The field of Artificial Life examines systems composed of many interacting components that exhibit behaviors that are not predictable from their individual parts. Models inspired by artificial life have been used to study swarm behavior in animals, urban growth, traffic flow, and even financial markets. These simulations help researchers identify the underlying principles of system dynamics, allowing for better predictions and optimizations.

Artificial life contributes to both theoretical understanding and practical applications. In synthetic biology, principles derived from live models are applied to engineering new forms of life or to modify existing organisms for tasks like drug production and environmental remediation. In computer science, artificial life inspires algorithms for optimization and decision-making, such as genetic algorithms and swarm intelligence.

Philosophically, artificial life challenges our understanding of what constitutes \textit{life}. By demonstrating life-like properties in non-biological systems, we must reconsider the boundaries between natural and artificial, raising questions about the essence of life and consciousness. It also invites us to reflect on the ethical and societal implications of creating synthetic systems that mimic or replicate living organisms.

\section{Artificial Intelligence}

Human fascination with creating intelligent machines is not new. Ancient myths, such as the Greek tale of Talos reflect humanity's enduring desire to animate the inanimate. The Enlightenment era brought more systematic approaches to mechanizing thought, culminating in inventions like Charles Babbage's Analytical Engine—a precursor to modern computing.

However there is a fundamental problem in definition of the \textit{intelligence} itself.
Intelligence remains a mystery, a concept without a universally accepted definition. Philosophers, psychologists, scientists, and engineers all hold varying views on what intelligence is and how it arises. We observe signs of intelligence in nature, such as the collaborative behaviour of animal groups, and in human thought and behaviour. Generally, intelligence is attributed to entities that exhibit autonomy and adaptability. 

Autonomy implies the ability to operate without constant external instructions, while adaptability refers to the capacity to modify behaviour in response to changes in the environment or problem. The fundamental driver of intelligent actions is data. The images we see, sounds we hear, and measurements of surrounding objects are all forms of data. We process these inputs to make decisions. Therefore, understanding data is crucial for grasping artificial intelligence (AI) algorithms.

Improper data selection, flawed representation, or missing data can hinder the performance of algorithms. Data can be quantitative, representing measurable values like temperature, or qualitative, capturing subjective observations such as a flower’s scent or a political view. While quantitative data is easier to interpret, qualitative data reflects individual perceptions rather than absolute truths. Data itself is raw and unbiased, but in practice, it is collected, recorded, and contextualized by humans, often shaped by specific purposes or assumptions. When data is analysed to derive meaningful conclusions, it becomes information. Applying information in practical and conscious ways generates knowledge—a quality AI aims to simulate.

The 20th century marked a transformative era for research on Computational Intelligence, with milestones such as the creation of programmable digital computers and the development of early AI algorithms. The Dartmouth Conference in 1956, often considered the birth of AI as a formal field, brought together pioneers who envisioned machines capable of reasoning, learning, and problem-solving. Since then, advances in computational power, data availability, and algorithmic sophistication have propelled AI into the forefront of scientific and technological innovation.

Today's AI systems excel in narrow domains, achieving remarkable feats in fields such as natural language processing, computer vision, and robotics. However, these systems are far from achieving the general intelligence exhibited by humans. The interplay between cognitive science, neurobiology, and AI development continues to drive progress, but challenges remain in understanding and replicating the full complexity of human thought and emotion.

The journey towards creating artificial systems with human-like intelligence intersects with cognitive science, neurobiology, and the philosophy of intelligence.
Cognitive science investigates the processes underlying thought, learning, perception, and problem-solving. By modelling human cognition, researchers aim to uncover principles that could guide the development of artificial systems capable of similar intellectual feats. Concepts such as memory, attention, and decision-making are studied not only to understand human intelligence but also to design systems that replicate these capabilities. For instance, neural networks in AI take inspiration from the way neurons in the brain process information, emphasizing pattern recognition and adaptability.

A complimentary approach to explore the human brain is Neurobiology. Neurobiology considers the workings of the human brain from a biological lens (rather than a computational one). However, the human remains the most advanced "intelligent machine" we know. Hence, efforts to mimic the brain's architecture have heavily influenced AI development. Neural networks, which form the basis of deep learning, are inspired by the synaptic connections between neurons. Moreover, emerging fields like neuromorphic computing seek to replicate the efficiency and parallelism of the brain's computational processes. By understanding the mechanisms of neuroplasticity, learning, and memory, AI research aims to develop systems that not only process information but also learn and evolve over time.

In parallel to cognitive and biological sciences, the philosophy of AI asks more fundamental questions: What does it mean to be intelligent? Can machines possess consciousness or self-awareness? Philosophical inquiries into the nature of mind and intelligence shape both the ethical and theoretical frameworks of AI development. Concepts like the Turing Test, proposed by Alan Turing, provide benchmarks for assessing machine intelligence, while debates about free will, creativity, and moral responsibility continue to challenge our understanding of what it means to build ``intelligent'' systems.

\section{Landscape of Artificial Intelligence Research}

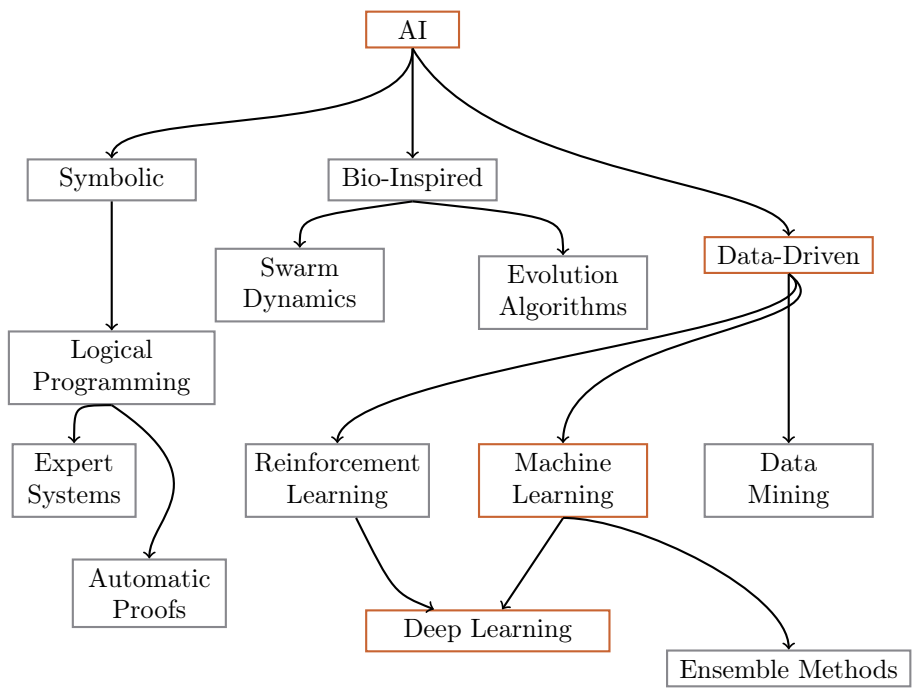
\begin{figure}
    \centering
    \begin{tikzpicture}

        \node[draw={rgb,255:red,200; green,100; blue,50}, thick, rectangle, text width=1cm, align=center] (parent) at (-1, 0) {AI};
        \node[draw={rgb,255:red,136; green,136; blue,141}, thick, rectangle, text width=2cm, align=center] (child1) at (-5, -2) {Symbolic};
        
        \node[draw={rgb,255:red,136; green,136; blue,141}, thick, rectangle, text width=2cm, align=center] (child2) at (-1, -2) {Bio-Inspired};
        
        \node[draw={rgb,255:red,200; green,100; blue,50}, thick, rectangle, text width=2cm, align=center] (child3) at (4, -3) {Data-Driven};

        \node[draw={rgb,255:red,136; green,136; blue,141}, thick, rectangle, text width=2.5cm, align=center] (grandchild1) at (-5, -4.5) {Logical \\ Programming};

        \node[draw={rgb,255:red,136; green,136; blue,141}, thick, rectangle, text width=2cm, align=center] (grandchild2a) at (-2.5, -3.4) {Swarm \\ Dynamics};
        
        \node[draw={rgb,255:red,136; green,136; blue,141}, thick, rectangle, text width=2cm, align=center] (grandchild2b) at (1, -3.5) {Evolution \\ Algorithms};

        \node[draw={rgb,255:red,136; green,136; blue,141}, thick, rectangle, text width=2.2cm, align=center] (grandchild3a) at (-2, -6) {Reinforcement \\ Learning};
        
        \node[draw={rgb,255:red,200; green,100; blue,50}, thick, rectangle, text width=2cm, align=center] (grandchild3b) at (1, -6) {Machine \\ Learning};
        
        \node[draw={rgb,255:red,136; green,136; blue,141}, thick, rectangle, text width=2cm, align=center] (grandchild3c) at (4, -6) {Data \\ Mining};
        
        \node[draw={rgb,255:red,136; green,136; blue,141}, thick, rectangle, text width=1.4cm, align=center] (greatgrandchild1a) at (-5.5, -6) {Expert \\ Systems};
        
        \node[draw={rgb,255:red,136; green,136; blue,141}, thick, rectangle, text width=1.8cm, align=center] (greatgrandchild1b) at (-4.5, -7.5) {Automatic \\ Proofs};

        \node[draw={rgb,255:red,200; green,100; blue,50}, thick, rectangle, text width=3cm, align=center] (greatgrandchild3b1) at (0, -8) {Deep Learning};
        
        \node[draw={rgb,255:red,136; green,136; blue,141}, thick, rectangle, text width=3cm, align=center] (greatgrandchild3b2) at (4, -8.5) {Ensemble Methods};


        \draw[->, thick] (parent.south) .. controls (-1, -1.5) and (-5, -1) .. (child1.north);
        \draw[->, thick] (parent.south) .. controls (-1, -1) and (-1, -1.5) .. (child2.north);
        \draw[->, thick] (parent.south) .. controls (0, -2) and (4, -2) .. (child3.north);


        \draw[->, thick] (child1.south) .. controls (-5, -2.5) .. (grandchild1.north);

        \draw[->, thick] (child2.south) .. controls  (-2.5, -2.5) .. (grandchild2a.north);
        \draw[->, thick] (child2.south) .. controls (1, -2.5) .. (grandchild2b.north);

        \draw[->, thick] (child3.south) .. controls (5, -4) and (-2, -4.5) .. (grandchild3a.north);
        \draw[->, thick] (child3.south) .. controls (5, -4) and (1, -4) .. (grandchild3b.north);
        \draw[->, thick] (child3.south) .. controls (4, -4) and (4, -5) .. (grandchild3c.north);

        \draw[->, thick] (grandchild1.south) .. controls (-5.5, -5) .. (greatgrandchild1a.north);
        \draw[->, thick] (grandchild1.south) .. controls (-3.5, -6) and (-4.5, -6.5) .. (greatgrandchild1b.north);

        \draw[->, thick] (grandchild3b.south) .. controls (1,-6.5) .. (greatgrandchild3b1);
        \draw[->, thick] (grandchild3a) .. controls (-1.25, -7.5) ..(greatgrandchild3b1);

        \draw[->, thick] (grandchild3b.south) .. controls (2,-6.5) and (4, -7.5) .. (greatgrandchild3b2);

    \end{tikzpicture}
    \caption{Landscape of artificial intelligence (AI). Nowadays, the most important paradigm in AI is the data-driven approach, where large datasets are utilized, and statistical methods are applied. In these notes, we primarily focus on Deep Learning, the subfield of Machine Learning.}
    \label{fig:landscape_mindmap}
\end{figure}

AI is typically categorized into three types. The first one,  {Narrow AI}, is designed to solve specific problems, narrow AI systems excel within a particular context or domain but cannot generalize their knowledge to other areas. The second, called General AI, mirrors human capabilities. Humans can learn from diverse experiences and interactions, applying knowledge from one domain to solve problems in others. General AI encompasses memory, spatial reasoning based on visual inputs, and the ability to utilize knowledge adaptively. The last one,  {Superintelligence}, is a theoretical concept referring to an intelligence surpassing human capabilities. Philosophers debate whether humans can create something more intelligent than themselves and, if so, whether they would recognize it. For now, superintelligence remains speculative and undefined.

Research in the field of AI can also be divided into three approaches. 
The first one, historically older, are {  Symbolic approaches to AI}, which are rule-based expert systems,
automatic theorem provers, and
intelligent forms of search
constraint-based approaches. The second one is based on constructing biologically inspired algorithms capable of emergent behaviour. Finally, { Statistical Approaches to AI}, are where knowledge is extracted from the massive amount of data. This paradigm is currently the main paradigm of approaching AI, with data being the main driver behind statistics. From there, we see a number of subfields which stem from these three categories, shown in Fig.~\ref{fig:landscape_mindmap}.

The {Symbolic approaches to AI} rely on rule-based systems where humans define rules and behaviours based on deep knowledge or trial and error. These systems, known as expert systems, often involve decision trees meticulously crafted by hand.
On the other hand, the {Statistical Approaches to AI} focus on algorithms and models that learn from data to generate rules autonomously, often outperforming manually defined approaches. These systems identify patterns in data that might elude human observation, enabling advanced insights and efficiencies.

To the Traditional AI we include 
{fuzzy logic}, {expert systems}, and {declarative reasoning}. 
Fuzzy logic extends classical binary logic by allowing reasoning with degrees of truth. Unlike binary logic, which categorizes information as entirely true or false, fuzzy logic handles imprecise, uncertain, or ambiguous data. For example, instead of describing temperatures as either "hot" or "cold," fuzzy logic introduces terms like "moderately warm" or "slightly cool," enabling more nuanced decision-making. This flexibility makes fuzzy logic invaluable in scenarios where sharp boundaries between categories are impractical.

Applications of fuzzy logic span various domains. In control systems, devices such as air conditioners and washing machines rely on fuzzy logic to adjust operations based on real-time data, providing smooth and adaptive performance. In medical diagnostics, fuzzy logic supports doctors by analysing ambiguous patient data to aid in disease prognosis and treatment planning. Similarly, in finance, it helps model uncertainties, enhancing decision-making in risk assessments and investment strategies.

Expert systems simulate the decision-making capabilities of human experts. These systems rely on a knowledge base of predefined rules and an inference engine that applies logical reasoning to solve problems or draw conclusions. For example, in medical diagnosis, expert systems analyse symptoms to suggest possible conditions and treatments. In technical support, they provide automated troubleshooting for IT systems and engineering tasks. Additionally, in regulatory compliance, expert systems interpret complex legal and organizational standards to ensure adherence.
Despite their strengths, expert systems are most effective in domains with well-structured and codifiable knowledge. They often struggle in unstructured or highly ambiguous situations where flexibility and adaptability are required.

Finally, \textit{declarative reasoning} focuses on specifying the desired outcome rather than detailing the steps to achieve it. This paradigm emphasizes relationships and logical structures over procedural instructions, making it particularly well-suited for tasks requiring flexibility and high-level problem-solving. Prolog, a leading language in declarative programming, exemplifies this approach. Prolog allows users to define relationships through facts and rules, enabling the system to solve problems by answering logical queries. Unlike traditional programming languages, which require detailed step-by-step instructions, Prolog empowers developers to focus on what the solution should accomplish.

Prolog finds applications across various fields. In natural language processing, it parses and interprets language structures, facilitating tasks like syntax analysis and semantic understanding. In knowledge representation, Prolog excels at managing complex hierarchies and ontologies, enabling efficient organization and retrieval of information. Furthermore, Prolog is widely used in constraint-solving tasks, such as scheduling, optimization, and logic puzzles, where defining relationships and constraints is more effective than procedural coding.

By abstracting problem-solving logic from implementation details, declarative reasoning and tools like Prolog offer a powerful and efficient framework for designing adaptable AI systems. This approach not only simplifies the development process but also enhances the system's ability to address a wide range of challenges.

Another group of algorithms considered as a Traditional AI are {Bio-Inspired Algorithms}. Biology-inspired algorithms draw from natural phenomena observed in the world around us. {Evolutionary Algorithms} are inspired by the theory of evolution. These algorithms mimic natural selection. Reproduction introduces genetic variation, including mutations, leading to offspring better suited to their environment. Genetic algorithms, a subset of evolutionary algorithms, use mechanisms such as selection, crossover, and mutation to optimize solutions to complex problems. These techniques are particularly effective for problems where the solution space is vast, such as route optimization or scheduling tasks. The other family of bio-inspired algorithms are {Swarm Intelligence algorithms} taking inspiration from swarms behaviour like ant foraging or bird migration. Swarm intelligence demonstrates how simple individual rules can lead to complex collective behaviours. For example, ant colony optimization algorithms emulate how ants find the shortest paths to food sources, and particle swarm optimization leverages social dynamics to solve optimization problems in multi-dimensional spaces. These methods are widely applied in robotics, logistics, and network optimization.

\section{Machine Learning: Data Driven approach to Artificial Intelligence}

The Modern AI is closely related to access to large amount of data. This field of AI solely based on parsing massive data sets is called Machine Learning.  At its core, machine learning algorithms aim to identify patterns, extract meaningful features, and learn from data to make predictions or decisions. However, the quality and representation of the data play a crucial role in the effectiveness of these algorithms. Feature extraction and feature engineering are critical steps in machine learning. Feature extraction involves selecting the most relevant characteristics from raw data that contribute to solving the given problem. For example, in image classification tasks, features might include edges, textures, or shapes extracted from images. Similarly, in text processing, features could include word frequencies, semantic meanings, or sentence structures. Feature engineering, on the other hand, is a more manual and creative process, where humans design and construct new features from the raw data to enhance the algorithm's performance. For instance, in predicting house prices, combining individual features like the number of rooms and square footage into a single feature representing the overall size of the property could improve the model's ability to make accurate predictions.

Traditional machine learning methods such as decision trees, random forests, and support vector machines often rely heavily on these engineered features. Algorithms like logistic regression and gradient boosting are also effective in problems where well-designed features capture the underlying structure of the data. Traditional feature engineering remains indispensable in many contexts where data is sparse or domain-specific knowledge is essential. Feature extraction and engineering underscore the importance of understanding data's nature and context, bridging the gap between raw information and actionable insights. These processes, though time-intensive, are often the foundation of successful machine learning applications.

Machine learning (ML) leverages statistical methods to build models from data. ML encompasses diverse algorithms that help uncover relationships in data, make decisions, and generate predictions. Its main approaches include:
{Supervised Learning} which involves training data where outcomes are already known. For example, a system might predict a fruit’s type based on features like weight, colour, and texture. This approach is widely used in classification tasks, such as spam detection, and regression tasks, where continuous outcomes like house prices are predicted. The next approach called  {Unsupervised Learning:} focuses on discovering hidden patterns and structures in data, often aiding exploratory analysis. For instance, it might cluster similar fruits based on shared features without predefined labels. Applications include customer segmentation and anomaly detection.
Finally, the {Reinforcement Learning} is inspired by behavioural psychology, this approach rewards desired behaviours and penalizes undesirable ones. For example, a robot might receive a reward for successfully opening a door and a penalty for failing. Over time, the robot  learns the optimal sequence of actions to achieve the goal. Reinforcement learning is particularly successful in gaming, robotics, and dynamic decision-making environments.

\section{Deep Learning}

Deep learning is a subset of machine learning which employs a vast array of techniques and algorithms to build advanced systems, often aimed at achieving narrow AI with aspirations toward general AI. Deep learning typically addresses problems requiring generalization, such as image recognition and speech processing.

Deep learning has revolutionized the field of machine learning by automating the process of feature extraction. Unlike traditional machine learning methods that depend on hand-crafted features, deep learning algorithms use layered neural networks to learn hierarchical representations of data directly from raw inputs. This capability has bridged a significant gap, reducing the reliance on human expertise for feature engineering.
Neural networks, the backbone of deep learning, consist of multiple layers of interconnected nodes. Each layer processes the data in increasingly abstract ways, enabling the system to identify complex patterns and relationships. For example, in image recognition tasks, initial layers might detect edges and textures, while deeper layers recognize higher-level features such as shapes and objects. This layered architecture allows deep learning models to autonomously extract features that are often more effective than those manually designed by humans.

The ability to automate feature extraction has made deep learning particularly impactful in fields with high-dimensional data, such as computer vision, natural language processing, and speech recognition. Tasks like object detection, language translation, and voice-to-text conversion have seen unprecedented advancements due to the powerful representation learning capabilities of deep neural networks.
Moreover, the integration of deep learning into machine learning pipelines has enabled the handling of unstructured data, such as images, audio, and text, which were traditionally challenging to process. By combining the strengths of traditional machine learning algorithms with the automated feature extraction of deep learning, modern AI systems can achieve superior performance across a wide range of applications.

While deep learning reduces the need for manual feature engineering, it requires large datasets and substantial computational resources to train effectively. Nevertheless, its ability to learn directly from raw data and generalize to complex, real-world problems has established deep learning as a cornerstone of modern artificial intelligence.

Deep learning’s foundation lies in the {Universal Approximation Theorem}, which states that a sufficiently large neural network can approximate any function. This theorem enables deep learning to model complex relationships and patterns in high-dimensional data. By stacking multiple layers of artificial neural networks, deep learning systems incrementally extract features from raw inputs. Each layer specializes in processing certain aspects of the data, such as edges, shapes, or textures in an image.

Applications of deep learning include:
Image Recognition, Speech Recognition, Converting spoken language into text,
Natural Language Processing (NLP), tasks like translation, summarization, and sentiment analysis, as well as Generative Modelling, where neural networks create new data, such as deep-fake images or realistic audio.

\subsection*{Deep Learning: \textit{why now?}}
Deep learning has been known since late 1980s and early 1990s. There arises natural question: why we experience boom in the field in last few years? There are five factors, which combined together, allowed boom in the field.

The first one, is the widespread availability of inexpensive data and affordable computational power has its full potential been realized. The success of deep learning, like any other subfield of machine learning, relies on access to vast amounts of high-quality training data. This success has been made possible by the development of the Internet and the proliferation of mobile devices capable of capturing inexpensive yet high-quality photographs, videos, and audio recordings. In 2024 alone, \href{https://edgedelta.com/company/blog/how-much-data-is-created-per-day#:~:text=An%20internet%20user%20generates%20an%20average%20of%20146%2C880%20MB%20daily.,-(Domo%2C%20World%20Population&text=Researchers%20have%20stated%20that%20the,create%20about%20506%2C736%20MB%20daily.}{Facebook generates 4,000 terabytes of data}. 

The second essential factor in the success of deep learning is access to (relatively) affordable computational power. This has been facilitated by advancements in graphics cards designed for numerical computations, particularly for fast single-precision matrix multiplications.

The third key factor enabling the rapid development of deep learning is the advancement of programming languages that allow for writing efficient prototype code with a low entry/learning threshold. Programming languages such as Python or Julia allow everyone to start training simple models without spending too much time learning a complicated API.

The fourth factor, no less important in the rapid development of deep learning research, is the automation of the neural network training process. The workhorse of neural network training is the computation of neural network gradients. A neural network is, in fact, a parametrised non-linear function, and the training process involves the optimal selection of its parameters. This selection is based on calculating the gradients of the objective \textit{loss function} with the respective neural network's parameters. In past decades, each new network architecture required manual gradient computation, which was done by researchers, and prone to errors. With the advancement of scientific software and programming languages, an automatic differentiation algorithm was implemented, relieving developers of the need to manually compute gradients each time. As a result, prototyping a wide range of network architectures became simpler and faster, enabling easy testing, rejection, or acceptance of new neural network architectures.

The fifth factor, is related to  the proper labelling of training data—a task often performed by low-cost labour in developing countries. This aspect is frequently overlooked but deserves recognition as the contributions of these workers are crucial, even though their efforts often go unacknowledged\footnote{To read more about this, check out this post by MIT Technology Review\href{https://www.technologyreview.com/2020/12/11/1014081/ai-machine-learning-crowd-gig-worker-problem-amazon-mechanical-turk/}{\textit{AI needs to face up to its invisible-worker problem}}, or \href{https://www.sciencefocus.com/future-technology/artificial-intelligence-quietly-relies-on-workers-earning-2-per-hour}{this article by BBC Science Focus, titled ``\textit{Artificial intelligence quietly relies on workers earning \$2 per hour}''}. A commentary on the economics behind the AI Labour market is offered by Venture Beat \href{https://venturebeat.com/ai/low-wage-workers-drive-the-global-ai-labor-market/}{here}. Even frontier models from OpenAI rely on this form of labour, \href{https://time.com/6247678/openai-chatgpt-kenya-workers/}{check out this article in the Times}, titled  \textit{``OpenAI Used Kenyan Workers on Less Than \$2 Per Hour to Make ChatGPT Less Toxic.''}}.

\section{Chronology of Artificial Intelligence Development}

\subsubsection*{Early Foundations (1940s-1950s)}
\begin{itemize}
    \item \textbf{1943}: Warren McCulloch and Walter Pitts publish \textit{A Logical Calculus of Ideas Immanent in Nervous Activity}, proposing the first mathematical model of a neural network.
    \item \textbf{1950}: Alan Turing publishes \textit{Computing Machinery and Intelligence}, introducing the Turing Test and discussing machine intelligence.
    \item \textbf{1956}: The Dartmouth Conference, organized by John McCarthy and others, is held, where the term \textit{Artificial Intelligence} is coined, marking the formal beginning of AI research.
\end{itemize}

\subsubsection*{Early Progress (1950s-1960s)}
\begin{itemize}
    \item \textbf{1956}: Allen Newell and Herbert A. Simon develop the Logic Theorist, considered the first AI program.
    \item \textbf{1958}: John McCarthy develops the Lisp programming language, which becomes a primary language for AI research.
    \item \textbf{1966}: Joseph Weizenbaum creates ELIZA, an early natural language processing program simulating conversation.
    \item \textbf{1969}: Marvin Minsky and Seymour Papert publish \textit{Perceptrons}, analysing the capabilities and limitations of neural networks, particularly in handling non-linear problems.
\end{itemize}
 
\subsubsection*{Expert Systems Era (1970s-1980s)}
\begin{itemize}
    \item \textbf{1972}: PROLOG, a programming language based on logic, is developed and becomes a key tool in AI research.
    \item \textbf{1979}: The Stanford Cart, an early autonomous vehicle, successfully crosses a room avoiding obstacles, demonstrating early robotics and computer vision capabilities.
    \item \textbf{1980}: The rise of expert systems, such as MYCIN for medical diagnosis, showcases practical applications of AI.
    \item \textbf{1987-1993}: The \textit{AI Winter} occurs due to reduced funding and interest, as expectations for AI progress are not met.
\end{itemize}

\subsubsection*{Machine Learning Renaissance (1990s-2000s)}
\begin{itemize}
    \item \textbf{1997}: IBM's Deep Blue defeats world chess champion Garry Kasparov, showcasing the potential of AI in strategic games.
    \item \textbf{1998}: Yann LeCun and colleagues develop LeNet-5, a convolutional neural network for handwritten digit recognition, advancing image processing.
    \item \textbf{2000}: Cynthia Breazeal develops Kismet, a robot capable of recognizing and simulating emotions, advancing human-robot interaction.
    \item \textbf{2006}: Geoffrey Hinton and colleagues revive deep learning through the development of deep belief networks, sparking renewed interest in neural networks.
\end{itemize}
 
\subsubsection*{Modern Deep Learning Era (2010s-Present)}
\begin{itemize}
    \item \textbf{2012}: AlexNet, developed by Alex Krizhevsky, Ilya Sutskever, and Geoffrey Hinton, wins the ImageNet competition, revolutionizing computer vision with deep convolutional neural networks.
    \item \textbf{2014}: Ian Goodfellow and colleagues introduce Generative Adversarial Networks (GANs), enabling AI to generate realistic data.
    \item \textbf{2015}: Google DeepMind's AlphaGo defeats professional Go player Fan Hui, marking a significant achievement in AI.
    \item \textbf{2016}: AlphaGo defeats Lee Sedol, one of the world's best Go players, in a landmark victory.
    \item \textbf{2017}: The Transformer model is introduced by Vaswani et al., leading to breakthroughs in natural language processing (e.g., BERT, GPT series).
    \item \textbf{2020}: OpenAI releases GPT-3, setting a new standard for large language models and applications.
    \item \textbf{2020}: DeepMind's AlphaFold AI system demonstrates the ability to predict protein structures with remarkable accuracy, solving a 50-year-old grand challenge in biology.
    \item \textbf{2022}: OpenAI's ChatGPT showcases advanced conversational AI capabilities to a broad audience.
    \item \textbf{2024}: Demis Hassabis and John Jumper of DeepMind, along with David Baker, are awarded the Nobel Prize in Chemistry for their work on protein structure prediction using AI.
    \item \textbf{2024}: Geoffrey Hinton and John Hopfield receive the Nobel Prize in Physics for their foundational contributions to artificial neural networks and machine learning.
\end{itemize}

\section{Literature}

For readers interested in the philosophical aspects of artificial intelligence research, we recommend  
\cite{Turing_1950, Copeland_1993,Penrose_1999,Larson_2021}. Those wishing to explore the history of AI research are encouraged to refer to the resources listed in 
\cite{Sejnowski_2018}.
Readers interested in the research on Artificial Life are encourage to follow references
\cite{Steels_1995, Langton_1995, Adami_1999}

Next, readers interested in the technical aspects of broad family of artificial intelligence algorithms are encouraged to consider the following literature: algorithms inspired by biology \cite{Chambers_1995, Haupt_2003, Rutkowski2008, Kramer_2017}, machine learning algorithms and knowledge discovery from data \cite{Bishop_2006,  Shalev_Shwartz_2014, Witten_2016}, and, finally, deep learning algorithms and neural networks \cite{Prince_2023, Bishop_2024}

Lastly, we recommend the following references for students and researchers intrigued by the applications of machine learning and deep neural networks in quantum physics \cite{Marquardt_2019, Carleo_2019, Mehta_2019, Neupert_2022, dawid2022modern, Sadegh_2023}.

%% file: chapters/chapter_2_NN_fundamentals.tex
\chapter{Fundamentals of Deep Learning}
\label{CH:FUNDAMENTALS}
\section{Neural Network Anatomy}

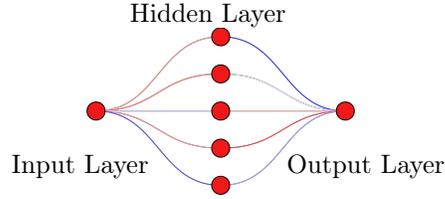
\begin{figure}[ht]
    \centering
     \begin{tikzpicture}
        \node [anchor=south west, inner sep=0] (image) at (0,0) {\includegraphics[width=0.3\textwidth]{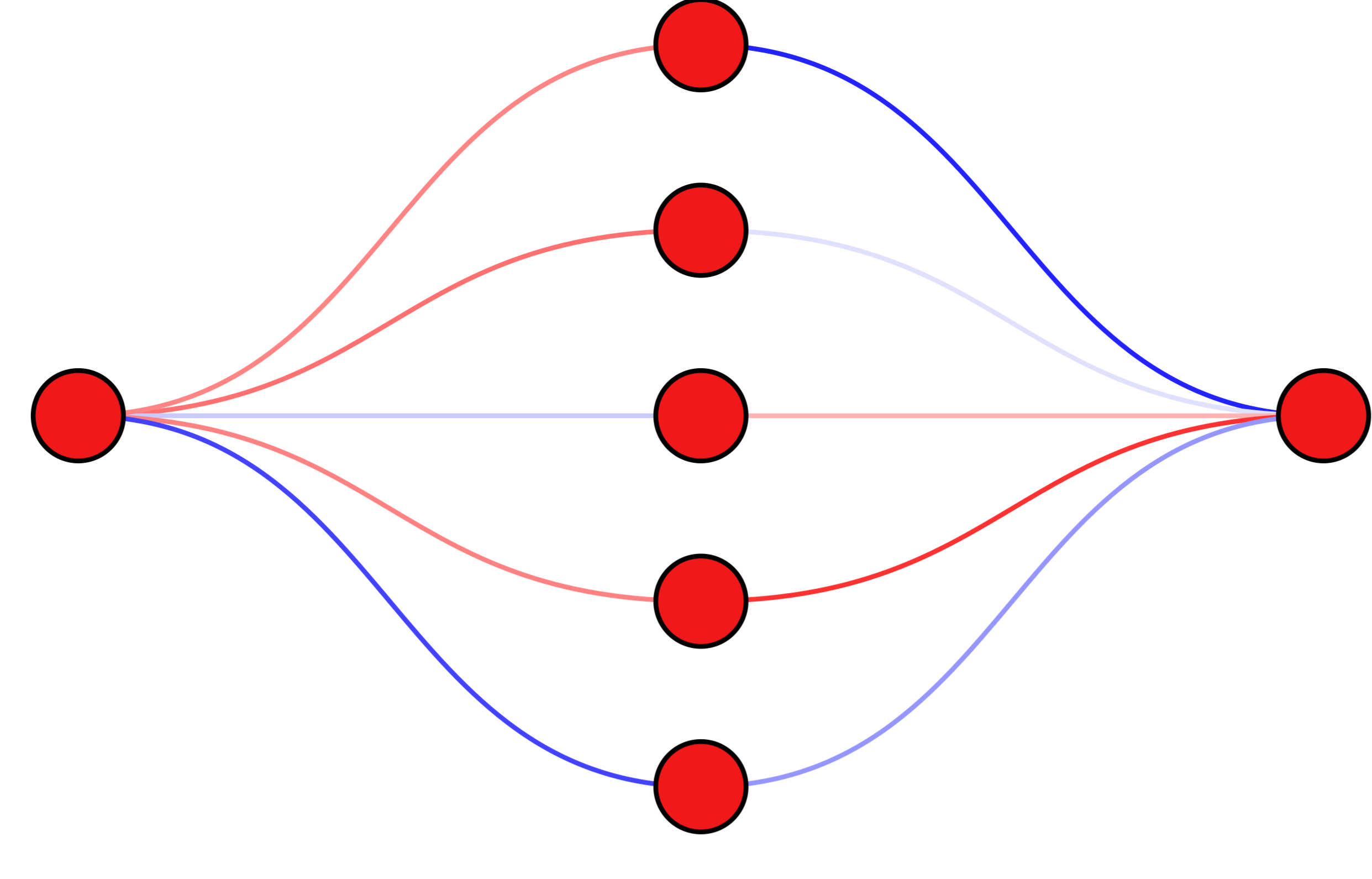}};
        \begin{scope}[shift={(0,0)}]
            \node at (0.0, 0.5) {Input Layer};
            \node at (1.7, 2.5) {Hidden Layer};
            \node at (3.8, 0.5) {Output Layer};
            
        \end{scope}
    \end{tikzpicture}
    \caption{Pictorial representation of a neural network. For a precise definition of its components, see Eq.\eqref{eq:NN_approximator}.}
    \label{fig:nn_basic_anatomy}.
\end{figure}

At the heart of deep learning is neural networks. A neural network is a special type of \textit{tunable} non-linear function. What makes it special is its structure, which bears a remarkable resemblance to how we previously believed human neurons function\footnote{The term neural network (or Artificial Neural Network) historically appeared in 1950s,  inspired by modelling of biological neurons. Nowadays, with the advances in neurobiology research, we know that this analogy does not hold any more, however \textit{artificial neural networks} as a term has persisted, and is used ubiquitously in the field.}. Their structure is best understood visually, with a simple neural network diagram shown in Fig.~\ref{fig:nn_basic_anatomy}. We read these diagrams from left to right, so the \textit{input} node is the leftmost node, and the \textit{output} node is the rightmost node. Between the input and output, we see two things; there are edges representing connections between nodes, and a \textit{hidden} layer of five nodes. 

In general, the number of nodes on a given layer specifies how large its input and output are. For example, the input layer has one node, meaning it receives a single (possibly real or binary) number as input. Whereas the hidden layer having five nodes means that we can think of this hidden layer holding a vector of five numbers. 

Within a given neuron, there is is a non-linear function which we call an \textit{activation function}, whose form specifies (one part of) our choice of \textit{network architecture}. Some common activation functions are shown in \hyperlink{box:activation}{Box~2.2}. On a given layer of a neural network, activation functions are element-wise, meaning each neuron independently applies its activation to the input value it receives.

Between layers, the edges connect the nodes mathematically through a linear transformation, usually matrix-vector multiplication. For example, the edges that connect the hidden layer to the output neuron specify the product of a $5\times5$ matrix with the 5-component vector on the hidden layer. Since the output layer contains only one neuron, we then sum element-wise over this transformed vector. Let us now make these ideas more precise.

\begin{figure}[h]
    \centering
    \begin{mybox}[\hypertarget{box:NN_definitions}{Box 2.1: Neural Network Anatomy - see Fig.~\ref{fig:nn_basic_anatomy}}]
    \textbf{Neural Network} - a structured, tunable, non-linear function consisting of nodes (called \textit{neurons}) and edges which connect them (called weights). \\
    
    \textbf{Input Layer} - the first layer of a neural network which receives some input. \\

    \textbf{Output Layer} - the last layer of a neural network. \\

    \textbf{Hidden Layer} - any layer of a neural network contained between the input and output layer. \\

    \textbf{Activation function} - the non-linear function contained in each neuron . \\

    \textbf{Neural Architecture} - our choice of shape, and activation functions for a neural network.

    \end{mybox}
\end{figure}

\begin{figure}[h]
    \centering
    \begin{mybox}[\hypertarget{box:activation}{Box 2.2: Activation functions}]
    A given neuron contains a function which \textit{must} be non-linear should we want to harness the full power of deep learning. Some common activation functions are the sigmoid,
    \begin{equation}
    \sigma(x) = \frac{1}{1 + e^{-x}},
    \end{equation}
    the tanh function,
    \begin{equation}
    \text{tanh}(x) = \frac{e^x - e^{-x}}{e^x + e^{-x}},
    \end{equation}
    the relu (rectified linear unit),
    \begin{equation}
    f(x) = \text{max}(0, x).
    \end{equation}
    We might think that non-linear means the Taylor series of the function (about any point) has \textit{at least} quadratic order that is non-vanishing, which is not the case for the relu function. From a deep learning perspective, we non-linear really means ``not a straight line'' over the whole domain $x$. In this sense, relu is non-linear. 
    \end{mybox}
\end{figure}

\subsection{Neural Networks - Mathematical Structure}

From the mathematical point of view any neural network is a non-linear function $f_{\vec{\theta}}: \mathbf{x} \in \mathbb{R}^m \to \mathbf{y}\in\mathbb{R}^n$, $m,n\in \mathbb{N}$, endowed with a set of tuneable parameters denoted by $\vec{\theta}$. We call these the \textit{variational parameters} of a neural network function $f_{\vec{\theta}}(\vec{x})$, see \hyperlink{box:VarParams}{Box~2.3}. 
The layers in a neural network are constructed by function composition,
\begin{equation}
    f_{\vec{\theta}}(\mathbf{x}) = h^{(L)} \circ h^{(L-1)} \circ \cdots \circ h^{(2)} \circ h^{(1)} (\mathbf{x})
    \label{eq:NN_layers},
\end{equation}
where $(l) = 1\dots L$ is the index that specifies which layer we are considering, and $\circ$ indicates function composition, i.e. $ h^{(2)} \circ h^{(1)} (\mathbf{x}) = h^{(2)} \big( h^{(1)} (\mathbf{x}) \big)$. This means the \textit{depth} of a neural network is given by $L$, the total number of layers (i.e. number of function compositions) in the network's architecture.

Within a given layer, $h^{(l)}$, the edges (representing connections) apply a linear transformation, followed by the application of a non-linear, differentiable\footnote{As we shall see, differentiability is very important when it comes to training!} \textit{activation function}. Clearly, we must describe this in two stages. First, the linear transformation of an input $\mathbf{z}$ is usually given by
\begin{equation}
    \mathbf{z} \gets \mathbf{W}^{(l)} \mathbf{z} + \mathbf{b}^{(l-1)}
\end{equation}
where $\mathbf{W}^{(l)}$ is a $N^{(l-1)}\times N^{(l)}$ matrix, and  vector $\mathbf{b}^{(l)}$ is a vector of so-called $\textit{biases}$. We use the left-arrow notation $\leftarrow$ to signify that $\mathbf{z}$ \textit{becomes} the RHS of this expression. This is to avoid making too many new variables, as neural networks apply \textit{lots} of transformations!
We can think of $\mathbf{W}^{(l)},\; \mathbf{b}^{(l)}$ as a way of allowing \textit{every} neuron on a given layer to receive a ``complete'' picture of the input $\mathbf{z} \in \mathbb{R}^{\text{dim}(l)}$ on that layer. This is because the matrix $\mathbf{W}^{(l)}$ can in-principle couple every element in $\mathbf{z}$ to every other, with the strength of that coupling given by the matrix value. Whilst $\mathbf{b}^{(l)}$ can \textit{independently} bias this coupling. As such, this matrix and vector really do represent \textit{couplings} between two neighbouring layers $h^{(l-1)}$ and $h^{(l)}$, each with $N_{l-1}$ and $N_{l}$ neurons respectively. Whenever we draw edges between neighbouring layers, we are representing these couplings.

Onto the second stage, where we apply a non-linear differentiable activation function. We can represent this using an element-wise non-linear function applied to the neurons of a given layer, $\sigma: \mathbb{R}^{\text{dim}(l)} \rightarrow \mathbb{R}^{\text{dim}(l)}$. Explicitly, we may write for an input $\mathbf{z}$ to layer $l$,
\begin{equation}
h^{(l)}(\mathbf{z}) = \sigma(\mathbf{W}^{(l)} \mathbf{z} + \mathbf{b}^{(l-1)}) 
\end{equation}
where $N_l = \text{dim}(l)$ is the number of neurons on layer $l$, and the argument of $\sigma$ is the first stage of our layer.  Be careful not to get confused here by the $(l)$ index and its meaning. The layer index $(l)$ only specifies \textit{which coupling} we are on, and not the values of a particular connection. To see a layer in its full form, we can write down the matrix-vector multiplication explicitly, giving,
\begin{equation}
     h^{(l)}(\mathbf{z}) = \sigma\big(\sum_{i=1}^{N_{l}} \sum_{j=1}^{N_{l-1}} \mathbf{W}^{(l)}_{i,j}\mathbf{z}_j + \sum_{i=1}^{N_l}\mathbf{b}_i^{(l)}\big).
\end{equation}
Now the $i,j$ indices specify the elements of the matrix, and $(l)$ specifies \textit{which} matrix we apply depending on the layer $l$. See \hyperlink{box:activation}{Box~2.2} for example activation functions.

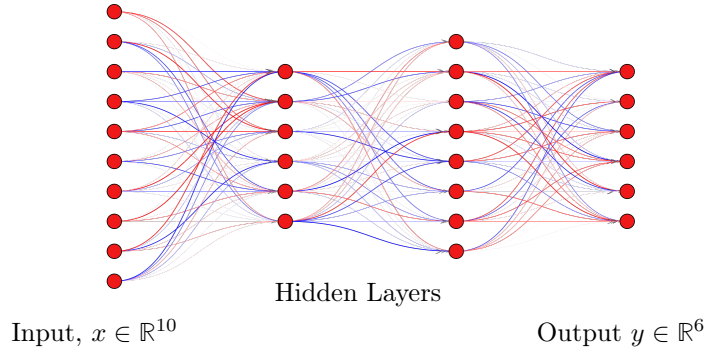
\begin{figure}[h]
    \centering
     \begin{tikzpicture}
        \node [anchor=south west, inner sep=0] (image) at (0,0) {\includegraphics[width=0.6\textwidth]{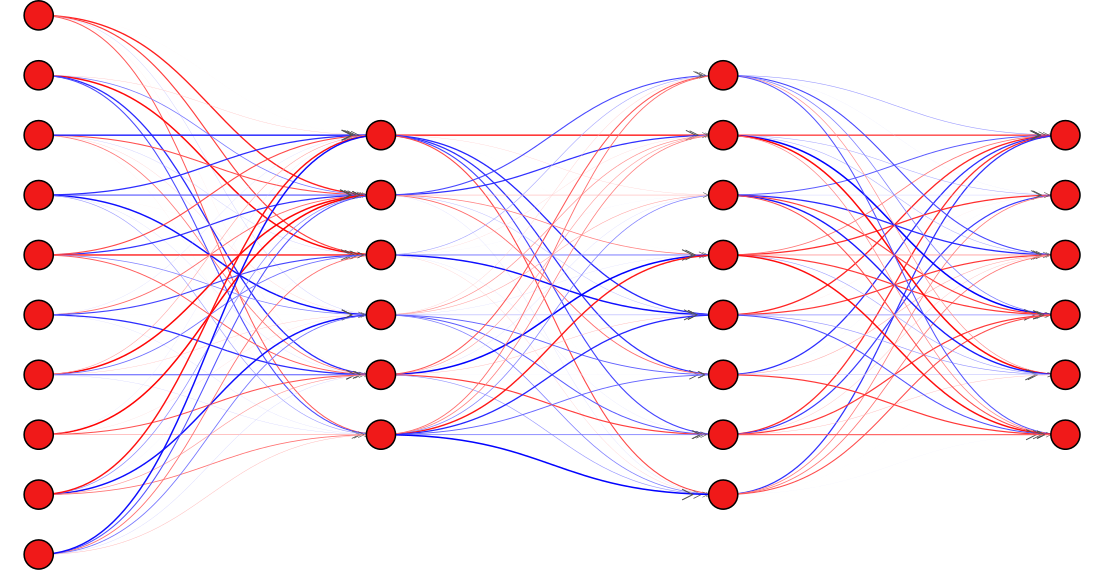}};
        \begin{scope}[shift={(0,0)}]
            \node at (0.0, -0.5) {Input, $x \in \mathbb{R}^{10}$};

            \node at (3.5, 0.0) {Hidden Layers};

            \node at (7.0, -0.5) {Output $y \in \mathbb{R}^6$};
        \end{scope}
    \end{tikzpicture}
    \caption{Graphical representation of a non-linear function $f_{\vec{\theta}}: \vec{x}\in \mathbb{R}^{10}\to \vec{y}\in \mathbb{R}^{6}$ with two hidden layers, each with $N^{(2)} = 6$, and $N^{(3)}=8$ neurons respectively. The horizontal coloured lines represent weight matrices $\vec{W}^{(l)}$ and biases vectors $\vec{b}^{(l)}$. Each node in the layer $h^{(l)}$, $l>1$, represents action of the non-linear activation function $\sigma$. Number of layers $L$, number of nodes in each layer $N^{(l)}$, and activation function $\sigma$ (see \protect\hyperlink{box:activation}{Box~2.2}) defines the neural network \textit{architecture}.}
    \label{fig:another_NN}
\end{figure}

Let's now consider another example of a neural network in Fig.\ref{fig:another_NN}. 
Recall that we defined a neural network to be a \textit{tunable} non-linear function. So far, the layer transformations cover the non-linear bit, but we are yet to make our construction tunable. Earlier, we said that tuneability comes from this elusive term ``variational parameters''. Let us now make this more precise.

To do this, we must \textit{choose} a set of variational parameters $\vec{\theta}$ of our neural network. In our construction, this will be  the layers' matrices $\mathbf{W}^{(l)}$, sometimes called \textit{weights}, and \textit{biases}, $\mathbf{b}^{(l)}$ (see \hyperlink{box:NN_definitions}{Box~2.1})
\begin{equation}
    \vec{\theta} = \{\mathbf{W}^{(l)}, \mathbf{b}^{(l)}| l = 1,\ldots,L\}.
\end{equation}
is a set variational parameters.  For example, in Fig.\ref{fig:another_NN}, the variational parameters $\vec{\theta} = \{\mathbf{W}^{(l)}, \mathbf{b}^{(l)}|l = 1,2,3\}$. 

The power and popularity of neural networks comes from the Universal Approximation Theorem. In the next section, we will explain this theorem, as well as the steps we can take to use it in practise. For now, make sure you are happy with the terms in \hyperlink{box:NN_definitions}{Box~2.1}, \hyperlink{box:VarParams}{Box~2.2} and \hyperlink{box:activation}{Box~2.3}.


\begin{figure}[h]
    \centering
    \begin{mybox}[\hypertarget{box:VarParams}{Box 2.3: Variational Parameters}]
    For a given neural network architecture, its variational parameters, $\vec{\theta}$, is a set of tunable parameters that modify how neural network acts on its inputs. The choice of values for $\vec{\theta}$ is in principle free for us to choose. You can think of this as having lots of dials on our neural network which we can turn. Every configuration of the dials sets new values for the variational parameters, in turn changing the behaviour of the neural network. However, in deep learning we are usually interested in an \textit{optimal} value for variational parameters, $\vec{\theta}^*$, which allow the neural network to operate best. The variation of $\vec{\theta}$ towards the optimum $\vec{\theta}^*$ is done by \textit{training} a neural network.
    \end{mybox}
\end{figure}

\subsection[Universal Approximation Theorem]{Universal Approximation Theorem \\ \textbf{- Deep Learning \textit{is} non-linear curve fitting}}

The Universal Approximation Theorem  states \cite{Cybenko1989ApproximationBS} that any continuous function on a compact domain\footnote{i.e a domain which has no holes or missing endpoints.} can be approximated to an arbitrary level of accuracy with a sufficiently deep neural network. We can understand this simple-yet-powerful result by considering the composition structure of the layers and non-linear activation functions in a deep neural network that give rise to $N \gg 1$ variational parameters, $\text{dim}(\vec{\theta}) = N$.

To that end, let $g: \mathbb{R} \to \mathbb{R} $ be a continuous function with domain $x$ and range $y$ defined on a compact set $ x,y \in K \subset \mathbb{R}$. Also, let's define $\sigma: \mathbb{R} \to \mathbb{R}$ be a non-constant, bounded, and continuous non-linear function\footnote{This is a formal way of describing an activation functions. The non-linearity is the ingredient that makes deep learning models \textit{expressive}. That is, they can be used to represent a large (in fact universal!) class of functions.}. Next, let us consider a weighted mixture of non-linear functions with linearly rescaled argument $x$:
\begin{equation}\label{eq:NN_approximator}
f(x; \vec{W}^{(1)}, \vec{b}, \vec{W}^{(2)}) = \sum_{i=1}^N \vec{W}^{(2)}_i \sigma(\vec{W}^{(1)}_i x + \vec{b}_i).
\end{equation}
This is just the first neural network architecture $f(x; \vec{W}^{(1)}, \vec{b}, \vec{W}^{(2)})$, which we encountered in Fig.\ref{fig:another_NN}. Then, for any $\epsilon > 0$, there exists an integer $N \in \mathbb{N}$, and parameters $\vec{W}^{(1)}_i, \vec{b}_i, \vec{W}^{(2)}_i$, $i = 1, \dots, N$ such that:
\begin{equation}
\sup_{x \in K} \left| f(x; \vec{W}^{(1)}, \vec{b}, \vec{W}^{(2)}) - g(x) \right| < \epsilon.
\end{equation}

Whilst this result looks rather technical, it is in fact \textit{remarkable}. It says that with enough depth, a neural network can approximate \textit{any} continuous, compact function.
\begin{figure}[h]
    \centering
    \begin{mybox}[\hypertarget{box:UAT}{Box 2.4: Universal Approximation Theorem}]
    With enough variational parameters, $\text{dim}(\vec{\theta}) \gg 1$, and training data, neural networks can approximate any continuous, compact function to arbitrary accuracy.
    \end{mybox}
\end{figure}

\begin{figure}[h]
    \centering
     \begin{tikzpicture}
        \node [anchor=south west, inner sep=0] (image) at (0,0) {\includegraphics[width=\textwidth]{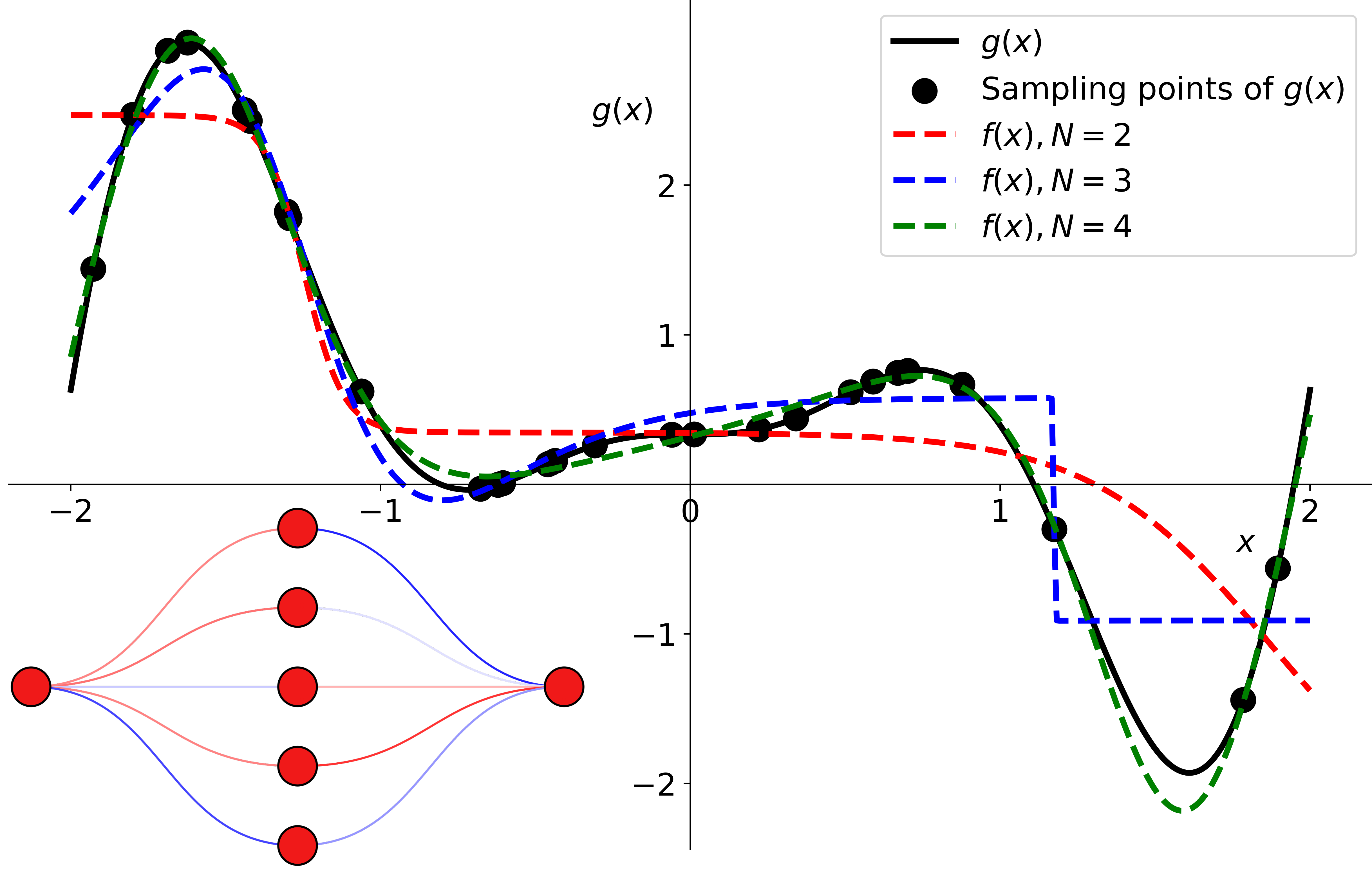}};
        \begin{scope}[shift={(0,0)}]
            \node at (0.0, 0.5) {Input, $x \in \mathbb{R}$};

            \node at (2.5, -0.5) {Hidden Layer};

            \node at (5.0, 0.5) {Output $y \in \mathbb{R}$};
        \end{scope}
    \end{tikzpicture}
     \caption{Approximation of function $g(x)=x^2\sin(\pi x) - \frac{1}{2+\cos(x)}$ with neural network ansatz, Eq.\ref{eq:NN_approximator}, with a single hidden layer with $N = 2,3,4$ neurons, schematically presented on the bottom left panel.}
    \label{fig:fig_1_2}.
\end{figure}

As an example let us consider function $g(x)=x^2\sin(\pi x) +\frac{1}{2+\cos(x)}$, shown as a black line in Fig.~\ref{fig:fig_1_2}. According to the universal approximation theorem, we should be able to numerically approximate this function with a sufficiently deep neural network. To that end, let's assume we have \textit{some} access to data relating to $g(x)$ in order to train our neural network\footnote{a full description of training will be given below - for now you can think of training as the strategy we use to tune the ``dials" that are the variational parameters. See \hyperlink{box:VarParams}{Box~2.3}}. The total information we use to train the neural network constitutes its \textit{training dataset}, $\mathcal{D}$, see \hyperlink{box:dataset}{Box~2.4}.

\begin{figure}[h]
    \centering
    \begin{mybox}[\hypertarget{box:dataset}{Box 2.5: Datasets}]
    The variational parameters, $\vec{\theta}$, of a neural network can be trained to find the \textit{optimal} values $\vec{\theta}^*$ that approximate an a compact, continuous function  per the universal approximation theorem. To train a neural network, we need some data about the function we wish to approximate. Any data that we use to train our neural network constitutes \textit{training data}. The total set of data we have available to perform training constitutes a \textit{training dataset}, denoted by $\mathcal{D}$.

    Our \textit{choice} of data is super important; if we expect a neural network to learn some desired behaviour, we must have good examples of what we seek. Our \textit{representation} of data is also critically important. Presented in the right way, it is easy to extract meaning from patterns. Presented in the wrong way, it can be almost impossible!
    \end{mybox}
\end{figure}

Since our example is trying to approximate $g(x)$, let $\mathcal{D}= \{x_i, g(x_i)\}_{i=1}^{K}$, be the countable set of values of function $g(x_i)$ at specific points, $x_i$. We can use this data to train our neural network model because $g(x_i)$ are the \textit{true} values of the function $g(x)$ at $x_i$. Therefore we would expect a well trained model, $f_{\vec{\theta}}(x)$ that approximates $g(x)$ to have outputs that are \textit{as close as possible} to these true values. This is, after-all, what the UAT tells us is possible when $\text{dim}(\theta) \gg 1$. Therefore, the difference between  $f_{\vec{\theta}}(x)$ and $g(x)$ for training points $x_i \in \mathcal{D}$ gives us an estimate of how well our model is doing; the better our approximation, the smaller $|f_{\vec{\theta}}(x_i) - g(x_i)|$ should be. Mathematically, the idea of measuring the performance of a neural network model using its training dataset gives rise to a \textit{loss function}, see \hyperlink{box:loss}{Box.~2.6}.

Our goal is to find a set of optimal variational parameters,
\begin{equation}
    \vec{\theta}^* = \{\vec{W}^{(1)*}, \vec{b}^*, \vec{W}^{(2)*}\},
\end{equation}
that minimise $|f_{\vec{\theta}}(x_i) - g(x_i)|$. We can do this with a commonly-used loss function known as the \textit{mean square error},
\begin{equation}
    {\cal L}(\vec{\theta}) = \frac{1}{K}\sum_{i=1}^{K}|g(x_i)-f(x_i; \vec{\theta})|^2,
    \label{eq:MSE}
\end{equation}
which gets its name from the fact that it calculates the arithmetic mean of the squared difference between the true values $g(x_i)$ and the neural network approximation  $f(x_i; \vec{\theta})$. Notice here that we have included the implicit dependence of $\mathcal{L}$ on $\vec{\theta}$ through $f(x_i; \vec{\theta})$. We can see the UAT in action in Fig.\ref{fig:fig_1_2}, which uses neural networks of varying depth to approximate $g(x)$ via Eq.\eqref{eq:NN_approximator} for a fixed training set ${\cal D}$. As we can see, increasing $N$ increases quality of our approximation. How then, do we actually \textit{vary} $\vec{\theta}$ in order to minimise a loss function when given a dataset? The answer lies in \textit{training}, which will be the main focus of the next section. We will also build a deeper understanding of how loss functions can encode desired behaviour or objectives for a neural network.

\section{Building Blocks for Deep Learning Models}

So far, we have seen the anatomy of a neural network, and its mathematical structure. This allowed us to define a network's \textit{variational parameters}, as well as a \textit{loss function} which quantifies how well a neural network is performing in a given task. Together, these two concepts allowed us to see the \textit{Universal Approximation Theorem}, which said that a neural network can approximate any compact, continuous function with enough variational parameters. Using the \textit{mean squared error} loss function, we saw this theorem in action, tuning a function to arbitrary accuracy in Fig.~{\ref{fig:fig_1_2}}.

In this section, we will seek to understand two key ingredients that will allow us to build our own models from scratch. First, we will better understand loss functions in their totality. In the last section, we saw a single example of a loss function, but there are many more ways to construct them. Since the loss function is our way of encoding how well a neural network is performing, it is essential for us to have a wide range of loss functions available should we want neural networks to be able to solve a diverse set of problems.

We will then proceed to understand how to train neural networks with gradient descent. This training algorithm is the simplest in a family of gradient-based training strategies. Once we understand this one, we will be able to upgrade our training to more recent and sophisticated strategies.

\subsection{Defining the Loss Function: Regression vs Classification}

The universal approximation theorem allows us to construct two main tasks called \textit{regression} and \textit{classification}.

A \textit{regression task} involves predicting a continuous numeric value, meaning the output can take any value within a range. The model learns to map input features to a real-valued output, where the relationship between the inputs $\vec{x}$ and the target output $\vec{y}$ can be expressed as $f_{\vec{\theta}}(\vec{x}) = \bar{\vec{y}} \approx \vec{y}$. The learning process optimizes a loss function, such as the Mean Squared Error (MSE):
\begin{equation}
{\cal L}_{\text{MSE}} = \frac{1}{N} \sum_{i=1}^N \left( \vec{y}_i - \bar{\vec{y}}_i \right)^2,
\label{eq:MSE_general}
\end{equation}
where $\vec{y}_i$ is the true target value, $\bar{\vec{y}}_i$ is the predicted value, and $N$ is the number of samples in the dataset.
In regression task, the model’s output layer typically has one node for single-target regression or multiple nodes for multi-output regression, with no activation function (linear output) applied.
During inference, the model predicts a continuous value $\bar{\vec{y}}$, which for example, in predicting house prices, the output could be $\hat{y} = 250,000$ dollars. 

A \textit{classification task} involves predicting discrete categories or class labels. The model maps the input $x$ to a probability distribution over $K$ classes, where $K$ is the number of classes. For example, in binary classification, the model predicts probabilities $P(\text{class}=1 \mid \vec{x})$ and $P(\text{class}=0 \mid \vec{x})$, and the final class $k$ is chosen based on the maximum probability:
\begin{equation}
k = \arg\max_{i \in \{1, 2, \dots, K\}} P(\text{class}=i \mid \vec{x}).
\end{equation}

During training, the model adjusts its weights to maximize the likelihood of the correct class, while during inference, the model predicts the most probable class $k$. For example, in an image classification task, the output might indicate that an image is $95\%$ likely to be a dog and $5\%$ likely to be a cat, so the final prediction is ``dog.'' 

\begin{figure}[t!]
    \centering
    \begin{mybox}[\hypertarget{box:loss}{Box 2.6: Loss function}]
    A loss function is a way of quantifying the performance of a neural network model using its training dataset. In general, a loss function is problem dependent. Nevertheless its interpretation is always the same: it is minimized for well trained neural network for a specific problem. In general, a loss function $\mathcal{L}: \mathbb{R}^{\text{dim}(y_T)} \times \mathbb{R}^{\text{dim}(y)} \rightarrow \mathbb{R}$, takes two objects as its input: (i) some true value, $y_T$, coming from the model's training dataset, and the output from the neural network $y = f_{\vec{\theta}}(x)$. Notice that since $y$ is the output of a neural network, $\mathcal{L}$ is implicitly dependent on the variational parameters $\vec{\theta}$. In this sense, a perfect neural network model should solve the following equation
    \begin{equation}
        \frac{\partial {\cal L}(\vec{\theta})}{\partial\vec{\theta}} = 0
    \end{equation}
    obtaining optimal set of parameter $\vec{\theta}^*$. Equivalently
    \begin{equation}
        \vec{\theta}^* = \arg\min_{\vec{\theta}} {\cal L}(\vec{\theta}),
    \end{equation}
    since $\vec{\theta}^*$ is the location of $\mathcal{L}$'s minimum.
    In more involved tasks, for example involving images, the function space is high dimensional. For example, when we consider black-and-white image classification we deal with function $g: \vec{x}\in\mathbb{R}^n\times\mathbb{R}^n\to \vec{y}\in\mathbb{N}$, where $\vec{x}$ is a numerical representation of an image with $n\times n$ pixels, and $y\in \mathbb{N}$ enumerates the class of a given image. Let's assume we are given the dataset denoted as ${\cal D} \equiv \{ (\bar{x}_i, \bar{y}_i), i=1\dots N_{\cal D}\}$, where images are $\{x_i\}_{i=1}^{N_{\cal D}}$ and labels are $\{y_i\}_{i=1}^{N_{\cal D}}$. The objective of the training process is to find set of trainable parameters $\vec{\theta}$ minimizing loss function ${\cal L}$ 
    \begin{equation}
        {\cal L}(\vec{\theta}) = \frac{1}{N}\sum_{i=1}^N d(f_{\mathbf{\theta}}(\vec{x}_i),\bar{\vec{y}}_i).
    \end{equation}
    where $d(\vec{y}_i,\vec{y}_j)$ is a task specific distance function fulfilling\footnote{As an observant reader you may notice this means the distance function is a well defined \textit{metric}.}  $d(\vec{y}_i,\vec{y}_j)  \ge 0$ and $d(\vec{y}_i,\vec{y}_j) = 0$ when $\vec{y}_i = \vec{y}_j$. 
    \end{mybox}
\end{figure}

We can see, that in the multiclass classification, the output of the neural network is in fact a discrete probability density - and for this output we should construct proper loss function. Here, with help comes the concept known as \textit{one-hot-encoding},  See 
\hyperlink{box:one_hot_encoding}{Box~2.7}. Each class in the dataset can have assigned natural number from zero to number of all classes $k$. Without loss of generality, let us assume our dataset contains labelled images, each image $\vec{x}$ has assigned single enumerated label $\vec{y} \in 1\dots K$, where $K$ is total number of labels. In the one-hot-encoding process, instead of assigning to a given image $\vec{x}$ a given class number $\vec{y}$, we assign to an image a vector $\vec{p}$ with elements $\vec{p}_i = \delta_{i, \vec{y}}$, $i=1\dots k$, where $\vec{y}$ is a class number. Vector $\vec{p}$ we can interpret as a discrete probability density of an image $\vec{x}$ representing class $\vec{y}$. Now, to interpret output of our NN $(f_{\vec{\theta}}(\vec{x}))$, as a probability distribution, we have to 
make sure that it has only positive elements, and all outcomes are normalized to one. It can be done via applying the SoftMax function to each neuron of the output layer, i.e. $\vec{q} = \text{SoftMax}(f_{\vec{\theta}}(\vec{x}))$, where the SoftMax function acts element-wise to a vector as
\begin{equation}
\text{SoftMax}(\vec{z}) =  \{ e^{\vec{z}_i}/\sum_{j=1}^{K}e^{\vec{z}_j}, i=1\dots K \}.
\end{equation}
This allows to interpret $\vec{q}$ as a discrete probability density that image $\vec{x}$ belongs to class $\vec{y} \in [0\dots K]$.

The quantity allowing comparing two discrete density probabilities is Kullback-Leibler divergence defined as
\begin{equation}
    \textit{KL}(\vec{p}||\vec{q}) =   \sum_{i=1}^K p_i \log\left(\frac{p_i}{q_i}\right) = {\cal H} + {\cal L}_{\rm ce},
\end{equation}
where ${\cal H}$ is Shannon entropy of $\vec{p}$, and ${\cal L}_{\rm ce}$ is a so-called \textit{categorical cross-entropy}:
\begin{equation}
    {\cal L}_{\rm ce} = -\sum_{i=1}^{K} p_i\log q_i.
\end{equation}
The Shannon entropy does not depend on the neural network parameters, thus only the categorical cross-entropy ${\cal L}_{\rm ce}$ serves as a loss function for multiclass classification tasks. Categorical cross-entropy penalizes incorrect predictions by assigning higher penalties when the predicted probability $q_i$ for the true class is low.

\begin{figure}[h]
    \centering
    \begin{mybox}[\hypertarget{box:one_hot_encoding}{Box 2.7: One-hot encoding technique}]
One-hot encoding is a method used to represent categorical data numerically in multiclass classification. Each class label is transformed into a binary vector where:
\begin{itemize}
    \item The length of the vector equals the total number of classes, denoted as $K$.
    \item All elements are $0$, except for a $1$ at the index corresponding to the class label.
\end{itemize}
For example, consider three classes: $A$, $B$, and $C$, indexed as:
$A \to 0, \quad B \to 1, \quad C \to 2$. The one-hot encoded vectors for these classes are:
\begin{equation}
A \to [1, 0, 0], \quad B \to [0, 1, 0], \quad C \to [0, 0, 1]
\end{equation}

If the original class labels are:
$[A, C, B, A]$
Their one-hot encoded representation becomes:
\begin{equation}
\begin{bmatrix}
1 & 0 & 0 \\ 
0 & 0 & 1 \\ 
0 & 1 & 0 \\ 
1 & 0 & 0
\end{bmatrix}
\end{equation}
One-hot encoding ensures there is no ordinal relationship between class labels. It is compatible with machine learning algorithms that require numerical input. For a large number of classes $K$, one-hot encoding results in high-dimensional sparse vectors, which can increase memory usage and computational cost. As such, the one-hot encoding of words from dictionary is not optimal way of numerical encoding human language. Alternative approaches, such as embeddings, may be more efficient when $K$ is very large.
     \end{mybox}
\end{figure}

\begin{figure}
    \centering
\begin{mybox}[\hypertarget{box:loss2}{Box 2.8: Common Loss Functions}]
    In a multiclass classification problem, the objective is to predict the class of an input instance among $K$ possible categories. The model outputs a probability distribution $\vec{q} = [q_1, q_2, \ldots, q_K]$, where $q_i$ is the predicted probability of the $i$-th class, with $\sum_{i=1}^K q_i = 1$. The true class distribution is represented by a one-hot encoded vector $\vec{p} = [p_1, p_2, \ldots, p_K]$, where $p_i = 1$ for the true class and $p_j = 0$ for $j \neq i$.

The \textbf{Kullback-Leibler (KL) divergence} quantifies the difference between the true distribution $\vec{p}$ and the predicted distribution $\vec{q}$. It is given by 
\begin{equation}
    \text{KL}(\vec{p} \parallel \vec{q}) = \sum_{i=1}^K p_i \log\left(\frac{p_i}{q_i}\right)
\end{equation}
Since $\vec{p}$ is a one-hot vector, $p_i = 1$ for the true class and $p_j = 0$ for $j \neq i$. Therefore, the KL divergence simplifies to $ \text{KL}(\vec{p} \parallel \vec{q}) = -\log(q_i) $, where $i$ is the index of the true class. This is equivalent to the \textbf{categorical cross-entropy loss}. The KL divergence can be expressed as sum of Shannon entropy ${\cal H}$ of probability distribution $\vec{p}$, and a categorical cross entropy ${\cal L}_{\rm ce}$, i.e.
\begin{equation}
    \text{KL}(\vec{p} \parallel \vec{q}) =   \sum_{i=1}^K p_i\log p_i - \sum_{i=1}^K p_i\log q_i = {\cal H} + {\cal L}_{\rm ce}.
\end{equation}
Because the Shannon entropy ${\cal H}$ contains information only about true labels, it does not depends on the NN parameters. Thus only the categorical cross-entropy ${\cal L}$ is used as a loss function in multiclass classification tasks. ${\cal L}$ penalizes incorrect predictions by assigning higher penalties when the predicted probability $q_i$ for the true class is low. It encourages the model to output higher probabilities for the correct class, making it an effective choice for multiclass classification tasks due to its probabilistic foundation.

Please note, that KL divergence is a semi-distance measure: while it is bounded from below $\text{KL}(\vec{p} \parallel \vec{q})\ge0$, vanishing when $\vec{p}=\vec{q}$ it is asymmetric $\text{KL}(\vec{p} \parallel \vec{q})\ne \text{KL}(\vec{q} \parallel \vec{p})$.

\end{mybox}\label{box:common_loss_functions}
\end{figure}

\subsection{Neural Network Training (Backpropagation)}
\label{CH_FUNDAMENTAls_sec:NN_training}

The central concept for training neural networks is the \textit{gradient descent} algorithm. This method solves for $\vec{\theta}^*$ for a given loss function $\mathcal{L}(\vec{\theta})$ (see \hyperlink{box:loss}{Box~2.6}). Gradient descent is \textit{iterative}, meaning each step is performed sequentially, depending only on the previous step(s). To this end, we will use a time-like index $t$ to denote which step of the algorithm we are on. In the first step, $t=0$, we randomly initialize $\vec{\theta}$ by selecting its values from the uniform distribution\footnote{This might seem like a poor strategy, and in fact it is. Mention warm starts here.}. Next, we calculate gradient of the loss function with the respect to initialized parameters, allowing us to update $\vec{\theta}_{t+1}$ via
\begin{equation}
    \vec{\theta}_{t+1} \leftarrow \vec{\theta}_t - \eta\frac{{\cal L}(\vec{\theta}_t)}{\partial\vec{\theta}_t},
    \label{eq:GD_vanilla}
\end{equation}
where $\eta \in \mathbb{R}^+$ is a so-called \textit{hyperparameter} of the training process, see \hyperlink{box:hyperparameters}{Box~2.9}. It was for \textit{this} reason that we required the layer's transformations to be differentiable.
\begin{figure}[h]
    \centering
    \begin{mybox}[\hypertarget{box:hyperparameters}{Box 2.9: Hyperparameters}]
    Hyperparameters are pre-set parameters that govern the training process and architecture of a neural network. You can think hyperparameters as anything that can be tuned, just not with the neural network training algorithm. That is, unlike model parameters (weights, biases), they are not learned during training. Hyperparameters can be divided into three categories:
\begin{itemize}
    \item \textbf{Architecture-Related:} such as number of layers,  activation function, dropout rate or weight initialization.
    \item \textbf{Optimization-Related:} such as learning rate, batch size, regularization of the loss function, or optimization algorithm.
    \item \textbf{Training-Related:} such as number of epochs
\end{itemize}
     Those that can be tuned with training are \textit{variational} parameters, see \hyperlink{box:VarParams}{Box~2.3}.
     \end{mybox}
\end{figure}

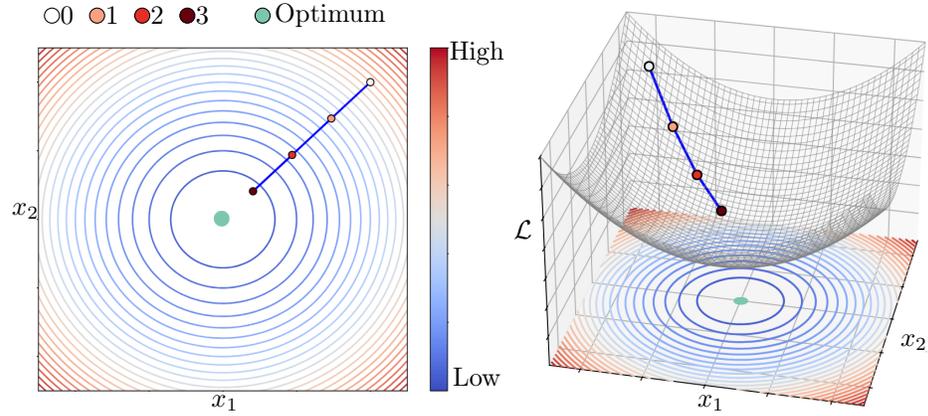
\begin{figure}[h]
    \centering
    \begin{subfigure}[b]{0.45\textwidth}
        \centering
        \begin{tikzpicture}
            \definecolor{customcolor}{HTML}{FCA082}
            \definecolor{customcolor2}{HTML}{E32F26}
            \definecolor{customcolor3}{HTML}{67000D}
            \definecolor{customcolor4}{HTML}{7CC7AC}
            \node [anchor=south west, inner sep=0] (image) at (0,0) {\includegraphics[width=\textwidth]{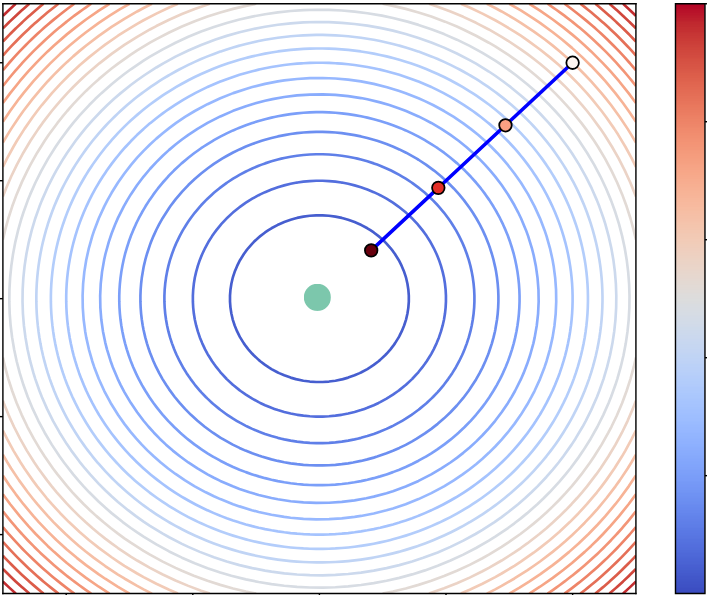}};
            \begin{scope}[shift={(0,0)}]
                \node at (5.85,4.5) {High};
                \node at (5.85, 0.2) {Low};
                \draw[fill=white, draw=black] (0.2,5.0) circle (0.1cm);
                \node at (0.4, 5.0) {0};
                \draw[fill=customcolor, draw=black] (0.8,5.0) circle (0.1cm);
                \node at (1.0, 5.0) {1};
                \draw[fill=customcolor2, draw=black] (1.4,5.0) circle (0.1cm);
                \node at (1.6, 5.0) {2};
                \draw[fill=customcolor3, draw=black] (2.0,5.0) circle (0.1cm);
                \node at (2.2, 5.0) {3};

                \draw[fill=customcolor4, draw=black] (3.0,5.0) circle (0.1cm);
                \node at (3.9, 5.0) {Optimum};

                \node at (2.5, -0.15) {$x_1$};
                \node at (-0.15, 2.4) {$x_2$};

            \end{scope}
        \end{tikzpicture}
    \end{subfigure}
    \hfill
    \begin{subfigure}[b]{0.45\textwidth}
        \centering
        \begin{tikzpicture}
            \node [anchor=south west, inner sep=0] (image) at (0,0) {\includegraphics[width=\textwidth]{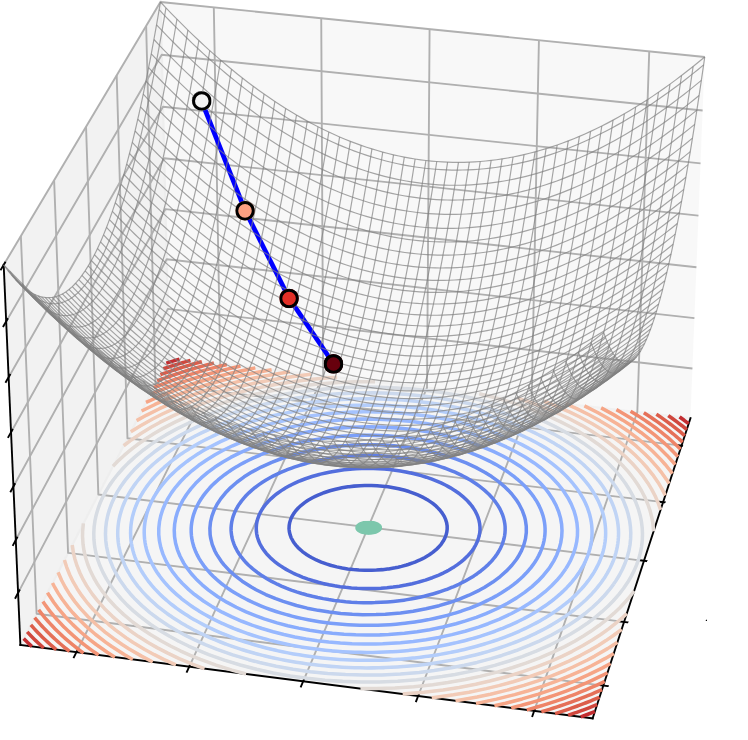}};
            \begin{scope}[shift={(0,0)}]
                \node at (2.3, 0.2) {$x_1$};
                \node at (5.0, 1.0) {$x_2$};
                \node at (-0.2, 2.5) {$\mathcal{L}$}; 
            \end{scope}
        \end{tikzpicture}
    \end{subfigure}
    \caption{Three steps of gradient descent visualised for a two-component quadratic loss function $\mathcal{L} = (\tilde{x}_1 - x_1)^2 + (\tilde{x}_2 - x_2)^2$. The blue straight lines joining the dots show the step \textit{size}, whilst their \textit{direction} points towards the optimum (light green). The algorithm initialises at step 0 (white circle) and makes its way towards the optimum. On the left we see this path on the contour plot, whilst the right shows what this looks like on the surface (grey mesh), with the contour underneath for comparison.}
    \label{fig:GD_basic}
\end{figure}

Visually, you can think of each iteration in gradient descent as a step around the landscape of $\mathcal{L}(\vec{\theta})$ \textit{in its direction of steepest descent}. In this way, $\eta$ tells you the step \textit{size}, whilst $\frac{\partial {\cal L}(\vec{\theta}_t)}{\partial\vec{\theta}_t} = \nabla_{\vec{\theta}}\mathcal{L}(\vec{\theta})$ tells you the step \textit{direction}. Hence, gradient descent is a way of navigating a loss function's landscape to find the optimum $\vec{\theta}^*$ which minimises it, see Fig.~\ref{fig:GD_basic}.
In an idealised setting, gradient descent converges to the optimum $\vec{\theta} = \vec{\theta}^*$ after a sufficiently large number of iterations, $t\gg 1$. At this stage, we will see $\vec{\theta}_{t} = \vec{\theta}_{t+1}$, since by definition $\nabla_{\vec{\theta}} \mathcal{L}(\vec{\theta)} = 0$ at the optimum. 

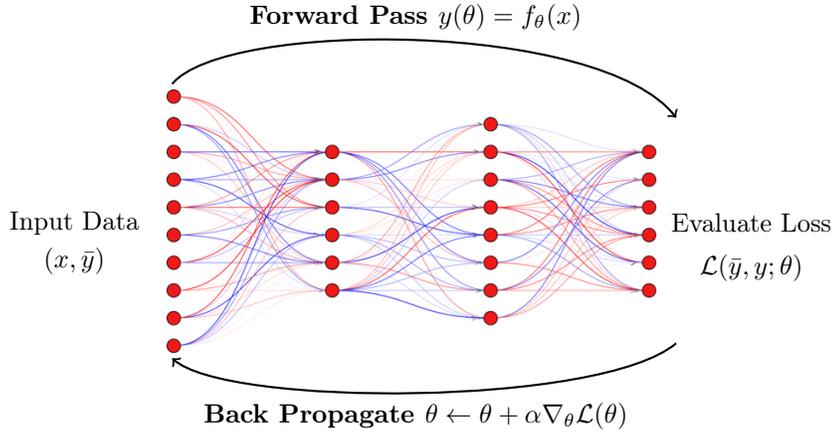
\begin{figure}[t!]
    \centering
     \begin{tikzpicture}
        \node [anchor=south west, inner sep=0] (image) at (0,0) {\includegraphics[width=0.6\textwidth]{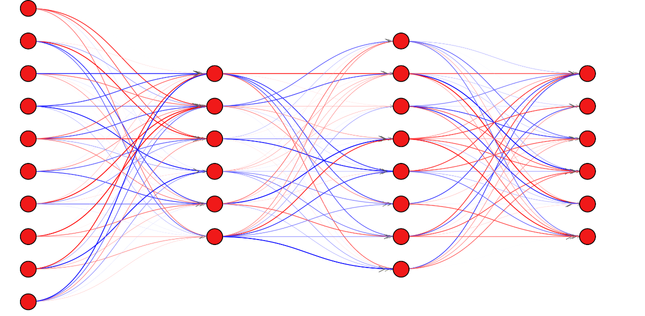}};
        \begin{scope}[shift={(0,0)}]

        \node at (-1.0, 1.8) {Input Data };  
        \node at (-1.0, 1.3) {$(x, \bar{y})$};
        \draw[->, thick] 
        (0.3,3.65) .. controls (1,4.5) and (6,4.5) .. (7,3.2)
        node[midway, above] {\textbf{Forward Pass $y(\theta) = f_{\theta}(x)$}};
        \node at (8.0, 1.8) {Evaluate Loss};
        \node at (8.0, 1.2) {${\cal L}(\bar{y},y;\theta)$};
        \draw[->, thick] 
        (7,0.2)  .. controls (6,-0.65) and (1,-0.65) .. (0.3,0.0)
        node[midway, below] {\textbf{Back Propagate $\theta \gets \theta + \alpha \nabla_{\theta} {\cal L}(\theta)$}};
        \end{scope}
    \end{tikzpicture}
    \caption{Visual representation of neural network training. Data, $x$, with true label $\bar{y}$ is received by the input layer, then a forward pass is executed to yield an output $y(\theta) = f(x)$. The output is then compared to some true value (or \textit{label}), $\bar{y}$, via a loss function, $L(\bar{y},y;\theta)$. Finally, the loss function (and its derivatives!) are used to update the neural network's variational parameters via backpropagation. Back propagation can be achieved through simple means like vanilla gradient descent (shown in the figure), or with more sophisticated optimisers - see Sec.~\ref{subsec:adv_optimisers}. For example using a \textit{mean square error} loss (Eq.~\ref{eq:MSE_general}) on this neural network would look like performing gradient descent in a landscape like Fig.~\ref{fig:GD_basic}.}
    \label{fig:fig_Ch2_training_infographic}.
\end{figure}

In more practical settings like those encountered in deep learning, our model usually high-dimensional and the dataset is large\footnote{i.e. the number of items, $N_{\mathcal{D}} \gg 1$ in a dataset $\mathcal{D}$.}. This creates a bottleneck in the computational memory when doing gradient descent; we cannot calculate loss function for the \textit{whole} dataset at once. Even if we could, this would be a huge amount of data that is not practical to handle. Consider for example the \textit{mean square error} function we saw in Eq.~(\ref{eq:MSE}). Having more data means our arithmetic mean draws from a larger set. In would be both impractical and necessary to take this mean over $K = N_{\mathcal{D}} \gg 1$ points, since often a subset of $\mathcal{D}$ gives us a ``good enough'' estimate with a more reasonable computational overhead. 
Doing gradient descent over a uniformly sampled subset $\mathcal{D}^{(j)}~\subset \mathcal{D}$ solves this obstacle, giving the \textit{stochastic gradient descent} algorithm. In this version of gradient descent, the dataset ${\cal D}$ is divided into \textit{batches} of data, ${\cal D}~=~\cup_{j=1}^{N_{b}}{\cal D}^{(j)}$, ${\cal D}^{(j)} =\{ (\vec{x}^{(j)}_i, \vec{y}^{(j)}_i), i = 1\dots N_{{{\cal D}^{(j)}}}$. 
Each batch is separately used to execute a step of the update Eq.~(\ref{eq:GD_vanilla}). In the literature, each step in the process of iterative optimization of neural network parameters is called a \textit{training step} (sometimes called train-step). The single $t$-th training step of the training constitutes doing three things for each batch of data ${\cal D}^{(j)}$:
\begin{enumerate}
\item {\bf Forward Pass} 
\begin{equation}
    \bar{\vec{y}}^{(j)}_i = f_{\vec{\theta}_t}(\vec{x}^{(j)}_i)
\end{equation}

\item {\bf Loss calculation}  
\begin{equation}
    {\cal L}^{(j)}(\vec{\theta}_t)= \frac{1}{N_{{{\cal D}^{(j)}}}}\sum_{i=1}^{N_{{{\cal D}^{(j)}}}}
    {\cal L}_i(\vec{y}^{(j)}_i, \bar{\vec{y}}^{(j)}_i)
\end{equation}

\item {\bf Backpropagation }  
\begin{equation}\label{eq:theta_update}
    \vec{\theta}_{t+1} \leftarrow \vec{\theta}_t - \eta\frac{\partial{\cal L}^{(j)}(\vec{\theta}_t)}{\partial\vec{\theta}_t}.
\end{equation}
\end{enumerate}
Steps $1$-$3$ are performed for each batch ${\cal D}^{(j)}$ - when whole dataset is used, then we say that $t$-th step of the training procedure, i.e. \textit{one training epoch}, has been performed, see Fig.~\ref{fig:fig_Ch2_training_infographic}. 

Usually, a neural network model takes many epochs before it converges to the optimum. The rate of convergence generally depends on our choice of update rule (based on gradient descent), our architecture, and the model's hyperparameters. So far, we have seen two examples for an update rule which were (vanilla\footnote{In the ML lingo, vanilla usually refers to the na\"ive way of doing something. For example, the first (and simplest) gradient descent strategy is commonly referred to as \textit{vanilla gradient descent}.}) gradient descent and stochastic gradient descent, based on batching data. However, these are not the only strategies. In fact, there is a whole research field for optimization algorithms which \textit{best} train neural networks. We will partially cover some key milestones of this field in the Sec.~\ref{sec:training_nn_mathematical}.

\newpage
\section[Deep Learning Hello World]{Deep learning \texttt{Hello world} - MNIST dataset}

\begin{figure}[h]
    \centering
    \includegraphics[scale=0.4]{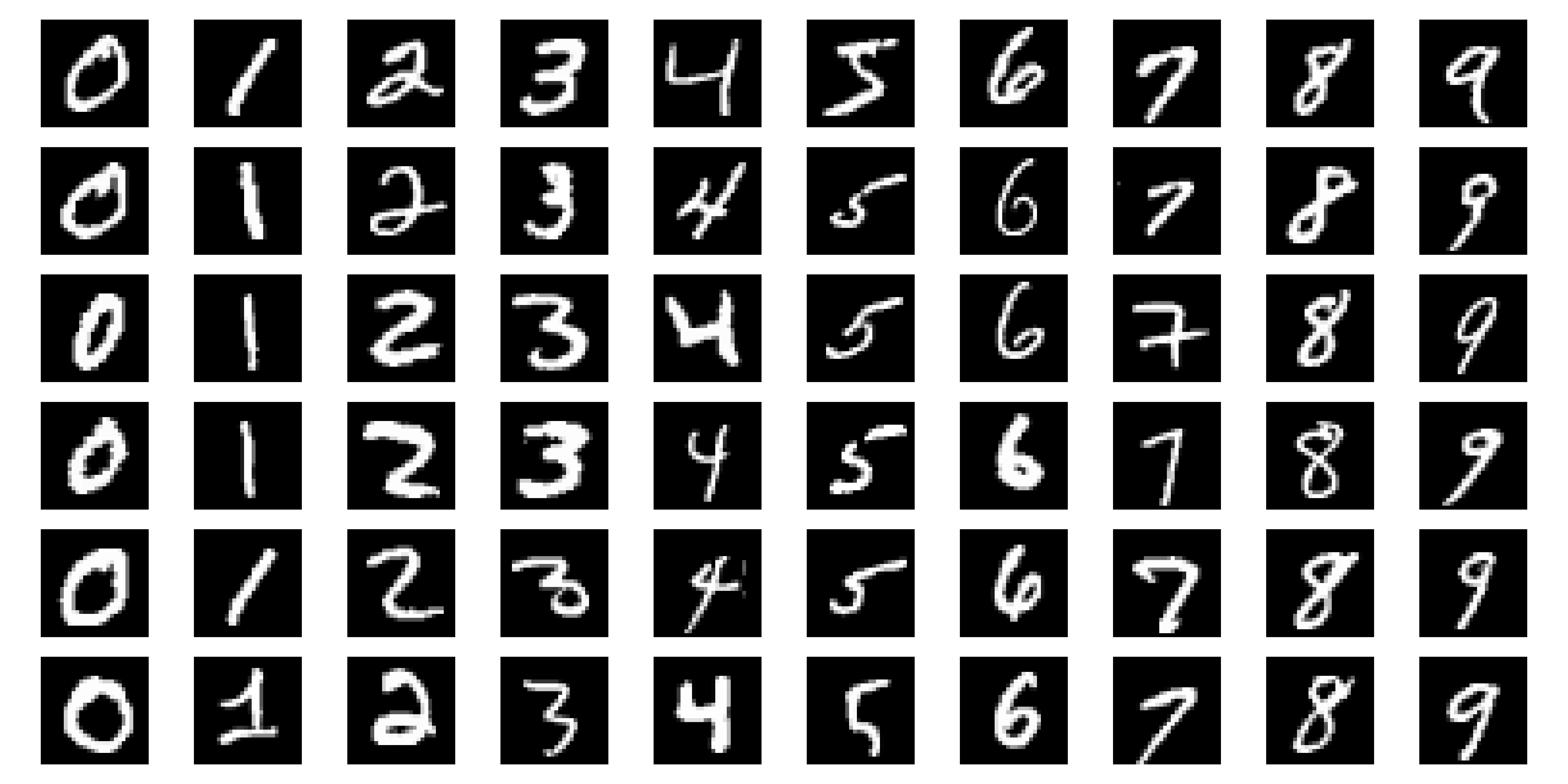}
    \caption{Examples of hand-written digits from the MNIST dataset. Each image is a  $28\times28\times 1$ tensor, where first two dimension describe picture resolution, while the last one, number of colour channels. One chanel corresponds to black-white images.}
    \label{fig:fig_MNIST_examples}
\end{figure}

As a concrete example, let us consider \textit{Hello World} deep learning problem, i.e. hand-written digit classification task. The \href{https://yann.lecun.com/exdb/mnist/}{MNIST dataset} (Modified National Institute of Standards and Technology) is a widely used benchmark for evaluating machine learning algorithms, particularly in image classification. It consists of grayscale images of handwritten digits ranging from 0 to 9, each represented as a $28\times28$ pixel grid. With $6\times 10^4$ training samples and $10^4$ test samples, it provides a straightforward setup for training and evaluating models. Each image is labeled with an integer corresponding to the digit it represents, with pixel values ranging from $0$ (black) to $255$ (white), Fig.\ref{fig:fig_MNIST_examples}.

\begin{figure}[h]
    \centering
    \begin{mybox}[\hypertarget{box:dataset2}{Box 2.10: Splitting dataset}]

 In machine learning, dividing a dataset into \textbf{training} and \textbf{test} sets is essential to ensure that a model learns patterns effectively and can make accurate predictions on new, unseen data. The training set is used to teach the model how to identify patterns, while the test set evaluates how well the model performs on data it has never seen before. This helps confirm that the model is not just memorizing the training data.
 
Often, the dataset is further split into \textbf{training}, \textbf{validation}, and \textbf{test} sets. The validation set is used to fine-tune the model's settings (such as parameters or architecture), while the test set is used only once to assess the final model’s quality. As an example of a dataset with $10^4$ samples, we might split it as:
\begin{itemize}
    \item \textbf{80\% (8,000 samples):} Training set
    \item \textbf{10\% (1,000 samples):} Validation set
        \item \textbf{10\% (1,000 samples):} Test set
\end{itemize}
This kind of data splitting ensures that the model is evaluated on data it has not been exposed to during training, providing a realistic measure of its ability to make predictions on future, unseen data.

    \end{mybox}
\end{figure}

\subsection{Image Classification with a Single Output Node} 

We will consider simple neural network architecture allowing for classification hand-written digit represented by an image via single output neuron.

We can imagine that there exist a function $G: \vec{x}\in\mathbb{R}^{28}\times\mathbb{R}^{28}\to \vec{y}\in\mathbb{N}$ which is mapping between particular configurations of pixels to a digit $0\dots 9$. We can build an analogy from example on Fig.\ref{fig:fig_1_2}. The unknown function $G(\vec{x})$, mapping intensity of $m=28^2=784$ pixels to a discrete set $\{0\dots 9\}$, is a high-dimensional equivalent of continuous function $g(x)$ from Fig.\ref{fig:fig_1_2}. Each hand-written digit from MNIST dataset corresponds to sampling point on $x$, and value of digit $y$ corresponds to value of $g(x_i)$ from Fig.\ref{fig:fig_1_2}. However, we have access only to finite set of images, being equivalent of black dots in Fig.\ref{fig:fig_1_2}, while the true $G$ is unknown to us. We will consider \textit{regression} problem, i.e. the output layer will contain single neuron which value we will take $y \in (0,1)$, and we will interpret given digit rescaled by factor $1/10$.

In \hyperlink{box:dataset}{Box~2.5}, we stated the importance of our \textit{choice} and \textit{representation} of data. The MNIST dataset already contains good choices of data for recognising hand-written digits, as it contains images of them. However, our \textit{representation} of data is yet to be set. Therefore, let's find a way of representing the data as an input that makes it as easy as possible for a neural network to extract meaningful patterns.
Our input data will be vectorized images $\vec{x}$, rescaled to have mean $0$, and standard deviation $1$, i.e. $\vec{x}_i \leftarrow (\vec{x}_i - \textit{mean}(\vec{x}_i)/ \textit{std}(\vec{x}_i)$, while label $y_i$ will be number $0\dots9$. This rescaling means we can invoke the UAT and argue that our task is now learning a standard multi-variate normal.

Next, our loss function will be the mean squared error
\begin{equation}
\begin{split}
{\cal L} & = \frac{1}{N}\sum_{i=1}^N {\cal L}_i(\vec{y}_i, \bar{\vec{y}}_i)\\
{\cal L}_i(\vec{y}_i, \bar{\vec{y}}_i) & = \frac{1}{2}(\vec{y}_i - \bar{\vec{y}}_i)^2,
\end{split}
\end{equation}
where $\bar{\vec{y}}_i$ is our prediction and $\vec{y}_i$ the ground truth. 

We update the weights $\vec{W}$ and biases $\vec{b}$ with a gradient descent procedure (i.e. we calculate value of the loss function for the \textit{whole} training dataset at each parameters update iteration):
\begin{equation}
\begin{split}
\vec{W}^{(1)} & \leftarrow {\vec W}^{(1)} - \eta \frac{\partial {\cal L}}{\partial \vec{W}^{(1)}} = \vec{W}^{(1)} - \frac{\eta}{N} \sum_{i=1}^N \frac{\partial {\cal L}_i}{\partial \vec{W}_1}\\
\vec{b} & \leftarrow \vec{b} - \eta \frac{\partial {\cal L}}{\partial \vec{b}} = W - \frac{\eta}{N} \sum_{i=1}^N \frac{\partial {\cal L}_i}{\partial \vec{b}},\\
\vec{W}^{(2)} & \leftarrow {\vec W}^{(2)} - \eta \frac{\partial {\cal L}}{\partial \vec{W}^{(2)}} = \vec{W}^{(2)} - \frac{\eta}{N} \sum_{i=1}^N \frac{\partial {\cal L}_i}{\partial \vec{W}_2}
\end{split}
\end{equation}
where $\eta$ is the learning rate, and $N$ is total number of images in the dataset.

As a neural network architecture for MNIST dataset recognition task we will consider a single hidden layer and sigmoid activation functions, i.e.
\begin{equation}
    \sigma(x) = \frac{1}{1+e^{-x}}.
\end{equation}
The input layer has the same size as the number of features in our data, i.e., $m=784$ neurons. Then, the hidden layer has $N^{(1)}$ neurons, and the output layer has $n$ neurons. The first weight  matrix $\vec{W}^{(1)}$ has shape $m\times N^{(1)}$, and the second weight matrix $\vec{W}^{(2)}$ has shape $N^{(1)}\times 1$. In this case, we only consider biases in the hidden layer $\mathbf{b}_1$ which is a vector with $N^{(1)}$ entries.
Finally, we consider $N_{{\cal D}_i}$ images in each batch of data. The training process contains the following steps: \\

\noindent {\bf 1. Feed-forward pass}
Now, let's go through the feed-forward pass of the training data, witch contains $i$-th batch of $N_{{\cal D}_i}$ vectors with $m=784$ pixels, 
$X_i\in\mathbb{R}^{ N_{{\cal D}_i} \times m}$ .
\begin{itemize}
    \item The input goes through the first linear layer $\vec{h} \leftarrow X \vec{W}^{(1)} + \vec{b}_1$ with shapes $[N_{{\cal D}_i}, m] \times [m, N^{(1)}] = [N_{{\cal D}_i}, N^{(1)}]$
    
    \item We apply the activation function $\bar{\vec{h}} \leftarrow \sigma(\vec{h})$ with shape $[N_{{\cal D}_i}, N^{(1)}]$
    
    \item Then, we apply the second linear layer $\vec{g} \leftarrow \bar{\vec{h}}\vec{W}^{(2)}$ with shapes $[N_{{\cal D}_i}, N^{(1)}] \times [N^{(1)}, n] = [N_{{\cal D}_i}, n]$

    \item Finally, we apply the activation function $\bar{\vec{y}} \leftarrow \sigma(\vec{g})$ with shape $[N_{{\cal D}_i}, n]$

\end{itemize}

\noindent {\bf 2. Parameter update} We will use the mean-squared error loss function denoted in matrix representation as
${\cal L} = \frac{1}{2N_{{\cal D}_i}}||\vec{Y} - \bar{\vec{Y}}||^2$,
where $\vec{Y}$ and $\bar{\vec{Y}}$ represent predicted and true classes in a considered batch of data, respectively. The parameter update rule is
\begin{equation}
\begin{split}
  \vec{W}_1 & \leftarrow \vec{W}_1 - \frac{\eta}{N_{{\cal D}_i}}\frac{\partial {\cal L}_i}{\partial \vec{W}_1} \\
  \vec{b}_1 & \leftarrow \vec{b}_1 - \frac{\eta}{N_{{\cal D}_i}}\frac{\partial {\cal L}_i}{\partial \vec{b}_1} \\
  \vec{W}_2 & \leftarrow \vec{W}_2 - \frac{\eta}{N_{{\cal D}_i}}\frac{\partial {\cal L}_i}{\partial \vec{W}_2}.
\end{split}
\end{equation}
The loss function with respect to  $\vec{W}^{(1)}$, $\vec{b}^{(1)}$ and $\vec{W}^{(2)}$ is calculated using the chain rule:
\begin{equation}
\begin{split}
\frac{\partial {\cal L}_i}{\partial \vec{W}^{(2)}}  & = \frac{\partial {\cal L}_i}{\partial \bar{Y}}\frac{\partial \bar{Y}}{\partial \mathbf{g}}\frac{\partial \vec{g}}{\partial \vec{W}^{(2)}} \\
\frac{\partial {\cal L}_i}{\partial \vec{W}^{(1)}} & = \frac{\partial {\cal L}_i}{\partial \bar{Y}}\frac{\partial \bar{Y}}{\partial \vec{g}}\frac{\partial \vec{g}}{\partial \bar{\vec{h}}}\frac{\partial \bar{\vec{h}}}{\partial \vec{h}}\frac{\partial \vec{h}}{\partial \vec{W}^{(1)}}\\
\frac{\partial {\cal L}_i}{\partial \vec{b}_{1}} & = \frac{\partial {\cal L}_i}{\partial \bar{Y}}\frac{\partial \bar{Y}}{\partial \vec{g}}\frac{\partial \vec{g}}{\partial \bar{\vec{h}}}\frac{\partial \bar{\vec{h}}}{\partial \vec{h}}\frac{\partial \vec{h}}{\partial \vec{b}_{1}}
\end{split}
\end{equation}

We can write down every term:
\begin{equation}
\begin{split}
 \frac{\partial {\cal L}_i}{\partial \bar{Y}}   &= \bar{Y} - Y, 
 \frac{\partial \bar{Y}}{\partial \vec{g}}  = \bar{Y}(1-\bar{Y}), 
 \frac{\partial \vec{g}}{\partial \vec{W}^{(2)}}   = \bar{\vec{h}},   
 \frac{\partial \vec{g}}{\partial \bar{\vec{h}}}   = \vec{W}^{(2)},\\  
 \frac{\partial \bar{\vec{h}}}{\partial \vec{h}}  & = \bar{\vec{h}}(1-\bar{\vec{h}}),  
 \frac{\partial \vec{h}}{\partial W_1}   = X,
 \frac{\partial \vec{h}}{\partial \vec{b}_1}   = \mathbb{1}.
\end{split}
\end{equation}
Finally, by defining 
\begin{equation}
\begin{split}
 Q_2 & \equiv \frac{\partial {\cal L}_i}{\partial \bar{Y}}\frac{\partial \bar{Y}}{\partial \vec{g}} = (\bar{Y}-Y)\bar{Y}(1-\bar{Y}) \\
 Q_1 & \equiv \frac{\partial {\cal L}_i}{\partial \bar{Y}}\frac{\partial \bar{Y}}{\partial \vec{g}}\frac{\partial \vec{g}}{\partial \bar{\vec{h}}}\frac{\partial \bar{\vec{h}}}{\partial \vec{h}} = Q_2 \vec{W}^{(2)}\bar{\vec{h}}(1-\bar{\vec{h}}),
 \end{split}
\end{equation}
the update rule, in the $i$-th batch of data, reads
\begin{equation}
\begin{split}
    \vec{W}^{(2)} &\leftarrow \vec{W}^{(2)} - \frac{\eta}{N_{{\cal D}_i}}\bar{\vec{h}}^TQ_2\\
\vec{b} &\leftarrow \vec{b} - \frac{\eta}{N_{{\cal D}_i}}Q_1,\\
\vec{W}^{(1)} &\leftarrow \vec{W}^{(1)} - \frac{\eta}{N_{{\cal D}_i}}X^TQ_1\\
\end{split}
\end{equation}
As we can see, even for simple single layer architecture with sigmoid activation function calculating parameters update rule requires substantial calculations.

\begin{figure}[t!]
     \centering
    \includegraphics[scale=0.5]{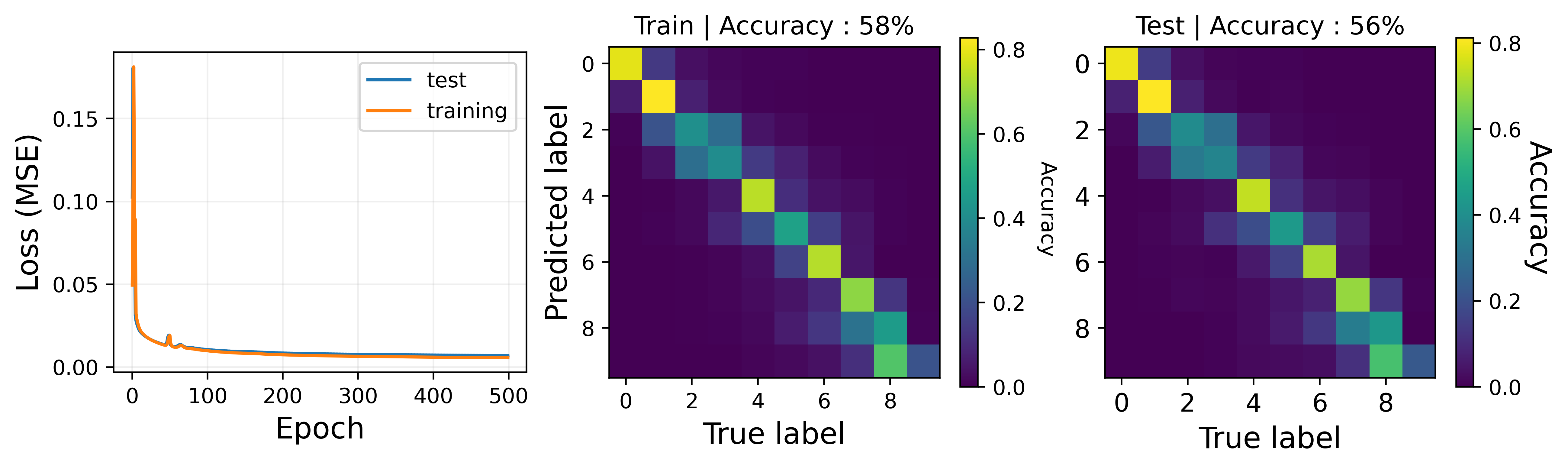}
    \caption{Performance of the simple NN architecture for MNIST dataset prediction. Input layer contains $28^2$ nodes, while single layer contains only $N^{(1)}=21$ nodes. Output layer is a single node. All nodes have sigmoid activation function. }
\label{fig:fig_MNIST_simple_NN_results}
\end{figure}

Now, we can analyse how loss function changes with training iterations, i.e. epochs, on the \textit{training} dataset, and how the optimized parameters $\{\vec{W}^{(1)}, \vec{b}, \vec{W}^{(2)}\}$ at each training epoch, perform against \textit{test} dataset, i.e. how our model predicts digits from images which were not used to optimize parameters. We consider $N^{(1)}=21$ neurons in the hidden layer, $N_{\rm epoch}=500$ training epochs and different values of the learning rate parameter $\eta = 50$. These parameters are not standard one, however for this simple architecture and classification task with the help of regression approach are sufficient. In Fig.\ref{fig:fig_MNIST_simple_NN_results}, left column, we present how loss function on training and tests datasets changes with training epochs, while second and third columns present predictions of trained model in the form of \textit{confusion matrix}, see , and quantify the quality of these predictions as a average value of diagonal elements of the confusion matrix, defining \textit{accuracy}, see \hyperlink{box:confusion_matrix}{Box~2.11}. As we can see model can recognize hand-written digits with almost $60\%$ accuracy on test dataset. As we can see, the training process is a simple nonlinear fitting of a high-dimensional function $G: \mathbb{R}^{784}\to \mathbb{R}$, based on finite dataset.

\begin{figure}[h]
    \centering
    \begin{mybox}[\hypertarget{box:confusion_matrix}{Box 2.11: Model evaluation with Confusion matrix}]
To evaluate model prediction for classification task of $K$ classes, a convenient tool is a \textit{confusion matrix} $C\in \mathbb{R}^{K\times K}$. To construct the confusion matrix $C$ we initialize its elements as $0$. Next, for each instance $\vec{x}$ from dataset having lass $y \in 1\dots K$, we prepare prediction $f_{\vec{\theta}^*}(\vec{x}) \equiv \bar{y} \in 1\dots K$, and update confusion matrix elements as $C_{\bar{y},y} \leftarrow C_{\bar{y},y}+1$. After the whole dataset is analysed we normalize the confusion matrix to sum up each row to one. Now, the confusion matrix represent normalized histogram of predicted classes. Accuracy of model predictions is defined as a mean value of diagonal of $C$.
     \end{mybox}
\end{figure}

\subsection{Multiclass Classification and Categorical Cross-Entropy}\label{sec:multiclass_classification}

In the previous example our NN architecture had only one neuron in the output layer. However, for a many-classes classification problem it is more natural to have as many neurons as classes in the output layer. For digits, we have $k = 10$ in the dataset. Therefore, in the ideal scenario of perfectly working model with an output layer of $k=10$ neurons, only one neuron will fire up when in the input we have image representing the proper class. 

\begin{figure}[t!]
    \centering
    \includegraphics[scale=0.5]{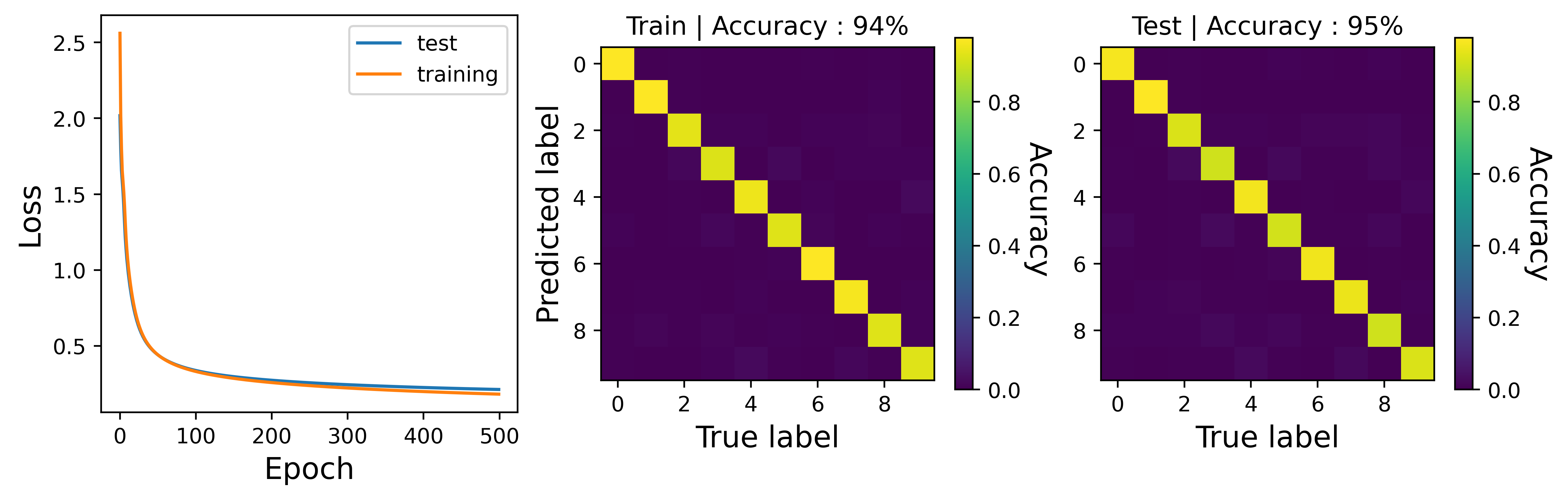}
    \caption{Performance of the simple NN architecture for MNIST dataset prediction. Input layer contains $28^2$ nodes, while single layer contains only $N^{(1)}=21$ nodes. Output layer is a single node. All nodes have sigmoid activation function. }
\label{fig:fig_MNIST_categorical_cross_entropy}
\end{figure}

Since we have modified the neural network architecture, we need to construct a different type of loss function. Namely, one which receives an array of 10 numbers. Here, we will show how to use cross entropy ${\cal L}_{\rm ce}$ as a loss function, and check how our model performs in the MNIST classification task. As in previous example, the number of hidden layer neurons $N^{(1)}=21$ and will consider same learning rate $\eta = 1$. Fig.\ref{fig:fig_MNIST_categorical_cross_entropy} presents how our simple architecture deals with MNIST classification with categorical cross-entropy loss function. Predictions accuracy on test dataset is $95\%$ (which we can compare with $56\%$ from our previous example). This example shows us how proper construction of the loss function can increase predictive power of our neural network.

\subsection{Convolutional Neural Networks}

It turned out that classification of hand-written digits from MNIST dataset can be done with extremely simple neural network architecture being the non-linear function mapping an black-white $28\times 28$ pixels image to a discrete probability function $f_{\vec{\theta}}:\vec{x}\in\mathbb{R}^{28\times 28}\to \vec{y}\in \mathbb{R}^{10}$, with such that $\sum_{k=1}^{K}\vec{y}_k=1$, where the loss function was categorical cross-entropy. 

\begin{figure}[t!]
    \centering
    \includegraphics[width=\linewidth]{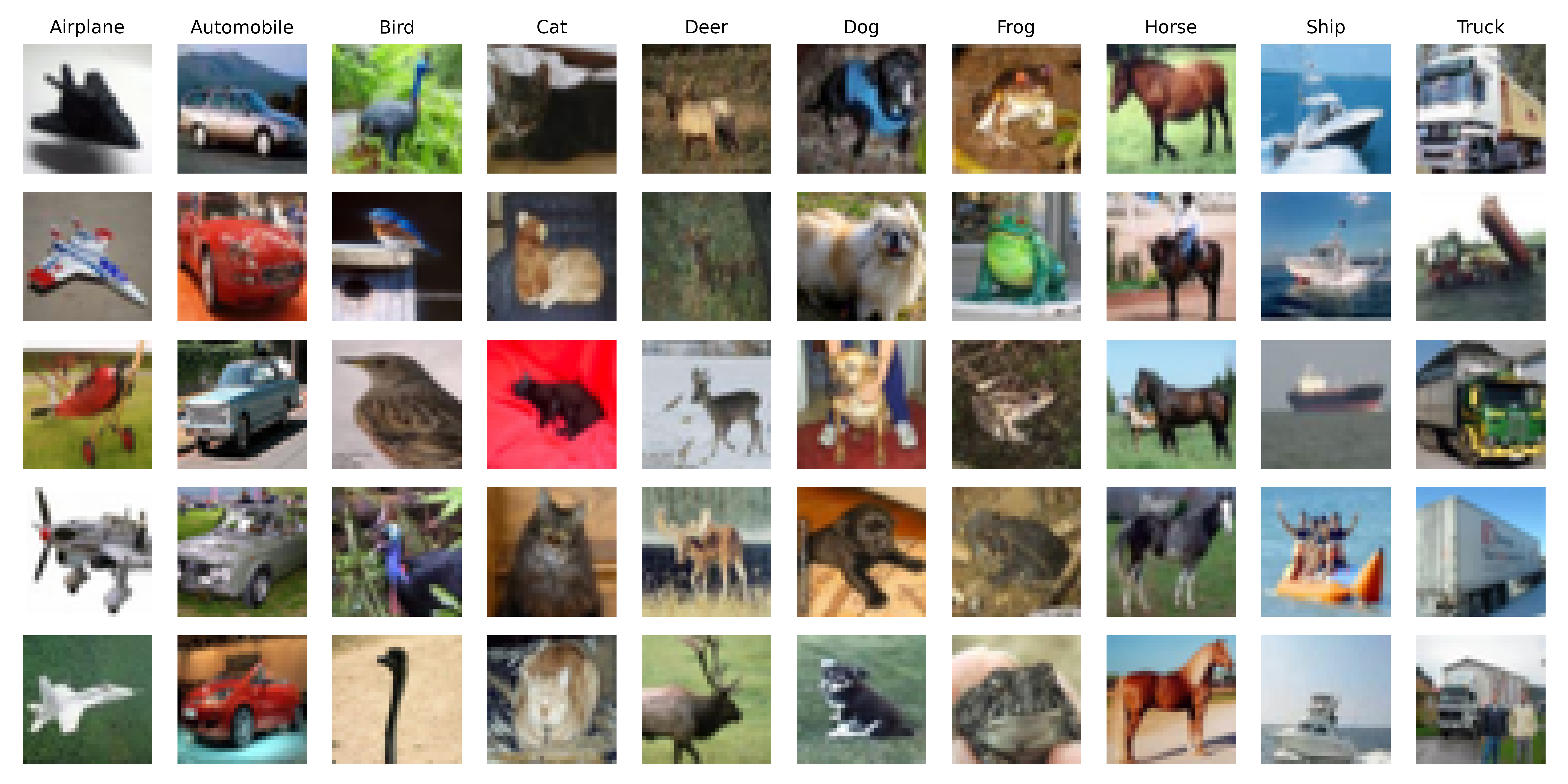}
    \caption{Examples of images of $10$ classes from CIFAR-10 dataset.}
\label{fig:CIFAR10_examples}
\end{figure}

Let us consider a more complicated problem: image classification for the \href{https://www.cs.toronto.edu/~kriz/cifar.html}{CIFAR-10} dataset, a popular benchmark dataset for computer vision tasks. It consists of {60,000 colour images}, each of size $32 \times 32$ pixels, divided into {10 classes}. The dataset is split into {50,000 training images} and {10,000 test images}, with each class containing {6,000 images}.
The classes in CIFAR-10 are:
\textit{Aeroplane, Auto-mobile, Bird, Cat, Deer, Dog, Frog, Horse, Ship}, and \textit{Truck}, see Fig.\ref{fig:CIFAR10_examples}
Each image is labelled with one of these classes, and the dataset has a balanced distribution of samples across all categories. The images presents much more complicated objects comparing to hand-written digits, and are coloured. Each image is a $32\times32\times3$ tensor, where each $32\times32$ matrix of pixels encodes colours: red, blue, and  green respectively, known as RGB colouring. 

We are interested in finding an architecture for the non-linear function  $32\times 32$ pixels and $3$ colour channels to a discrete probability density $f_{\vec{\theta}}:\vec{x}\in\mathbb{R}^{32\times 32\times 3}\to \vec{y}\in \mathbb{R}^{10}$, with such that $\sum_{k=1}^{K}\vec{y}_k=1$, with categorical cross-entropy loss function.
As a starting point, let us consider a simple architecture, similar to the last example with a single hidden layer with $21$ neurons.

\begin{figure}[t!]
    \centering
\includegraphics[width=\linewidth]{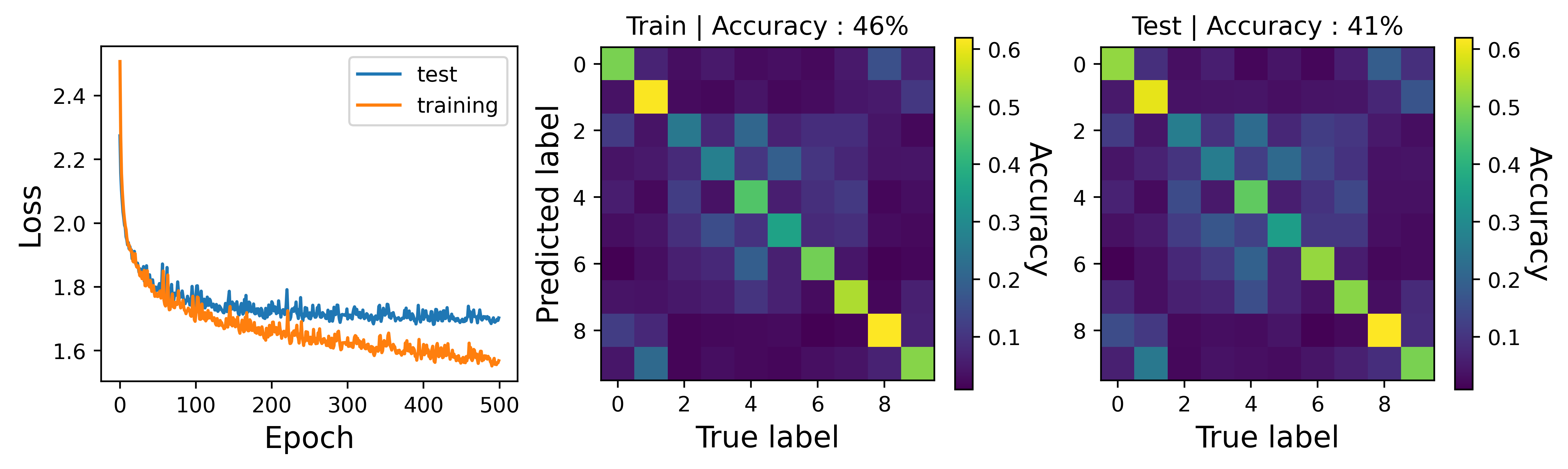}
    \caption{Simple neural network performance on CIFAR-10 dataset. Neural network has input layer with vectorized image of size $32^2\times 3$, single hidden layer with $N^{(1)}=21$ neurons, and output layer with $10$ neurons and softmax function.}
\label{fig:CIFAR10_simple_NN}
\end{figure}
Fig.\ref{fig:CIFAR10_simple_NN} presents change of loss function during training and confusion matrix. We see that prediction is much worse, below $50\%$ accuracy for both train and test dataset. Such a behaviour is known as \textit{underfitting}, which  occurs when a neural network is too simple to capture the underlying patterns in the data. As a result, the model performs poorly on both the training and test datasets because it fails to generalize or even fit the training data adequately.
The opposite to \textit{underfitting} is \textit{overfitting}, when when a neural network learns the training data too well, including its noise and irrelevant details, rather than capturing underlying patterns. This makes it generalize poorly to new data. As a result, the model performs well in the training dataset but poorly in the test datasets; see \hyperlink{box:underfitting_overfitting}{Box~2.12}. The solution to this problem is given by the architecture known as \textit{Convolutional Neural Networks} (CNNs) introduced by Yan Lecunn in 1989 \cite{Lecun1988}.

\begin{figure}[t!]
    \centering
    \begin{mybox}[\hypertarget{box:underfitting_overfitting}{Box 2.12: Overfitting vs Underfitting}]
       {\bf Overfitting} occurs when a model learns the training data too well, including noise and irrelevant details, leading to excellent performance on the training set but poor generalization to unseen data. It often arises from excessive model complexity, or small datasets. 

{\bf Underfitting} happens when a model is too simple to capture the underlying patterns in the data, resulting in poor performance on both the training and test sets. This typically stems from overly simplistic models, insufficient training time, or inadequate feature representation. 

Striking a balance between overfitting and underfitting ensures a model generalizes well to new data. 
    \end{mybox}
\end{figure}

\begin{figure}[t!]    \includegraphics[width=0.9\linewidth]{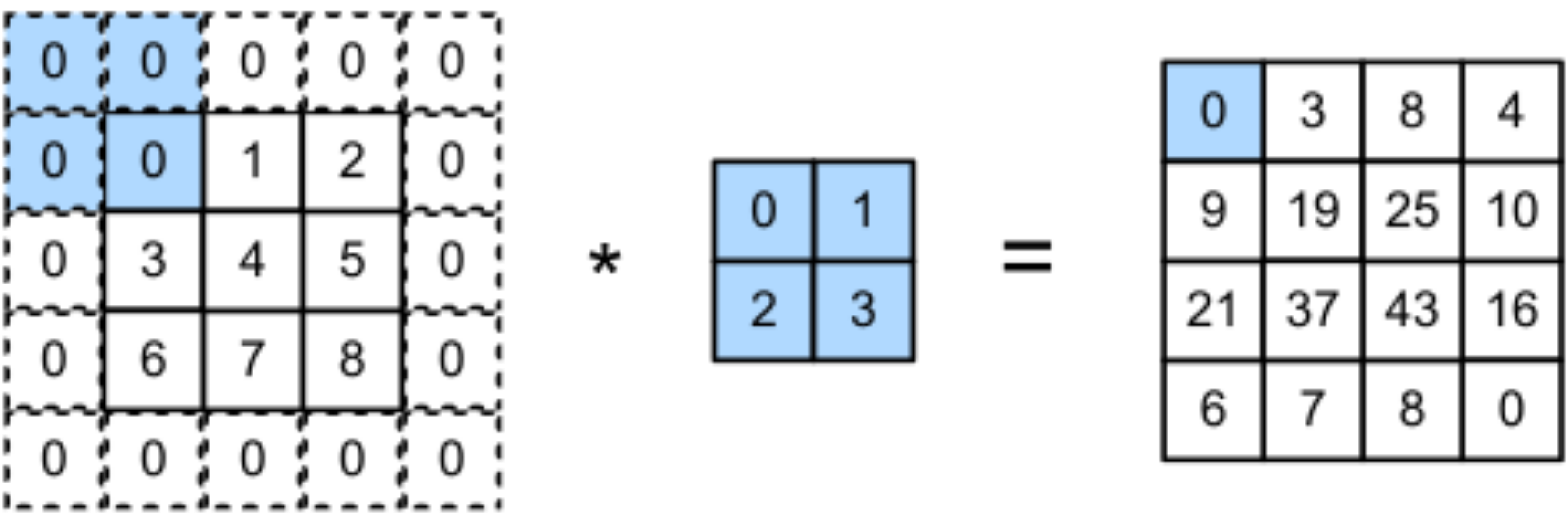}
    \caption{Example of convolution operation, Eq.\eqref{eq:convolution}. Kernel of size $2\times 2$ acts on an matrix of size $5\times 5$ producing the feature map of size $4\times 4$.}
    \label{fig:fig_Conv2d}
\end{figure}

CNNs are specialized neural networks designed for grid-like data, such as images. At the core of a CNN is the convolution operation, where a small, tuneable \textit{filter} (sometimes called a \textit{kernel}) slides over the input image and computes a dot product at each position. The  outcome of convolution operation of a given image $\vec{x}\in \mathbb{R}^{K\times K \times C}$, where $K$ is number of pixels, and $C$ number of channels, and given a kernel $\vec{G} \in \mathbb{R}^{k\times k}$ is called a \textit{feature map}, see Fig.\ref{fig:fig_Conv2d}
For an input image $\vec{x}$ and a kernel $\vec{G}$, the convolution in the $c$-th colour channel, is given by:
\begin{equation}\label{eq:convolution}
\vec{S}(i, j, c) = \sum_{m=1}^{k} \sum_{n=1}^{k} \vec{G}(m, n)\cdot\vec{x}(i - m, j-n, c) \equiv (\vec{G}*\vec{x})(i,j,c),
\end{equation}
where $\vec{S}(i, j, c)$, is the so-\textit{feature map} value at position $(i, j)$ and colour channel $c\in 1\dots C$, while $k$ is the filter size. The movement of the filter across the image is controlled by the \textit{stride}, which determines the step size.

\begin{figure}[t!]
    \includegraphics[width=0.99\linewidth]{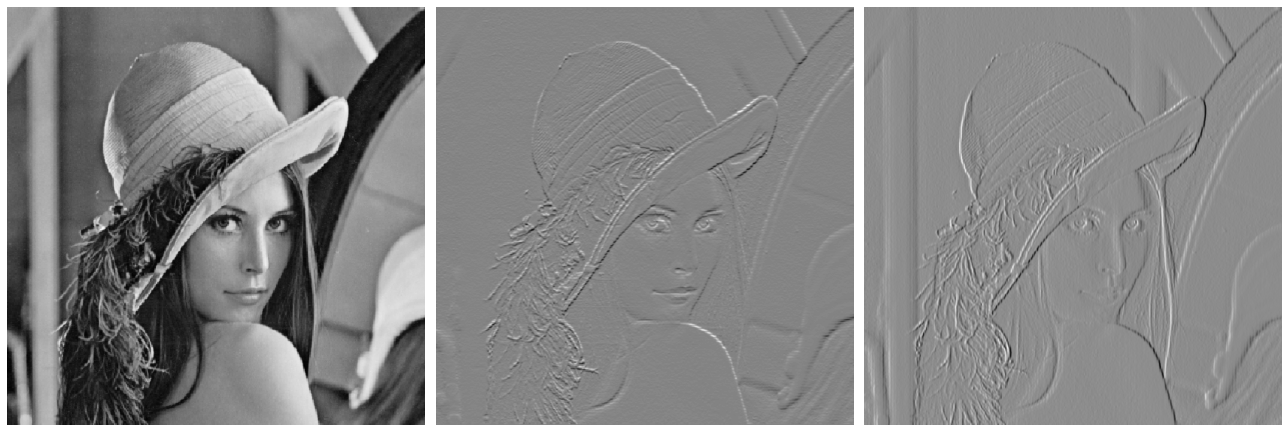}
    \caption{Action of Sobel filters for horizontal and vertical edge detection}
    \label{fig:fig_edge_detection}
\end{figure}

In image processing, kernels play a fundamental role in transforming and analysing images. Kernels are used to apply effects such as blurring, sharpening, or for edge detection. 
Let us consider problem of edge detection, which is done usually via \textit{Sobel filters}, i.e. 
pair of $3 \times 3$ convolution kernels used to approximate the gradient of the image intensity. They are designed to detect edges in both horizontal and vertical directions by highlighting regions with high spatial frequency changes. The Sobel kernel for horizontal edge detection, and vertical edge detection have form
\begin{equation}
\vec{G}_x = \begin{bmatrix}
-1 & 0 & +1 \\
-2 & 0 & +2 \\
-1 & 0 & +1 \\
\end{bmatrix},
\vec{G}_y = \begin{bmatrix}
+1 & +2 & +1 \\
0 & 0 & 0 \\
-1 & -2 & -1 \\
\end{bmatrix}.
\end{equation}
These kernels emphasizes changes in the vertical, and horizontal direction, respectively, effectively highlighting edges within the image, see Fig\ref{fig:fig_edge_detection}.

From the point of view on deep learning, the trainable parameters in convolutional neural networks are in fact kernels, which during training learn to extract key features of analysed image. Filters in the first layers of a CNN often detect simple features like edges by responding to changes in pixel intensity, such as vertical or horizontal gradients.
Each convolutional layer applies multiple filters, each capturing different features. For an input with $C_{\text{in}}$ channels (e.g., RGB images with three channels), the filters have dimensions $k \times k \times C_{\text{in}}$. Using $C_{\text{out}}$ filters results in an output with $C_{\text{out}}$ channels, providing a diverse set of feature maps.

Additionally to kernels, CNN contains pooling layers, which downsample feature maps, reducing spatial dimensions while retaining essential information. \textit{Max pooling} is a widely used method that selects the maximum value within a small region of the feature map. For instance, given a $2 \times 2$ pooling window, the following example illustrates max pooling:
\begin{equation}
    \text{Input:} 
\begin{bmatrix}
1 & 3 & 2 & 4 \\
5 & 6 & 7 & 8 \\
4 & 2 & 1 & 3 \\
9 & 8 & 6 & 5
\end{bmatrix}
\quad \xrightarrow{\text{Max Pooling}}
\quad 
\text{Output:}
\begin{bmatrix}
6 & 8 \\
9 & 6
\end{bmatrix}
\end{equation}
In this example, the $2 \times 2$ pooling operation with stride 2 selects the maximum value from each $2 \times 2$ sub-region of the input matrix, effectively reducing the size while preserving dominant features. Pooling layers improve computational efficiency and help make the network robust to small translations of the input.

CNNs learn to detect features: in the early layers, filters capture simple patterns such as edges, textures, or gradients. For example, an edge-detecting filter may respond strongly to regions with sharp changes in pixel intensity, allowing the network to identify object boundaries. As the network deepens, filters in later layers learn to detect more complex features, such as textures, shapes, and even specific objects.
Its hierarchical nature of CNNs, and ability to automatically learn spatial features, makes them great tools for capturing spatial correlations between pixels in images.

\begin{figure}[t!]
    \centering
    \includegraphics[width=\linewidth]{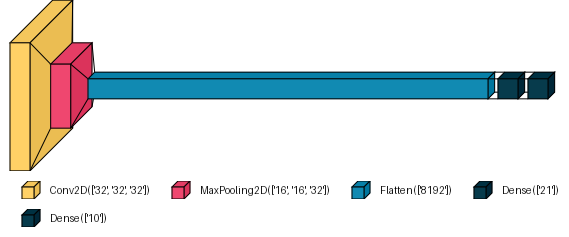}
\includegraphics[width=\linewidth]{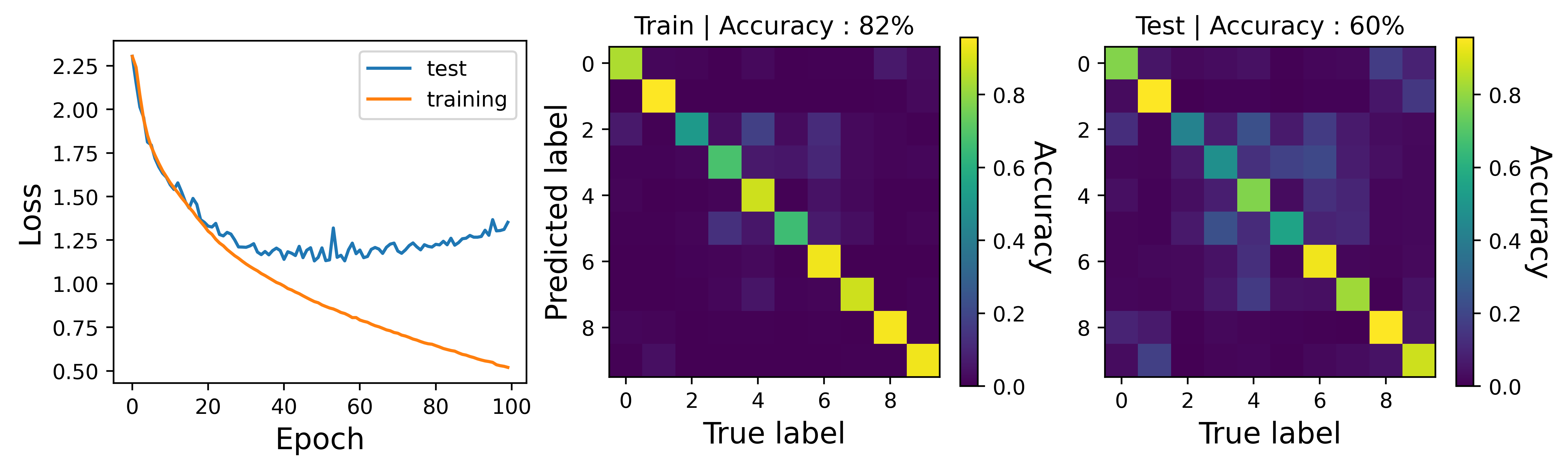}
    \caption{Simple Convolutional Neural Network Simple neural network performance on CIFAR-10 dataset. Neural network has input layer with shape $(32,32,3)$, followed by CNN layer with $32$ trainable kernels, each size $3\times 3$ with sigmoid activation function, with padding keeping input shape. Next layer is a MaxPooling of size $2\times 2$. Next, dense layer with $21$ nodes, sigmoid activation function, followed by output layer with $10$ nodes and softmax function. Example of \textit{overfitting}.}
\label{fig:CIFAR10_CNN_simple_NN}
\end{figure}

Fig.\ref{fig:CIFAR10_CNN_simple_NN} presents training and inference of simple CNN architecture.  Neural network has input layer with shape $(32,32,3)$, followed by CNN layer with $32$ trainable kernels, each size $3\times 3$ with sigmoid activation function, with padding keeping input shape. Next layer is a MaxPooling of size $2\times 2$. Next, dense layer with $21$ nodes, sigmoid activation function, followed by output layer with $10$ nodes and softmax function. As we can see, while the loss function on the training dataset has very nice decaying behaviour, the loss function vs epochs on test dataset starting $40$-th epoch increases. This behaviour is example of \textit{overfitting}, introduced earlier. Model performance on training dataset has $83\%$, which is quite good, however it has only $61\%$ on test dataset. Model learned features from the training dataset, however it was not able to generalize well, thus performance on the unknown data is much worse. 

\begin{figure}[t!]
    \centering

\includegraphics[width=0.8\linewidth]{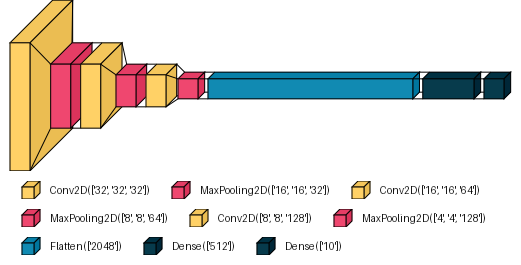}
\includegraphics[width=\linewidth]{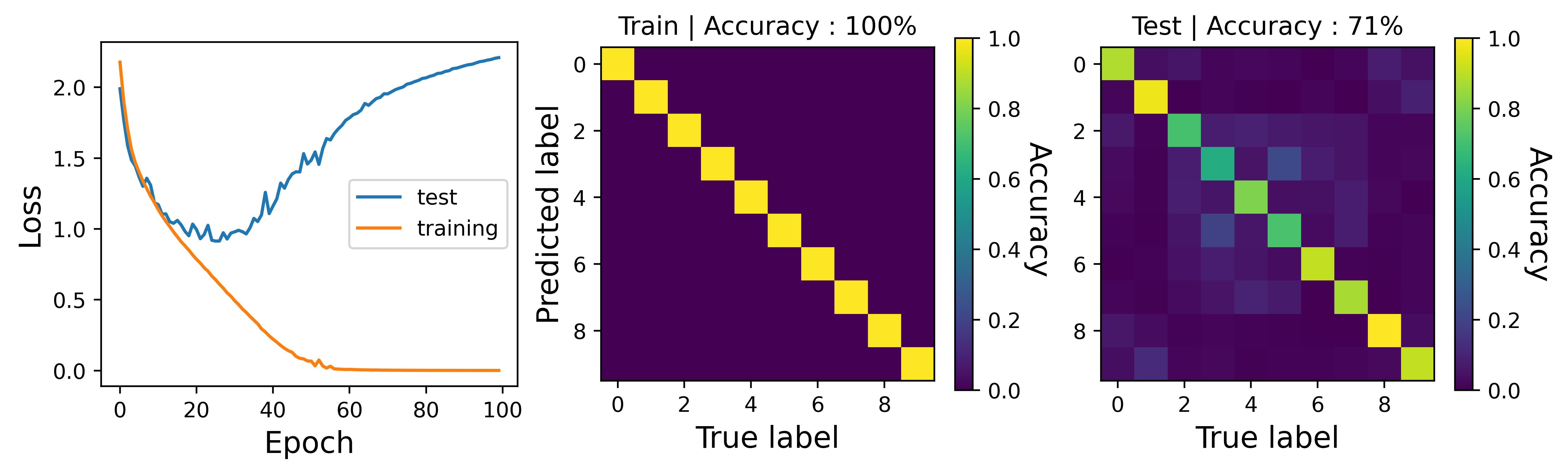}
    \caption{Convolutional Neural Network performance on CIFAR-10 dataset with relu activation function. Neural network has input layer with shape $(32,32,3)$, followed by three block of CNN layer with $32, 16, 8$ trainable kernels, each size $3\times 3$ with relu activation function. After each Conv2D layer, the maxpooling layer of size $2\times 2$ is applied. Next, dense layer with $512$ nodes, followed by output layer with $10$ nodes and softmax function. Example of \textit{overfitting}.}
\label{fig:1_8_CIFAR10_CNN_larger}
\end{figure}

Naturally we can think of increasing complexity of our architecture, increasing number of CNN blocks and changing activation function - hoping it will help with overfitting. Let us consider more complex NN: input layer with shape $(32,32,3)$, followed by three block of CNN layer with $32, 16, 8$ trainable kernels, each size $3\times 3$ with relu activation function. After each Conv2D layer, the maxpooling layer of size $2\times 2$ is applied. Next, dense layer with $512$ nodes, followed by output layer with $10$ nodes and softmax function. Fig.\ref{fig:1_8_CIFAR10_CNN_larger} shows example of overfitting - trained model has $99\%$ accuracy on the train dataset - however, despite increasing number of trainable parameters, overfitting is still visible. 
To overcome this difficulties, we should \textit{regularize} our architecture, which we will discuss in the next section.

\section[Training Neural Networks - Mathematical Structure]{Training Neural Networks \\ Mathematical Structure}
\label{sec:training_nn_mathematical}

The process of training a neural network is not a precise science, but rather a heuristic one. There is no single recipe or method which guarantees the best performance of a given architecture. This section covers a set of techniques which can help in training neural networks. Combining these techniques together forms the \textit{art} of training strategies. We will begin with regularization, then revisit backpropagation to understand how to perform \textit{meaningful} updates to a model. This will allow us to understand the concept of a computational graph, and the more sophisticated training strategies that underpin their optimisation, known as \textit{optimisers}.

\subsection{Neural Networks' Regularization}

Regularization is a key component in training neural networks, ensuring that models generalize well to unseen data and do not overfit to the training dataset. It involves techniques that constrain the learning process or model capacity to achieve better generalization. Below, we describe some of the most popular regularization techniques.

\paragraph{L1, L2 regularization} One of the simplest and most widely used regularization techniques is {L1 and L2 regularization}. These methods introduce penalty terms into the loss function to constrain the magnitude of model weights. L1 regularization adds a term proportional to the sum of the absolute values of the weights, leading to sparse solutions where some weights are driven to zero. On the other hand, L2 regularization (often referred to as weight decay) adds a term proportional to the sum of the squares of the weights. This discourages large weight values and promotes smoother models. These are mathematically expressed as:

\begin{equation}
\begin{split}
{\cal L}_{\text{regularized}} & = {\cal L}_{\text{original}} + \lambda \sum_i |\vec{\theta}_i| \quad \text{(L1)},\\
{\cal L}_{\text{regularized}} &= {\cal L}_{\text{original}} + \lambda \sum_i \vec{\theta}_i^2 \quad \text{(L2)},
\end{split}
\end{equation}
where $\lambda$ is the regularization strength and $\vec{\theta}_i$ are the model weights. This is nothing more than a \textit{Lagrange multiplier} for those that have done a little bit of statistical physics. It presents a \textit{soft} constraint, in the sense that the weights are \textit{encouraged} to take smaller values, but are not \textit{forced} to.

\paragraph{Dropout regularization} Dropout method randomly deactivates a fraction of neurons during training, forcing the network to learn redundant representations. By doing so, dropout reduces reliance on specific neurons and acts as an ensemble of sub-networks. During inference, all neurons are activated, but their outputs are scaled by the dropout rate used during training.

\paragraph{Early stopping} This technique is based on monitoring the performance of the model on a validation dataset during training. If the validation loss stops improving or begins to increase, training is halted to prevent overfitting. This technique is particularly useful when computational resources are limited.

\paragraph{Data augmentation} This technique is widely used  particularly in computer vision. By applying transformations such as rotations, translations, cropping, and colour adjustments to the training data, data augmentation increases the effective size of the training dataset. This prevents the model from overfitting by making it invariant to these transformations.

\paragraph{Learning rate scheduling} This is a dynamical regularization technique, where the learning rate is decreased as training progresses. It ensures that model converges to a smoother solution by taking smaller optimization steps later in training. This technique is implemented in optimization algorithms such as Adam, where the learning rate is adjusted for each parameter based on the observed gradients during training, allowing the network to converge more effectively without overfitting.

\begin{figure}[t!]
    \centering
\includegraphics[width=0.8\linewidth]{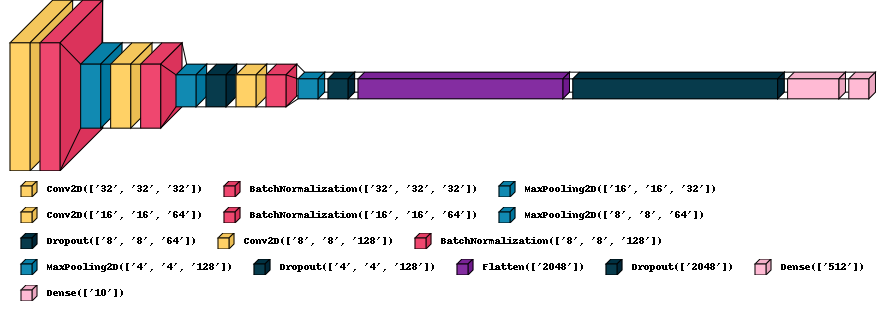}
\includegraphics[width=\linewidth]{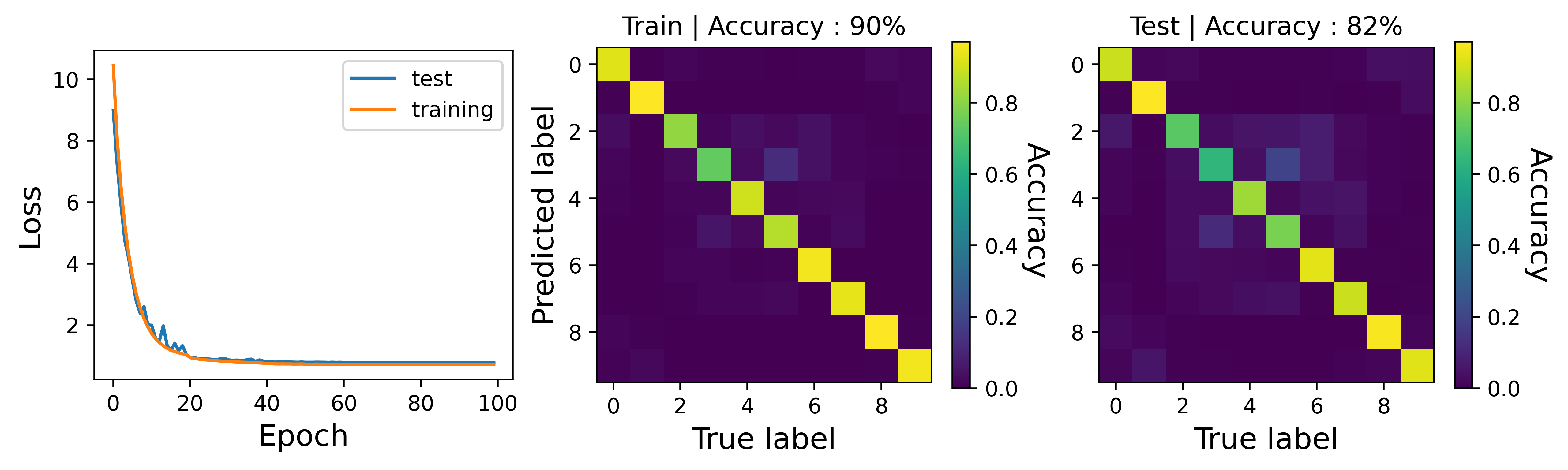}
    \caption{Regularized Convolutional Neural Network performance on CIFAR-10 dataset with ReLu activation function. Neural network has input layer with shape $(32,32,3)$, followed by three block of CNN layer with $32, 16, 8$ trainable kernels, each size $3\times 3$ with relu activation function. After each Conv2D layer, the maxpooling layer of size $2\times 2$ is applied. Next, dense layer with $512$ nodes, followed by output layer with $10$ nodes and softmax function.}
\label{fig:CIFAR10_CNN_larger_regularized}
\end{figure}

\paragraph{Batch Normalization (BatchNorm)} This technique normalizes the input to each layer in a neural network during training, helping to stabilize and accelerate training. This technique works as follows:
For a $i$-th mini-batch of inputs $\{\vec{x}_j\}_{j=1}^{N_{{\cal D}_i}}$ of size $N_{{\cal D}_i}$, mean and standard deviation is calculated, i.e. \begin{equation}
    \begin{split}
        \mu_B &= \frac{1}{N_{{\cal D}_i}} \sum_{j=1}^{N_{{\cal D}_i}} \vec{x}_j,\\ 
        \sigma_B^2 &= \frac{1}{N_{{\cal D}_i}} \sum_{j=1}^{N_{{\cal D}_i}} (\vec{x}_j - \mu_B)^2.
    \end{split}
\end{equation}
Next, each input is normalized via 
\begin{equation}
    \vec{x}_j \leftarrow \frac{\vec{x}_j - \mu_B}{\sqrt{\sigma_B^2 + \epsilon}},
\end{equation}
where $\epsilon$ is a small constant added for numerical stability. Finally, the input is linearly transformed as 
\begin{equation}
    \vec{x}_j \leftarrow \gamma \vec{x}_j + \beta,
\end{equation}
where $\gamma$ and $\beta$ are learnable parameters that allow the network to recover the original data distribution if necessary. BatchNorm  regularization method reduces the dependency of the network on specific initialization strategies. By normalizing the inputs to each layer BatchNorm reduces sensitivity to parameter changes and prevents exploding or vanishing gradients.

Equipped with regularization techniques, we can try to improve performance of our chosen NN architecture. Fig.\ref{fig:CIFAR10_CNN_larger_regularized} shows loss function vs epoch for model after implementing dropout, batch normalization and learning rate scheduler - as we can see, overfitting is removed. Loss function for training and test dataset has similar behaviour, and performance on the test dataset increased to $82\%$, while on training dataset is is reduced to $90\%$, comparing to Fig.\ref{fig:1_8_CIFAR10_CNN_larger}. Such trained model learned generalization well.

\subsection{Backpropagation Revisited}

As we have seen training any neural network is based on calculating derivative of the given loss function with respect to trainable parameters which are given in each layer. 
We can think of any neural network with $L$ layers as a many-level function composition
\begin{equation}
 f_{\vec{\theta}}(\vec{x}) = (h^{(L)}\circ h^{(L-1)} \circ \dots h^{(1)})(\vec{x}) = h^{(L)}(h^{(L-1)}(\cdots(h^{(1)}(\vec{x}))) = \vec{y},
\end{equation}
where $\vec{x}$ are inputs, while $\vec{y}$ are outputs (e.g. predicted labels), and $j = 0,\dots,L$ enumerates layers of the network.
The input data flow can be decomposed in the following steps:
\begin{equation}
\begin{split}
  h^{(0)} & \equiv \vec{x} \\
  h^{(1)} & = \sigma(\vec{W}^{(1)} h^{(0)} + \vec{b}^{(1)}) \\
  h^{(2)} & = \sigma(\vec{W}^{(2)} h^{(1)} + \vec{b}^{(2)}) \\
  \vdots\\
  h^{(j)} & = \sigma(\vec{W}^{(j)} h^{(j-1)} + \vec{b}^{(j)})\\
  \vdots\\
h^{(L-1)} & = \sigma_K(\vec{W}^{(L-1)} h^{(L-2)} + \vec{b}^{(L-1)}).
\end{split}
\end{equation}

Neural network training is done via an iterative process of passing the training data, and updating the weights according to gradient of the loss function with respect to the training parameters $\theta = \{W_0, b_0, W_1, b_1, \dots, A_{K-1}, b_k\}$. The gradient of the loss function with respect to the parameters set $\theta$ requires the partial derivatives with respect to each layer $h^{(j)}$ trainable parameters $\theta_j = \{W_j, b_j\}$, $j = 1,\dots, L$:
\begin{equation}
 \begin{split}
  \frac{\partial {\cal L}}{\partial \theta^{(L-1)}} & = \frac{\partial L}{\partial h^{(L)}}\color{black}{ \frac{\partial h^{(L)}}{\partial \theta^{(L-1)}}} \\
  \frac{\partial {\cal L}}{\partial \theta^{(K-2)}} & = \frac{\partial {\cal L}}{\partial h^{(L)}}\color{black}{ \frac{\partial h^{(K)}}{\partial h^{(L-1)}}}\color{black}{ \frac{\partial h^{(L-1)}}{\partial \theta^{(L-2)}}} \\
  \frac{\partial {\cal L}}{\partial \theta^{(L-3)}} & = \frac{\partial {\cal L}}{\partial h^{(L)}}\color{black}{ \frac{\partial h^{(L)}}{\partial h^{(L-1)}}}\color{black}{ \frac{\partial h^{(L-1)}}{\partial h^{(L-2)}}}\color{black}{ \frac{\partial h^{(L-2)}}{\partial \theta^{(L-3)}}} \\
  \frac{\partial {\cal L}}{\partial \theta^{(L-4)}} & = \frac{\partial {\cal L}}{\partial h^{(K)}}\color{black}{ \frac{\partial h^{(K)}}{\partial h^{(L-1)}}}\color{black}{ \frac{\partial h^{(L-1)}}{\partial h^{(L-2)}}}\color{black}{ \frac{\partial h^{(L-2)}}{\partial h^{(L-3)}}}\color{black}{ \frac{\partial h^{(L-3)}}{\partial \theta^{(L-4)}}} \\
  \vdots
 \end{split}
\end{equation}
The crucial element of the training process is an efficient way to calculate derivatives of the loss function with the respect of model parameters, which we discuss in the following.

Let us start by considering a function $f: \mathbb{R}^m \to \mathbb{R}$. We can distinguish three methods for calculating (partial) gradients $\nabla f = \big(\frac{\partial f}{\partial x_1},\cdots,\frac{\partial f}{\partial x_m}\big)$,
in a computer program.

\paragraph{Finite-difference method} is mainly used in numerical simulations. It is based on approximation the derivative of a function as 
\begin{equation}
 \frac{\partial f(\vec{x})}{\partial x_i} \sim \frac{ f(\vec{x} + h\vec{e}_i) - f(\vec{x})}{h},
\end{equation} 
where $\vec{e}_i \in \mathbb{R}^m$ is the $i$-th unit vector and $h$ is small step size.
However, approximating $\nabla f$ requires $\mathcal{O}(m)$ evaluations of f. Additionally round-off errors due 
to floating-point arithmetic dominate the errors as $h\to 0$.

\paragraph{Symbolic differentiation} is based on calculating derivatives analytically, though by automated manipulation of mathematical expressions allowing us to obtain explicit results (with help of computer algebra systems such us Mathematica, Maple, Maxima). For example for two given functions $f(x)$ and $g(x)$ derivative of their product is calculated explicitly following
\begin{equation}
\frac{d}{dx}(f(x)g(x)) = \frac{df(x)}{dx}g(x) + f(x)\frac{dg(x)}{dx}   
\end{equation}
The benefit of symbolic expression is that results are interpretable and allow us to find analytical solutions. However, symbolic derivatives generated through symbolic differentiation typically do not allow for efficient calculations of derivative values.

\paragraph{Automatic differentiation (AD)} This method is most important from the the deep learning point of view. AD allows to obtain exact numerical value of the derivatives without the
need of the symbolic expression - this method lies between symbolic differentiation and numerical differentiation. The logic lying behind AD is that all numerical computations are compositions of a finite set of elementary operations for which derivatives are known. By combining the derivatives of the constituent operations through the computational diagram, and applying the chain rule for derivatives, we can obtain exact numerical values of the derivative of the overall expression for a given input. This is the subject of the next section.

\subsection{Differentiable Computation Graphs}
A computational graph is a graph-based representation of a mathematical computation. It is a way of visually representing the operations performed in a computation, and the dependencies between these operations.

\begin{figure}[t!]
    \centering
    \includegraphics[width=\linewidth]{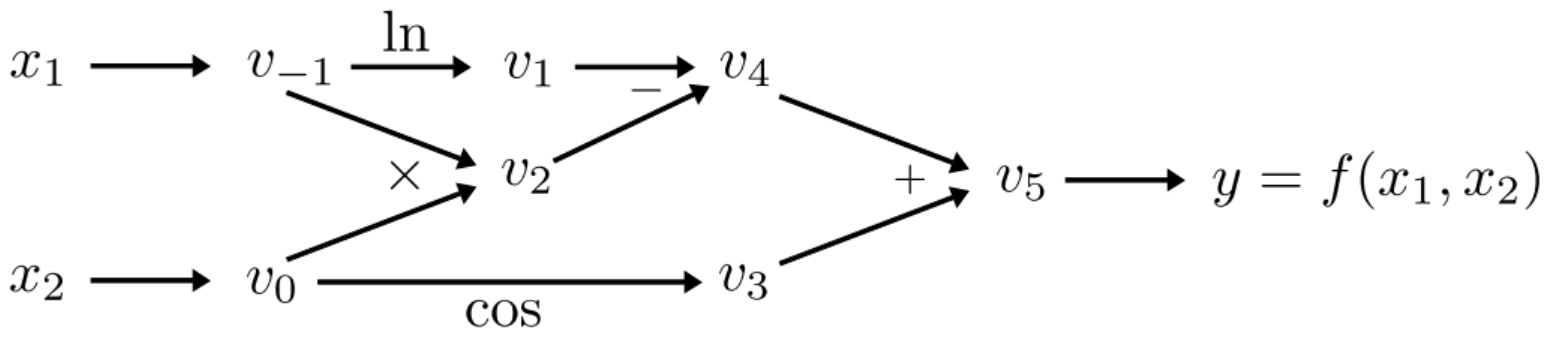}
    \caption{Computational graph representing two dimensional function $f(x_1,x_2) = \ln x_1 + \cos x_2 - x_1 x_2$.}
    \label{fig:computational_graph}
\end{figure}

Let us consider expression $f(x_1,x_2) = \ln x_1 + \cos x_2 - x_1 x_2$, 
Function $f$ is a mapping $f: \mathbb{R}^n \to \mathbb{R}^m$ with $n = 2$, $m=1$. Following \cite{dawid2022modern}, we introduce the notation for the computational graph as follows:
\begin{enumerate}
    \item Input variables are denoted as $v_{1-i}$, where $i = 1,\dots,n$.
    \item Intermediate variables are denoted as $v_i$, $i = 1,\dots,l$.
    \item Output variables are denoted as $v_{l+i}$, $i = 1,\dots,m$.
\end{enumerate}
The computational graph related to considered function $f(x_1,x_2)$ is presented in Fig.\ref{fig:computational_graph}
As an exercise, let us calculate value of $f(x_1,x_2)$ at $(x_1, x_2) = (2,1)$ via passing the diagram from left to right:
\begin{equation}
\begin{array}{|l|r|}
\hline
v_{-1}  = x_1  & = 2 \\
v_0  = x_2 & = 1 \\ \hline
v_1  = \ln v_{-1} & = \ln 2 \\
v_2  = v_{-1} v_0 & = 2 \\
v_3  = \cos v_0 & = \cos 1 \\
v_4  = v_1 - v_2  &= -1.307 \\
v_5  = v_3 + v_4  &= -0.767 \\ \hline\hline
v_5  = y  &= -0.767 \\ \hline\hline
\end{array}
\end{equation}

Automatic differentiation allows calculation of an exact value of the gradient at a given point. In our example, we are interested in value of $\frac{\partial f}{\partial x_1}$ at given point $(x_1, x_2) = (2,1)$. This can be obtain in two modes. 

\paragraph{The forward-mode AD} is implemented by complementing each intermediate variable $v_i$ with a derivative:
\begin{equation}
\dot{v}_i = \frac{\partial v_i}{\partial x_j},
\end{equation}
where $j=1\dots n$, when we are interested in derivative with respect to $x_j$. Next, by applying chain rule for differentiation we can obtain desired gradient.
Derivatives are propagated forward in sync with the function evaluation. As an exercise, let us calculate value of $\frac{\partial f}{\partial x_1} = \dot{v}_5$ at $(x_1, x_2) = (2,1)$ through passing the diagram:
\begin{equation}
\begin{array}{|l|r||l|r|}
\hline
v_{-1} = x_1  &= 2 & \dot{v}_{-1} = \dot{x}_1 & \quad = 1 \\
v_0 = x_2  &= 1 & \dot{v}_0 = \dot{x}_2 &\quad = 0 \\ \hline
v_1 = \ln v_{-1}  &= \ln 2 & \dot{v}_1 = \dot{v}_{-1}/v_{-1} &\quad = 1/2 \\
v_2 = v_{-1}v_0  &= 2 & \dot{v}_2 = \dot{v}_{-1}v_0 + v_{-1}\dot{v}_0 &\quad = 1 \\
v_3 = \cos v_0  &= \cos 1 & \dot{v}_3 = -v_0 \sin v_0 &\quad = 0 \\
v_4 = v_1 - v_2  &= -1.307 & \dot{v}_4 = \dot{v}_1 - \dot{v}_2 &\quad = -1/2 \\
v_5 = v_3 + v_4  &= -0.767 & \dot{v}_5 = \dot{v}_3 + \dot{v}_4 &\quad = -1/2 \\ \hline\hline
v_5 = y  &= -0.767 & \dot{v}_5 = \dot{y} &\quad = -1/2 \\ \hline\hline
\end{array}    
\end{equation}

In practice, forward-mode is implemented by extending the algebra of real numbers via introducing $\textit{dual}$ numbers, defined as
\begin{equation}
 \tilde{z}_1 = a_1 +\epsilon b_1,
\end{equation}
where $a,b \in \mathbb{R}$, and $\epsilon^2 = 0$. Next, addition and multiplication of dual numbers is defined as:
\begin{enumerate}
    \item Addition: $z_1 + z_2 = (a_1 + a_2) + \epsilon(b_1 + b_2)$
    \item Multiplication: $z_1z_2 = a_1a_2 + a_1b_2\epsilon +b_1a_2\epsilon + b_1b_2\epsilon^2 = a_1a_2 + \epsilon(a_1b_2+a_2b_1)$
\end{enumerate}
Next, when we consider Taylor series expansion around $\epsilon$, we have
  \begin{equation}
 f(z) = f(a+\epsilon) = f(a) + f'(a)\epsilon + \frac{1}{2}f''(a)\epsilon^2 + \dots,
 \end{equation}
 we see that this simplifies to
 \begin{equation}
 f(a+\epsilon) = f(a) + \epsilon f'(a),
 \end{equation}
 which means that operations on dual number $a$ automatically provides numerical value for $f(a)$ and derivative $f'(a)$.
In numerical implementation in libraries like PyTorch, dual numbers are handled by operator overloading where all mathematical operators are working appropriately on the new algebra of dual numbers.

\paragraph{Reverse-mode (backpropagation) AD} In a reverse mode we calculate gradients backwards. We are interested in calculating derivative of $y$ with respect to $v_i$, i.e. $\frac{\partial y_j}{\partial v_i} $. For a computational graph we can write the chain rule as
\begin{equation}
 \frac{\partial y_j}{\partial v_i} = \frac{\partial y_j}{\partial v_k}\frac{\partial v_k}{\partial v_i},
\end{equation}
where $v_k$ is a parent of a $v_i$ in a computational graph. When $v_i$ has more than one parent we sum up the chain rule as:
\begin{equation}
 \frac{\partial y_j}{\partial v_i} = \sum_{p\in parents(i)} \frac{\partial y_j}{\partial v_p}\frac{\partial v_p}{\partial v_i}.
\end{equation}
In the literature the above expression is called as $\textit{adjoint}$ and denoted as
\begin{equation}
\bar{v}_i \equiv \frac{\partial y_i}{\partial v_i}.
\end{equation}
Next, we can rewrite the adjoint in term of the adjoints of the parents, i.e.
\begin{equation}
 \bar{v}_i = \sum_{p\in \text{parents(i)}} \bar{v}_p \frac{\partial v_p}{\partial v_i}
\end{equation}
which gives us a recursive algorithm node $y$ with setting starting point as $\bar{y} = 1$. Let us write parents of each node in our example:
\begin{equation}
\begin{split}
 \text{parents}(i=5) &\to  y \\
 \text{parents}(i=4) &\to \{v_5\} \\
 \text{parents}(i=3) &\to \{v_5\} \\
 \text{parents}(i=2) &\to \{v_4\} \\
 \text{parents}(i=1) &\to \{v_4\} \\
 \text{parents}(i=0) &\to \{v_2, v_3\} \\
 \text{parents}(i=-1) &\to \{v_1, v_2\}\\
\end{split}
\end{equation}

Now we can write adjoints:
 \begin{equation}
 \begin{split}
  \bar{v}_5 & = \bar{y} \\
  \bar{v}_4 & = \bar{v}_5\frac{\partial v_5}{\partial v_4}\\
  \bar{v}_3 & = \bar{v}_5\frac{\partial v_5}{\partial v_3}\\
  \bar{v}_2 & = \bar{v}_4\frac{\partial v_4}{\partial v_2}\\
  \bar{v}_1 & = \bar{v}_4\frac{\partial v_4}{\partial v_1}\\
  \bar{v}_0 & = \bar{v}_2\frac{\partial v_2}{\partial v_0} + \bar{v}_3\frac{\partial v_3}{\partial v_0} \\
  \bar{v}_{-1} & = \bar{v}_1\frac{\partial v_1}{\partial v_{-1}} + \bar{v}_2\frac{\partial v_2}{\partial v_{-1}} \\
 \end{split}
 \end{equation}
Finally, we notice that
\begin{equation}
\begin{split}
 \bar{v}_0    & = \bar{x}_2 = \frac{\partial y}{\partial x_2}\\
 \bar{v}_{-1} & = \bar{x}_1 = \frac{\partial y}{\partial x_1}.
\end{split}
\end{equation}
As we can see, with a single backward pass we have both $\frac{\partial y}{\partial x_1}$ and $\frac{\partial y}{\partial x_2}$ (in forward mode we can obtain $\frac{\partial y}{\partial x_1}$ only in one pass). 

As an exercise, let us calculate the value of $\frac{\partial f}{\partial x_1} = \dot{v}_5$ at $(x_1, x_2) = (2,1)$ by passing the diagram.
\begin{equation}
    \begin{array}{|l|r||l|r|}
\hline
v_{-1} = x_1       & = 2         & \bar{v}_5 = \bar{y} \quad   & = 1 \\
v_0 = x_2          & = 1         &                             &        \\ \hline
v_1 = \ln v_{-1}   & = \ln 2 & \bar{v}_4 = \frac{\partial v_5}{\partial v_4} \bar{v}_5 &= 1 \\
v_2 = v_{-1} v_0   & = 2     & \bar{v}_3 = \frac{\partial v_5}{\partial v_3} \bar{v}_5 &= 1 \\
v_3 = \cos v_0     & = \cos 1  & \bar{v}_2 = \frac{\partial v_4}{\partial v_2} \bar{v}_4 &= -1 \\
v_4 = v_1 - v_2    & = -1.307 & \bar{v}_1 = \frac{\partial v_4}{\partial v_1} \bar{v}_4 &= 1 \\
v_5 = v_3 + v_4    & = -0.767 & \bar{v}_0 = \frac{\partial v_2}{\partial v_0}\bar{v}_2 + \frac{v_3}{v_0} \bar{v}_3 &= -2.841 \\
                   &       & \bar{v}_{-1} = \frac{\partial v_1}{\partial v_{-1}} \bar{v}_1 + 
                           \frac{\partial v_2}{\partial v_{-1}} \bar{v}_2 &= -1/2 \\ \hline\hline
v_5 = y  &= -0.767         & \bar{v}_{0} = \bar{x}_2 &= -2.841 \\
                   &       &  \bar{v}_{-1} = \bar{x}_2 &= -1/2 \\ \hline\hline
\end{array}
\end{equation}

An important question is how to choose between forward-mode and reverse-mode? The answer depends on the dimensionality of the function. Let us consider function $f:\mathbb{R}^m \to \mathbb{R}^n$, then if $m \ll n$, i.e. the number of inputs is much smaller than the number of outputs, from a computational point of view it is more favorable to use forward-mode automatic differentiation. On the other hand, if $m \gg n$, that is, the number of inputs is much larger than the number of outputs (and this is the case of neural networks), from a computational point of view, it is more favorable to use backward-mode automatic differentiation.

\subsection{Advanced Optimization algorithms}
\label{subsec:adv_optimisers}

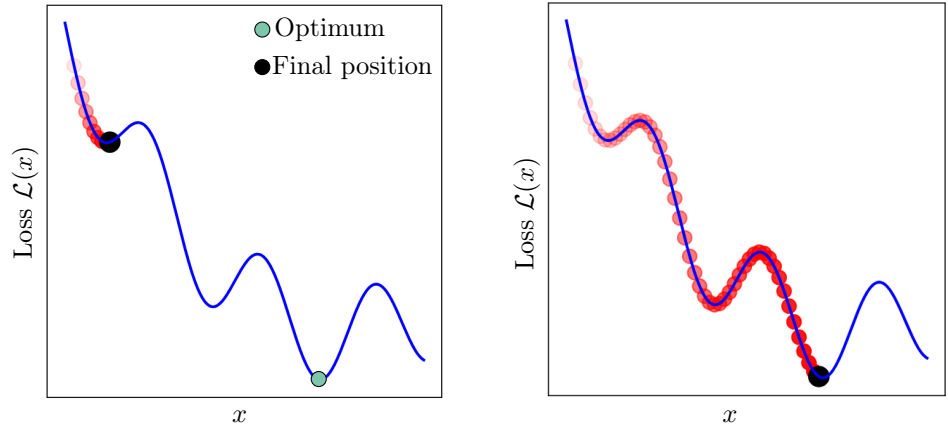
\begin{figure}[h]
    \centering
    \begin{subfigure}[b]{0.45\textwidth}
        \centering
        \begin{tikzpicture}
            \definecolor{customcolor}{HTML}{FCA082}
            \definecolor{customcolor2}{HTML}{E32F26}
            \definecolor{customcolor3}{HTML}{67000D}
            \definecolor{customcolor4}{HTML}{7CC7AC}
            \node [anchor=south west, inner sep=0] (image) at (0,0) {\includegraphics[width=\textwidth]{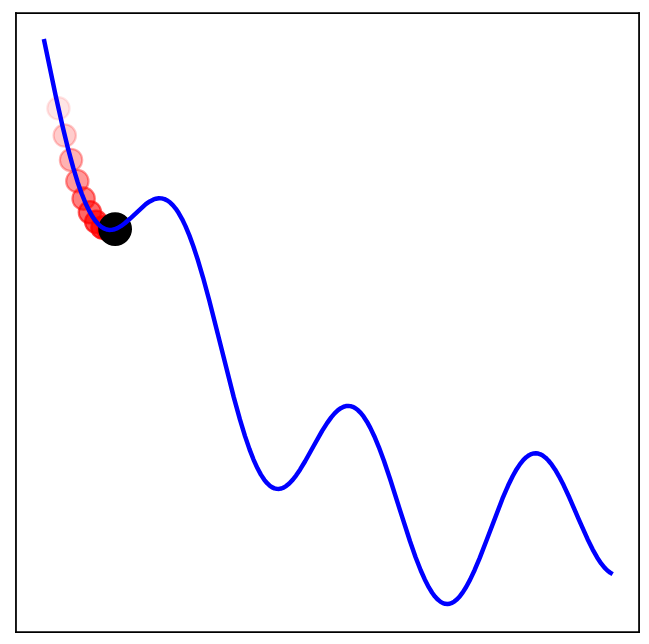}};
            \begin{scope}[shift={(0,0)}]
                \draw[fill=customcolor4, draw=black] (3.0,5.0) circle (0.1cm);
                \node at (3.9, 5.0) {Optimum};
                \draw[fill=black, draw=black] (3.0,4.5) circle (0.1cm);
                \node at (4.2, 4.5) {Final position};

                \node at (2.7, -0.15) {$x$};
                \node[rotate = 90] at (-0.15, 2.7) {Loss $\mathcal{L}(x)$};

                \draw[fill=customcolor4, draw=black] (3.75,0.35) circle (0.1cm);

            \end{scope}
        \end{tikzpicture}
        \label{fig:fig_Ch2_optimiser_vanilla}
    \end{subfigure}
    \hfill
    \begin{subfigure}[b]{0.45\textwidth}
        \centering
        \begin{tikzpicture}
            \node [anchor=south west, inner sep=0] (image) at (0,0.19) {\includegraphics[width=\textwidth]{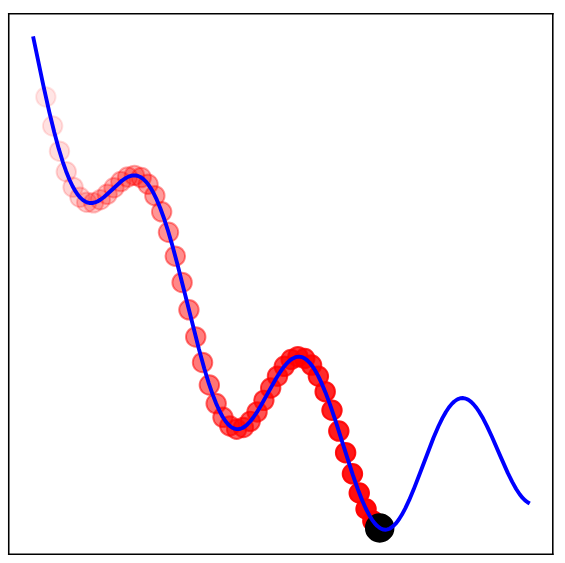}};
            \begin{scope}[shift={(0,0)}]
                \node at (2.5, 0.0) {$x$};
                \node[rotate = 90] at (-0.2, 2.7) {Loss $\mathcal{L}(x)$}; 
            \end{scope}
        \end{tikzpicture}
        \label{fig:fig_Ch2_optimiser_momentum_sub2}
    \end{subfigure}
    \caption{ Visual comparison between two gradient descent strategies. Steps are shown with increasing intensity from step 0 (most translucent, red) to final step (black, opaque).
    (left) A simple vanilla gradient descent strategy, where $x~\gets~x~-~\alpha~\nabla_x~\mathcal{L}(x)$. In this strategy, the step size, $\alpha \in \mathbb{R}$ is fixed, hence the constant \textit{horizontal} spacing between each step of gradient descent. With a fixed step size, and no other information beyond the local gradient at the current step of the optimiser, this simple strategy can get stuck in local mimuma with no way of getting out. This means the vanilla strategy often does not converge to the global optimum in its final position. The global optimum is indicated with a green circle. 
    (right) A more sophisticated gradient descent strategy which includes momentum. You can think of this as using the \textit{history} of past steps gathering momentum to carry the ball over each local maximum until it reaches the optimum. Other even more clever strategies exist as are discussed in the main text.}
    \label{fig:GD_advanced}
\end{figure}

In Sec.~\ref{CH_FUNDAMENTAls_sec:NN_training} we saw gradient descent as the foundation for neural network training methods. It formed the backbone of the backpropagation algorithm that allows us to iteratively tune the variational parameters toward an optimum $\vec{\theta}^*$ that minimizes a loss function $\mathcal{L}(\theta)$. We alluded to the fact that gradient descent and its closely related variant, stochastic gradient descent, were two examples of \textit{simple} training strategies. In this section, we will show how improvements can be made beyond SGD, leading us towards the current state-of-the-art optimization family: ADAM. To begin with, consider the visual comparison of two gradient descent strategies shown in Fig.~\ref{fig:GD_advanced}.

In stochastic gradient descent, we use mini-batches comprised of a few training samples, and the model's parameters are updated based on the average loss across the samples in each mini-batch. An improved version of SGD algorithm is SGD with \textit{momentum}, which is based on adding a fraction $\gamma$ of the previous parameter update to the current one, which helps the model make faster progress in the right direction and avoid getting stuck in local minima. This fraction is called the momentum coefficient, and it is a hyperparameter that can be adjusted according to the problem. The momentum algorithm accumulates a history of the past gradients and continues to move in their direction:
\begin{equation}
\begin{split}
 g_t &=  \frac{\partial {\cal L}(\vec{\theta}_{t-1})}{\partial \vec{\theta}_{t-1}}\\
 v_k &= \gamma v_{t-1} - \eta g_t \\
 \vec{\theta}_{t+1} &= \vec{\theta}_t + v_t,
 \end{split}
\end{equation}
where $t$ enumerates training epoch, $\theta$ are the trainable parameters of the Neural Network, $\gamma$ is the momentum coefficient and $\eta$ is the learning rate.
The velocity $v$ accumulates the gradient of the loss function ${\cal L}$; the larger $\gamma$ with respect to $\eta$, the more previous gradients affect the current direction.
In the standard SGD algorithm, the update size depended on the gradient and the learning rate. With momentum, it also depends on how large and how aligned consecutive gradients are.
In addition to speeding up training, momentum optimization can also help the model to generalize better to new data. 

The next improved optimization algorithm is called
\href{https://jmlr.org/papers/v12/duchi11a.html}{Adaptive Gradient algorithm}, which is based on the idea of adapting the learning rate to the parameters, performing larger updates for infrequent and smaller updates for frequent parameters. The Adagrad algorithm works by accumulating the squares of the gradients for each parameter, and then scaling the learning rate for each parameter by the inverse square root of this sum. This has the effect of reducing the learning rate for parameters that have been updated frequently, and increasing the learning rate for parameters that have been updated infrequently. The update rule for Adagrad  reads
\begin{equation}
 \begin{split}
 \vec{\theta}_{t+1} & = \vec{\theta}_t + \Delta\vec{\theta},\\
 \Delta \vec{\theta} &= - \frac{\eta}{\sqrt{diag( \epsilon\mathbb{1} + G_t )}} \odot g_t,\\
 g_t &= \frac{\partial {\cal L}(\vec{\theta}_{t-1})}{\partial \vec{\theta}}\\
 G_t &= \sum_{\tau = 1}^{t} g_\tau g_\tau^T.
 \end{split}
\end{equation}
where $\odot$ means element-wise multiplication. The $\epsilon \ll 0$ is a regularizing parameter, preventing from division by 0. Adagrad eliminates the need to manually tune the learning rate, i.e. initially $\eta \ll 1$, and it is effectively adapted during training process. Algorithm can be sensitive to the choice of the initial learning rate, and it may require careful tuning to achieve good results.

The next optimization algorithm is \href{https://arxiv.org/pdf/1212.5701.pdf}{Adadelta}, based on the idea of adapting the learning rate to the parameters, similar to Adagrad, but it does not require the specification of a learning rate as a hyperparameter. Adadelta uses an Exponentially Weighted Moving Average (EWMA) of the squared gradients to scale the learning rate. 
The Exponentially Weighted Moving Average (EWMA) for $x_t$ is defined recursively as:

\begin{equation}
 E[x]_t = \gamma E[x]_{t-1} + (1-\gamma) x_t
\end{equation}

In general Adadelta algorithm uses EMWA for $g_t^2$ instead  $G_t = \sum_{\tau = 1}^t g_\tau g_\tau^T$, as in Adagrad, i.e.:

\begin{equation}
 E[g^2]_t = \gamma E[g^2]_{t-1} + (1-\gamma)g_t^2,
\end{equation}
and we update parameters as
\begin{equation}
\begin{split}
 \vec{\theta}_{t+1} & = \vec{\theta}_t + \Delta\vec{\theta}_t \\
 \Delta\vec{\theta}_t & = - \frac{\eta}{\sqrt{E[g^2]_t + \epsilon}}\odot g_t.
\end{split}
\end{equation}

However, authors of Adadelta algorithm noticed that the parameter updates $\Delta\theta$ being applied to $\theta$ should have matching units. Considering, that the parameter had some hypothetical units $[\theta]$, the changes to the parameter should be changes in those units as well, i.e. $[\theta] = [\Delta\theta]$. However, assuming the loss function is unitless, we have $[\Delta\theta] = \frac{1}{[\theta]}$, thus units do not match. This is the case for SGD, Momentum, or Adagrad algorithms.

The second order methods such as Newton’s method that use the second derivative information preserve units for the parameter updates. For function $f(x)$, we have
\begin{equation}
 \Delta x = \frac{\frac{\partial f}{\partial x}}{\frac{\partial^2 f}{\partial x^2}},
\end{equation}
thus units $[\Delta x] = [x]$ are preserved. Keeping this in mind, the update rule in Adadelta algorithm is re-defined as:
\begin{equation}
 \begin{split}
  \vec{\theta}_{t+1} & = \vec{\theta}_t + \Delta\vec{\theta}_t \\
  \Delta\vec{\theta}_t &= -\frac{RMS[\Delta\theta]_{t-1}}{RMS[g]_t}\odot g_t,
 \end{split}
\end{equation}
where  $RMS$ stands for root-mean-square, $RMS[g]_t = \sqrt{E[g^2]_t + \epsilon}$.

This has the effect of automatically adapting the learning rate to the characteristics of the problem, which can make it easier to use than other optimization algorithms that require manual tuning of the learning rate.

Finally, we comment on 
\href{https://arxiv.org/pdf/1412.6980.pdf}{Adaptive Moment Estimation - Adam}, which combines the ideas of momentum optimization and Adagrad to make more stable updates and achieve faster convergence. Like momentum optimization, Adam uses an exponentially decaying average of the previous gradients to determine the direction of the update. This helps the model to make faster progress in the right direction and avoid oscillations. Like AdaGrad, Adam also scales the learning rate for each parameter based on the inverse square root of an exponentially decaying average of the squared gradients. This has the effect of reducing the learning rate for parameters that have been updated frequently, and increasing the learning rate for parameters that have been updated infrequently. Adam uses Exponentially Modified Moving Average for gradients and its square:
\begin{equation}
\begin{split}
 g_t &= \frac{\partial {\cal L}(\vec{\theta}_{t-1})}{\partial \vec{\theta}_{t-1}}\\
 m_t &= \beta_1 m_{t-1} + (1-\beta_1)g_t \\
 v_t &= \beta_2 v_{t-1} + (1-\beta_2)g_t^2.
\end{split}
\end{equation}
The update rule for the parameters reads:
\begin{equation}
\Delta \vec{\theta}_{t+1} = \vec{\theta}_t - \frac{\eta}{\sqrt{\bar{v}_t} + \epsilon}\bar{m}_t,
\end{equation}
where
\begin{equation}
\begin{split}
\bar{m}_t &= \frac{m_t}{1-\beta_1}\\
\bar{v}_t &= \frac{v_t}{1-\beta_2},
\end{split}
\end{equation}
are bias-corrected first and second gradient moments estimates. Authors suggest to set $\beta_1 = 0.9$, $\beta_2 = 0.999$, $\eta = 10^{-8}$.

\subsubsection{Loss function sensitivity}

The stability of training and convergence depends on factors like the sensitivity of the loss function and the behaviour of gradients during backpropagation. These issues can significantly affect the model's ability to learn effectively, especially in deep architectures.

Loss function sensitivity problem arises when small changes in model parameters cause large fluctuations in the loss function, making optimization unstable. This is particularly problematic when the loss function has steep or flat regions. For example, in regions where gradients are large, the optimizer might overshoot the minimum, while in flat regions, training becomes slow. Proper normalisation of the input data, such that it has zero mean and unit variance, can help reduce this sensitivity. Additionally, choosing loss functions suited to the task, such as mean squared error for regression or cross-entropy for classification, ensures smooth optimization. A practical approach is to apply learning rate scheduling, where the learning rate decreases over time, making the optimization process more stable.

\subsubsection[Vanishing Gradients]{Vanishing Gradients - the Achilles Heel of Deep Learning}
Exploding gradients occur when the gradients grow excessively large as they propagate backward through the network. This issue often arises in deep networks due to repeated multiplication of large weights. The gradients can become so large that they overflow, causing numerical instability. One common solution is \textit{gradient clipping}, where the gradient vector $g$ is rescaled to ensure its magnitude remains below a threshold. Mathematically, this is expressed as:
\begin{equation}
g = \frac{g}{\max(1, \|g\| / \text{threshold})}.
\end{equation}
 Reducing the learning rate also helps control the size of updates and prevents gradients from exploding.

Vanishing gradients, on the other hand, occur when gradients become exponentially smaller as they are propagated backward through the network. This happens in activation functions like sigmoid or tanh, where the gradients diminish in magnitude as the input moves away from zero. As a result, the earlier layers in the network receive negligible updates, slowing down training. A commonly used solution is the ReLU (Rectified Linear Unit) activation function, defined as:
\begin{equation}
f(x) = \max(0, x).
\end{equation}
ReLU avoids saturation for positive inputs, ensuring that gradients do not vanish. Additionally, batch normalisation can be used to normalize the inputs to each layer during training, stabilizing activations and gradients. Architectures like residual networks (ResNets) incorporate skip connections, allowing gradients to bypass certain layers. This helps maintain the flow of gradients through the network, as shown by:
\begin{equation}
y = F(x) + x,
\end{equation}
where $F(x)$ is the transformation applied by the layer.

Both exploding and vanishing gradients are especially problematic in very deep networks. A unified approach involves using adaptive optimizers like Adam, which adjust the learning rate dynamically based on the gradient magnitudes. Monitoring gradient magnitudes during training can also help detect and address these issues early. Techniques like layer normalisation or weight regularization can further stabilize the optimization process.

\subsubsection[Initializing Weights]{Initializing Weights of a Neural Network}

Proper weight initialization is crucial in neural networks to ensure stable and efficient training. Poor initialization can lead to exploding or vanishing gradients, causing training to become unstable or excessively slow. The act of initialising a neural network's weights \textit{intelligently} is known a \textit{warm start}. This sits in contrast with random strategies sometimes referred to as \textit{cold stats}. This subsection gives a brief overview for warm start techniques.

One commonly used technique is {Xavier initialization}, which is suited for activation functions like sigmoid or tanh. In Xavier initialization, weights are drawn from a distribution with zero mean and variance $\frac{2}{n_{\text{in}} + n_{\text{out}}}$, where $n_{\text{in}}$ and $n_{\text{out}}$ are the number of input and output neurons, respectively. Another approach is {He initialization}, designed for ReLU activation functions, which scales the variance to $\frac{2}{n_{\text{in}}}$. These methods ensure that the variance of activations remains consistent across layers, preventing gradients from exploding or vanishing. Biases are typically initialized to zero to allow the network to learn from the data distribution without introducing additional bias. Uniform and normal distributions can also be used, but the choice of initialization depends on the activation function and architecture of the network. Proper initialization accelerates convergence, stabilizes training, and improves the overall performance of the model.

\section{Deep Learning in the Classical 2D Ising Model}

Now, let us use deep learning to study our first physical model. We will consider a system of classical spins on a regular lattice at finite temperature. Specifically, our goal will be to pinpoint critical temperature $T_c$ in the two-dimensional Ising model. We will begin with introducing two-dimensional Ising model, then proceed to understand what techniques can be employed for the numerical (computational) study of the Ising model. This will allow us to construct a neural network model that can estimate the critical temperature of a two-dimensional Ising model. To se how deep learning can be employed to find the critical temperature, we will first recall the key ideas underpinning the classical Ising model. We will then briefly describe an analytical solution for the critical temperature. This will provide us with two things. First, it gives us a \textit{ground truth}; something to which we can compare a neural network model. Second, it will allow us to appreciate the role of neural networks in physics research in Sec.~\ref{sec:chinese_thought_experiment}. We emphasise here that you don't need to fully understand \textit{how} Onsager found his solution. Instead, we wish to highlight that he found one, and we can compare this solution to neural network estimates.

\subsection{Classical Ising Model}
The Ising model, proposed by Ernst Ising in 1925, is one of the most fundamental models in statistical mechanics describing classically  ferromagnets, giving insights into phase transitions and critical phenomena \cite{Huang2009, Honig2018-lp}.

The model is described on a regular lattice of size $L$ in $d$ dimensions ($d = 1, 2, 3$).  At each lattice site, there is a classical spin variable $s_i$ that can take values $+1$ (spin up) or $-1$ (spin down). The Hamiltonian of the system, i.e. energy of a given spin configuration $\{s_i\}$, is described by the Hamiltonian:
\begin{equation}\label{eq:Ising_model}
H(\{s_i\}) = -J \sum_{\langle i,j \rangle} s_i s_j - h \sum_i s_i,
\end{equation}
where $J$ is the exchange interaction constant, and its positive value $J > 0$ favours alignment of neighbouring spins (ferromagnetic interaction). In the following, we set $J = 1$. The notation $\langle i,j \rangle$ indicates that the sum runs over all pairs of nearest neighbours, while $h$ describes strength the external magnetic field. In the following we set $h = 0$, focusing on the spontaneous magnetization arising from interactions alone. The full information about thermodynamical properties is encoded in the partition function $Z = \sum_{\{s_i\}} e^{-H / T}$,
where, the sum runs over all possible spin configurations, and $e^{-H / T}$ is the Boltzmann factor weighting each configuration according to its energy $H$ and temperature $T$ (we set the Boltzmann constant $k_B = 1$). Let us denote the spin configurations at given temperature $T$ as $\{s_i\}$, where $i$ index each lattice site.

Historically, the model was initially suggested by Wilhelm Lenz, Ising's doctoral advisor, as a simplified framework to study ferromagnetism. Ising solved the one-dimensional (1D) version of the model and found that it does not exhibit a phase transition at finite temperatures, leading him to conclude (incorrectly) that the model could not explain ferromagnetism. In 1944, Lars Onsager provided an exact solution for the two-dimensional (2D) Ising model without an external magnetic field, revealing a phase transition at a finite critical temperature~$T_c$.

The Ising model captures the essence of ferromagnetic interactions, where spins tend to align due to exchange interactions. In two-dimensional geometry, the model provides a paradigmatic example of a system undergoing a second-order (continuous) phase transition, characterized by divergent correlation lengths and critical exponents. The critical behaviour near $T_c$ is universal, meaning it is independent of microscopic details and depends only on general features like dimensionality and symmetry.

\begin{figure}[t!]
\includegraphics[width=\linewidth]{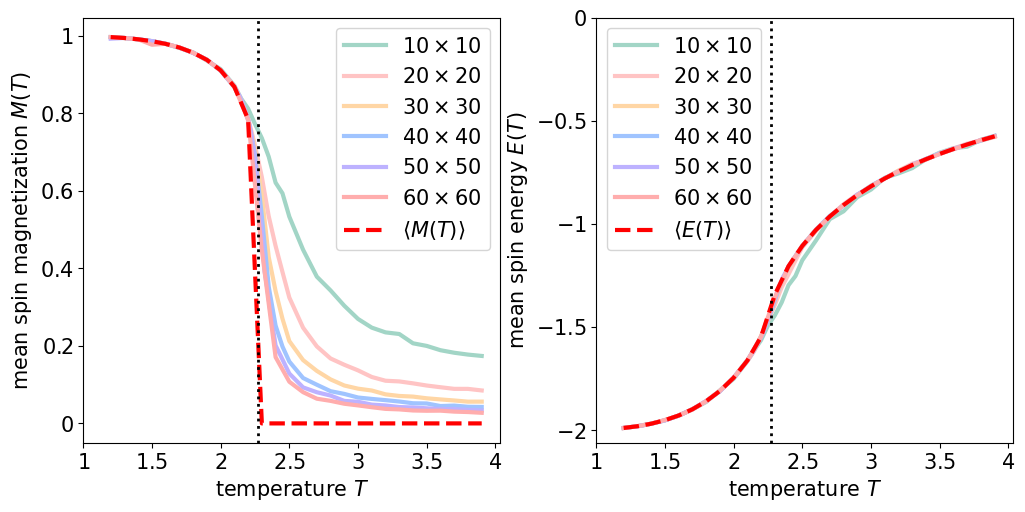}
    \caption{Mean magnetization pers spin, and mean energy per spin, averaged over $1000$ spin configurations obtained via the metropolis algrotihm. Solid lines presents the average, while shaded areas present standard deviation. Dashed red lines present Onsager's analytical solution for thermal averages, Eq.\eqref{eq:Magnetization_analytical} and Eq.\eqref{eq:Energy_analytical}. Vertical dotted line indicates critical temperature $T_c = \frac{2}{\ln(1+\sqrt{2})}\approx 2.269$}
    \label{fig:magnetization_vs_T}
\end{figure}

Important concept in physics of many-body system is the order parameter, i.e. quantity which distinguishes different phases of the physical system, based on some observable. In two-dimensional Ising model, for the given spin configuration $\{s_i\}$, the order parameter  is given by the mean magnetization, which measures the net alignment of spins in the lattice:
\begin{equation}
M(\{s_i\}) = \frac{1}{N} |\sum_i s_i|,
\end{equation}
where  $N = L^d$ is number of spins on a $d$ dimensional regular lattice. The magnetization $M$ distinguishes between ordered ($M \neq 0$) and disordered ($M = 0$) phases. It varies with temperature, dropping to zero at the critical temperature $T_c$, see Fig.\ref{fig:magnetization_vs_T}. The other important observable is the mean energy per spin
\begin{equation}
E(\{s_i\}) = \frac{1}{N} \sum_{\langle i,j\rangle}s_i s_j.
\end{equation}

Onsager provided an exact solution for the thermal average of magnetization $\langle M(T)\rangle$:
\begin{equation}
\langle M \rangle = \frac{1}{Z} \sum_{\{s_i\}} M(\{s_i\}) \, e^{-\beta H(\{s_i\})}
\end{equation}
and thermal average of energy $\langle E(T)\rangle$
\begin{equation}
\langle E \rangle = \frac{1}{Z} \sum_{\{s_i\}} E(\{s_i\}) \, e^{-\beta H(\{s_i\})}
\end{equation}
where the sums run over all possible spin configurations $\{s_i\}$. He indicated the existence of the critical temperature $T_c$
\begin{equation}
T_c = \frac{2}{\ln(1 + \sqrt{2})}
\end{equation}
at which thermal fluctuations overcome the spin alignment tendency, leading to a phase transition from an ordered to a disordered state.

At temperatures $T<T_c$, the magnetization $M(T)$ is non-zero, i.e. most of spins are aligned in the same direction. For $T>T_c$  magnetization vanishes, $M(T)=0$, because spins are randomly pointing up or down. For temperatures below the critical temperature ($T < T_c$), the spontaneous magnetization has analytical form:
\begin{equation}\label{eq:Magnetization_analytical}
\langle M(T)\rangle = \left[1 - \left( \sinh\left( \frac{2}{T} \right) \right)^{-4} \right]^{1/8}
\end{equation}

The exact expression for the average energy per spin as a function of temperature reads
\begin{equation}\label{eq:Energy_analytical}
\langle E(T)\rangle = -\, \coth\left( \frac{2}{T} \right) \left[1 + \frac{2}{\pi} \left( \frac{2 \tanh^2\left( \frac{2}{T} \right) - 1}{\tanh\left( \frac{2}{T} \right)} \right) K\left( k \right) \right]
\end{equation}
where, $K(k)$ is the complete elliptic integral of the first kind
defined as:
\begin{equation}
K(k) = \int_{0}^{\pi/2} \frac{d\phi}{\sqrt{1 - k^2 \sin^2 \phi}},
\end{equation}
while the $k$ is defined as:
\begin{equation}
k = \frac{2 \sinh\left( \frac{2}{T} \right)}{\cosh^2\left( \frac{2}{T} \right)}
\end{equation}

For temperatures well below $T_c$,  the energy per spin approaches:
$E(T) \approx -2$, while at high temperatures, $T \gg T_c$, the energy per spin tends toward zero $E(T) \approx 0$. The specific heat $C(T) = \frac{d\langle E(T)\rangle}{dT}$ shows a logarithmic divergence at $T_c$, consistent with the nature of the second-order phase transition.

The two-dimensional Ising model near critical temperature $T_c$ has critical behaviour, i.e. power law dependence on $T$. In particular, near $T_c$, the magnetization exhibits critical behaviour, following the relation $M(T) \sim (T_c - T)^\beta$ with a critical exponent $\beta = 1/8$. There are two more physical quantities that characterize the critical behaviour near the phase transition: the magnetic susceptibility $\chi$ and the correlation length $\xi$.

The magnetic susceptibility $\chi$, measures how the magnetization of the system responds to an external magnetic field. Even in the absence of an external field, the susceptibility can be defined in terms of the fluctuations of the total magnetization
\begin{equation}
\chi = \frac{1}{N T} \left( \langle M^2 \rangle - \langle M \rangle^2 \right).
\end{equation}
Near the critical temperature $T_c$, the susceptibility diverges as:
\begin{equation}
\chi \propto |T - T_c|^{-\gamma}
\end{equation}
where the $\gamma = \frac{7}{4}$ is a critical exponent.

The correlation length $\xi$, characterizes how quickly spin correlations decay with distance in the system, i.e. the typical size over which spins are correlated. The spin-spin correlation function $G(r)$ quantifies how the spins at two sites separated by a distance $r$ are correlated:
\begin{equation}
G(r) = \frac{1}{N} \sum_{i} \left[ \langle s_i s_{i+r} \rangle - \langle s_i \rangle \langle s_{i+r} \rangle \right],
\end{equation}
where $s_{i+r}$ denotes the spin at a distance $r$ from site $i$.
At large distances and away from $T_c$, the correlation function decays exponentially $G(r) \propto e^{-r / \xi(T)}$, where $\xi(T)$ is a temperature dependent correlation length. Near the critical temperature, the correlation length diverges as:
\begin{equation}
\xi \propto |T - T_c|^{-\nu}
\end{equation}
For the two-dimensional Ising model, the critical exponent is $\nu = 1$

\subsubsection{Onsager's solution}

Let us start be recalling properties of on a central concept of the statistical mechanics, the \textit{partition function}. The partition function encodes the essential thermodynamic properties of a system in thermal equilibrium. For a given Hamiltonian $H$ it is defined as:
\begin{equation}
Z = \sum_{\{s_i\}} e^{-\beta H},
\end{equation}
where the sum is over all possible spin configurations $\{s_i\}$, and $\beta = \frac{1}{T}$ is the inverse temperature. 
The partition function $Z$ allows calculation thermodynamic observables such as the mean energy, its variance, the mean magnetization, and its variance. 
The mean energy $\langle E \rangle$ is given by:
\begin{equation}
\langle E \rangle = -\frac{\partial \ln Z}{\partial \beta}.
\end{equation}
The energy variance $\mathrm{Var}(E)$, which characterizes energy fluctuations, is expressed as:
\begin{equation}
\mathrm{Var}(E) = \langle E^2 \rangle - \langle E \rangle^2 = \frac{\partial^2 \ln Z}{\partial \beta^2} - \left( \frac{\partial \ln Z}{\partial \beta} \right)^2.
\end{equation}

The mean magnetization $\langle M \rangle$, measuring the average alignment of spins, is:
\begin{equation}
\langle M \rangle = \frac{1}{\beta} \frac{\partial \ln Z}{\partial h}.
\end{equation}
The variance of the magnetization $\mathrm{Var}(M)$, which quantifies magnetization fluctuations, is:
\begin{equation}
\mathrm{Var}(M) = \langle M^2 \rangle - \langle M \rangle^2 = \frac{\partial^2 \ln Z}{\partial h^2} - \left( \frac{\partial \ln Z}{\partial h} \right)^2.
\end{equation}

The central result of the Onsager's solution is the explicit expression of the partition function at zero magnetic field $h=0$. From this solution, all of these physically observable properties can be calculated. Onsager showed that in the limit $L\to \infty$, the logarithm of the partition function reads
\begin{equation}
    \ln Z = \ln(2\cosh(2\beta J)) + \frac{1}{2\pi} \int_{0}^{\pi}d\phi \ln\left(\frac{1}{2}(1+\sqrt{1-\kappa^2\sin^2\phi}\right),
\end{equation}
where $\kappa = \frac{2\sinh(2\beta J)}{\cosh^2(2\beta J)}$. From here it followed that specific heat capacity
\begin{equation}
    C=\beta^2\frac{\partial^2\ln Z}{\partial \beta^2},
\end{equation}
diverges logarithmically at $T\to T_c$.

The importance of the Onsager's result lies in the fact that spins can spontaneously form an ordered phase at vanishing magnetic field $h=0$ below their critical temperature. For details and historical background of the model we refer to \cite{bhattacharjee1995ising,Baxter2011}.

\subsection{Numerical Studies of the Ising Model}

Simulating the Ising model numerically is based on the Metropolis algorithm, which is a stochastic process that generates a sequence of spin configurations according to the Boltzmann distribution. This ensures that the system evolves toward thermal equilibrium.

\begin{algorithm}[H]
\caption{Metropolis Algorithm for the 2D Ising Model}
\begin{algorithmic}[1]
\State Initialize a $L \times L$ lattice with random spins $s_{i,j} = \pm 1$
\State Set temperature $T$
\For{each Monte Carlo step (MCS)}
    \For{each lattice site $(i, j)$}
        \State Calculate energy change:\\$\Delta E = 2 s_{i,j} \left( s_{i+1,j} + s_{i-1,j} + s_{i,j+1} + s_{i,j-1} \right)$
        \If{$\Delta E \leq 0$}
            \State Flip the spin: $s_{i,j} \leftarrow -s_{i,j}$
        \Else
            \State Generate a random number $r \in [0,1)$
            \If{$r < \exp(-\Delta E / T)$}
                \State Flip the spin: $s_{i,j} \leftarrow -s_{i,j}$
            \EndIf
        \EndIf
    \EndFor
    \If{Equilibration is reached}
        \State Measure observables (e.g., magnetization $M$, energy $E$)
    \EndIf
\EndFor
\end{algorithmic}
\end{algorithm}

\begin{figure}
 \centering
\includegraphics[width=0.9\linewidth]{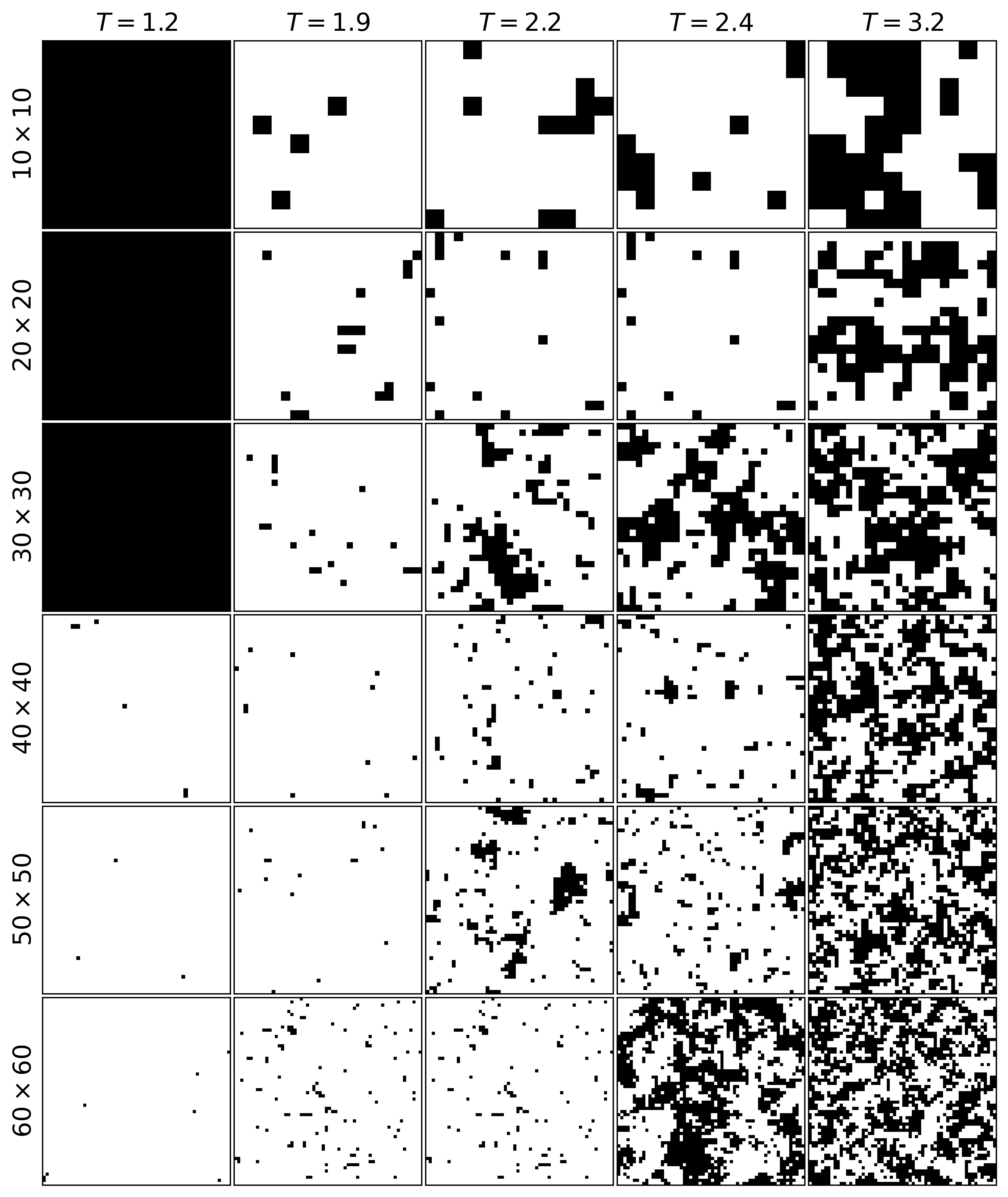}
 \caption{Example of spins configurations on square lattice $L\times L$,  $L = 5, 10, 20, 30, 40, 50$ at  different temperatures $T$. Spins configurations in the third, and fourth column, at temperatures in the vicinity of $T_c \approx 2.269$ are hardly distinguishable, however belong to different phases of the two-dimensional Ising model. At high temperatures, $T>3$ spins configurations have random structure with vanishing net magnetization.}
\label{fig:Ising_spin_configuration}
\end{figure}

In this algorithm, we start with a random spin configuration to avoid bias. One Monte Carlo step (MCS) involves attempting to update all spins in the lattice. For each spin, we calculate the change in energy $\Delta E$ if the spin were to be flipped. The change in energy due to flipping a single spin $S_{i,j}$ is given explicitly by considering its interactions with its four nearest neighbours. This is because the Ising model assumes nearest neighbour interactions only, and the 2D geometry of a square lattice has four neighbours. If flipping lowers the energy ($\Delta E \leq 0$), we accept the flip. If $\Delta E > 0$, we accept the flip with probability $\exp(-\Delta E / T)$, using the sampled outcome $r$ to decide whether to accept energetically unfavorable flips. After the system reaches equilibrium, we measure the physical quantities to compute the averages.
Fig.\ref{fig:Ising_spin_configuration} presents examples of spin configurations for square lattices of different sizes $L\times L$ at temperatures $T = 1.2, 1.9, 2.2, 2.4, 3.2$. Below critical temperatures almost all spins are aligned giving rise to finite magnetization in the ordered phase, whereas above critical temperature $T_c \approx 2.269$ system is in disordered phase with vanishing magnetization.

Now we can define our task: we seek to train a neural network model in such a way that it can provide a numerical approximation, $\bar{T}_c$, for the critical temperature, $T_c$. However, before starting, one might ask a valid question: \textit{"Why do we need to implement a neural network to find the critical temperature, when we can simply examine the net magnetization versus temperature and pinpoint \(T_c\) from there?"} To answer this, we need to refer to the Ginzburg-Landau theory of phase transitions and the concept of the \textit{order parameter}, i.e. a quantity describing the state of a system. It measures the degree of order in the system and typically changes its value (often from zero to non-zero) as the system undergoes a transition between two phases. For a given physical system, the existence of a phase transition is monitored by a properly constructed order parameter.

In the case of the classical two-dimensional Ising model, the order parameter is simply the average spin magnetization. However, in general, defining the correct order parameter is one of the most challenging steps in model analysis.  There are very few physical systems for which analytically its order parameter is known. It is here that deep learning can be a great tool to uncover new behaviour in physical systems, around which more theoretical, or structure predictions can be made. In this sense, deep learning serves as a tool to reveal that \textit{something is happening} around a specific temperature, even if we do not know \textit{exactly what} is occurring. In the following, we will demonstrate how neural networks can help capture the dependence of spin configurations on temperature, without referring to the concept of an order parameter.

\subsection{Estimating the Critical Temperature with Neural Networks}

In the following, we will explore the confidence-based approach to find estimation $\bar{T}_c$ for the critical temperature $T_c$ of the two-dimensional Ising model. This method leverages the intrinsic uncertainty of a classifier trained to distinguish between ordered and disordered phases, and we will analyse its prediction confidence across temperatures. The critical temperature $\bar{T}_c$ is identified as the point where the classifier exhibits the greatest uncertainty in its predictions.

Our dataset consists of spin configurations represented as $L \times L$ grids, each corresponding to a specific temperature $T$. To ensure effective training, we divide the configurations into two distinct classes: ordered ($T < T_{\text{min}}$, labelled as $y=0$) and disordered ($T > T_{\text{max}}$, labelled as $y=1$). Configurations near the expected critical temperature $T_c$ are excluded from the training process. Specifically, we omit configurations within the range $T_{\text{min}} \leq T \leq T_{\text{max}}$, as these are likely to contain ambiguous features due to critical fluctuations. By excluding this region, we allow the model to learn clear distinctions between the ordered and disordered phases.

We employ a Convolutional Neural Network (CNN) to classify these configurations into ordered and disordered phases. We consider simple architecture with two stacks of convolutional layers with $32$ kernels of size $3\times 3$, and maxpooling layer, followed by fully connected dense layer with $64$ nodes with ReLU activation function, while the output layer has $2$ nodes and softmax function. We consider categorical cross-entropy loss function, and use ADAM optimizer for training. Our model is trained on configurations from well-separated temperature ranges, focusing on identifying the spatial patterns characteristic of each phase. Once the model is trained, we evaluate its performance on configurations across the entire temperature range, including those from the critical region.

\begin{figure}  
\includegraphics[width=0.99\linewidth]{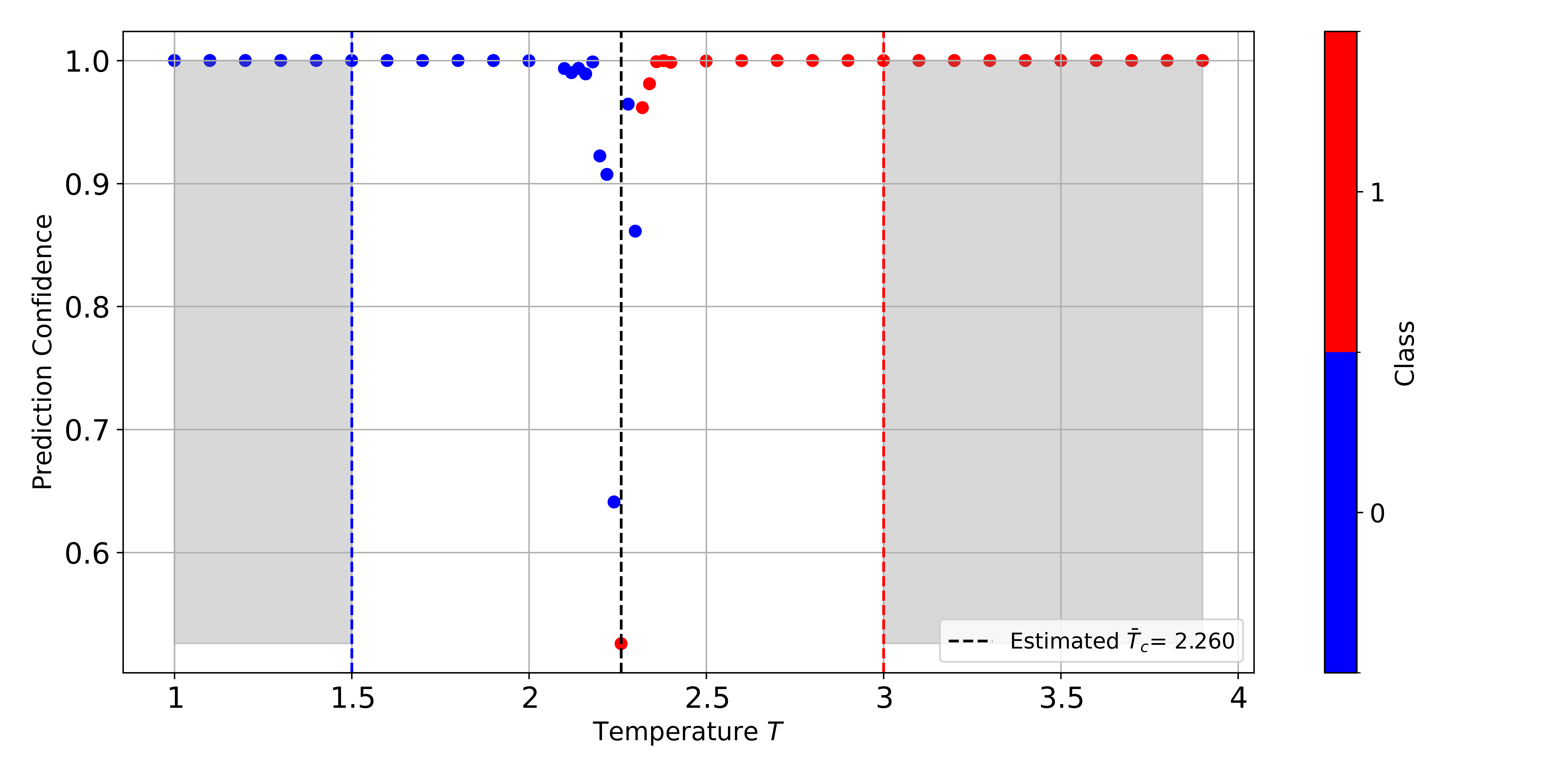}
    \caption{Class prediction confidence vs temperature $T$ for $L\times L$ lattice, $L = 50$. For temperatures $T$ near the critical temperature, the confidence of class predictions for classifier trained on data with  $T<T_{\rm min} = 1.5$, and $T>T_{\rm max} = 3$ (shaded areas) significantly drops. Temperature for the lowest confidence prediction is our estimation for critical temperature in the two dimensional classical Ising model, $\bar{T}_c = 2.26 \approx T_c \approx 2.269$  }    \label{fig:fig_Ising_prediction_confidence}
\end{figure}

To analyse the classifier's behaviour, we examine its prediction confidence, which is defined as the softmax probability associated with the predicted class. Far from $T_c$, where the phases are well-separated, the model’s confidence is high, as classification is relatively straightforward. However, as the temperature approaches $T_c$, critical fluctuations blur the distinction between phases, causing a significant drop in confidence. By plotting prediction confidence against temperature, we can identify $\bar{T}_c \approx T_c$ as the temperature where the model’s confidence is minimized, see Fig.\ref{fig:fig_Ising_prediction_confidence}

This confidence-based approach is firmly grounded in the physical properties of the Ising model.  This ambiguity is naturally reflected in the classifier’s reduced confidence when tasked with classifying configurations in the critical region.  Near $T_c$, the model exhibits critical behaviour, marked by large fluctuations and the coexistence of ordered and disordered features. The order parameter, which measures the net magnetization, approaches zero, blurring the distinction between ordered and disordered phases. Similarly, the divergence of the correlation length causes spin configurations to appear similar over large spatial scales, further complicating classification. The drop in prediction confidence near $T_c$ reflects the intrinsic ambiguity in phase classification arising from critical fluctuations. 

Our estimation is not free from finite size effects. The lattice size $L$ plays a significant role in the performance of the confidence-based method. Larger lattices exhibit sharper phase transitions, as the thermodynamic limit is approached, making it easier to pinpoint $T_c$ from the confidence curve. For smaller lattices, finite-size effects smooth out the transition, potentially broadening the confidence dip near $T_c$. By analysing multiple lattice sizes and extrapolating to the thermodynamic limit, we can refine our estimate of $\bar{T}_c$ and study how critical behaviour depends on system size.

While the current approach classifies configurations into two phases (ordered and disordered), it can be extended to multi-class classification by dividing the temperature range into multiple bins. Each class would correspond to a specific temperature interval, allowing the model to predict not only the phase but also the approximate temperature. This approach could provide a more detailed view of the phase transition and how the system evolves as temperature changes.

 \section[Understanding Deep Learning - Chinese Room Experiment]{Understanding Deep Learning through the Chinese Room Thought Experiment}
 \label{sec:chinese_thought_experiment}

As we have seen, neural networks and deep learning can decipher complex patterns in data. This was a consequences of Universal Approximation Theorem. Inspired by the human brain's structure, these networks learn from vast amounts of data through training processes. Even a simple neural network has the capability to learn how to recognize hand written digits in the MNIST dataset, or even much more complicated objects from CIFAR10 dataset. In the last section, we even managed to use deep learning to predict the critical temperature of the two-dimensional classical Ising model (to a good accuracy too!). 

We might therefore be tempted to say that indeed, neural networks were capable of \textit{discovering} this critical temperature. Can we therefore conclude that we understood the physics of the Ising model by employing a neural network in this way? Was any real insight into criticality and phase transitions gathered here? More generally, this raises the question of whether neural networks \textit{understand} the processes they model. In this section, we will consider a simple thought experiment to demonstrate that the answer is \textit{no}; deep learning does not provide an understanding of the physical process it was used to model. In this sense, deep learning is simply a tool amongst others in physics; it is useful but yet replaces human understanding and insight at the time of writing these notes.

\begin{figure}[t!]

    \includegraphics[width=0.9\linewidth]{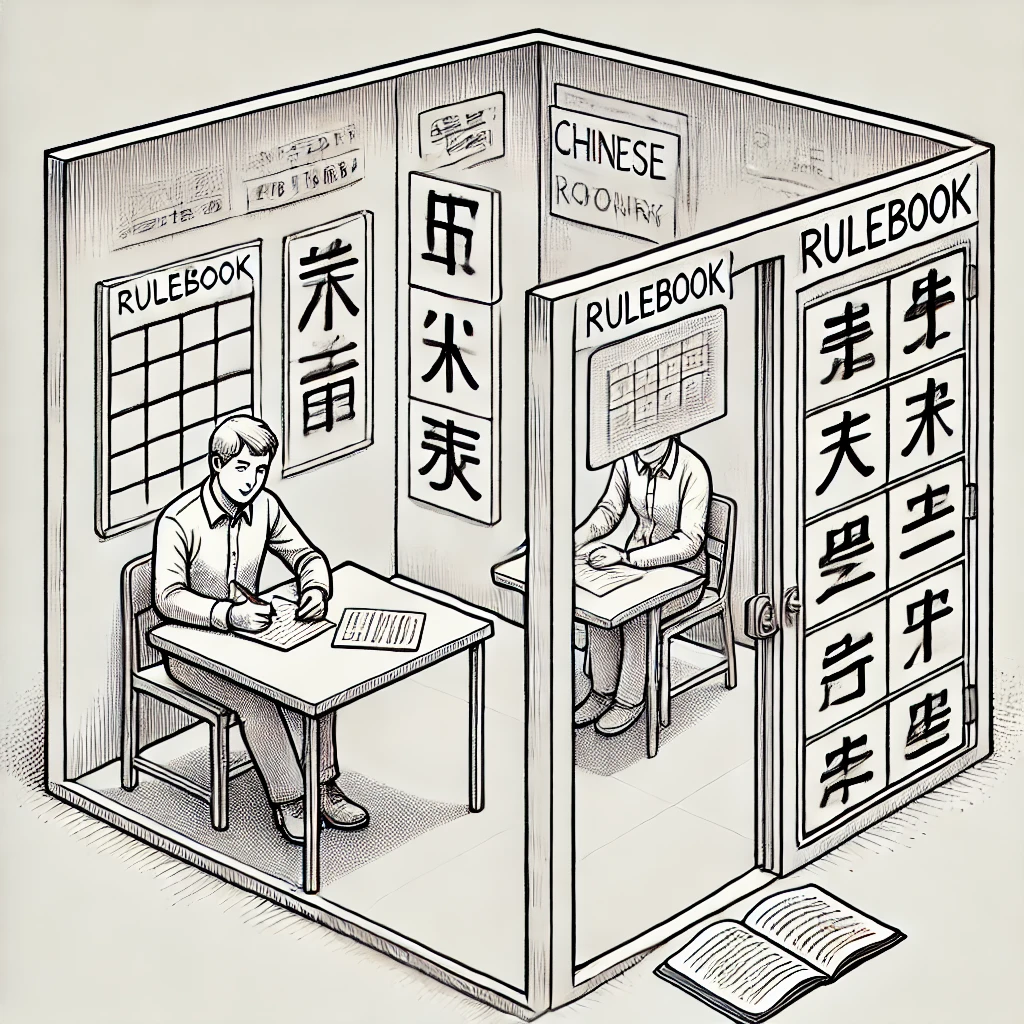}
    \caption{Inside a sealed Chinese Room, a person who speaks no Chinese receives slips of paper covered in Chinese characters. Armed only with an enormous rule-book that lists symbol-to-symbol transformation procedures, the person painstakingly looks up each pattern, copies the matching response, and pushes the new slip back through the slot. From the outside, native speakers see coherent answers and naturally assume the room “understands” Chinese. Yet everything inside is mechanical symbol shuffling, devoid of meaning. The illustration captures John Searle’s point: executing perfect syntactic rules can mimic intelligence without producing genuine semantic comprehension.}
    \label{fig:chinese_room}
\end{figure}

The question about understanding via data, can be summarised by the Chinese Room Thought Experiment \cite{Searle1980}, first proposed by John Searle in 1980. The Chinese Room Thought Experiment challenges the notion that a computer running a program can possess a "mind" or "understand" language merely by processing symbols. It provides a compelling framework for examining the nature of deep learning and its capabilities, and goes as follows...

Let us imagine a person who does not understand Chinese sitting inside a room. This person receives Chinese characters through a slot and uses a rulebook (written in a language they understand, such as English) to manipulate these symbols, see Fig~\ref{fig:chinese_room}. Based on these rules, the person sends out appropriate Chinese characters as responses. To an external observer, it appears as though the room understands Chinese, even though the person inside does not comprehend the language. Searle argues that this scenario illustrates that syntactic manipulation of symbols (the basis of how computers process information) is insufficient for semantic understanding or consciousness. That being said, deep learning is a rapidly evolving field, and much attention is currently being given (at the time of writing) to protocols which allow \textit{semantic interpretability} of a model.

Drawing parallels between deep learning and the Chinese Room highlights intriguing similarities and distinctions. Both deep learning models and the person in the Chinese room engage in symbol manipulation without inherent understanding. Deep learning systems process inputs like images produce outputs such as classifications or regression that may appear intelligent to observers. Similarly, the Chinese room uses a rulebook to process symbols without any comprehension of their meaning. Both rely on rule-based processing: the Chinese room follows explicit rules, while deep learning models rely on learned weights and activation patterns to process information.
However, there are more significant differences. The Chinese room exemplifies a lack of understanding and consciousness, operating purely mechanically. Deep learning models, while also lacking consciousness, can simulate understanding by recognizing and leveraging complex patterns in data. They can generalize from training data to perform tasks on new, unseen data, albeit through statistical methods rather than cognitive understanding. This generalization allows deep learning models to excel in specific domains, performing tasks that seem to require intelligence without possessing true understanding or awareness. The implications of this comparison are profound, spurring a multitude of different fads where members of the public called these models ``AI''. 

Deep learning models achieve behavioural equivalence in certain tasks, performing them with a level of proficiency comparable to humans. However, despite their impressive capabilities, these types models cannot possess genuine understanding or consciousness. This distinction aligns deep learning with the concept of Narrow AI, which refers to machines designed to simulate intelligent behaviour without possessing true consciousness or understanding. This is in contrast to General AI, which posits that machines could possess consciousness and understanding comparable to humans. Philosophically, this invites questions about the nature of intelligence and consciousness in machines. Will machines ever truly understand, or will they always be limited to simulating understanding through sophisticated pattern recognition and symbol manipulation? The Chinese Room Thought Experiment serves as a valuable lens through which to examine deep learning and AI. While deep learning models demonstrate impressive capabilities by simulating aspects of human intelligence, they are yet to reach a conscious understanding that characterizes human cognition. This is all, of course, the subject of much debate, since the very notion of consciousness is ill-defined and hard to pin down.

%% file: chapters/chapter_3_unsupervised_learning.tex
\chapter[Unsupervised Learning]{Unsupervised Learning}
\label{CH:UNSUPERVISED}
\section{Introduction}
Despite its tremendous success, supervised learning comes with its own limitations. First, supervised learning algorithm requires labelled data. This can be intensive or difficult to gather as it requires "knowing" the answer to enough examples of a problem so that we can train a model. This leads to the second point; supervised learning is data \textit{intensive}. The performance of a model is directly affected by the amount of training data made available to it. Too few examples means a model is likely to overfit to those examples, rather than infer a general rule about how to classify or approximate. Finally, many physics problems cannot be phrased in terms of classification or regression. Often, we are instead interested in learning something about an underlying distribution or rule that generates data and observations. Whilst this can be related to classification or regression, a good prediction for a label or function value can come from a "wrong" model, which may or may not tell us about the underlying physics of a problem.

In this chapter, we will explore how techniques from unsupervised learning can alleviate these limitations. The core idea behind unsupervised learning is to extract meaningful features from data \textit{without labels}. Typically, we are given a dataset where each piece of data is high-dimensional\footnote{A $64 \times 64$ pixel coloured image is already $\mathbb{R}^{64 \times 64 \times 3}$ dimensional. Think about the size of high-resolution images!}, and our task is to extract the key characteristics, or \textit{features}, of the data in a low-dimensional representation. The success of unsupervised learning can be summarised by the Manifold Hypothesis:

\textit{\textbf{The Manifold Hypothesis}: All of the important features of a high-dimensional dataset are contained in a smaller, lower-dimensional space.}

A central task for us will be to infer an invertible map to this low-dimensional space, formally known as \textit{dimensionality reduction}. We will begin by looking at NN-free methods like PCA, t-SNE to build some intuition for how unsupervised processes look, and then explore how deep learning strategies like Auto Encoders (AEs) can be applied to this task. This will allow us to take a more probabilistic view of unsupervised deep learning based on sampling, leading us to Variational Auto Encoders (VAEs), General Adversarial Networks (GANs) and generative models. Before exploring these methods, we lay out a few definitions:

\begin{itemize}
    \item A \textbf{feature space} is a vector space which contains all the possible data that we could possibly observe.  For example, the space of $64 \times 64$ pixel coloured images is $\mathbb{R}^{64 \times 64 \times 3}$. You can think of this as a $64 \times 64$ grid of tuples $(R,G,B)$.
    \item A \textbf{datapoint} is any vector in this space. For example, every different image in a dataset would correspond to a unique vector in this high-dimensional space. 
    \item A \textbf{feature vector} is a unit vector in feature space corresponding to a particular feature of interest. For example if we were representing black vs white pixel data on a 2D grid, a the black feature vector could be, $\mathbf{\hat{i}}$, and the white feature vector $\mathbf{\hat{j}}$, with an arbitrary (grey) pixel being a sum over this basis.
\end{itemize}

\noindent The key problem in unsupervised learning is how to provide feedback to a model without having labels for your data. How then, do we train a model?

\begin{figure}[h!]
    \centering
     \begin{tikzpicture}
        \node [anchor=south west, inner sep=0] (image) at (0,0) {\includegraphics[width=\textwidth]{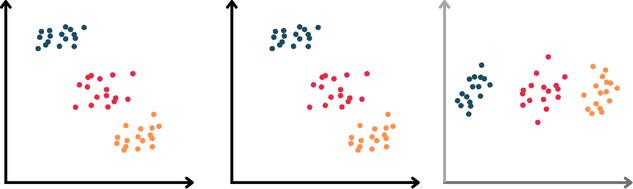}};
        \begin{scope}[shift={(0,0)}]
        \node at (-0.2,1.9) {$x_1$};
        \node at (2.0,-0.2) {$x_2$};

        \node at (4.1,1.9) {$x_1$};
        \node at (6.5,-0.2) {$x_2$};

        \node[gray] at (10.3, -0.2) {$w_2$};
        \node[gray] at (8.2, 2.0) {$w_1$};

        \draw[thick, gray] 
        (6.5, 2) ellipse [x radius=1.8cm, y radius=0.9cm, rotate=-45];
        \draw[->, gray, thick] (6.5,2) -- (7.2,2.5) node[left] {$w_2$};     
        \draw[->, gray, thick] (6.5,2) -- (5.3,3.2) node[left] {$w_1$}; 
            
        \end{scope}
    \end{tikzpicture}
    \caption{Visual representation of Principle Component Analysis (PCA) divided into three stages. In the first stage (left), we see the raw  input data with no labels. Taking this data, we can find an ellipsoid which contains all points in the dataset. Notice that when this ellipse is \textit{tight}, i.e. there is little we can do to make its area smaller whilst still containing all data, its principle axes ($w_1$, $w_2$) provide us with a new vector basis to represent the data. To transform between bases, we can apply a rotation matrix to each data-point, see the main text for further detail. The point of PCA is that in higher dimensions, many of these principle components are vanishingly small. This means ignoring them, or working in a \textit{subspace} spanned by the \textit{principle} components of an ellipse, allows us to find a compact representation of the data which still captures its essential features.}
    \label{fig:PCA_fig}.
\end{figure}
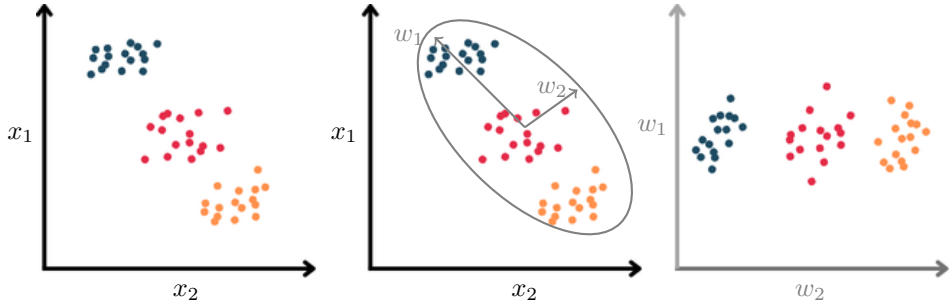

In order to train, we must change our perspective a little. We can consider unsupervised learning as task of finding efficient representations of data from a given dataset in line with the \textit{Manifold Hypothesis}. Given a piece of data in latent space, we can therefore ask \textit{how likely it is that this data belongs do the true training dataset?} By maximising this likelihood, we can create a model which maximises the probability of observing a given dataset. This technique is known as Maximum Likelihood Estimation, see~\hyperlink{box:MLE}{Box 3.1}.

\begin{figure}[h]
    \centering
    \begin{mybox}[\hypertarget{box:MLE}{Box 3.1: Maximum Likelihood Estimation (MLE)}]
    \label{box:MLE}
    Given $n$ independent observations of data $X = \{\mathbf{x}_1,\ldots,\mathbf{x}_n\}$, the likelihood function represents the joint probability of observing all the data. Should this data have come from a model with trainable parameters $\theta$, we may write
    \begin{equation}
        L(\theta; X) = P(X \mid \theta) = \prod_{i=1}^{n} P(\mathbf{x}_i \mid \theta).
    \end{equation}
    MLE is the task of varying $\theta$ such that we \textit{maximise} this likelihood function. That was, we are most likely to observe the data so our model parameters $\theta$ are optimal. Maximizing this product directly can be challenging because multiplying many probabilities (especially when they are small) can lead to numerical underflow. To overcome this, we take the logarithm of the likelihood function to obtain the log-likelihood function:
    \begin{equation}
        \ell(\theta; x) = \log L(\theta; x) = \log \left( \prod_{i=1}^{n} P(x_i \mid \theta) \right) =  \sum_{i=1}^{n} \log P(x_i \mid \theta).
    \end{equation}
    This is a common maths trick in deep learning - logarithms are monotonic so they preserve the direction in which we need to optimise. By converting the product of probabilities into a sum of logarithms, we can efficiently apply gradient-based optimization techniques (see Chapter~\ref{CH:FUNDAMENTALS}) to find the parameters, $\theta^*$, that make the observed data most probable under our model.
    \begin{equation}
        \theta^* = \arg \max_{\theta} \ell(\theta; x) = \arg \max_{\theta} \sum_{i=1}^{n} \log P(x_i \mid \theta).
    \end{equation}
    \end{mybox}
\end{figure}
\subsection{Principle Component Analysis (PCA)}

Best understood geometrically, PCA is a method to identify the most significant directions of a collection of points in a vector space. PCA works by containing all the dataset of different vectors inside an ellipsoid, and computing the principle axes of this ellipsoid. The principle components of a dataset can then be identified with the priniple axes of this ellipsoid. This is because the least prominent components of a set of vectors are ones that do not change. Therefore we can find which ones are most important by computing the principle axes of the ellipsoid which minimally contains the data. We can see how this works in 2D in Fig.~\ref{fig:PCA_fig}. 
\begin{algorithm}
\caption{Principle Component Analysis (PCA) algorithm}\label{alg:PCA}
    \begin{algorithmic}[1]
        \State \textbf{Input:} dataset $\mathcal{D}=\{\mathbf{x}_j: j = 1,\ldots N\}$ 
        \Comment{ feature vectors comprising a dataset \(\mathcal{D}\)}
        
        \State \(
        \mathbf{x}_j \leftarrow \mathbf{x}_j - \frac{1}{N} \sum_{i = 1}^N \mathbf{x}_i
        \)
        \Comment{Centre the data on the origin of the feature space}

        \State Compute \(\sigma = \sum_{j = 1}^N \mathbf{x}_j \mathbf{x}_j^T\)
        \Comment{Covariance matrix}

        \State Compute Singular Value Decomposition (SVD) \(\sigma = U \Lambda U^{T}\) 
        \Comment{Each element of the diagonal matrix $\Lambda$ is an eigenvalue of $\sigma$ with associated eigenvector in the rows of $U$.}

        \State Select eigenvectors with $M$ largest eigenvalues to construct, and place them in an (decreasing) ordered list        
        
        \State \textbf{Output:} Principle vectors \(\mathcal{V} = \{\mathbf{w_k}: k = 1, \ldots M \leq N\}\)
	\end{algorithmic} 
\end{algorithm}

\begin{figure}[h]
    \centering
    \begin{mybox}[\hypertarget{box:MLE}{Box 3.2: What it Means to Sample From \(p(x)\)}]
    \label{box:sampling}
    Sampling from a probability distribution \(p(x)\) means producing a random variable
    \(X\) whose probability of falling in any measurable set \(A\subset\mathbb{R}\)
    matches the area of \(p\) over that set:
    \(\mathbb{P}[X\in A]=\int_A p(x)\,dx\).
    Repeated draws therefore turn the \emph{formula} \(p(x)\) into a tangible
    empirical histogram that converges to \(p\) itself. 
    
    In practise, we can generate samples using a procedure called Inverse-transform sampling. This uses the cumulative distribution function
    \(F(x)=\int_{-\infty}^{x}p(t)\,dt\) to convert an easily generated
    uniform variate into one that obeys \(p\), because \(F(X)\sim\mathcal{U}(0,1)\)
    whenever \(X\sim p\). This is done with the following sub-routine which samples from a given discretized probability distribution with support
    \(\{x_1,\dots,x_k\}\) and probabilities \(p_j=\Pr[X=x_j]\)
    (\(\sum_{j=1}^k p_j = 1\)). First, we
    \textbf{(i)} draw the uniform \(u\),
    \textbf{(ii)} accumulate probabilities in \(r\) until \(r\) exceeds \(u\),
    and \textbf{(iii)} return the first value whose cumulative slice
    swallows \(u\).  Over many repeats, the emitted values match the target
    probabilities exactly. This is summarised by the following pseudocode:

    \begin{algorithmic}[1]
      \Require probability array \((p_1,\dots,p_k)\), sample count \(N\)
      \For{$i\gets1$ \textbf{to} $N$}
         \State \(u \sim \mathcal{U}(0,1)\)      \Comment{step 1: pick a random number}
         \State \(r \gets 0\)                    \Comment{running cumulative probability}
         \For{$j \gets 1$ \textbf{to} $k$}       \Comment{step 2: walk through the bins}
            \State \(r \gets r + p_j\)           \Comment{\(r\) now equals \(\sum_{\ell=1}^j p_\ell\)}
            \If{\(u < r\)}                       \Comment{step 3: stop when \(u\) fits}
               \State \textbf{output} \(x_j\)    \Comment{\(x_j\) is the sample}
            \EndIf
         \EndFor
      \EndFor
    \end{algorithmic}
    \end{mybox}
\end{figure}

Executing Algorithm~\ref{alg:PCA}, we have a new orthonormal basis which contains the principle components of our dataset. All that is left is to transform our dataset into this basis by applying a matrix to each datapoint
\begin{equation}
    W = \begin{pmatrix}
        \mathbf{w}_1 \\
        \mathbf{w}_2 \\
        \vdots \\
        \mathbf{w}_M
    \end{pmatrix},
\end{equation}
where each $\mathbf{w}_l$ is one of the principle eigenvectors. Since we have truncated this list of eigenvectors, we now have a feature space $\mathbb{R}^M \subset \mathbb{R}^N$. Notice that the basis in our reduced feature space is not some selection of the original basis in $\mathbb{R}^N$. Rather, we have selected some linear combination of this basis, whose unit vectors are automatically selected by PCA.

PCA comes with the advantage of simplicity - there are features that we wish to extract which are a linear combination of the basis in our feature space. What might happen then, if the dataset's features are a \textit{non-linear} combination of the old basis? 

Intuitively, we would expect a \textit{non-linear} transformation is needed to extract these kinds of features. There are lots of ways to engineer a non-linearity here, but we will focus primarily on t-distributed Stochastic Neighbourhood Embedding (t-SNE). Other methods based on introducing non-linearity with Neural Networks are reserved for Sec.~\ref{sec:AEs}.

\subsection{t-SNE}
t-distributed Stochastic Neighborhood Embedding (t-SNE) is a technique which introduces nonlinearity to feature extraction \cite{van2008visualizing}. It is particularly useful for two reasons; (i) it allows us to visualize high-dimensional dataset in low-dimensional spaces and (ii) it can perform more general feature extraction than PCA and friends. The algorithm is stochastic, meaning it can generate different results each time, which perform well \textit{on average}.

Consider embedding each datapoint in a feature space $\mathbf{x}_j \in \mathbb{R}^N$. Then if two pieces of data, $\mathbf{x}_i,\;\mathbf{x}_j$ represent something similar, they should be close in this embedding according to the euclidean distance, $|\mathbf{x}_i - \mathbf{x}_j|$. Next, let's define a map from the high-dimensional space, to a lower one; $f: \mathbb{R}^N \rightarrow \mathbb{R}^M$ for $M < N$ such that $\mathbf{y}_j = f(\mathbf{x}_j) \in \mathbb{R}^M$. The key structure we want to keep in the low-dimensional space is \textit{closeness} of similar points in the domain. That is when $|\mathbf{x}_i - \mathbf{x}_j| \leq \epsilon$, we have $|f(\mathbf{x}_i) - f(\mathbf{x}_j)| \leq \delta$. To achieve this, we could phrase our problem in terms of maximising that the nearest neighbour of $\mathbf{y}_i$ is $\mathbf{y}_j$ whenever the nearest neighbour of $\mathbf{x}_i$ is $\mathbf{x}_j$. 

Inspired by classical statistical physics, we can write the conditional probability in the domain as
\begin{equation}
    p(i|j) = \frac{e^{-|\mathbf{x}_i - \mathbf{x}_j|^2/\sigma_i^2}}{\sum_{k \neq j} e^{-|\mathbf{x}_j - \mathbf{x}_k|^2/\sigma_j^2}},
\end{equation}
where $\sigma^2$ is some variance to be chosen\footnote{There are strategies for making an \textit{intelligent choice}. But for now, we know that eventually we are going to do some tuning via gradient descent. So it is possible to simply include this as one of the variational parameters, and tune it later.} and $p(i|j)$ encodes that $\mathbf{x}_i$ is the nearest neighbour given we start at $\mathbf{x}_j$, and $p(i|i) =0$. We can see this is just a Boltzmann distribution over the Euclidean distances between our data. Notice that $ p(i|j)$ is a conditional probability. To make this symmetric\footnote{i.e. that we have the same outcomes no matter which way around $i$ and $j$ are.}, we can use the joint probability
\begin{equation}
    p_{ij} = \frac{1}{2} \bigg( p(i|j) +  p(j|i) \bigg).
\end{equation}

Next, we need to define a probability distribution in the image of $\mathbf{y}_j = f(\mathbf{x}_j)$. We could go for another Gaussian via the Boltzmann distribution but there is a more cunning and strategic distribution which works better for is: the t-distribution\footnote{dun dun DUUUUNNNN statistics classes really were for something weren't they! This is is a nice place to observe a key difference between machine learning and \textit{deep} learning. In machine learning, we need an underlying statistical model like the t-distribution. This sits in contrast with deep learning in unsupervised tasks, which seeks to capture the underlying correlations in an automated way.}.
\begin{equation}
    q_{ij} = \frac{(1 + \| \mathbf{y}_i - \mathbf{y}_j \|^2)^{-1}}{\sum_{k \neq l}(1 + \| \mathbf{y}_k - \mathbf{y}_l \|^2)^{-1}}.
\end{equation}
The mathematical details of this distribution don't actually matter that much, only that it is computationally efficient and that it is a fat-tailed distribution. This allows us to tune our map quickly and efficiently. 

Given that we now have two distributions for closeness in the domain and image of $f$, we are now in a position to minimise the distance between the distributions $p_{ij}$ and $q_{ij}$. To do this, we need another object; something which allows us to measure the similarity between two probability distributions. A popular choice is the so-called Kullback-Leibler (KL) divergence\footnote{which for those studying quantum information theory should feel the same as the \textit{relative entropy} measure between two density matrices. This is not a coincidence!},
\begin{equation}
    L(\{\mathbf{y}_j\}) = \sum_{ij} p_{ij} \log \frac{p_{ij}}{q_{ij}},
\end{equation}
When these distributions are identical ($L = 1)$, we get our closeness property perfectly! 

All that remains is to \textit{tune} the function $f$ with respect to the images $\{\mathbf{y}_j\}$. We can do this with gradient descent using $L$ as a loss function. 

We have seen two of many different but effective strategies\footnote{Some others of note are clustering and kernel-based methods. Although we won't introduce them here, a fantastic introduction can be found in \cite{Prince_2023}.}. Therefore, we might ask what is the \textit{most} effective strategy to reduce the dimensionality of our data? This question forms the premise of Auto Encoders (AEs) which automate the process of learning which embedding of data is best.

\section{Auto Encoders (AEs)}
\label{sec:AEs}

\begin{figure}
    \centering
    \includegraphics[angle = 90, width=0.8\linewidth]{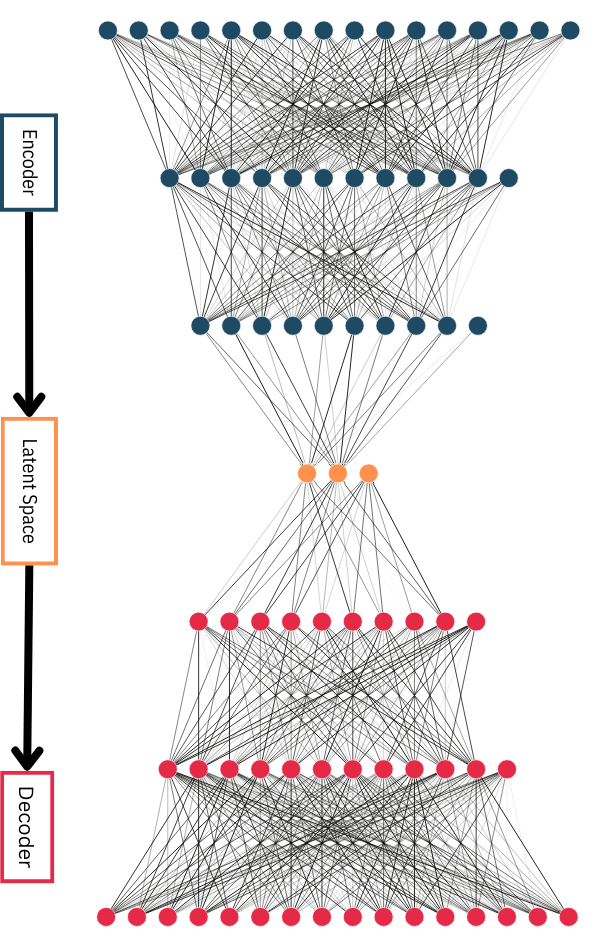}
    \caption{The Auto Encoder (AE) architecture of three components (left to right, blue, orange, and red). In the first component, the model receives some piece of data $x \in \mathbb{R}^N$ and passes it through the encoder to a \textit{latent space} vector, $z \in \mathbb{R}^L$ (centre, orange). At this stage, a well-trained model gives you a compressed version of your input. To recover the original input, simply pass the latent space vector through the decoder.}
    \label{fig:AEs}
\end{figure}

An Auto Encoder (AEs) is a type of neural network designed to learn efficient, low-dimensional representations (encodings) of input data by compressing it into a latent space (encoder) and then reconstructing the original data from this representation (decoder) \cite{bengio2009learning}. It is trained to minimize the reconstruction error between the input and output. AE can reduce the dimension of data by acting as a tunable information funnel or compression scheme (see Fig~\ref{fig:AEs}). 

There are three components to an AE:
\begin{itemize}
    \item Encoder - This is a map $E: \mathbb{R}^N \rightarrow \mathbb{R}^L$ from datapoints $\mathbf{x}_j \in \mathbb{R}^N$ to a low-dimensional space, $E(\mathbf{x}_j) = \mathbf{z}_j \in \mathbb{R}^L$ called a \textit{latent space}, $L < N$.
    \item Latent space - this is the "central" layer in an AE which has lower dimension $L < N$. 
    \item Decoder - This is a map from the latent space, back to the input space, $D: \mathbb{R}^L \rightarrow \mathbb{R}^N$.
\end{itemize}
The aim of AEs is to be able to map any piece of data $\mathbf{x}_j$ to itself through this information funnel. That is, when an AE functions perfectly,
\begin{equation}
\label{eq:perfect_encoder}
    \mathbf{x}_j = D \big( E (\mathbf{x}_j) \big),
\end{equation}
for all data $\mathbf{x}_j$ in the feature space. \textit{Why do this}, one might ask? When we satisfy Eq.~\ref{eq:perfect_encoder}, it means it was possible to pass $\mathbf{x}_j \in \mathbb{R}^N$ through a \textit{latent} layer with $L < N$ dimensions, compressing it down with a neural network! Therefore, if we stop the process in the middle, we end up with encoded data in a latent space, reducing the dimension of our feature space. Whenever we want to retrieve the original vector, we simply pass it as an input through the Encoder.

To make an AE, we can construct the encoder, $E$, and decoder, $D$, as neural network functions, $E:\mathbb{R}^N \xrightarrow{\varphi} \mathbb{R}^L$, 
$D:\mathbb{R}^L \xrightarrow{\phi} \mathbb{R}^N$, where $\theta = \{\phi, \varphi\}$ are the trainable parameters of our model. By executing a forward pass, 
\begin{equation}
    \tilde{\mathbf{x}}_j =  D \big( E (\mathbf{x}_j) \big),
\end{equation}
we can compare the estimated reconstruction $\tilde{\mathbf{x}}_j$ with the original, true, datapoint $\mathbf{x}_j$ with a MSE loss,
\begin{equation}
    L_{\text{AE}}(\theta) = \frac{1}{M} \sum_{j} \| \mathbf{x}_j - \tilde{\mathbf{x}}_j\|^2 = \frac{1}{M} \sum_{j} \| \mathbf{x}_j -  D \big( E (\mathbf{x}_j) \big)\|^2.
\end{equation}
Geometrically, this is simply comparing average euclidean distance between $\mathbf{x}_j$ and its reconstruction $\tilde{\mathbf{x}}_j$. 

The latent space data, $\mathbf{z}_j \in \mathbb{R}^L$ give us a NN analogue to PCA and t-SNE. However, AEs come with some very useful advantages:
\begin{itemize}
    \item They can be trained sequentially over batches of data, rather than dealing with functions of an \textit{entire} dataset. This is particularly useful when we have lots of data available!
    \item They contain non-linearity like t-SNE.
    \item It is easy to control the size of the latent space - just choose a different number of neurons on this layer. \textit{Done}.
\end{itemize}
On the other hand, AEs make it very likely to overfit to data. We don't necessarily have a way of testing how an AE can \textit{generalise} to new instances of data.  A highly complex\footnote{i.e. with sufficient depth and lots of trainable parameters $\text{dim}(\theta) \gg 1$.} encoder-decoder network could simply memorize the entire training dataset by assigning each input a unique scalar value. In this way, the information bottleneck could be compressed to a single neuron, $\mathbf{z} \in \mathbb{R}$, giving the \textit{illusion} of lossless compression. Given a new, unseen piece of data, this kind of network would perform terribly, as this new point has no scalar encoding. We need therefore need to be careful when regularising AEs to make sure they are extracting meaningful features, rather than overfitting to their training data.

From Chapter~\ref{CH:FUNDAMENTALS}, we have seen lots of regularisation techniques for handling overfitting. Whilst we could unpack these and see how well they work, notice that no amount of regularisation will truly handle this generalisation problem at its core. First, consider that the data we are provided likely came from some underlying process in physics\footnote{or some other process depending on the use case at hand. As this is a course in deep learning for classical and quantum physics, we are mainly focussed on this perspective.}, and they are merely observations of this process.
This means the data, $X = \{\mathbf{x}_1, \ldots,\mathbf{x}_M\}$ we are given are actually samples from some underlying data distribution $X \sim \mathcal{D}$. We might therefore wonder what an AE is doing in terms of this underlying distribution. Since it only ever attempts to correlate the samples themselves, it is acting in the space that connects data to the latent space, $\text{AE}: X \rightarrow_{\theta} Z$. However, we have just established that these data could be connected to some underlying process, namely a probability distribution. The existence of a perfect encoder defined in Eq.~\ref{eq:perfect_encoder} is deterministically mapping the points $X$ over into $Z$, which means if we were to resample $X$ from $\mathcal{D}$, the AE would fail, despite the data coming from the same distribution. It is this fact that requires a change in perspective. We could instead try to use the universal approximation theorem to learn a map which transforms the \textit{distribution} $X \sim \mathcal{D}$ to a lower-dimensional distribution $Z \sim \mathcal{Z}$. By seeking this latent-space distribution, we are changing our perspective from points to probabilities. 

This subtle change in perspective has profound consequences that has helped create a young and flourishing sub-field of generative AI, and deserves its own discussion, which we reserve for Sec.~\ref{sec:points_to_probabilities}. For the AE, we can now define a new type of encoder on probability distributions, instead of points. We call this a \textit{Variational Auto Encoder} (VAE). 

\section{Variational Auto Encoder (VAE)}
In the above, we alluded to a change in perspective from points to probabilities, motivating the VAE. VAEs are a way of compressing the underlying distribution of a dataset through an information bottleneck such that the input and output distributions of the model are the same. Analogously to an AE, if we can do this perfectly, it implies we have found a latent probability distribution which is low-dimensional representation of our input distribution, in line with the \textit{manifold hypothesis} \cite{kingma2013auto}. To accomplish this, we can construct probabilistic encoders and decoders that map data to and from a latent probability space, and train them.

Training aligns the input-output distributions, enabling the VAE to learn a meaningful, low-dimensional representation of the input distribution. This is what can generate new, \textit{similar} data\footnote{Similar in the sense that we would compare how samples from the two distributions compare, rather than a fixed deterministic set of points.}. A probabilistic construction also conveniently solves our overfitting problem from AEs - a perfect VAE has learned a latent \textit{distribution} that constructs our dataset, so it cannot overfit to samples! What remains is to understand the mathematical structure of neural networks that can learn probability distributions. To that end, it is useful to understand step-by-step what happens during a forward pass of a VAE.

At the input, the encoder receives a sample from the underlying true data distribution. For simplicity, let's focus on the simple but powerful\footnote{thanks to the Central Limit Theorem} case of a Normal distribution.
We must now parametrise the latent space distribution by recasting our interpretation of the latent vector differently according to our change in perspective. We can \textit{interpret} the Encoder's output as the parameters that form a Gaussian distribution. Explicitly, the output of the encoder $\mathbf{z}(\varphi) = E(\mathbf{x};\varphi)$ with trainable parameters $\varphi$, can be used to construct a Normal distribution\footnote{As an observant reader, you might notice that this constrains the size of the latent space. If we want to construct a multi-mode Gaussian, the number of elements in $\mathbf{z}(\varphi)$ is fixed by the size of $\mu$ and $\sigma$. Generically, we see an $n$-mode Gaussian needs $n^2 + n$ parameters, meaning $\mathbf{z}(\varphi)$ must have $n^2 + n$ elements or we must have repeats.} 
\begin{equation}
    \mathbf{z}(\varphi) :\rightarrow \mathcal{N}(\mu_{\varphi}, \sigma_{\varphi}).
\end{equation}
Notice we are keeping track of the trainable NN parameters $\varphi$ in order to back-propagate through this process during training. In order to proceed to the decoder, we now sample from  $\mathcal{N}(\mu_{\varphi}, \sigma_{\varphi})$ and pass these samples through the decoder. Let's pause here and notice that this is the step that prevents overfitting. A VAE cannot assign a unique number to each point in the dataset because the decoder receives \textit{samples} from  distribution constructed by the encoder, not the points themselves. This allows us to compare the output and input in the same way as an AE, for example with MSE loss, minus the overfitting problem.

An interesting and subtle problem does however come from interpreting $\mathbf{z}$ in the way we have. Namely, $\mathbf{z}$ defining a distribution means there is no continuity that connects the trainable parameters $\varphi$ of the encoder to the output estimator, $\tilde{\mathbf{x}}_j(\varphi)$, of $\mathbf{x}_j$. To see why this is the case, lets consider the simplest case of sampling from the 1D Gaussian $x \sim \mathcal{N}(\mu_{\theta}, \sigma_{\theta})$, where $\mu_{\theta},\; \sigma_{\theta} \in \mathbb{R}^2$ are the two-component outputs of a NN. Let's assume we have some loss function $L: \mathbb{R}^2 \rightarrow \mathbb{R}$ such that $L(x)$ is differentiable. In order to perform back propagation, we need to connect the sample $x$ to the neural network parameters, $\theta$, that gave rise to its distribution, $\mathcal{N}(\mu_{\theta}, \sigma_{\theta})$. But a sample is just a point. It has no continuity properties. Even sets of samples are just points with no continuity between them! We need a method that allows us to differentiate through a sampling process should we wish to perform back propagation over distributions. To remedy this, Kingma et al. in their seminal work noticed that we can create a distribution which is equivalent to $\mathcal{N}(\mu_{\theta}, \sigma_{\theta})$, but differentiable with respect to $\theta$. This is called the \textit{reparametrisation trick}, see \hyperref[box:reparam]{Box 3.3}.

\begin{figure}
    \centering
    \includegraphics[angle = 90, width=0.8\linewidth]{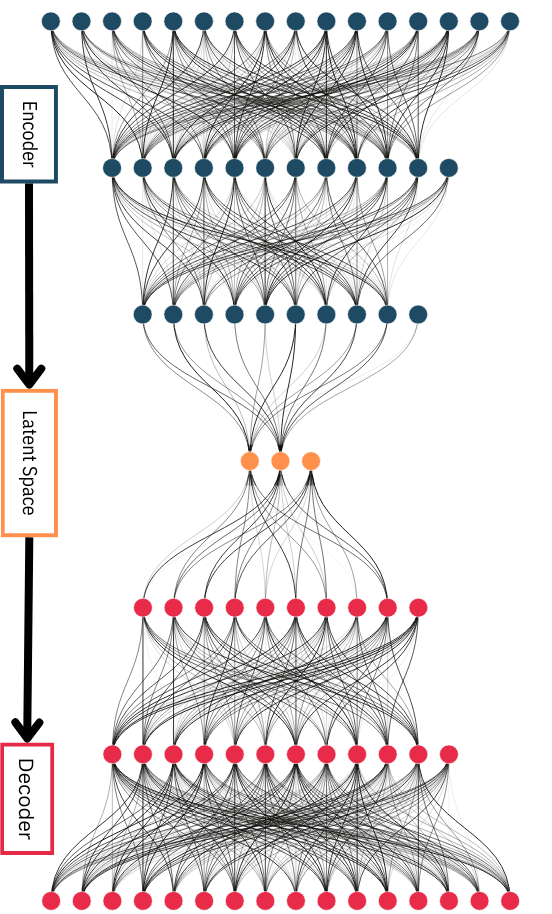}
    \caption{An example architecture for a Variational Auto-Encoder (VAE). This examples uses a Multi-Layer Perceptron architecture introduced in Chapter~\ref{CH:FUNDAMENTALS}, with bias units shown as the top neuron in each layer.
    The left (blue) side shows the Encoder $E(\mathbf{x};\phi)$ receiving a 16-feature input $x \in \mathbb{R}^{16}$ and compressing it to a two-component \textit{latent space}, $\mathbf{z} = (z_1,z_2) \in \mathbb{R}^2$.
    In the latent space, we can construct a normal distribution $\mathcal{N}(\mu = z_1, \sigma = z_2)$, from which we can take samples. These samples are then fed into the Decoder (red) shown on the right, $D(\mathbf{z};\phi)$. 
    From the output, we yield a forward pass $\tilde{x} = D(E(\mathbf{x})$ which is differentiable thanks to the reparametrisation trick 
    - see Box.~\protect\hyperref[box:reparam]{Box 3.3}}
    \label{fig:enter-label}
\end{figure}

\begin{figure}
    \centering
    \begin{mybox}[\hypertarget{box:reparam}{Box 3.3: Reparametrisation Trick}]\label{box:reparam}
     Kingma et al. in their seminal work \cite{kingma2013auto} noticed that we can create a distribution which is equivalent to $\mathcal{N}(\mu_{\theta}, \sigma_{\theta})$, but differentiable with respect to $\theta$. Instead of sampling $x$ directly from $\mathcal{N}(\mu_{\theta}, \sigma_{\theta})$, let
    \begin{equation}
        x' = \mu_{\theta} + \sigma_{\theta} \cdot n,
    \end{equation}
    where $n \sim \mathcal{N}(0,1)$. Now the first and second moments of $x$ and $x'$ are identical. Since these are the only unique moments of a Gaussian, the two distributions are \textit{indistinguishable}. That is, given two "black boxes" containing $x$ and $x'$, no amount of sampling will ever let you tell them apart.
    Therefore, we may use one in place of the other! The technique designed by Kingma et al. has been revolutionary, unlocking the ability to do deep learning in probability spaces. As we will see later, this trick is highly applicable in quantum processes, since quantum mechanics is itself a probabilistic theory. Much attention has been given to constructing this trick on other more exotic distributions, see for example \cite{rezende2020normalizing}.
    \end{mybox}
\end{figure}

Under the reparametrisation trick, we now have an end to end method for performing forward passes with a VAE in a way that is differentiable with respect to the encoder's neural parameters, $\varphi$. This lets us properly define back-propagation, constructing Algorithm~\ref{alg:VAE_trainstep}. 

An interesting and open problem in VAEs lies in providing a meaningful interpretation to the latent space, see for example \cite{schoulepnikoff2025interpretable}. There is also the problem of \textit{disentangling} features in the latent space by encouraging the VAE's latent layer neurons to fire independently of one another. This is in attempt to establish a \textit{causal} relationship between different data-classes and neural activation patterns, and forms the conceptual premise of the $\beta$-VAE \cite{higgins2017beta}.

\begin{algorithm}
\caption{Variational Auto Encoder (VAE) Train Step  }\label{alg:VAE_trainstep}
    \begin{algorithmic}[1]
        \State \textbf{Input} $E(\varphi)$, $D(\phi)$ and values $\theta = (\varphi, \phi)$
        \Comment{Encoder and Decoder with current values $\theta = \{\varphi, \phi\}$}
        \State Calculate $\mathbf{z} = E(\mathbf{x};\varphi)$ 
        \Comment{Latent space vector}
        \State Construct $(\mu_{\varphi},\sigma_{\varphi})$ from $\mathbf{z}$
        \State Sample $\tilde{\mathbf{z}} = \mu_{\varphi} + \sigma_{\varphi} \cdot n$ \Comment{$n \sim \mathcal{N}(\mathbf{0},\mathbb{1})$}
        \State Calculate $\tilde{\mathbf{x}} = D(\tilde{\mathbf{z}};\phi)$
        \State Calculate $L(\mathbf{x},\tilde{\mathbf{x}};\varphi, \phi)$ 
        \Comment{Loss function $L: \mathbb{R}^N \times \mathbb{R}^N \rightarrow \mathbb{R}$}
        \State Back propagate $\frac{\partial L}{\partial \theta}$ \Comment{$\theta = (\varphi, \phi)$ are the trainable parameters of $E$ and $D$}
        \State Update $\theta \gets \theta + \alpha \nabla_{
        \theta} L(\mathbf{x},\tilde{\mathbf{x}};\varphi, \phi)$
        \Comment{Learning rate $\alpha \in \mathbb{R}$}
	\end{algorithmic} 
\end{algorithm} 
\section{Points to Probabilities make Infinite Possibilities - Generative AI}
\label{sec:points_to_probabilities}
The reparametrisation trick (\hyperref[box:reparam]{Box 3.3}) has given us a new way use neural networks over \textit{distributions}, whilst training with samples. Armed with a sufficiently well trained VAE, we might wonder what happens if we use \textit{only} the decoder component for some inputs of our choosing. Then the output of the decoder will be samples in the input space that correspond to the features we engineered in the latent space. We can hence think about using the decoder to \textit{generate} new samples that are typical of the distribution that our training data is sampled from. This gives rise to a whole new strategy to use neural networks; \textit{we may use them in generative tasks}. More formally, we can define a generative model as follows: \\

\textit{\textbf{Deep Generative Models} are defined as neural networks employed to represent (typically high-dimensional) probability distributions such that samples from them are indistinguishable from the true underlying distribution} \\

Here, we will focus on three types of deep generative model called Generative Adversarial Networks (GANs), (Restricted) Boltzmann Machines, and Normalising Flows. We study GANs as the natural extension of a VAE to illustrate how to use our change in perspective from points to probabilities to generate new data. We will then proceed to the other two as further examples with pertinent use-cases in the quantum sciences.

\section{Generative Adversarial Networks (GAN)} 

To realise our new strategy, we can set up an adversarial game between the two neural networks for encoding and decoding, hereby referred to as the \textit{discriminator} and the \textit{generator} \cite{goodfellow2014generative}. The generator seeks to capture the underlying distribution of the dataset by generating samples that resemble real data. The discriminator, on the other hand, aims to distinguish between genuine data samples and those produced by the generator. During training, both networks are refined simultaneously—the generator improves its ability to produce realistic samples to fool (sometimes called spoof) the discriminator, while the discriminator enhances its capability to detect generated samples. This adversarial training process adjusts the generator's output distribution to closely align with the input data distribution. By iteratively tuning the input-output relationships through this competition, we can train a model to generate new data points that are statistically similar to the original dataset. This forms the conceptual premise of a Generative Adversarial Network (GAN).

To make these ideas more precise, let's define a generator network 
\begin{equation}
    G: L \xrightarrow{\varphi} \chi,
\end{equation}
which maps samples in the latent space distribution $z \sim \mathcal{Z}$ to points the full data space $\mathbf{x} \in \chi$. That is, the generator produces samples $\tilde{\mathbf{x}} = G(\mathbf{z})$. Next, we can define a discriminator network
\begin{equation}
    D: \chi \xrightarrow{\phi} [0,1] \subset \mathbb{R},
\end{equation}
which we interpret as the probability of point in the dataspace, $\mathbf{x} \in \mathbb{R}^N$ belonging to the true dataset, i.e. that $\mathbf{x}$ was not made by the generator, $G$. This gives us a total trainable parameter set $\theta = \{\varphi, \phi\}$. We can formally place these networks in an adversarial setting by defining a value function which maps all possible pairs of discriminator and generator to a score, i.e.
\begin{equation}
    V: D \times G \rightarrow \mathbb{R}.
\end{equation}
Then an optimal generator and discriminator will solve
\begin{equation}
\label{eq:GAN_objective}
    \min_G \max_D V(D, G) = \mathbb{E}_{\mathbf{x} \sim \chi} \left[ \log D(\mathbf{x}) \right] + \mathbb{E}_{\mathbf{z} \sim \mathcal{Z}} \left[ \log \left( 1 - D(G(\mathbf{z})) \right) \right],
\end{equation}
where $\mathbb{E}_{a \sim \mathcal{A}}$ denotes expectation over samples $a$ with distribution $\mathcal{A}$.

Whilst this may look quite intimidating, we can unpack Eq.~\ref{eq:GAN_objective} to build some intuition. The first term represents the expected log-likelihood that the discriminator correctly identifies real data samples. This is natural; we should encourage $D$ to assign high probabilities to real data. The second term is the expected log-likelihood that the discriminator correctly identifies fake data, coming from the generator network, $G$. This makes sure that $D(\mathbf{x})$ is small whenever $\mathbf{x} = G(\mathbf{x})$. We want these two factors to be simultaneously true, which means their log-likelihoods are added (per usual log rules).

Since two networks participate in solving Eq.~(\ref{eq:GAN_objective}), we should understand how they can be trained. For the discriminator, let
\begin{equation}
    \mathcal{L}_D = - \frac{1}{m} \sum_{i=1}^{m} \left[ \log D\left( \mathbf{x}_i \right) + \log \left( 1 - D\left( G\left( \mathbf{z}_i \right) \right) \right) \right],
\end{equation}
since $\mathcal{L}_D$ is minimised when $\left[ \log D\left( \mathbf{x}_i \right) + \log \left( 1 - D\left( G\left( \mathbf{z}_i \right) \right) \right) \right]$ is at a maximum on average over a batch of samples $\{\mathbf{x}_i\} \sim \chi, \; \{\mathbf{z}_i\} \sim \mathcal{Z}$, taken respectively from the true distribution $\chi$ and the generator's latent space distribution $\mathcal{Z}$. Similarly for the generator, we can define
\begin{equation}
    \mathcal{L}_G = - \frac{1}{m} \sum_{i=1}^{m} \log D \left( G\left( \mathbf{z}^{(i)} \right) \right),
\end{equation}
since minimising this loss maximises the chances that the discriminator incorrectly classifies samples from $\mathcal{Z}$ as coming from $\chi$.

\begin{figure}
    \centering
     \begin{tikzpicture}
            \node[anchor=south west, inner sep=0] (image) at (0,0) {\includegraphics[width=0.9\textwidth]{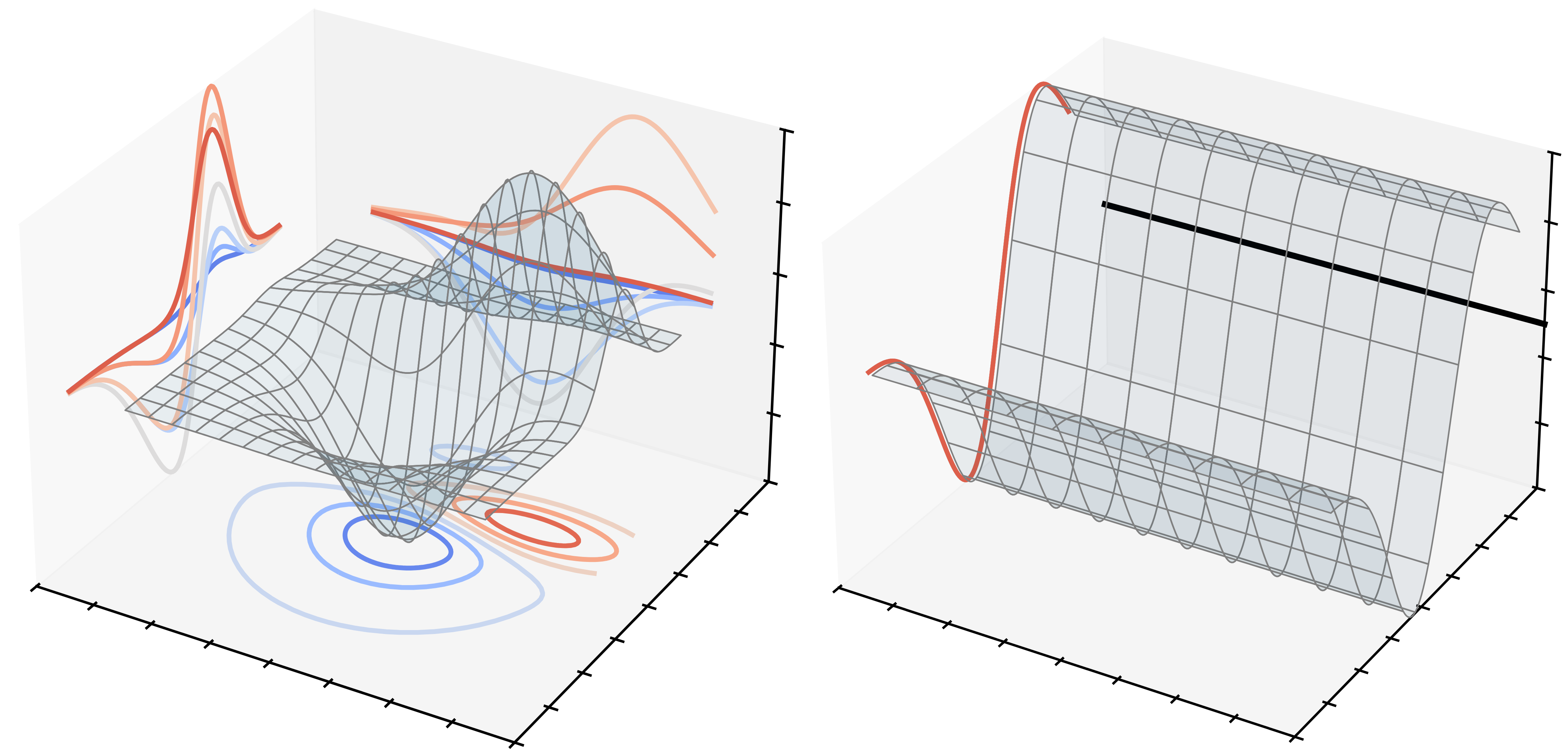}};
            \begin{scope}[x={(image.south east)}, y={(image.north west)}]
                \node at (0.15, 0.06) {$x$}; 
                \node at (0.44, 0.15) {$y$}; 
                \node at (0.5, 0.87) {$p(x,y)$};
                
                \node at (0.65, 0.06) {$x$}; 
                \node at (0.94, 0.15) {$y$};
                \node at (1.05, 0.8) {$\tilde{p}(x,y)$};

            \end{scope}
        \end{tikzpicture}
    \caption{An example visualisation for mode collapse. In both plots, the height represents probability density $p(x,y)$ with respect to two independent variables $x$ and $y$. The marginal distributions $p(x|y)$ and $p(y|x)$ are projected respectively in the $x-z$ and $y-z$ planes, with a contour plot of the left distribution displayed in the $x-y$ plane. Consider trying to use a GAN to learn the left joint distribution $p(x,y)$, hereby referred to as the \textit{ground truth}. If we are not careful with the order of \texttt{min} and \texttt{max} when defining the problem, there is a loophole for the generator to exploit. Namely, the GAN can capture the main characteristics of the $p(x|y)$ marginal (right figure, $x-z$ plane) whilst ignoring the $p(y|X)$ marginal (right figure, $y-z$ plane). Notice that in both of these distributions, we see the $p(x|y)$ marginal has a similar shape. As such, the learned joint distribution $\tilde{p}(x,y)$ is incorrect, but the marginal $\tilde{p}(x|y)$ which lives in a \textit{subspace} of $\tilde{p}(x,y)$ is still ``good enough". In fact, even when we get the ordering right (see Eq.~(\ref{eq:GAN_objective}), mode collapse can still be a problem. For this reason (amongst others), GANs are typically no longer the go-to for generative tasks. They are however of great importance for understanding how to learn high-dimensional distributions! So let's not belittle their importance too much.}
    \label{fig:mode_collapse}
\end{figure}
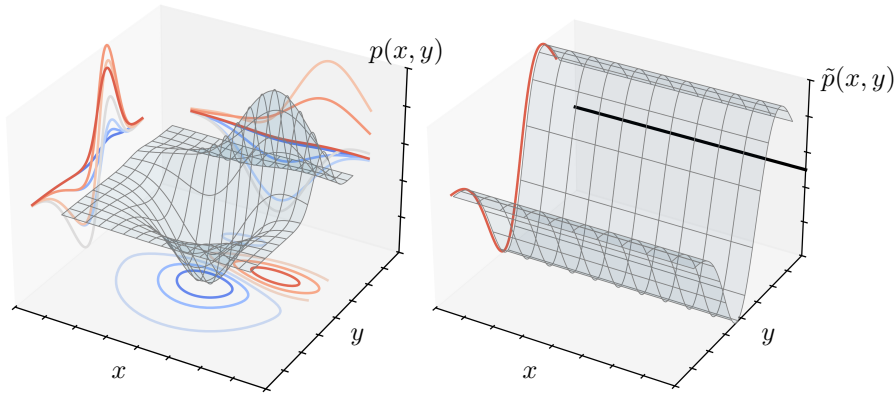

When training a GAN, we have to simultaneously train the generator, $G$, and discriminator, $D$. Here, a subtle point arises. Notice that if we switch the order of $\texttt{min}$ and $\texttt{max}$ in Eq.~{\ref{eq:GAN_objective}}, that is, we attempt to solve
\begin{equation}
    \max_D \min_G  V(D, G),
\end{equation}
there is a resultant loophole. Now, the generator $G$ can solve this problem by always producing a few samples $\mathbf{\tilde{x}}$ which the discriminator classifies as genuine with high probability. That is, there are sets of points $\{\mathbf{z}_j\} \in \mathcal{Z}$ that get mapped to the same generated samples $\mathbf{\tilde{x}}$. Instead of getting it right over the whole distribution, the generator only gets it right over a subspace. This phenomenon is known as \textit{mode collapse}, shown in Fig.~\ref{fig:mode_collapse}. 

In fact, mode collapse can even happen when we get the ordering right! That is, when we attempt to solve Eq.~(\ref{eq:GAN_objective}), as well as with other more sophisticated pipelines, mode collapse can still happen. It is worth taking a moment to understand how we can mitigate mode collapse.
To mitigate mode collapse, we can apply a few tricks inspired from regularisation techniques. First, we can employ batched data. To see how this helps fight mode collapse, consider the generator outputting samples $\{\mathbf{\tilde{x}}_j\ \in \chi$ satisfying
\begin{equation}
    |\mathbf{\tilde{x}}_j - \mathbf{\tilde{x}}_j'| \leq \delta,
\end{equation}
where $\delta \ll 1 \in \mathbb{R}$ is small. From the perspective of the generator $G$, this data is \textit{unusually} close together given that it should have been sampled from some true data distribution. Therefore, using mini-batches forces the generator away from mode collapse by ensuring that outputs $\mathbf{\tilde{x}}_j$ are sufficiently diverse.

In spite of this nice trick, GANs are typically not the go-to for generative tasks at the time of writing these notes. This is because of other structural problems that can arise in the adversarial setting of GANs. Some problems of note include
\begin{itemize}
    \item \textbf{Training Instability} - we have two neural networks that need to be simultaneously trained if they are to make progress of one another. Each neural network has independent trainable parameters, but they influence the behaviour and performance of each other (by construction). This creates a training instability because we have to balance genuine independence of the neural networks' parameters with their sensitivity to each other

    \item \textbf{Oscillations and Weak Convergence} - GANs have no stopping criterion. As such, we cannot definitively say \textit{when} a GAN has finished training. Again, the adversarial setting means that the loss functions $L_G$ and $L_D$ can often oscillate. 
\end{itemize}

\section{(Restricted) Boltzmann Machines}

Having seen one instance of our new strategy (points to probabilities) and its limitations, we might wonder what alternative methods can be employed to construct deep generative models. Recall that these are models which form a representation of an underlying (usually high-dimensional) distribution such that samples of our learned representation cannot be distinguished from the true distribution.
In their \href{https://www.nobelprize.org/prizes/physics/2024/press-release/}{Nobel-Prize-winning work}, Hinton and Hopfield showed that we can use a model inspired by statistical physics to solve this problem \cite{ackley1985learning, hopfield1985neural}! For a review on energy-based models, see \cite{DawidLeCun2024, Carbone2024}.

Consider the simple case where the distribution we wish to model has binary random variables, $p(\mathbf{z}): \{0,1\}^N \rightarrow \mathbb{R}$, of length $N$. In statistical physics, this corresponds to the configuration probability of a micro-state of a given system of $N$ spins. Motivated by this connection, we can construct a probability distribution for configurations (microstates) in our ensemble by assigning an energy to each configuration, and computing the Boltzmann distribution. To that end, let
\begin{equation}
    \label{eq:unrestricted_boltzmann_energy}
    E(\mathbf{z}) = -\sum_j h_j z_j - \sum_{jk} W_{jk} z_k z_j
\end{equation}
be the energy of a given configuration $\mathbf{z} = (z_1,\ldots,z_N)$. This is simply an Ising-like model energy where spins experience a local transverse field $h_j$. However, coupling here is between any two spins, unlike the Ising model which only connects nearest beighbours. Represented graphically in the language of deep learning, an energy function like Eq.~(\ref{eq:unrestricted_boltzmann_energy}) is a fully connected neural network where each node represents a site, and the edges that connect them represent the interaction strength. Given an energy function, we can now compute the probability of observing a configuration with energy $E$ via
\begin{equation}
    \label{eq:probability_rbm}
    p(\mathbf{z}) = \frac{1}{Z} e^{- E(\mathbf{z})},
\end{equation}
where 
\begin{equation}
    \label{eq:partition_fn}
    Z = \sum_{\mathbf{z}} E(\mathbf{z}),
\end{equation}
is the usual partition function. By defining variational parameters $\theta = \{h_j, W_{jk}\}$, we have a tuneable energy $E(\mathbf{z})$ that can be trained from samples of a target distribution via Maximum Likelihood Estimation (see \hyperlink{box:MLE}{Box 3.1}). In the language of neural networks, we can see the coupling $W_{jk}$ as summing over the edges of the network for a given input $\mathbf{z}$, and the transverse field $h_j$ as a bias term. The exponential function in Eq.~\ref{eq:probability_rbm} is then nothing more than a softmax function. Explicitly, the exponential map acts as our activation function, and the partition function, $Z$, renormalises so that the output $p(\mathbf{z})$ is indeed a proper probability distribution. This defines the working principles of a \textit{Boltzmann Machine}, whose training we will discuss after an important caveat.

\begin{figure}[t!]
    \centering
    \begin{subfigure}[b]{0.55\textwidth} 
        \includegraphics[angle = 90, width=\textwidth]{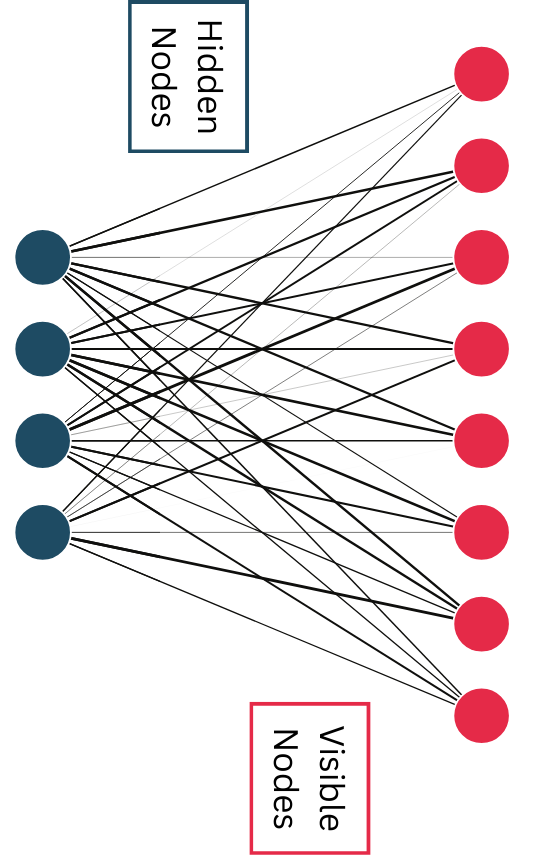} 
        \caption{Restricted Bolztmann Machine}
        \label{fig:RBM}
    \end{subfigure}
    \hfill
    \begin{subfigure}[b]{0.35\textwidth} 
        \centering
        \includegraphics[width=\textwidth]{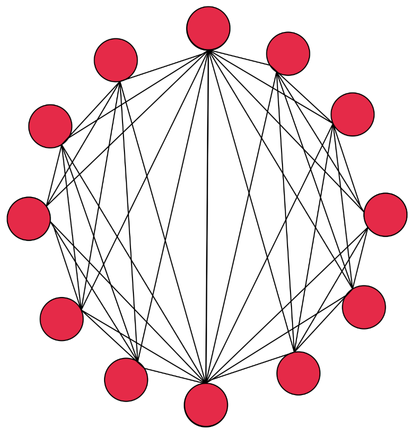} 
        \caption{Boltzmann Machine}
        \label{fig:BM}
    \end{subfigure}
    
    \caption{Two types of Boltzmann machine architectures. In the restricted Boltzmann machine we only sample from the visible (red) nodes, whereas we sample all nodes in an unrestricted Boltzmann machine. Boltzmann machines are generative in the sense that they receive no input, we use the energy based soft-max function to sample from their ``energy'' distribution.}
    \label{fig:RBM_main_figure}
\end{figure}

Notice that so far, we have placed no restrictions on the coupling matrix $W_{jk}$ despite the fact that our model is physically inspired. As things stand, we have a coupling matrix that enables arbitrary connections between our nodes. This comes with a significant training overhead; we have to train an all-to-all connected network in order to realise our target distribution. This is where our defining restriction comes in (hence the name Restricted Boltzmann Machine). We can partition the nodes in a Boltzmann machine into \textit{visible} and \textit{hidden} nodes, as shown in Fig.~\ref{fig:RBM}.

By demanding that visible nodes can only connect to hidden ones (and vice-versa) we can simplify a Boltzmann machine into a bipartite structure with fewer trainable parameters. We sample only from the visible nodes, and the hidden layers act as a sort of intermediary between the visible layers, allowing for coupling complexity. This is advantageous because it means a Reduced Boltzmann Machine remains \textit{expressive}, that is, it has a high capacity to represent high-dimensional probability distributions. 

To see how we can partition a Boltzmann Machine into a Restricted Boltzmann Machine, let $\mathbf{h} \in \{0,1\}^*$ represent hidden nodes, and $\mathbf{v} \in \{0,1\}^*$ represent visible nodes, and $\{0,1\}^*$ simply means a binary vector of arbitrary length. Then the energy equation from Eq.~(\ref{eq:unrestricted_boltzmann_energy}) becomes 
\begin{equation}
    E(\mathbf{h},\mathbf{v}) = -\sum_j a_j v_j -\sum_j b_j h_j  - \sum_{jk} W_{jk} h_k v_j,
\end{equation}
allowing us to apply a softmax (Eq.~(\ref{eq:probability_rbm})) as before. In an RBM, the variational parameters are therefore $\theta = \{a_j, b_j, W_{jk}\}$.  Thanks to this restricted architecture, the conditionals between visible and hidden layers factor beautifully. Let’s derive this step by step.

We start recalling meaning of conditional probability -  conditional probability formalises how our belief about one random variable updates when another is observed - this is given by \emph{Bayes’ theorem}: 
starting from the fundamental product rule
\begin{equation}
    P(\mathbf{v},\mathbf{h})=P(\mathbf{h})\,P(\mathbf{v}\mid \mathbf{h}),
\end{equation}
we isolate the posterior distribution of interest by normalising with the marginal \(P(\mathbf{v})\).
This delivers the conditional–probability form of Bayes’ rule:
\begin{equation}
    P(\mathbf{h} \mid \mathbf{v})=\frac{P(\mathbf{v},\mathbf{h})}{P(\mathbf{v})}.
\end{equation}
With this bridge from joint models to posterior inferences in place, we can now proceed to compute \(P(h\!\mid v)\) for the problem at hand. Next, let’s isolate the part of the energy that depends on \( h_j \),
\begin{equation}
E(\mathbf{v}, \mathbf{h}) = -\sum_i a_i v_i - \sum_j h_j \left( b_j + \sum_i v_i W_{ij} \right)
\end{equation}
Thus,
\begin{equation}
P(\mathbf{h} \mid \mathbf{v}) \propto \prod_j \exp\left( h_j \left( b_j + \sum_i v_i W_{ij} \right) \right)
\end{equation}
Now notice since \( h_j \in \{0,1\} \), the above is just a Bernoulli distribution,
\begin{equation}
P(h_j = 1 \mid \mathbf{v}) = \sigma\left( b_j + \sum_i W_{ij} v_i \right)
\end{equation}
\begin{equation}
P(h_j = 0 \mid \mathbf{v}) = 1 - \sigma\left( b_j + \sum_i W_{ij} v_i \right)
\end{equation}

\begin{equation}
P(\mathbf{h} \mid \mathbf{v}) = \prod_j P(h_j \mid \mathbf{v}) \quad \text{(fully factorized)}
\end{equation}

Similarly, we can isolate the part of the energy that depends on \( v_i \), and arrive at,
\begin{equation}
P(v_i = 1 \mid \mathbf{h}) = \sigma\left( a_i + \sum_j W_{ij} h_j \right)
\end{equation}
\begin{equation}
P(\mathbf{v} \mid \mathbf{h}) = \prod_i P(v_i \mid \mathbf{h})
\end{equation}
Now we want to compute the probability of a visible configuration,
\begin{equation}
P(\mathbf{v}) = \sum_{\mathbf{h}} P(\mathbf{v}, \mathbf{h})
\end{equation}
Substituting the energy function,
\begin{equation}
P(\mathbf{v}) = \frac{1}{Z} \sum_{\mathbf{h}} \exp\left( \sum_i a_i v_i + \sum_j h_j \left( b_j + \sum_i v_i W_{ij} \right) \right)
\end{equation}
We can pull out the part independent of \( \mathbf{h} \),
\begin{equation}
= \frac{1}{Z} \exp\left( \sum_i a_i v_i \right) \prod_{j=1}^m \sum_{h_j = 0}^1 \exp\left( h_j \left( b_j + \sum_i v_i W_{ij} \right) \right)
\end{equation}
Each sum over \( h_j \) evaluates to,
\begin{equation}
\sum_{h_j = 0}^1 \exp(h_j x_j) = 1 + e^{x_j}
\end{equation}
So we get,
\begin{equation}
P(\mathbf{v}) = \frac{1}{Z} \exp\left( \sum_i a_i v_i \right) \prod_{j=1}^m \left( 1 + \exp\left( b_j + \sum_i v_i W_{ij} \right) \right)
\end{equation}
This is the key formula that we can employ when we train based on maximum likelihoods. What remains is for us to find an efficient training scheme to tune our variational parameters.

Recall that MLE (see \hyperlink{box:MLE}{Box 3.1}) aims to maximise the log-likelihood of observing samples from the true distribution. Therefore, we can use negative log-likelihood as our loss function, since minimising this will maximise the probability of samples belonging to the true distribution. For a batch of $m$ samples of input data, $\mathbf{x}$, let
\begin{equation}
    \label{eq:log_likelihood_loss}
    L(\theta) = -\sum_{j = 1}^m \log p_j(\mathbf{x}_j),
\end{equation}
where $p_j(\mathbf{x}_j)$ is the probability of observing a binary vector, $\mathbf{x}_j$, calculated via (a forward pass of) Eq.~(\ref{eq:probability_rbm}). Notice however, that Eq~(\ref{eq:probability_rbm}) contains a sum over \textit{all} possible configurations. This means the sum is over $2^N$ terms for $N$ visible nodes. To get around this, we need a new trick based on Monte Carlo Markov chains called Gibbs Sampling, see Box.~\hyperlink{box:MCMC}{3.5}.

To understand why Gibbs sampling will help us, it is instructive to write down the log-likelihood loss function's (Eq.~(\ref{eq:log_likelihood_loss})) gradient as we will need this to perform updates via gradient descent. Recall that the variational parameters here are $\theta = \{a_j, b_j, W_{ij}\}$. Hence, 
\begin{equation}
\begin{split}
     \frac{\partial L}{\partial \theta} &= -\sum_{k = 1}^m \frac{\partial \log p(\mathbf{x}_k)}{\partial \theta} \\
     &= -\sum_{k = 1}^m \left( - \frac{\partial}{\partial \theta} \log (Z) 
        - \frac{\partial}{\partial \theta} E(\mathbf{x}_k)
    \right) \\
\end{split}
\end{equation}
If we focus for a second on the gradient with respect to the matrix-entries $W_{ij}$, we may write,
\begin{equation}
\frac{\partial \log P(\mathbf{v})}{\partial W_{ij}} = \mathbb{E}_{P(h \mid \mathbf{v})}[v_i h_j] - \mathbb{E}_{P(\mathbf{v}, \mathbf{h})}[v_i h_j],
\end{equation}
for which we identify a positive phase \( \mathbb{E}_{P(h|\mathbf{v})}[v_i h_j] \), driven by data, and the negative phase \( \mathbb{E}_{P(v,h)}[v_i h_j] \), which is an expectation over the model's joint distribution \( P(\mathbf{v}, \mathbf{h}) \) and serves to reduce associations that are not supported by the data. 

While the positive phase is tractable due to the factorization of \( P(\mathbf{h} \mid \mathbf{v}) \) which we derived above, the negative phase is intractable for large systems, as it involves a sum over all possible visible and hidden configurations. So in its current form, the gradient of our loss requires differentiating through the partition function, $Z$, from Eq.~(\ref{eq:partition_fn}).   

To address this, Hinton proposed the \textit{Contrastive Divergence} (CD) algorithm as an efficient approximation.
The idea of CD is to replace the expectation over the model distribution with samples generated from a short Markov chain initialized at the data. In the simplest version, known as CD-1, the algorithm proceeds as follows. Given a data point \( \mathbf{v}^{(0)} \), we first sample the hidden units from the conditional distribution:
\begin{equation}
\mathbf{h}^{(0)} \sim P(\mathbf{h} \mid \mathbf{v}^{(0)}).
\end{equation}
Using this sample, we then generate a reconstruction of the visible units:
\begin{equation}
\mathbf{v}^{(1)} \sim P(\mathbf{v} \mid \mathbf{h}^{(0)}),
\end{equation}
followed by a new hidden sample:
\begin{equation}
\mathbf{h}^{(1)} \sim P(\mathbf{h} \mid \mathbf{v}^{(1)}).
\end{equation}
These steps form a single Gibbs sampling step. The weight update is then given by:
\begin{equation}
\Delta W_{ij} \propto v^{(0)}_i h^{(0)}_j - v^{(1)}_i h^{(1)}_j.
\end{equation}
This provides an approximation to the true gradient, where the negative phase expectation is replaced by the statistics of the one-step reconstruction.
Despite its simplicity and bias, CD-1 works surprisingly well in practice. It captures the idea that the model should increase the likelihood of the data while reducing the likelihood of nearby, incorrect reconstructions. Extensions such as CD-\(k\), where the chain is run for \( k > 1 \) steps, can improve accuracy at the cost of more computation.

Now we can see more clearly why Gibbs sampling is useful. It is merely an extension of CD-1 to include many steps (see \hyperlink{box:MCMC}{Box 3.5}), which makes for a more unbiased estimator for the negative phase. Hence, instead of trying to get multiple samples and differentiate through an exponentially large partition function, we can do Gibbs sampling with $t = r$ steps, allowing us to estimate the costly sum.

We now have all the ingredients to create and train RBMs. This is summarised Algorithm~\ref{alg:RBM_trainstep}, allowing you to construct and train your own RBM. 
\begin{algorithm}
\caption{Contrastive Divergence (RBM) Train Step}\label{alg:RBM_trainstep}
\begin{algorithmic}[1]
\State \textbf{Input:} Dataset batch $S = \{\mathbf{x}_1, \mathbf{x}_2, \dots, \mathbf{x}_n\}$, learning rate $\alpha$, and current RBM weights $\theta = \{a, b, J\}$
\For{$x \in S$ do:}
    \State Calculate $P_{\theta}^{\text{rbm}}(h_j = 1 | x)$
    \State Perform $r$-step Gibbs sampling starting from $x$ to obtain a sample $v'$
    \State Compute the negative phase by calculating $P_{\theta}^{\text{rbm}}(h_j = 1 | v')$ \Comment{See \hyperlink{box:MCMC}{Box 3.5} for generating a Gibbs sample}
    \State Calculate $\Delta W_{ij} = x_i P_{\theta}^{\text{rbm}}(h_j = 1 | x) - v_i' P_{\theta}^{\text{rbm}}(h_j = 1 | v')$
    \State Calculate $\Delta a_i = (x_i - v_i')$
    \State Calculate $\Delta b_j = (P_{\theta}^{\text{rbm}}(h_j = 1 | x) - P_{\theta}^{\text{rbm}}(h_j = 1 | v'))$
\EndFor
\State Update $W_{ij} \gets W_{ij} + \eta \Delta W_{ij}$ (and similarly update $a_i$ and $b_j$)
\end{algorithmic}
\end{algorithm}

\begin{figure}[H]
    \centering
    \begin{mybox}[\hypertarget{box:MCMC}{Box 3.4: Key Ingredients in Restricted Boltzmann Machines (RBMs)}]

    Restricted Boltzmann Machines (RBMs) are simple generative models that learn a probability distribution over a set of binary variables. They consist of two layers:
\begin{itemize}
  \item A \textbf{visible layer} \( \mathbf{v} = (v_1, \dots, v_n) \), which represents observed data.
  \item A \textbf{hidden layer} \( \mathbf{h} = (h_1, \dots, h_m) \), which learns latent (hidden) features.
\end{itemize}
The word \emph{restricted} means that there are no connections within a layer; visible units don’t interact with each other, and neither do hidden units. This simplifies the mathematics significantly and leads to factorized conditional probabilities.
The behaviour of an RBM is governed by an \textbf{energy function}, defined for a joint configuration \( (\mathbf{v}, \mathbf{h}) \) as,
\begin{equation}
E(\mathbf{v}, \mathbf{h}) = -\sum_{i=1}^n a_i v_i - \sum_{j=1}^m b_j h_j - \sum_{i=1}^n \sum_{j=1}^m v_i W_{ij} h_j
\end{equation}
where \( a_i \) and \( b_j \) are bias terms for visible and hidden units respectively, and \( W_{ij} \) is the weight matrix connecting visible unit \( i \) to hidden unit \( j \). From this energy, we define the \textbf{joint probability} of a visible and hidden configuration as:
\begin{equation}
P(\mathbf{v}, \mathbf{h}) = \frac{1}{Z} \exp\left(-E(\mathbf{v}, \mathbf{h})\right)
\end{equation}
where \( Z \) is standard the partition function, \(Z = \sum_{\mathbf{v}} \sum_{\mathbf{h}} \exp\left(-E(\mathbf{v}, \mathbf{h})\right)\).
Given a dataset \( \{\mathbf{v}^{(n)}\}_{n=1}^N \), we train the RBM by maximizing the log-likelihood,
\begin{equation}
\log P(\mathbf{v}) = \log \sum_{\mathbf{h}} e^{-E(\mathbf{v}, \mathbf{h})} - \log Z
\end{equation}
Taking the gradient with respect to the weight \( W_{ij} \), we get:
\begin{equation}
\frac{\partial \log P(\mathbf{v})}{\partial W_{ij}} = \mathbb{E}_{P(h \mid \mathbf{v})}[v_i h_j] - \mathbb{E}_{P(\mathbf{v}, \mathbf{h})}[v_i h_j],
\end{equation}
for which we identify the e \textbf{positive phase} \( \mathbb{E}_{P(h|\mathbf{v})}[v_i h_j] \), driven by data, and the \textbf{negative phase} \( \mathbb{E}_{P(v,h)}[v_i h_j] \), which is an expectation over the model's joint distribution \( P(\mathbf{v}, \mathbf{h}) \) and serves to reduce associations that are not supported by the data. We can estimate this gradient with contrastive divergence and Gibbs sampling.
    \end{mybox}
\end{figure}

\begin{figure}[h]
    \centering
    \begin{mybox}[\hypertarget{box:MCMC}{Box 3.4: Gibbs Sampling and Monte Carlo Markov Chains}]
    Gibbs Sampling is a Markov Chain Monte Carlo (MCMC) algorithm used to generate samples from a joint probability $p(\mathbf{x}) = p(x_1,\ldots,x_N)$ distribution when direct sampling is challenging (e.g. when our joint distribution is high-dimensional!). It does this by iteratively sampling each variable from its conditional (posterior) distribution given the current values of the other variables. Let $p(x_i | x_{-i})$, denote the probability that the $i^{th}$ component of $\mathbf{x}$ takes a value $x_i$, conditioned on all other components of $\mathbf{x}$ except $i$. Then we can approach the target distribution as by iterating the $t$ (for time) index in the following,
    \begin{enumerate}
        \item Sample $x_1^{(t)}$ from $p\left(x_1|x_2^{(t-1)}, x_3^{(t-1)},\ldots,x_N^{(t-1)}\right)$
        \item Sample $x_2^{(t)}$ from $p\left(x_2|x_1^{(t-1)}, x_3^{(t-1)},\ldots,x_N^{(t-1)}\right)$
        \item Repeat until $x_N^t$ is sampled
        \item $t \gets t + 1$
    \end{enumerate}
    All we require is some initial  (preferably random) seed $\mathbf{x}^0$ to kick the process off.
    \end{mybox}
\end{figure}

\subsubsection*{Example: Approximating a continuous distribution with an RBM.}

As a concrete, low–dimensional demonstration, we train an RBM to model a smooth univariate distribution \(p(x)\) supported on \([0,1]\).  

We draw a finite dataset $\{x^{(n)}\}_{n=1}^N$, and present it to the RBM by \emph{quantizing} each real \(x\) into an \(L\)-bit binary vector of visible units, i.e. we  map the real numbers $x \in [0,1]$ to the $L$-bit string $\mathbf{v}(x) \in \{0,1\}^{L}$. We divide the distribution domain \([0,1]\), into $2^L$ half-open bins of width $\Delta=2^{-L}$:
\begin{equation}
B_i \;=\; [\,i\Delta,\,(i+1)\Delta\,), \qquad i=0,1,\dots,2^{L}-1.
\end{equation}
Then, we can define a deterministic encoding from real numbers  \(x \in [0,1]\) into \(L\)-bit vectors by
\begin{equation}
i(x) \;=\; \Bigl\lfloor \tfrac{x'}{\Delta} \Bigr\rfloor\in\{0,\dots,2^{L}-1\},
\qquad
\mathbf{v}(x) \;=\; \mathrm{bin}_L\bigl(i(x)\bigr)\in\{0,1\}^{L},
\end{equation}
where \(x' = \min\{x,\,1-\Delta\}\) clips the endpoint and
\(\mathrm{bin}_L(\cdot)\) returns the \(L\) bit binary expansion of the integer argument
(least significant bit convention unless stated otherwise).  
This induces a discrete target \textit{probability mass function} (pmf) on indices,
\begin{equation}
p^{\star}(i) \;=\; \int_{B_i} p(x)\,dx, \qquad i=0,\dots,2^{L}-1,
\end{equation}
i.e., the probability mass of \(p\) captured by each bin.  
The RBM now has \(n_{\mathrm{vis}}=L\) visible units and is trained to approximate \(q^{\star}\) on \(\{0,1\}^{L}\).

Let \(\theta=\{a,b,W\}\) denote RBM parameters and \(P_{\theta}(\mathbf{v})\) its model pmf on \(\{0,1\}^{L}\).
Maximum likelihood seeks to minimize \(\mathrm{KL}\!\left(p^{\star}\,\|\,P_{\theta}\right)\),
but exact gradients involve the partition function.
Following the discussion above, we use Gibbs transitions and Contrastive Divergence (CD-\(k\)) to form a tractable, biased gradient estimator that increases data-driven associations (positive phase) and decreases model-only associations (negative phase).
Under mild conditions the Markov chain over \((\mathbf{v},\mathbf{h})\) has a stationary distribution, and longer chains reduce the bias in the negative phase.

\begin{figure}[t!]
    \centering
    \includegraphics[width=0.99\linewidth]{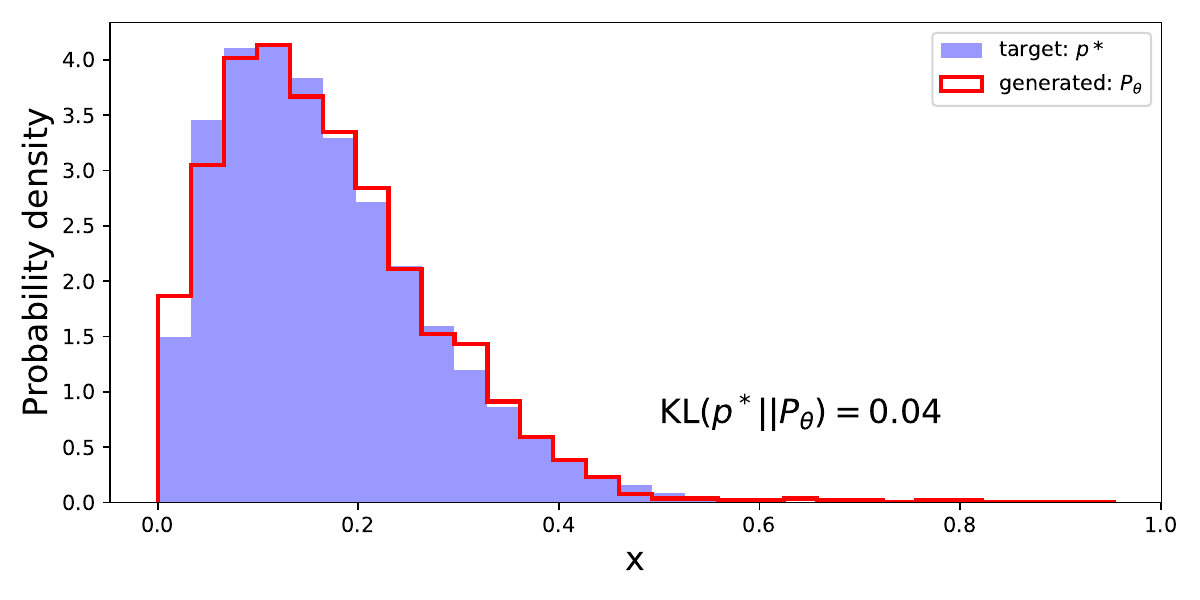}
    \caption{Comparison between the target density \(p^{\ast}\) (filled histogram) and the learned RBM model \(P_{\theta}\) (red step) on \([0,1]\). The distasnce between probability distributions can be quantified with the Kullback-Leibrer divergenec \(D_{\mathrm{KL}}(p^{\ast}\,\|\,P_{\theta})=0.04\) computed on the same discretization. The close overlap indicates that the RBM captures the continuous density up to the \(L\)-bit quantization.}
    \label{fig:RBM_example}
\end{figure}

After training, samples \(\mathbf{v}\sim P_{\theta}\) are mapped back to \([0,1]\) by reconstructing the index
\(i(\mathbf{v})=\sum_{\ell=0}^{L-1} v_\ell\,2^{\ell}\) and reporting the bin midpoint $\tilde{x}(\mathbf{v}) \;=\; \bigl(i(\mathbf{v})+\tfrac{1}{2}\bigr)\Delta$.
Thus the learned model defines a piecewise-constant density estimator on \([0,1]\) whose cell probabilities are given by the RBM’s discrete pmf.  To quantitatively compare target \(q^{\star}\) with learned \(P_{\theta}\) we calculate Kullback-Leibrer divergence, see Fig.\ref{fig:RBM_example}.

\section{Normalising Flows}

At the heart of normalising flows (NFs) is the concept of smoothly warping a simple distribution into a more intricate one by applying a chain of reversible transformations \cite{papamakarios2021normalizing}. More formally, NFs are a framework that transforms a simple probability distribution into a complicated one using a sequence of invertible, differentiable functions. We can use NFs as an estimation tool for the underlying probability density of a process, usually referred to as \textit{density estimation}.
Alternatively, we can use NFs for generative tasks, since they transform a simple ansatz distribution into one which creates data that looks like it was drawn from a given dataset, like in the RBM we saw above. Here however, we instead wish to directly modify the transformation between distributions, rather than modify the prior alone. In this sense, NFs are a very powerful technique as they enable us to do both. To see how we can do this, it is instructive to first recall how probability distributions transform under a change of variables.

Let $\mathbf{z} \in \mathbb{R}^d$ be a random variable with probability density $p_{\mathcal{Z}}(\mathbf{z}): \mathbb{R}^d \rightarrow \mathbb{R}$ under distribution $\mathbf{z} \sim \mathcal{Z}$. Consider transforming $\mathbf{z}$ by an invertible function $f: \mathcal{Z} \rightarrow \mathcal{Y}$,
\begin{equation}
    \mathbf{y} = f(\mathbf{z}),
\end{equation}
where $\mathbf{y} \sim \mathcal{Y} \in \mathbb{R}^d$ is the image of our transformation with distribution $\mathbf{y} \sim \mathcal{Y}$. Using the standard change of variables formula, we can compute the probability density $p_{\mathcal{Y}}(\mathbf{y})$,
\begin{equation}
    p_{\mathcal{Y}}(\mathbf{y}) = p_{\mathcal{Z}}(g(\mathbf{y})) |\det \, D g(\mathbf{y})|
    = p_{\mathcal{Z}}(g(\mathbf{y})) |\det \, Df(g(\mathbf{y}))|^{-1},
\end{equation}
where $g = f^{-1}$ is the inverse of $f$, and $D g(\mathbf{y}) = \frac{\partial g}{\partial \mathbf{y}} \in \mathbb{R}^{d \times d}$ is the Jacobian of the transformation $f$. In the language of transformations and differential geometry, we can think of $p_{\mathcal{Y}}(\mathbf{y})$ as the \textit{pushforward} density of $p_{\mathcal{Z}}(\mathbf{z})$, denoted by $f_* p_{\mathcal{Z}}$. In this way, $g = f^{-1}$ defines the \textit{pullback} direction. 

Whereas in the language of generative models, we can think of $f$ as the \textit{generative} direction from a simple base density to a final density.
That is, $f$ transforms simple (often noise!) distribution $\mathcal{Z}$ into a more intricate distribution $\mathcal{Y}$ by acting reversibly over probability density functions.  Then, $g = f^{-1}$ defines the \textit{normalising} direction, which takes a complicated distribution and transforms it so a simpler base distribution. Notice that in order to execute the transformations $f,\;g$, we require they be differentiable in order to compute their Jacobian. This means we require $f$ and $g$ to form a \textit{diffeomorphism}: \\

\textit{\textbf{Diffeomorphism} A function $f: \mathbb{R}^d \rightarrow \mathbb{R}^d$ is diffeomorphic when both $f$ and its inverse $g = f^{-1}$ are differentiable.} \\

Really, this is two conditions wrapped up as one: we require $f$ to have an inverse, \textit{and} that both $f$ and its inverse are differentiable. In fact, pretty much all generative models can be constructed from this perspective. If we think of $\mathcal{Z}$ and $\mathcal{Y}$ as measurable spaces\footnote{That is, the tuples $(\mathcal{Z}, \Sigma_{\mathcal{Z}})$ and $(\mathcal{Y}, \Sigma_{\mathcal{Y}})$ have well-defined sigma algebra which is Borel, and their measures are absolutely continuous with respect to the Lebesgue measure $\mu = p_{\mathcal{Z}}(\mathbf{z}) d\mathbf{z}$.}, and $f$ as a measurable mapping between them, then we can define a measure on $\mathcal{Y}$ as
\begin{equation}
    \mathbf{f}_* \mu(\mathbf{y}) = \mu(\mathbf{f}^{-1}(\mathbf{y})).
\end{equation}
From this perspective, data is simply a sample from the data-space distribution $\mathcal{Y}$ which we wish to infer. To learn $\mathcal{Y}$, we can start with a simple base distribution $\mathcal{Z}$ which is simpler to construct, and find a function $f: \mathcal{Z} \rightarrow \mathcal{Y}$ such that the induced measure is $\mu(\mathbf{y}) = f_{*} \mu(\mathbf{z})$. Then $f$ is the \textit{generator} of $\mathcal{Y}$ from a latent space $\mathcal{Z}$. What remains is to construct a method that can represent this function $f: \mathcal{Z} \rightarrow \mathcal{Y}$.

We can use neural networks for this task. We can motivate neural networks here thanks to the Universal Approximation Theorem (UAT). Recall that this theorem says we can approximate any arbitrary function given a neural network with sufficient depth and training. Combining this with our shift in perspective from points to probabilities, we can see how the UAT allows neural networks to approximate an arbitrary transformation $f: \mathcal{Z} \rightarrow \mathcal{Y}$ over probability distributions. You can find a formal proof of this fact in \textit{A Universal Approximation Theorem of Deep Neural
Networks for Expressing Probability Distributions} by Lu et. al, \cite{lu2020universal}. Just like in point-wise neural networks, we can compose transformations together in order to realise a complicated, expressive one. The base units that we can compose together are what we refer to as \textit{normalising flows}: \\

\textit{\textbf{Normalising Flows} - any bijective function where we can efficiently compute, invert and which has a Jacobian determinant which is efficient to compute.}\\

To see how this lets us compose functions together, consider applying $f$ as a composition,
\begin{equation}
    f = f_N \circ f_{N - 1} \circ \ldots \circ f_1,
\end{equation}
where each $f_j$ is a normalising flow. Then it can be shown that $f$ is bijective, with the inverse, $g$, defined as
\begin{equation}
    g = g_1 \circ g_2 \circ \ldots \circ g_N.
\end{equation}
This makes the overall Jacobian of our transformation have determinant,
\begin{equation}
    \det \, Dg(\mathbf{y}) =  \prod_{i=1}^{N} \det \, Dg_i(\mathbf{x}_i),
\end{equation}
where $\mathbf{x}_i = f_1 \circ \ldots f_i(\mathbf{z}) = g_{i + 1} \circ g_N (\mathbf{y})$ is the value $i^{\text{th}}$ intermediate flow, and $Dg_i(\mathbf{x}_i) = \frac{\partial g_i}{\partial x_i}$ is the Jacobian of $g_i$. This allows us to compose bijective functions between probability distributions, opening the avenue for neural network architectures based on layers and composition. Note that this comes with an important caveat - in order to make $f$ trainable with back-propagation, we must be able to reparameterise it! This is because the function $f$ must still be a differentiable computation graph, so if we sample from its output distribution to provide feedback via some loss function, we have to causally connect the parameters that construct $f$ to the samples that were generated from it. See Box.~\hyperref{box:reparam}{3.3} for more detail. \\

With a formal definition out of the way, let's now explore how we can use NFs for density estimation, as this has clear and well-defined use cases in quantum science (which we will see in the next chapter!). Since we have a clear composition rule for NFs, it is sufficient to understand how to do this with a \textit{single} transformation, and combine it with our composition rule to make more powerful models.

Let $\mathbf{y} = f_{\theta}(\mathbf{z})$ be the generative direction for a model of a single flow, with variational parameters $\theta$, with corresponding inverse $\mathbf{z} = g(\mathbf{y})$. That is, $g = f^{-1}$. Let's also assume we are given a dataset, $\mathcal{D} = \{\mathbf{y}_j\}_{j = 1}^M$, of $M$ observations of $\mathcal{Y}$, as this is usually accessible to us in practise\footnote{We saw this in the RBM, when the input dataset was a batch $S = \{\mathbf{x}_j\}_{j = 1}^n$ of samples from the \textit{true} distribution to be estimated.}. We can now use maximum likelihood estimation (MLE - see \hyperlink{box:MLE}{Box 2}) to find the optimal parameters $\theta^*$ that \textit{maximise} the likelihood of observing our data $\mathcal{D}$. Notice that the log-likelihood in this case can be written as
\begin{equation}
\begin{split}
    \log p(\mathcal{D} | \theta) &= \sum_{i=1}^{M} \log p_{\mathbf{Y}}(\mathbf{y}_i | \theta) \\
&= \sum_{i=1}^{M} \log p_{\mathbf{Z}}(g(\mathbf{y}_i | \theta)) + \log \left| \det Dg(\mathbf{y}_i | \theta) \right|,
\end{split}
\end{equation}
thanks to our usual log-rules. The first term, $\log p_{\mathbf{Z}}(g(\mathbf{y}_i | \theta))$, is the likelihood of the transformed data under our base (simpler) distribution, while the second term, $\log \left| \det Dg(\mathbf{y}_i | \theta) \right|$, corrects for the change in volume induced by the transformation. In this form, we can see the opportunity for extra expressivity; nothing stops us from constructing the base distribution out of variational parameters too. For example, a 1D Gaussian base distribution could have variational parameters $\{\mu, \sigma\}$. To that end, let $\Theta = (\theta, \phi)$ be a tuple of variational parameters for the transformation and base distribution respectively. Then we can write the log-likelihood as
\begin{equation}
\begin{split}
\log p(\mathcal{D} | \Theta) &= \sum_{i=1}^{M} \log p_{\mathbf{Y}}(\mathbf{y}_i | \Theta) \\
&= \sum_{i=1}^{M} \log p_{\mathbf{Z}}(g(\mathbf{y}_i | \theta) | \phi) + \log \left| \det Dg(\mathbf{y}_i | \theta) \right|.
\end{split}
\end{equation}
Notice now that the first term in the sum is conditioned on the base distribution's parameters $\phi$ whilst the second is not. This is because the structure of the transformation is independent of the values of the base distribution's variational parameters. Combined with the reparametrisation trick that we saw in VAEs, we are now in a position to train a NF model with gradient-based methods, allowing you to create and train your own NF model for density estimation. Pseudocode for this is provided in Algorithm~\ref{alg:NF_density}. For a more comprehensive introduction to NF models, refer to \cite{kobyzev2020normalizing}. This will cover more details on the various neural architectures that come from different ways of coupling together variables in a high-dimensional probability distribution.

\begin{algorithm}
\caption{Density Estimation with Normalizing Flows}
\label{alg:NF_density}
\begin{algorithmic}[1]
\State Dataset (batched) $\mathcal{D} = \{x_i\}$, number of iterations $N$, learning rate $\alpha$
\State Initialize parameters $\theta$ of the normalizing flow
\For{$\mathrm{iteration} = 1$ to $N$}
    \For{each data point $x$ in batch $B \subset \mathcal{D}$}
        \State $z = f_\theta^{-1}(x)$  \Comment{Map $x$ to latent space}
        \State $\log p_x = \log p_Z(z) + \log \left| \det \frac{\partial f_\theta^{-1}}{\partial x} \right|$  \Comment{Compute log-likelihood}
    \EndFor
    \State $\mathcal{L} = - \frac{1}{|B|} \sum_{x \in B} \log p_x$  \Comment{Negative log-likelihood}
    \State $\theta = \theta - \alpha \nabla_\theta \mathcal{L}$  \Comment{Update parameters}
\EndFor
\end{algorithmic}
\end{algorithm}

\begin{figure}[h]
    \centering
        \begin{tikzpicture}
            \definecolor{customcolor}{HTML}{FCA082}
            \definecolor{customcolor2}{HTML}{E32F26}
            \definecolor{customcolor3}{HTML}{67000D}
            \definecolor{customcolor4}{HTML}{7CC7AC}
            \node[anchor=north east, inner sep=0, rotate=90] (image) at (0,0) {\includegraphics[width=0.4\textwidth]{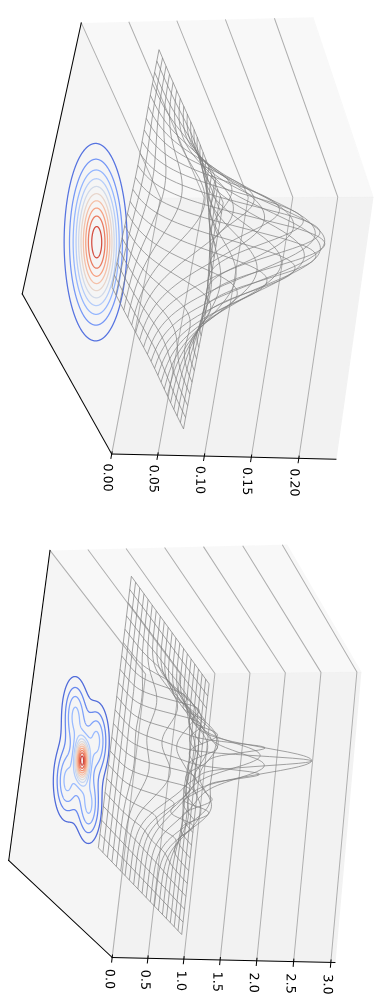}};

            \begin{scope}[shift={(0,0)}]
                \draw[thick, ->] (3,0.5) to[out=45, in=135] (7.5,0.5);
                
                \node at (5.25, 1.75) {$\mathbf{y} = f_{\theta}(\mathbf{z})$};
                
                \node at (5.25, 0.75) {Generative};
                
                \draw[thick, ->] (7.5,-5) to[out=-135, in=-45] (3,-5);
                
                \node at (5.25, -6.25) {$\mathbf{z} = f^{-1}(\mathbf{y})$};

                \node at (5.25, -5.25) {Normalising};
                
                \node[rotate = 90] at (0, -2.0) {$p_{\mathcal{Z}}(\mathbf{z})$};
                
                \node[rotate = 90] at (6.6, -2.0) {$p_{\mathcal{Y}}(\mathbf{y}|\theta)$};

                \node at (1.5, -4.5) {$z_1$};

                \node at (5.0, -4.2) {$z_2$};

                \node at (8.7, -4.6) {$y_1$};

                \node at (11.8, -4.5) {$y_2$};
            \end{scope}
        \end{tikzpicture}
    \caption{Visualisation of a normalising flow model over two-variable density functions. Here we start with a base distribution (left) $p_{\mathcal{Z}}(\mathbf{z})$, which can be mapped in the generative direction (top arrow), $\mathbf{y} = f_{\theta}(\mathbf{z})$, giving the new distribution, $p_{\mathcal{Y}}(\mathbf{y}|\theta)$. Since NFs are invertible, we can map back to the base distribution in the normalising direction (bottom arrow), $\mathbf{z} = f^{-1}(\mathbf{y})$.}
    \label{fig:GD_advanced_Ch3}
\end{figure}
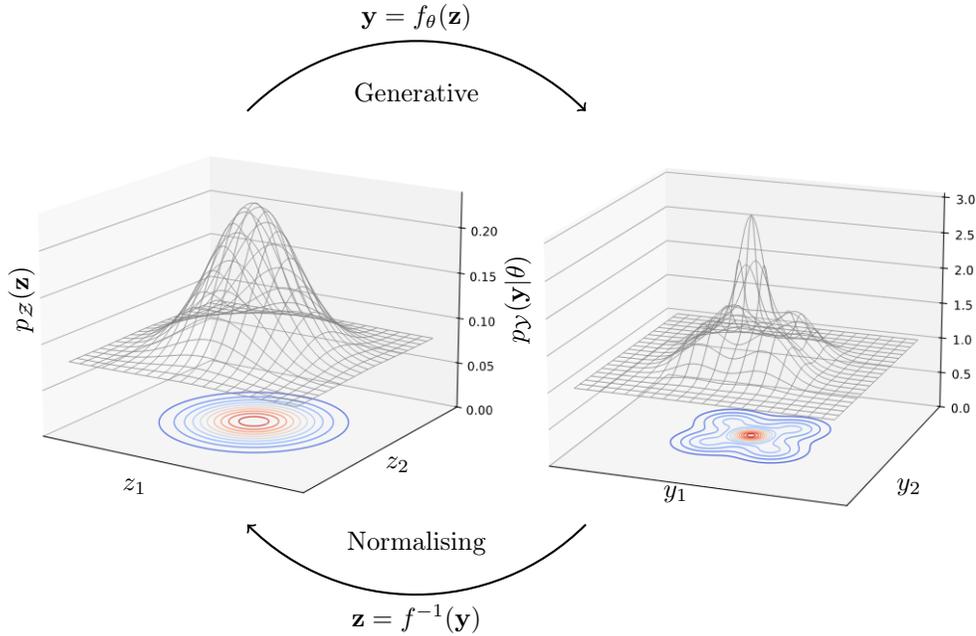

\section{Symmetries in Deep Learning}

What makes physics so powerful? Some might argue it is its conservation laws. These are, in some sense, the crowning achievements that different theories, from Newton and Maxwell, to Einstein and Schr\"odinger, have offered us. They allow us to make rigorous predictions about the future, and meaningful inference about the past. Given the diverse settings in which conservation laws have arisen, we might wonder if there is something that unifies them. 

As discovered and proven by Amalie Emmy Noether in 1918, there \textit{is}  a common structure that unifies all conservation laws. That structure is \textit{symmetry}. In brief, Noether's Theorem states that every symmetry present in a physical system must give rise to a conservation law, and vice versa. Although this may sound like a rather general statement, many consider it to be the \textit{fundamental theorem of physics}. It sets out the theoretical framework for constructing \textit{any} physical theory, based either on intuition for which quantities we might expect to be conserved or based on symmetries we have observed.

In the rest of this text, we will focus on how to use the deep learning methods covered so far in physical settings. It is therefore crucial for us to be able to enforce conservation laws and other types of physical intuition into our neural network models. The art of successfully doing this is known in the literature as giving a model an \textit{inductive bias}, see \hyperlink{box:inductive_bias}{Box 3.6}.

\begin{figure}[h]
    \centering
    \begin{mybox}[\hypertarget{box:inductive_bias}{Box 3.6: Inductive Bias}]
    An inductive bias is a way of trying to help a model learn better. Specifically, giving a model an inductive bias involves trying to encode everything you already know about a given task or problem. Generally, there are three ways to give an inductive bias to a model:
    \begin{itemize}
        \item With the loss function - we actually saw one example of this already in \hyperlink{box:underfitting_overfitting}{Box~2.12} when we were trying to prevent over and underfitting. By \textit{regularizing} our loss function, we gave the model an inductive bias - namely we encoded the fact that we don't want the model to simply memorise (or overfit) to training data.
        \item With the neural network architecture - we can change the shape, activation functions, and other hyperparameters to better match a given situation as we shall see in the rest of this section.
        \item With the training data - again we already saw one example of this in \hyperlink{box:one_hot_encoding}{Box~2.7}, where we used one-hot encoding to embed distinct classes into the model. 
    \end{itemize}
    \end{mybox}
\end{figure}

Given the result from Noether's Theorem, we will now seek to understand how to give an inductive bias to a model in the form of \textit{symmetry}. This will allow our models to respect the conservation laws that are cherished so dearly by physicists.

\subsection[Physics Informed Neural Networks]{Physics Informed Neural Networks - Conservation via Loss Function}

PINNs stand for Physics Informed Neural Networks \cite{raissi2019physics}. As their name suggests, PINNs have an inductive bias which helps them stay in the realm of physical solutions.  Instead of relying solely on training data to learn the optimal variational parameters of a model, PINNs also incorporate equations of motion as part of the loss function. This ensures that the neural network's predictions are tuned to fit both the training data and known physical principles.

For those familiar with statistical physics (or the regularization section from Chapter 2), we can incorporate equations of motions into a loss function using \textit{Lagrange Multipliers}. To see how this works, let's explore a simple worked example. Consider the 1d damped harmonic motion of a spring, whose equation of motion reads
\begin{equation}
m \frac{d^2x}{dt^2} + c \frac{dx}{dt} + kx = 0,
\label{eq:damped_oscillator_EOM}
\end{equation}
where $m,\; c,\;k \in \mathbb{R}$ correspond to mass, the damping coefficient, and the spring constant respectively. Our aim is to try to solve the regression task of predicting the trajectories with a neural network, $NN_{\theta}: \mathbb{R} \rightarrow \mathbb{R}$, with variational parameters $\theta$. The neural network will receive as input some times $t$, and its forward pass will produce an estimate $\tilde{x}(t;\theta)$ for the true positions $x_T(t)$. We can train our neural network with a simple mean square error loss function from the data we have available, say the true values for position at a collection of points $t_j$ in time, $\mathcal{D} = \{(t_j, x_T(t_j))\}_{j = 1}^{N} = (t_0, x_T(t_0)), (t_1, x_T(t_1)), \ldots (t_N, x_T(t_N))$. This means a vanilla model (with no knowledge of the underlying equation of motion) has a loss function,
\begin{equation}
L_{\text{data}}(\theta) = \frac{1}{N} \sum_{i=1}^{N} \left( x_{\theta}(t_i) - x_i \right)^2.
\label{eq:PINN_data_loss}
\end{equation}

\begin{figure}[t!]
    \centering
    \begin{subfigure}[b]{0.49\textwidth}
        \centering
        \begin{tikzpicture}
            \node [anchor=south west, inner sep=0] (image) at (0,0) {\includegraphics[width=\textwidth]{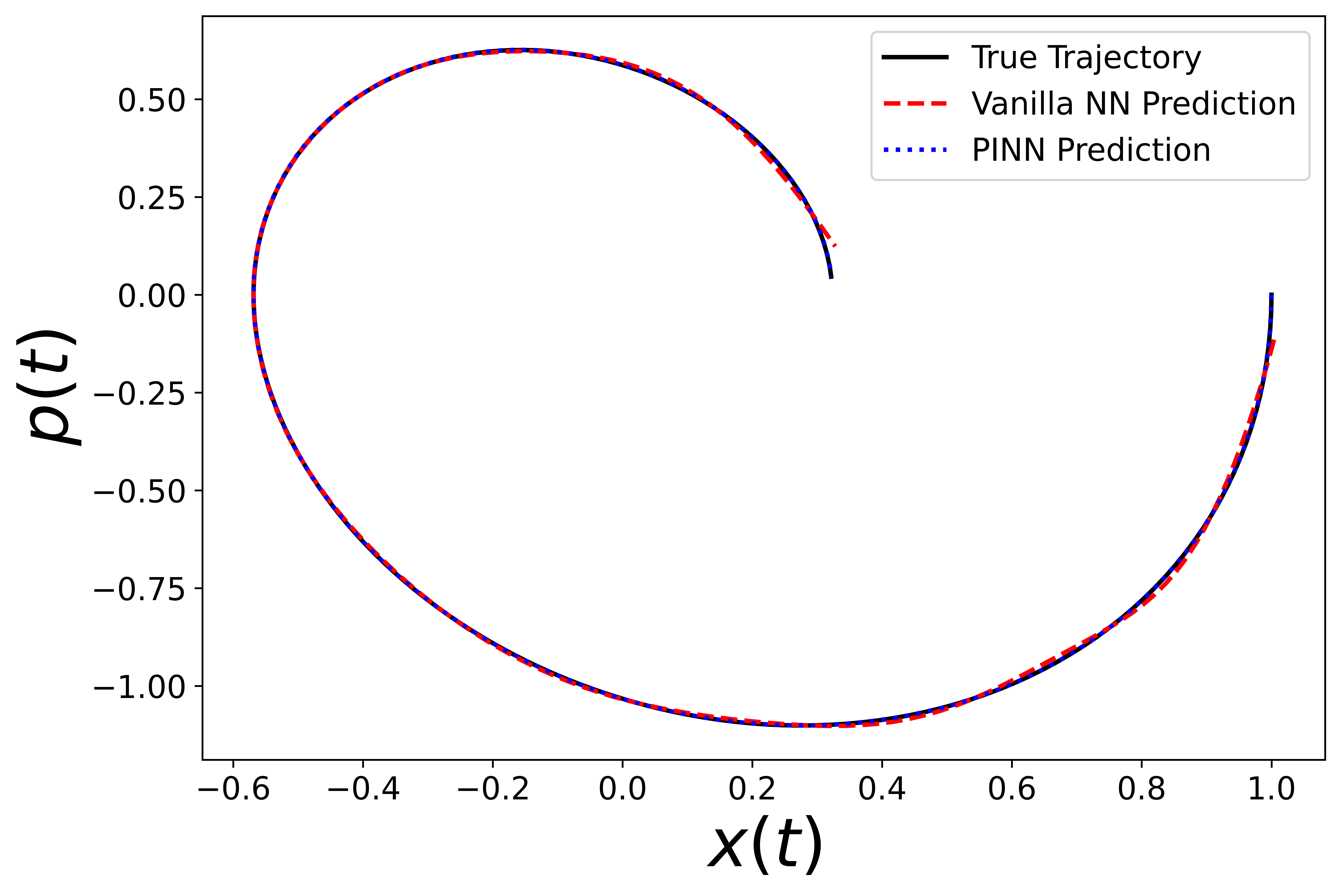}};

        \end{tikzpicture}
        \label{fig:sub1}
    \end{subfigure}
    \hfill
    \begin{subfigure}[b]{0.49\textwidth}
        \centering
        \begin{tikzpicture}
            \node [anchor=south west, inner sep=0] (image) at (0,0) {\includegraphics[width=\textwidth]{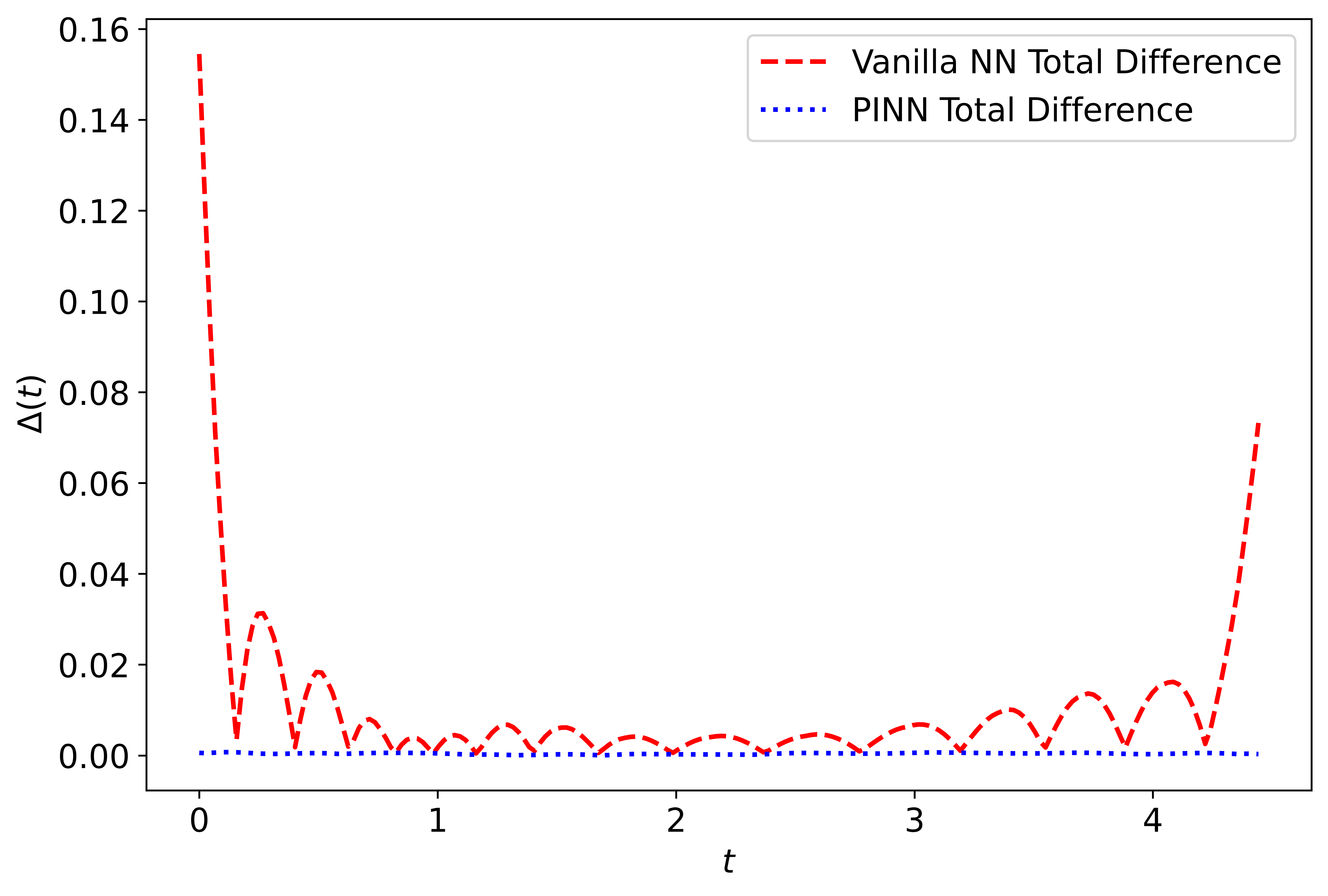}};

        \end{tikzpicture}
        \label{fig:sub2}
    \end{subfigure}
    \caption{Comparison of true trajectory $x_T(t)$ (black line) with vanilla NN (red, thick-dashed line) and PINN (blue dotted line). Times shown in this figure are in the training domain, whilst Fig. \ref{fig:PINN_comparison_generalisation} shows the generalisation of this outside training times.
    On the left is a phase space diagram for the damped oscillator, whilst the right plots the average residual error, $\Delta(t) = \sqrt{\Delta x (t)^2 + \Delta p(t)^2}$, defined as the euclidean distance to the true trajectory in the left plot at each point in time. Here, $\Delta x(t) = ||x_T(t) - x_{\theta}(t)||$ and $\Delta p(t) = ||\dot{x}_T(t) - \dot{x}_{\theta}(t)||$.
    Notice that the PINN is more accurate; it stays in the neighbourhood of the ground truth in phase space (on the right). Whereas the vanilla model oscillates on either side of the ground truth in phase space. Both models are based on the same neural network architecture, hyperparameters, and training dataset. Their only difference is the inclusion of $L_{\text{physics}}(\theta)$ for the PINN's loss function. Finally, note that since we are working in natural (arbitrary) units, the axes labels have been left blank.}
    \label{fig:PINN_comparison}
\end{figure}

Notice that Eq.~(\ref{eq:damped_oscillator_EOM}) gives 0 \textit{independently} of time. In this sense, the left-hand-side of Eq.~(\ref{eq:damped_oscillator_EOM}) is a conserved quantity for our system. Using a Lagrange multiplier, we can create a combined loss function,
\begin{equation}
L(\theta) = L_{\text{data}}(\theta) + \lambda L_{\text{physics}}(\theta),
\label{eq:PINN_combined_loss}
\end{equation}
where for a batch of points $M \leq N$,
\begin{equation}
L_{\text{physics}}(\theta) = \frac{1}{M} \sum_{j=1}^{M} \left( m \frac{d^2 x_{\theta}}{dt^2}(t_j) + c \frac{d x_{\theta}}{dt}(t_j) + k x_{\theta}(t_j) \right)^2,
\end{equation}
and $\lambda \in \mathbb{R}$ weights the importance of the physics-based loss function with respect to the data-based one. This is another instance of a \textit{hyperparameter}.
The combined loss function is zero only when the neural network makes predictions $\tilde{x}(\theta)$ that make Eq.~(\ref{eq:damped_oscillator_EOM}) conserved. In this sense, adding this extra term to the loss function has \textit{informed} the neural network of a quantity we wish to be conserved. This can make a difference in stability of solutions when training neural networks to model physical systems. See Fig.~\ref{fig:PINN_comparison} for an example which takes the same neural network (and hyperparameters) and trains it with Eq.~(\ref{eq:PINN_data_loss}) compared with Eq.~(\ref{eq:PINN_combined_loss}).

\begin{figure}[h!]
    \centering
    \begin{subfigure}[b]{0.49\textwidth}
        \centering
        \begin{tikzpicture}
            \definecolor{customcolor}{HTML}{FCA082}
            \definecolor{customcolor2}{HTML}{E32F26}
            \definecolor{customcolor3}{HTML}{67000D}
            \definecolor{customcolor4}{HTML}{7CC7AC}
            \node [anchor=south west, inner sep=0] (image) at (0,0) {\includegraphics[width=\textwidth]{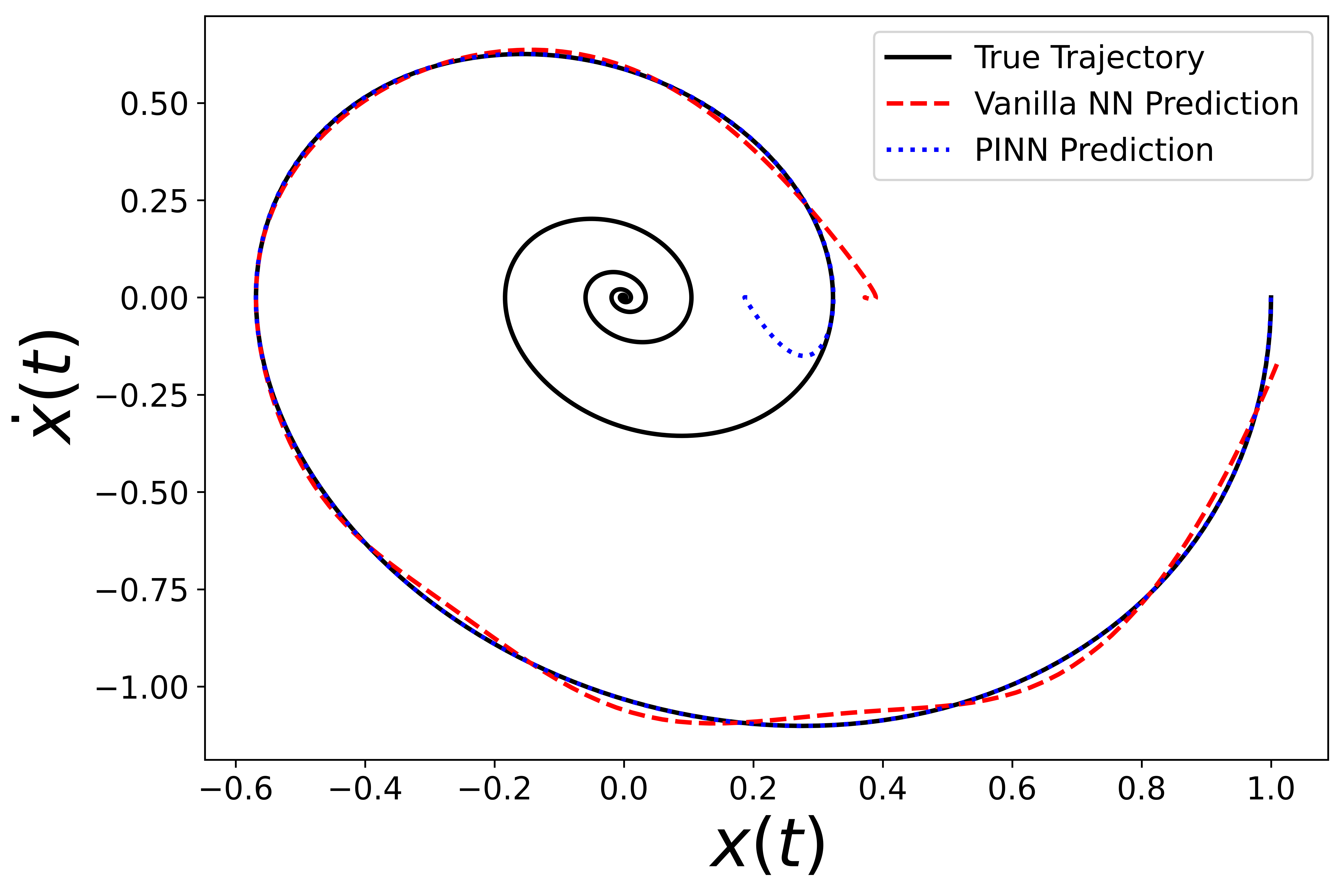}};
            \begin{scope}[shift={(0,0)}]
                \node at (2.75,3.25) {\textcolor{purple}{\Large \textbf{+}}};
            
            \end{scope}
        \end{tikzpicture}
    \end{subfigure}
    \hfill
    \begin{subfigure}[b]{0.49\textwidth}
        \centering
        \begin{tikzpicture}
            \node [anchor=south west, inner sep=0] (image) at (0,0) {\includegraphics[width=\textwidth]{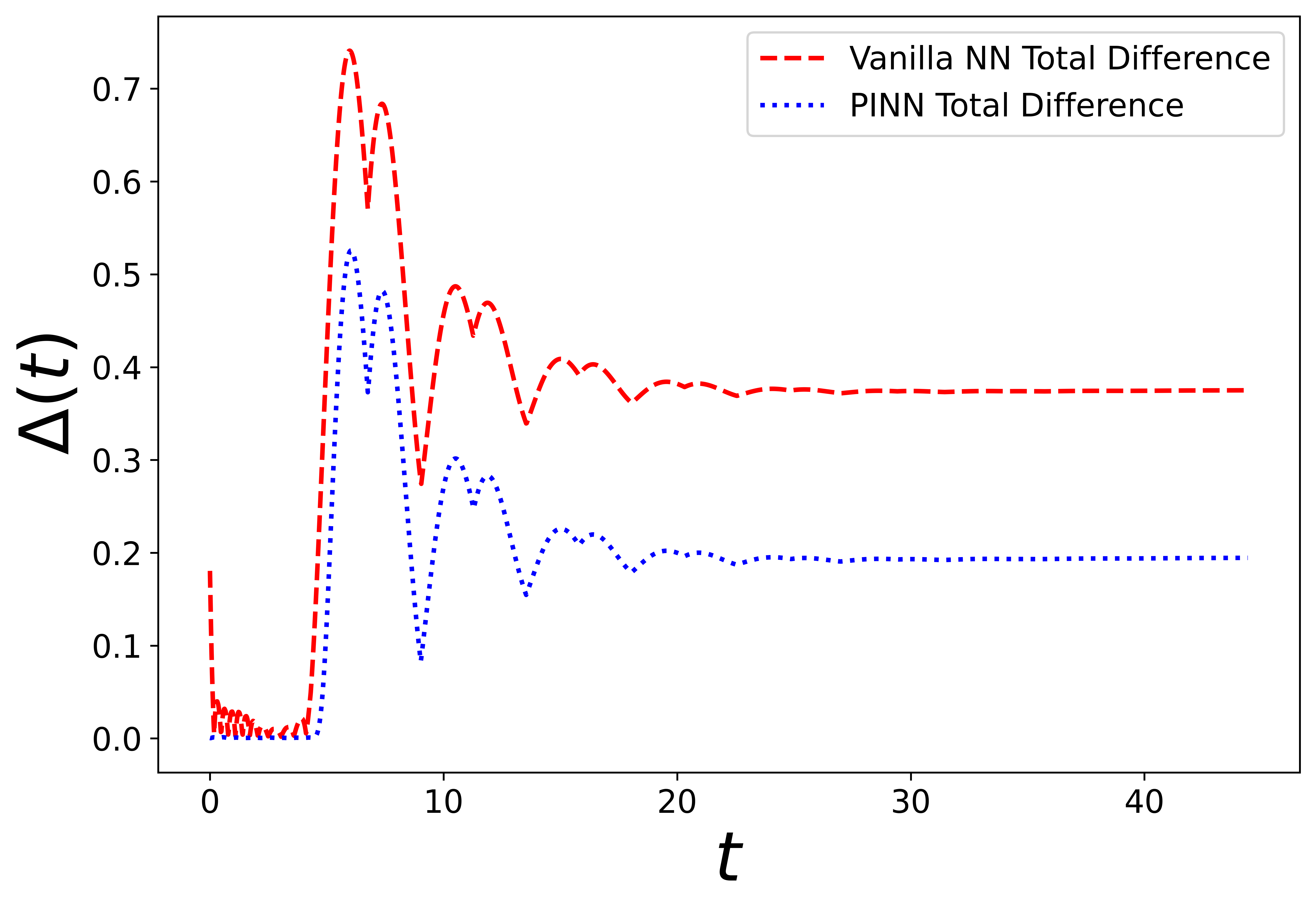}};
            \begin{scope}[shift={(0,0)}]
                \draw[purple, dashed, thick] (1.3,0.0) -- (1.3,4.0);
            \end{scope}
        \end{tikzpicture}
    \end{subfigure}
    \caption{(Left) Phase Space trajectories of the vanilla model (red dashed) and PINN model (blue dotted), with ground truth trajectory shown as a solid black line. Residuals were calculated in the same was as Fig.\ref{fig:PINN_comparison}, with the time elapsed increased to $10 \times$ the training time boundary, shown by a purple cross in the left figure, and a purple dashed line in the right.
    Notice that the PINN again performs better outside of training distribution, but both models are failing to properly generalise. This is a reflection of the fact that even a PINN is not hard-coded to respect the laws of physics. Finally, note that since we are working in natural (arbitrary) units, the axes labels have been left blank.}
    \label{fig:PINN_comparison_generalisation}
\end{figure}

\begin{figure}[h]
    \centering
        \begin{tikzpicture}
            \definecolor{customcolor}{HTML}{FCA082}
            \definecolor{customcolor2}{HTML}{E32F26}
            \definecolor{customcolor3}{HTML}{67000D}
            \definecolor{customcolor4}{HTML}{7CC7AC}
            \node[anchor=north east, inner sep=0] (image) at (0,0) {\includegraphics[width=\textwidth]{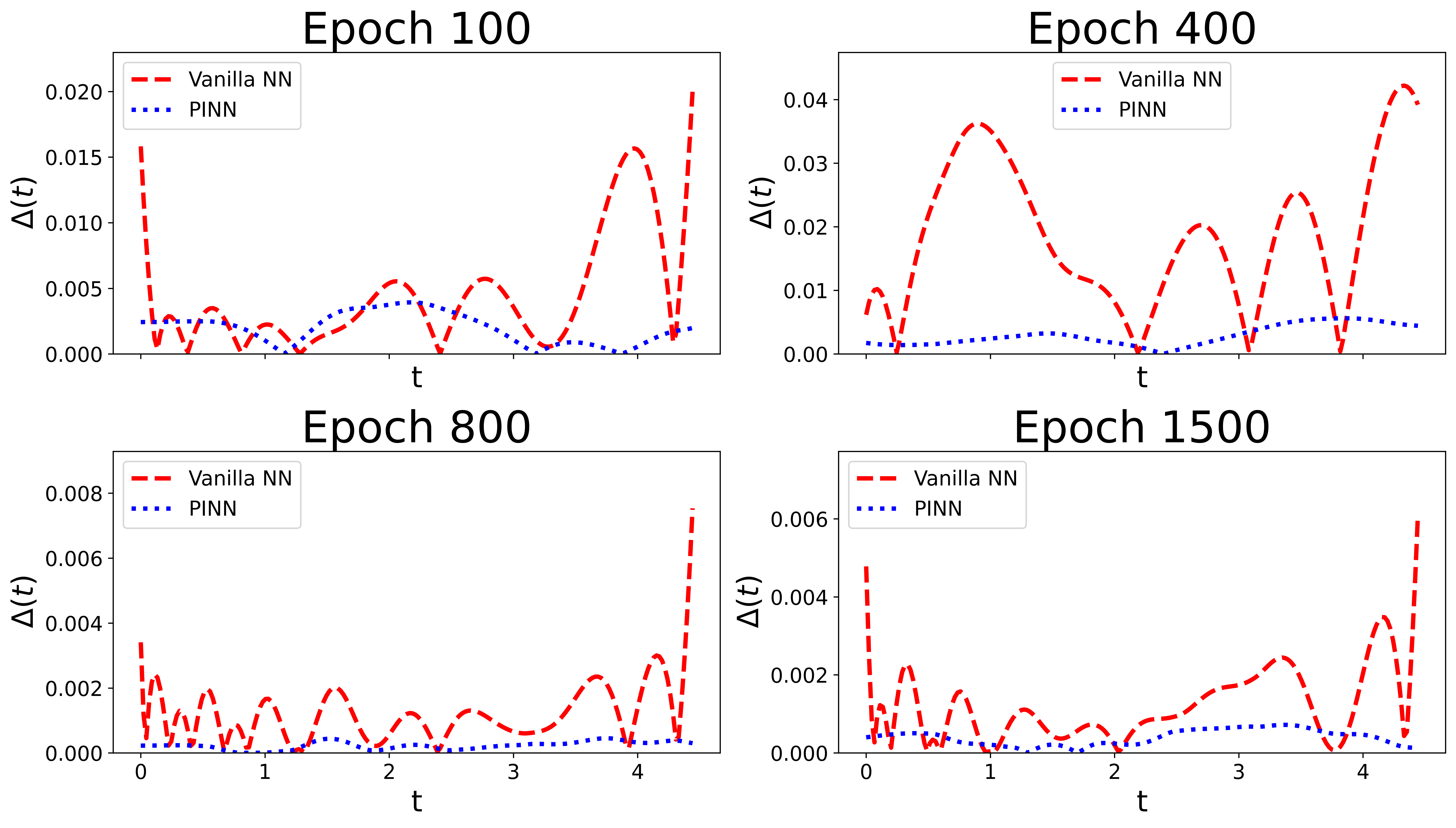}};

            \begin{scope}[shift={(0,0)}]
                \node at (1,1) {hi};
            \end{scope}
        \end{tikzpicture}
    \caption{Residual plots of the vanilla NN and its PINN counterpart at different epochs of a 2000 epoch training run. Notice that whilst the vanilla NN initially performs better thanks to a simpler loss landscape (it does not have the extra lagrange multiplier), the PINN quickly takes over in terms of performance before the half-way point of its training. For the same computational resources, we can therefore conclude that indeed PINNs boost performance of our neural models. Residuals were calculated per Fig.\ref{fig:PINN_comparison}. Finally, note that since we are working in natural (arbitrary) units, the axes labels have been left blank.}
    \label{fig:epochs_training_PINN}
\end{figure}

We see from Fig.~\ref{fig:PINN_comparison} that PINNs do not conserve quantities \textit{perfectly}. This is because adding a Lagrange multiplier presents our model with a soft constraint; the loss function value is improved when the model \textit{tends} towards solutions that conserve our quantity of interest. However, nothing is \textit{forcing} it to conserve energy, it merely comes (approximately) as a consequence of minimising loss - something which is not always guaranteed to be zero\footnote{If the loss were zero, then the PINN would conserve out quantity exactly. But we know this cannot be analytically possible with numerical techniques due to floating point error, training strategies, and a number of other factors that mean the optimal solution $\theta^*$ gives $L_{min} > 0$.}. How then can we build neural network models which \textit{analytically} conserve quantities of interest? The answer lies in two concepts known as \textit{equivariance} and \textit{invariance}.

\subsection[Invariance and Equivariance]{Invariance and Equivariance - Conservation via Data and Architecture}

So far, we have seen how to embed a physical equation of motion into the loss function of a neural network model. This produces a soft constraint, in the sense that nothing is forcing the neural network model to ensure this equation of motion is satisfied. 

In this section, we will look at how to create models that manifestly respect the physical symmetries present in a system we wish to model. There are two ways in which we can seek to embed symmetry into a model. The first is \textit{invariance}, defined as
\begin{equation}
    f_{\theta}(T \circ x) = f_{\theta}(x),
    \label{eq:invariance}
\end{equation}
where $f_{\theta}(x)$ is a neural network receiving $x \in \mathbb{R}^n$ as input, and $T \in \mathcal{G}$ is any element of a symmetry group present. For example, if $T$ is a permutation of the elements $x$, then we say that $f_{\theta}(x)$ is \textit{invariant} under the permutation group, $\mathbb{P}$ whenever $f_{\theta}(T \circ x) = f_{\theta}(x)$, for $T \in \mathbb{P}$. 

For those familiar with gauge theory, we can think of $\mathcal{G}$ as the gauge group of $f_{\theta}(x)$. Here, we use the notation $\circ$ to mean ``acts on'' whatever follows to the right. This is because depending on the group, its action could be multiplicative, additive, or something else entirely (e.g. permutation). Invariance is, in some sense, the strongest form of symmetry preserving property a neural network can have. Any neural network which is invariant under $T$ must conserve the physical quantity associated with it by Noether's Theorem. However, this can often be difficult to engineer as we shall see shortly. 

The second way to embed symmetry into a model is to use \textit{equivariance}, defined as,
\begin{equation}
    f_{\theta}(T \circ x) = T^{'} \circ f_{\theta}(x),
    \label{eq:equivariance},
\end{equation}
where $T^{'} = g(T)$ is related to $T$ and thus is predicable via the function $g: \mathcal{G} \rightarrow \mathcal{G}$. Equivariance comes from the idea that the image of the transformation $T$ is predicable, and thus in some sense, $f_{\theta}(x)$ is ``equally varying'' with $T$ as well as $x$. Since $T'$ can be anticipated, we can engineer invariance \textit{from} equivariance simply by dividing it out. Explicitly, transforming by $T$ means $T' = g(T)$ is known, as is its inverse\footnote{per the usual rules of Group Theory}, $(T^{'})^{-1}$, which means we can apply
\begin{equation}
    (T^{'})^{-1} f_{\theta}(T \circ x) = (T^{'})^{-1} \circ T^{'} \circ f_{\theta}(x) = \mathbb{1} \circ f_{\theta}(x) = f_{\theta}(x),
\end{equation}
leaving the overall result invariant with respect to $T$.

There are many advantages to using equivariance and invariance in deep learning for physical systems. First of all, we can see how a symmetry-endowed model will be more sample efficient than its symmetry-lacking counterpart. This is because models that respect symmetries do not need to learn redundant representations for transformed versions (i.e. $T \circ x$) of the same input. They also tend to generalise better, since physical systems often exhibit the same behaviour under certain transformations. We see this classically in the gauge fields of electromagnetism, and in quantum mechanics with basis transformations, or changing between the Schr\"odinger, Heisenberg and Dirac pictures of time dynamics. Models that incorporate symmetries also tend to be simpler in terms of the number of variational parameters. This is because when a symmetry group is present, fewer parameters need to be varied in order to search an entire space\footnote{A good mental picture of this is that by imposing symmetry constraints we are \textit{folding out} the unnecessary dimensions or degrees of freedom in a model, allowing us to vary over a lower-dimensional manifold. Another way of thinking about it is that we must be looking in a lower dimensional space since the symmetries have lowered the number of degrees of freedom. }.

Having seen what invariance and equivariance are, and why it is beneficial to employ them, we are now in a position to understand \textit{how} to make a model respect these constraints.

Let's start with invariance. In general, we can establish invariance by augmenting our representation of the input data to respect the symmetry group $\mathcal{G}$. That is, we find a one-hot encoding, $z$, of $x$ such that $T \circ z = z$. We can also augment the model itself to be invariant under $\mathcal{G}$. This involves fixing the structure of the architecture itself, i.e. the weights and biases, as well as the activation functions. Typically, the former (data) is easier than the latter (model) with deep learning, owing to the inherent complexity that comes with deep neural networks. The latter is however possible, and is the subject of ongoing research efforts in the community. 

One simple (vanilla) strategy to do establish invariance with discrete groups is to perform pooling over all the group actions, effectively averaging out the model's responses to all possible transformations. For instance, you can apply the group transformations to the input data, process each transformed version through the same network (sharing the weights), and then combine the outputs via operations like averaging or softmax. This ensures that regardless of how the input is transformed within $\mathcal{G}$, the final output remains constant, capturing the desired invariance. Essentially, the output is invariant because we have created an effective input that is the sum \textit{over all possible} transformations $T \in \mathcal{G}$. As an example, if $\mathcal{G} = \mathcal{S}_n$ is the permutation group of $n$ elements, instead of inputting $x \in \mathbb{R}^n$, we could input every permutation of $x$'s elements.

Equivariance follows a similar story. We can establish equivariance with data, or through the model architecture. To see how this works, it is instructive to again consider the permutation group $S_n$.  To construct layers that are permutation equivariant to the permutation group, we can design functions that commute with any permutation of the input elements. That is, $f_{\theta}(T\circ x) = T \circ f_{\theta}(x)$, for any $T \in \mathcal{S}_n$. By adapting how we construct layers we can make this possible. Lets consider a layer's function $f: \mathbb{R}^n \rightarrow \mathbb{R}^n$, where we assume the number of neurons per layer remains the same for simplicity. Instead of the usual weight matrix and bias that we see in a multi-layer perceptron, we could instead use
\begin{equation}
    [f(x)]_i = \sigma \left( W x_i + b + U \sum_{j=1}^{n} x_j \right),
\end{equation}
where $[f(x)]_i$ is the $i^{\text{th}}$ component of $f$, $W, U \in \mathbb{R}$ are scalar weights, $b$ is a bias, and $\sigma$ is any activation function. As an exercise, you can check that this layer function is indeed equivariant with $S_n$, satisfying $f(T\circ x) = T \circ f(x)$ for any $T \in \mathcal{S}_n$. Intuitively, we can see this must be the case since $\sigma$ receives a sum $\sum_{j=1}^{n} x_j$ which is invariant under $S_n$, and a single element $x_i$, which over the whole layer becomes permutation equivariant; changing \textit{which} $x_i$ is there by enacting a permutation just changes which node $x_i$ is pulled from in the previous layer. 

In unsupervised learning, we can therefore use our knowledge or anticipation of symmetry properties to guide the structure of the search for correlation, and vice versa. However, we must not forget the parable of the Chinese Room thought experiment from Chapter~\ref{CH:FUNDAMENTALS}. Namely, just because we have the ability to embed the symmetry properties of physical processes into a neural architecture, this does not imply that neural networks have actually provided real understanding of physical processes. The art of yielding an understanding from deep learning models comes from their explainability \cite{heese2025explaining, holzinger2019causability, fazi2021beyond} and interpretability \cite{pira2024interpretability, li2022interpretable, zhang2018visual}.

%% file: chapters/chapter_4_quantum_basics.tex
\chapter{Essential Quantum Mechanics}

\section{Introduction}

After exploring the foundations of deep learning, we now turn to the quantum mechanics. We emphasise here that the aim of this chapter is \textit{not} to reteach quantum mechanics from scratch. Rather, we will recall some essential parts of the theory and recast it in a form that makes them immediately usable for quantum information processing and for the deep-learning methods we will deploy in Chapter 5. We will focus on the Heisenberg (matrix) representation, because it meshes naturally with the linear-algebra viewpoint that underlies neural networks and other machine-learning models.

We begin with the notion of  qubits Hilbert space, introducing quantum gates as unitary matrices and illustrate how single-qubit rotations and two-qubit entangling gates form universal tool-kits for quantum algorithms.

Next, we briefly recall key properties of entanglement through the lens of resource theories showing clear algebraic criteria for detecting and quantifying entanglement in multi-qubit states. We will highlight its operational meaning through small-scale circuit examples.

Measurement is the bridge between the quantum and classical worlds, so we devote a full section to projective measurements and positive operator-valued measures (POVMs), emphasizing how information extraction inevitably disturbs a quantum state. 

Because many quantum-information tasks—from ground-state searches to quantum machine-learning models—rely on optimizing parameters in the presence of quantum uncertainty, we also introduce the variational principle and a basic example of it, leaving the hybrid quantum-classical algorithms that arise from it (the Variational Quantum Eigensolver and Quantum Approximation Optimization Algorithm) to Chapter 5.

Despite being structurally quite separate from qubits, we believe it is important to also offer some exposure to continuous-variable quantum systems, which lay the foundations for photonic technology. As such, we conclude this brief detour into quantum theory with 
a concise primer on quantum optics. 

By the end of the chapter you will have a compact, operator-centric toolkit that meshes naturally with deep-learning abstractions—setting the stage for Chapter 5, where we merge these two languages to tackle concrete problems in quantum technologies. 

Students with little to no familiarity with quantum mechanics are highly encouraged to supplement any sections not properly understood with extra reading. For a comprehensive introduction to quantum information theory, we recommend \cite{nielsen2010quantum}, for quantum mechanics, we recommend \cite{griffiths2018introduction}, whilst for quantum optics we recommend \cite{schleich2015quantum, klimov2009group}.

\section{Qubits, Circuits and spin  chains}

All the material below can be found in a standard quantum information textbook, we suggest \cite{nielsen2010quantum} or \cite{zygelman2018first}. Recall that a qubit is a two-level quantum system whose state lies in a two-dimensional Hilbert space $\mathcal{H}_2 \cong \mathbb{C}^2$,
\begin{equation}
    \ket{\psi} = c_0 \ket{0} + c_1 \ket{1},
    \label{eq:one_qubit}
\end{equation}
where $c_j \in \mathbb{C}$ are complex coefficients such that $|\!\braket{\psi|\psi}\!|^2 = 1$ is a well defined probability distribution. We can also opt for an explicit vector representation of $\ket{\psi}$ by defining
\begin{equation}
    \ket{0} = \begin{pmatrix}
        0 \\
        1
    \end{pmatrix} 
    \;\;\;\;\;\;\;\;
    \ket{1} = \begin{pmatrix}
        1 \\
        0
    \end{pmatrix}.
\end{equation}
 To incorporate multiple qubits, we can employ the tensor product. That is, a two-qubit system $\ket{\psi_1,\psi_2}$ lives in a bipartite Hilbert space $\mathcal{H}_2^{\otimes 2}$, with a four-component computational basis,
\begin{equation}
\begin{aligned}
\ket{00} &= \begin{pmatrix} 1 \\ 0 \end{pmatrix} \otimes \begin{pmatrix} 1 \\ 0 \end{pmatrix} = \begin{pmatrix} 1 \\ 0 \\ 0 \\ 0 \end{pmatrix}, \quad
\ket{01} = \begin{pmatrix} 1 \\ 0 \end{pmatrix} \otimes \begin{pmatrix} 0 \\ 1 \end{pmatrix} = \begin{pmatrix} 0 \\ 1 \\ 0 \\ 0 \end{pmatrix}, \\
\ket{10} &= \begin{pmatrix} 0 \\ 1 \end{pmatrix} \otimes \begin{pmatrix} 1 \\ 0 \end{pmatrix} = \begin{pmatrix} 0 \\ 0 \\ 1 \\ 0 \end{pmatrix}, \quad
\ket{11} = \begin{pmatrix} 0 \\ 1 \end{pmatrix} \otimes \begin{pmatrix} 0 \\ 1 \end{pmatrix} = \begin{pmatrix} 0 \\ 0 \\ 0 \\ 1 \end{pmatrix}.
\end{aligned}
\end{equation}
In the gate-based model of quantum computing, we can act on these states via operators or \textit{quantum gates}, represented as Unitary matrices. These are matrices, $\hat{U} \in \mathbb{C}^{d \times d}$ which satisfy
\begin{equation}
    \hat{U} \hat{U}^{\dag} = \mathbb{1} = U^{\dag} U,
\end{equation}
where $\mathbb{1}$ is the identity matrix, and $d = 2^N$ is the total dimension of our system of $N$ qubits. What distinguishes Unitary gates from classical gates is that they are \textit{reversible}. That is, when $\ket{\psi '} = \hat{U} \ket{\psi}$, we can reverse this process via $\ket{\psi} = U^{\dag}\ket{\psi '}$. This means that Unitary quantum gates form a group, $SU(d)$, for a quantum system of dimension $d$.
Just like states, we can represent quantum gates in a given basis. For example we say that
\begin{equation}
    \hat{U} = \sum_{i,j \in \{0,1\}^N} U_{ij} \ket{i}\!\bra{j},
\end{equation}
is a computational basis expansion of $U$. Another important basis we will employ is the Pauli basis, where
\begin{equation}
    \label{eq:pauli_basis}
    \hat{U} = \sum_{j_1,\ldots,j_N = 0}^3 U_{j_1,\ldots,j_N} \hat{\sigma}_{j_1} \otimes \ldots \otimes \hat{\sigma}_{j_N}.
\end{equation}
Here $\hat{\sigma}_j \in \{\mathbb{1}, \hat{\sigma}_x, \hat{\sigma}_y, \hat{\sigma}_z\}$ is a Pauli matrix, and the value of $j_k$ indexes which of these four is present on the $k^{\text{th}}$ tensor product.  We can also represent the action of operators $\hat{O}$ on $\ket{\psi}$ via matrices. For example, a one qubit operator in the computational basis reads
\begin{equation}
    \hat{O} \ket{\psi} = \begin{pmatrix}
        \braket{0|\hat{O}|0}& \braket{0|\hat{O}|1}\\
        \braket{1|\hat{O}|0} & \braket{1|\hat{O}|1}
    \end{pmatrix}
    \left(
    c_0 
    \begin{pmatrix}
        0 \\
        1
    \end{pmatrix}
    + c_1 
    \begin{pmatrix}
        1 \\
        0
    \end{pmatrix}
    \right),
\end{equation}
in the same basis as $\ket{\psi}$. For $N$-qubits this matrix would be $2^N \times 2^N$.

\begin{figure}[h]
    \centering
    \begin{tikzpicture}
        \node at (-1.5, 0) {$\ket{q}$};        
        \draw[thick] (-1, 0) -- (0, 0);
        \draw[thick] (0, -0.5) rectangle (1, 0.5);
        \node at (0.5, 0) {$\hat{U}$};
        \draw[thick] (1, 0) -- (1.5, 0);
        \node at (2, 0) {$\ket{q'}$};
    \end{tikzpicture}
    \caption{A unitary gate \( \hat{U} \) applied to state \( \ket{q} \), resulting in \( \ket{q'} \).  }
    \label{fig:unitary_gate}
\end{figure}
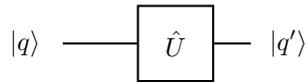

Quickly, we can see how keeping track of states and operators in Dirac and matrix notation can become cumbersome and counterproductive, especially when there are many qubits and gates we wish to model. It is for this reason that quantum circuit notation was introduced to the field. Quantum circuit notation is diagrammatic, with the simplest one-qubit example shown in Fig.~\ref{fig:unitary_gate}. We read circuit diagrams time-wise from left to right. In this circuit, the initial state is $\ket{q}$, and the output $\ket{q'} = \hat{U} \ket{q}$. Each wire in the circuit corresponds to a single qubit unless otherwise specified.

In classical computing, we rely on a small set of logical gates such as \texttt{AND}, \texttt{OR}, and \texttt{NOT}, to perform complex computations. Remarkably, \textit{any} Boolean function can be implemented using just a single type of gate, such as the \texttt{NAND} gate. As such, we say a gate set is classically \textit{universal} when any Boolean function can be computed in terms of this gate set\footnote{Note here that our definition does not require that this decomposition be efficient, rather that it simply exists in the first place!} This idea was crucial in the early development of classical computers because it simplifies the design and analysis of digital circuits, allowing us to build any classical algorithm from a minimal set of components.

In quantum computing, we face a similar challenge but with added complexity due to the non-classical properties of quantum systems. Quantum algorithms require operations that manipulate qubits in ways that exploit superposition and entanglement. To build any quantum algorithm, we need a set of quantum gates that can approximate \textit{any} unitary operation to arbitrary precision. This leads us to the concept of \textit{universal quantum gates}, which serve as the foundational building blocks for all quantum computations. Some important basic one-qubit gates which we will use (amongst others!) to construct universal gate sets are shown in Fig.~\ref{fig:four_basic_gates}.

\begin{figure}
    \centering
    \begin{tikzpicture}
    \draw[thick] (0,0) -- (1.0,0);
    \draw[thick] (2.0,0) -- (2.8,0);
    \draw[fill=white] (1.0,-0.5) rectangle (2.0,0.5);
    \node at (1.5,0) {\(H\)};
    \node at (4.0,0) {\( = \frac{1}{\sqrt{2}}\begin{pmatrix}1 & 1 \\ 1 & -1\end{pmatrix}\)};
    \end{tikzpicture}
    \hspace{1cm} 
    \begin{tikzpicture}
    \draw[thick] (0,0) -- (1.0,0); 
    \draw[thick] (2.0,0) -- (3.0,0); 
        \draw[fill=white] (1.0,-0.5) rectangle (2.0,0.5);
    \node at (1.5,0) {\(X\)};
    \node at (4.0,0) {\( = \begin{pmatrix}0 & 1 \\ 1 & 0\end{pmatrix}\)};
    \end{tikzpicture}

    \vspace{1cm}
    
    \begin{tikzpicture}
    
    \draw[thick] (0,0) -- (1.0,0);
    \draw[thick] (2.0,0) -- (3.0,0);
    \draw[fill=white] (1.0,-0.5) rectangle (2.0,0.5);
    \node at (1.5,0) {\(S\)};
    \node at (4.0,0) {\( = \begin{pmatrix}1 & 0 \\ 0 & i\end{pmatrix}\)};
    \end{tikzpicture}
    \hspace{1cm} 
    \begin{tikzpicture}
    \draw[thick] (0,0) -- (1.0,0); 
    \draw[thick] (2.0,0) -- (2.8,0); 
    \draw[fill=white] (1.0,-0.5) rectangle (2.0,0.5);
    \node at (1.5,0) {\(T\)};
    \node at (4.0,0) {\( = \begin{pmatrix}1 & 0 \\ 0 & e^{i \pi / 4}\end{pmatrix}\)};
    \end{tikzpicture}
    
    \caption{The Hadamard and \texttt{NOT} gate circuits with matrix form in the computational basis}
    \label{fig:four_basic_gates}
\end{figure}

Notice that these four gates are fixed, not variational. That is, they deterministically act on a one qubit input $\ket{\psi}$ from Eq.~(\ref{eq:one_qubit}). To create a variational gate, we can consider Unitary \textit{rotations} of Pauli matrices,
\begin{equation}
    \hat{R}(\theta) = e^{i \theta \hat{\sigma}} = \cos(\theta) 
    \mathbb{1} + i \sin(\theta) \hat{\sigma},
    \label{eq:euler_su(2)}
\end{equation}
where $\mathbb{1}$ is the one-qubit identity matrix, and $\sigma \in \{\sigma_x, \sigma_y, \sigma_z\}$ is a Pauli matrix\footnote{Actually, this relation holds for any $\sigma \in SU(2)$, but in quantum circuits, we are usually only concerned with Pauli matrices.}. Recall here that we are exponentiating matrices, which is well defined via their Taylor Expansion\footnote{If you have never seen this before, a good exercise is to show Eq.~(\ref{eq:euler_su(2)}) by taking the Taylor series of $e^{i \theta \sigma}$ about $\mathbb{1}$ and simplify using the fact that $\sigma^2 = \mathbb{1}$.}. In circuit notation, we denote this as
\begin{figure}
    \centering
    \begin{tikzpicture}
    \draw[thick] (0,0) -- (1.0,0);
    \draw[thick] (2.0,0) -- (2.7,0);
    
    \draw[fill=white] (1.0,-0.5) rectangle (2.0,0.5);
    \node at (1.5,0) {\(\hat{R}_j(\theta)\)};
    
    \node at (3.6,0.05) {\( = e^{i\frac{\theta}{2} \hat{\sigma}_j}\)};
    \end{tikzpicture}
    
    \caption{One-qubit circuit with rotation gate \(\hat{R}(\theta)\) and its Unitary form. Here $j \in \{x,y,z\}$ index which of the Pauli matrices we are rotating around, i.e. which of the three is a fixed axis of rotation.}
    \label{fig:rotation-gate-circuit}
\end{figure}

\begin{figure}
    \centering
    \includegraphics[width=0.5\linewidth]{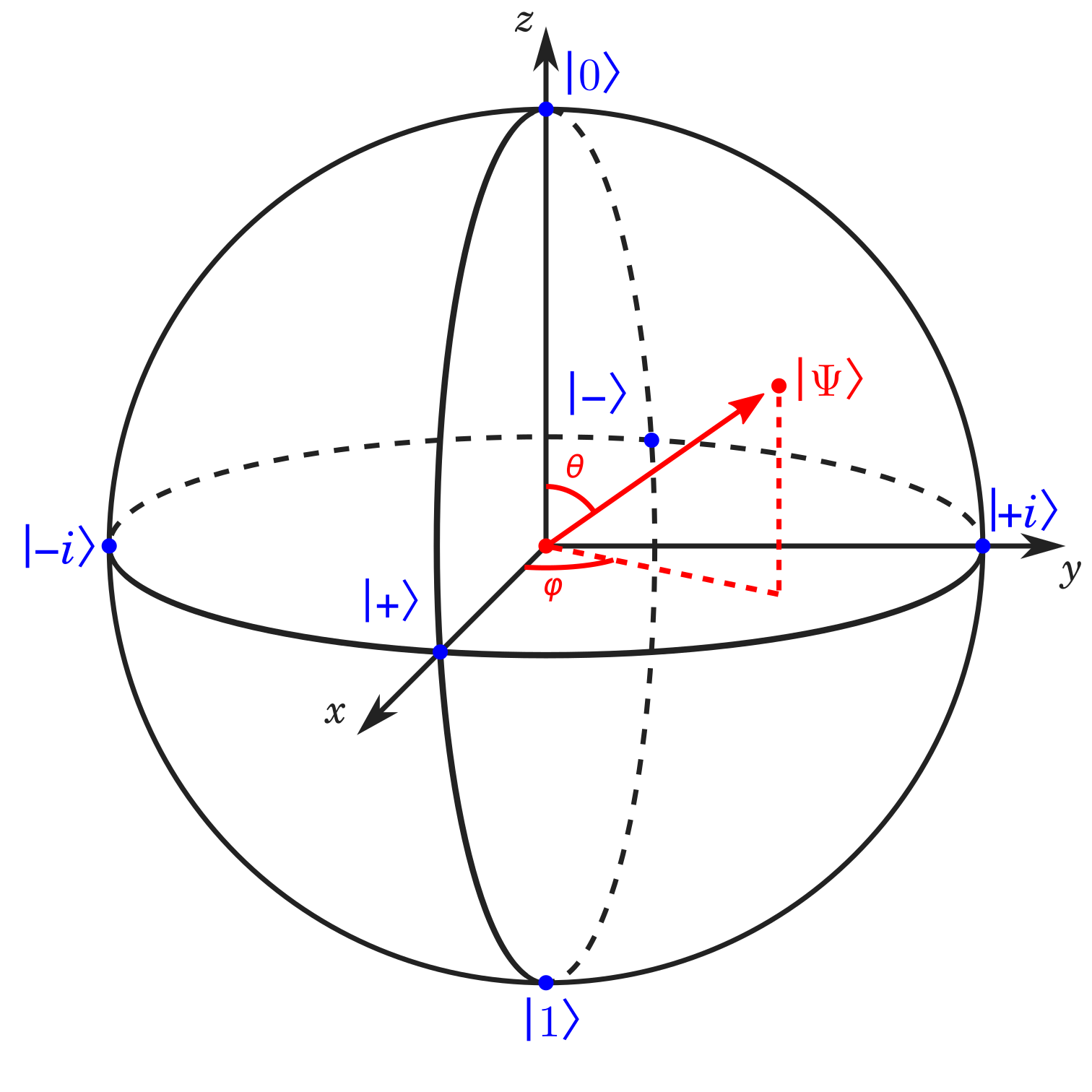}
    \caption{Bloch sphere representation of a single qubit pure state (red arrow). The $x,\;y,$ and $z$ axes represent coefficients of the density matrix $\hat\rho = \ket{\psi}\!\bra{\psi}$ in the Pauli basis, for example i.e. $x \leftrightarrow \sigma_x$. Here, the $\ket{\pm}$ states are the eigenstates of $\sigma_x$, whilst $\ket{\pm i}$ are the eigenstates of $\hat\sigma_y$.}
    \label{fig:bloch_sphere}
\end{figure}

There are many different sets of universal gates that are complete in the space of unitary operators. However, one-qubit gates alone are not sufficient to reach this goal. To see why this is the case, let us consider geometrically what one-qubit gates can do. Recall that a density matrix $\hat\rho = \ket{\psi}\!\bra{\psi}$, can be expanded in the Pauli basis,
\begin{equation}
    \hat{\rho} = \frac{1}{2}(\mathbb{1} + \mathbf{n} \cdot \mathbf{\hat{\sigma}}),
\end{equation}
where $\mathbf{n} = (n_x,n_y,n_z)^T$ are the Cartesian coefficients representing the amplitude of $\rho$ in each of the $\hat{\sigma}_x,\;\hat{\sigma}_y,\;\hat{\sigma}_z$ directions. Since $\hat{\rho}$ is normalized, we require $|\mathbf{n}| = 1$, meaning the space of one-qubit pure states is the surface of a sphere, called a \textit{Bloch Sphere} see Fig.~\ref{fig:bloch_sphere}. 

The action of any one-qubit unitary gate can be represented through motion of $\ket{\psi}$ on the Bloch sphere. For example, the Hadamard gate from Fig.~\ref{fig:four_basic_gates} rotates a given state by $\pi/2$ about the $x$-axis. More generally, all one-qubit gates can be thought of as rotations and reflections in this sphere. These gates are critical for placing qubits in superpositions and applying necessary phase shifts. 

However, in quantum computing, the power of quantum algorithms comes from their ability to exploit superposition \textit{and entanglement}; two features that cannot be fully realized by single-qubit operations alone. Whilst we can create superposition on a single qubit, entanglement is a property of many-body quantum systems. Creating an entangled system therefore requires us to apply operations to more than one qubit. Some important two-qubit gates are shown below .

\begin{figure}[h]
    \centering
    \begin{tikzpicture}
        \node at (1.3, 0.5) {$\ket{0}$};
        \draw[thick] (1.5, 0.5) -- (4, 0.5);  

        \node at (1.3, -0.5) {$\ket{0}$};
        \draw[thick] (1.5, -0.5) -- (4, -0.5); 
        
        \filldraw (2.5, 0.5) circle (3pt);
        \draw[thick] (2.5, 0.5) -- (2.5, -0.8);
        
        \draw[thick] (2.5, -0.5) circle (0.3); 
        
        \node at (6.0, 0) {$ = \begin{pmatrix}
        1 & 0 & 0 & 0 \\
        0 & 1 & 0 & 0 \\
        0 & 0 & 0 & 1 \\
        0 & 0 & 1 & 0
        \end{pmatrix}$};
    \end{tikzpicture}

     \begin{tikzpicture}
        \node at (1.3, 0.5) {$\ket{0}$};
        \draw[thick] (1.5, 0.5) -- (4, 0.5);  

        \node at (1.3, -0.5) {$\ket{0}$};
        \draw[thick] (1.5, -0.5) -- (4, -0.5);

        \draw[thick] (2.5, 0.5) -- (2.5, -0.5); 
        \node at (2.5, 0.5) {$\times$};
        \node at (2.5, -0.5) {$\times$}; 

        \node at (6.0, 0) {$ = \begin{pmatrix}
        1 & 0 & 0 & 0 \\
        0 & 0 & 1 & 0 \\
        0 & 1 & 0 & 0 \\
        0 & 0 & 0 & 1
        \end{pmatrix}$};
    \end{tikzpicture}

    \begin{tikzpicture}
        \node at (1.3, 0.5) {$\ket{0}$};
        \draw[thick] (1.5, 0.5) -- (4, 0.5);  

        \node at (1.3, -0.5) {$\ket{0}$};
        \draw[thick] (1.5, -0.5) -- (4, -0.5); 
        
        \filldraw (2.5, 0.5) circle (3pt); 
        \draw[thick] (2.5, 0.5) -- (2.5, -0.5);
        \draw[thick, fill = white] (2.3, -0.3) rectangle (2.7, -0.7); 
        \node at (2.5, -0.5) {$\hat{U}$};

       \node at (6.0, 0) {= $\begin{pmatrix}
            1 & 0 & 0 & 0 \\
            0 & 1 & 0 & 0 \\
            0 & 0 & \multicolumn{2}{c}{\raisebox{-0.5em}{\( \hat{U} \)}} \\
            0 & 0 & &
            \end{pmatrix}$};
    \end{tikzpicture}
    
    \caption{The \texttt{CNOT}, \texttt{SWAP} and \texttt{C-U} gates and their matrix form in the computational basis. As its name suggests, the \texttt{SWAP} gate swaps the two qubits it acts on. The \texttt{C-U} gate applies the one-qubit unitary $U$ if and only if the control qubit is in the state $\ket{1}$.}
    \label{fig:two_qubit_gates}
\end{figure}
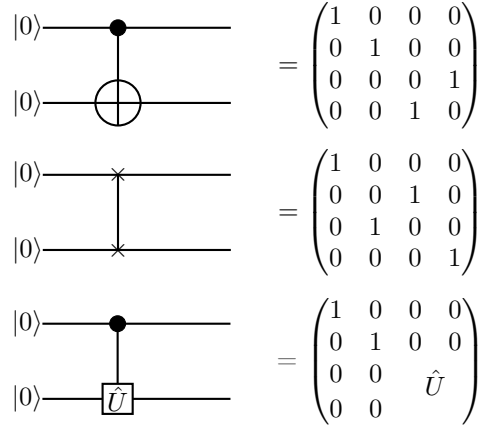

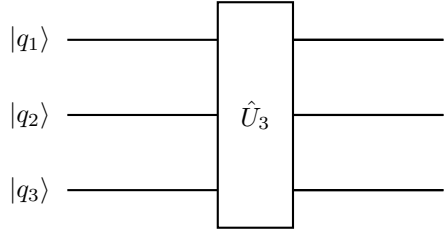
\begin{figure}
    \centering
    \begin{tikzpicture}
        \draw[thick] (0,0) -- (5,0);
        \draw[thick] (0,-1) -- (5,-1);
        \draw[thick] (0,-2) -- (5,-2);
        
        \node at (-0.5, 0) {\( \ket{q_1} \)};
        \node at (-0.5, -1) {\( \ket{q_2} \)};
        \node at (-0.5, -2) {\( \ket{q_3} \)};
        
        \draw[thick, fill = white] (2, 0.5) rectangle (3, -2.5);
        \node at (2.5, -1) {\( \hat{U}_3 \)};
        
        \draw[thick] (3,0) -- (5,0);
        \draw[thick] (3,-1) -- (5,-1);
        \draw[thick] (3,-2) -- (5,-2);

    \end{tikzpicture}
    \caption{A three-qubit unitary gate in quantum circuit notation.} \label{fig:3_qubit_unitary}
\end{figure}

In general, we can apply a Unitary gate to as many qubits as we like, for example Fig.~\ref{fig:3_qubit_unitary} is a generic three-qubit Unitary gate $U_3 \in SU(2^3) \cong SU(8)$.

\begin{figure}[h]
    \centering
    \begin{tikzpicture}
        \node at (-1.5, 0.5) {$\ket{0}$};
        \draw[thick] (-1, 0.5) -- (0, 0.5);  
        \draw[thick] (1, 0.5) -- (4, 0.5);   
        \node at (-1.5, -0.5) {$\ket{0}$};
        \draw[thick] (-1, -0.5) -- (4, -0.5); 
        \draw[thick] (0, 0) rectangle (1, 1);
        \node at (0.5, 0.5) {H};
        \filldraw (2.5, 0.5) circle (3pt);
        \draw[thick] (2.5, 0.5) -- (2.5, -0.8); 
        \draw[thick] (2.5, -0.5) circle (0.3);
        \node at (6.0, 0) {$\ket{\Phi^+} = \frac{1}{\sqrt{2}} (\ket{00} + \ket{11})$};
    \end{tikzpicture}
    \caption{Quantum circuit to create a Bell state \(\ket{\Phi^+}\). First a Hadamard is applied to qubit $1$, then a \texttt{CNOT} is applied to qubit $2$. In Dirac notation, this corresponds to $\ket{\Phi^+} = (\texttt{CNOT}_{12})(H \otimes \mathbb{1}) \ket{00}$.}
    \label{fig:bell_state}
\end{figure}
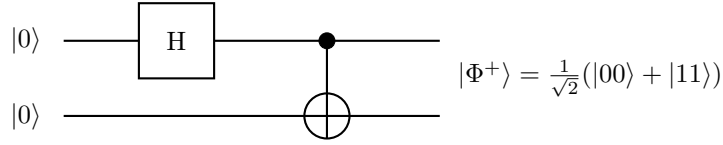

Quantum circuit notation also allows for time-wise composition of unitary gates, simply by placing them one after another. For example, Fig.~\ref{fig:bell_state} shows the quantum circuit for creating a Bell state.

We are now in a position to construct universal gate sets, the most common of which is one-qubit gates with \texttt{CNOT}.
In fact, many-qubit system is equivalent in its description to
spin-$1/2$ chain. This simple observation, allows to make a connection between \textit{analog} and \textit{digital} quantum computing, or quantum simulators. Let us start with defining Pauli string operator on a spin-$1/2$ chain with $L$ spins (qubits). We will use the following notation for Pauli spin operators acting on $i$-th spin (qubit): $\hat{X}_i,\hat{Y}_i,\hat{Z}_i$, defined as:
\begin{equation}
 \begin{split}
 \hat{X}_i  &= \mathbb{1}_1\otimes\dots\mathbb{1}_{i-1}\otimes\hat{\sigma}_x\otimes\mathbb{1}_{i+1}\dots\mathbb{1}_{L},\\
 \hat{Y}_i  &= \mathbb{1}_1\otimes\dots\mathbb{1}_{i-1}\otimes\hat{\sigma}_y\otimes\mathbb{1}_{i+1}\dots\mathbb{1}_{L},\\
 \hat{Z}_i  &= \mathbb{1}_1\otimes\dots\mathbb{1}_{i-1}\otimes\hat{\sigma}_z\otimes\mathbb{1}_{i+1}\dots\mathbb{1}_{L},
 \end{split}
\end{equation}
where $\hat{\sigma}^{x,y,z}$ are $2\times2$ Pauli operators.

It is worth noticing, that this representation of operators represent their decomposition in the eigenbasis of the  $\hat{Z} = \sum_{i=1}^{L} \hat{Z}_i$ operator, $\hat{Z}|v_l\rangle = v_l |v_l\rangle$, where the eigenvectors $|v_l\rangle$ are
\begin{equation}
 \begin{split}
   |v_1\rangle & = |\uparrow \uparrow \dots \uparrow \rangle \\
   |v_2\rangle & = |\uparrow \uparrow \dots \downarrow \rangle \\
   & \vdots \\
   |v_D\rangle & = |\downarrow \downarrow \dots \downarrow \rangle.
 \end{split}
\end{equation}

We can represent decomposition of  spin-chain Pauli operators in this basis as
\begin{equation}
\begin{split}
   \hat{X}_i & = \sum_{k,l} \langle v_k|\hat{X}_i|v_l\rangle|v_k\rangle\langle v_l| \\
   \hat{Y}_i & = \sum_{k,l} \langle v_k|\hat{Y}_i|v_l\rangle|v_k\rangle\langle v_l| \\
   \hat{Z}_i & = \sum_{k,l} \langle v_k|\hat{Z}_i|v_l\rangle|v_k\rangle\langle v_l| \delta_{k,l},
\end{split}
\end{equation}
where $\delta_{k,l}$ is Kronecker delta. In the following we denote: $\uparrow \equiv 0$, $\downarrow \equiv 1$.

The spin-$1/2$ many-body wave-function, or many-qubit system wave-function, $|\psi\rangle$, can be expressed in the composite Fock basis:
\begin{equation}
|\psi\rangle = \sum_{k} \psi_k |v_k\rangle,
\end{equation}
with $\sum_k |\psi_k|^2 = 1$. As such, the state of the system is encoded in a vector:
\begin{equation}
   |\psi\rangle    \Leftrightarrow \begin{pmatrix} \psi_1 \\ 
                               \psi_2 \\
                               \vdots \\ 
                               \psi_D 
                \end{pmatrix}
\end{equation}

Now, we can very directly define any qubit gate. For example 
\begin{enumerate}

    \item Rotational gates acting on $j$-th qubit 
\begin{equation}
 \begin{split}
 \hat{R}^x_j(\theta) = e^{-i\frac{\theta}{2} \hat{X}_j} \\
 \hat{R}^y_j(\theta) = e^{-i\frac{\theta}{2} \hat{Y}_j} \\
 \hat{R}^z_j(\theta) = e^{-i\frac{\theta}{2} \hat{Z}_j} \\
 \end{split}
\end{equation}

    \item Hadamard gate acting on $j$-th qubit

    \begin{equation}
    \hat{H}_j = \frac{1}{\sqrt{2}}(\hat{X}_j + \hat{Z}_j)
    \end{equation}

    In fact, Hadamard gate can be represented as a combination of two rotational gates, namely 
    \begin{equation}
        \hat{H}_j = \hat{R}_j^z(\pi)\hat{R}_j^y\left(\frac{\pi}{2}\right)
    \end{equation}

\item Phase gate acting on $j$-th qubit 

\begin{equation}
 \hat{P}_j(\theta) = e^{i\frac{\theta}{2}}\hat{R}^z_j(\theta)
\end{equation}

\item XX interaction gate between $i$-th and $j$-th qubit

 \begin{equation}
 \hat{R}^{xx}_{i,j}(\theta) = e^{-i\frac{\theta}{2}\hat{X}_i\hat{X}_j}
\end{equation}

\item YY interaction gate between $i$-th and $j$-th qubit

 \begin{equation}
 \hat{R}^{yy}_{i,j}(\theta) = e^{-i\frac{\theta}{2}\hat{Y}_i\hat{Y}_j}
\end{equation}

\item ZZ interaction gate between $i$-th and $j$-th qubit

 \begin{equation}
 \hat{R}^{zz}_{i,j}(\theta) = e^{-i\frac{\theta}{2}\hat{Z}_i\hat{Z}_j}
\end{equation}

\item Control-Z gate acting on $i$-th and $j$-th qubit:

\begin{equation}
    \text{CZ}_{ij} = \frac{1}{4} \left( \mathbb{1} + \hat{Z}_i + \hat{Z}_j - \hat{Z}_i\hat{Z}_j \right)
\end{equation}

In fact, $\text{CZ}_{ij}$ can be expressed using single-qubit rotational gates and the two-qubit ZZ interaction gate, namely
\begin{equation}
\text{CZ}_{ij} = \hat{R}^z_i(\pi)\hat{R}^z_j(\pi)\hat{R}^{zz}_{ij}\left(\frac{\pi}{4}\right)
\end{equation}

\item Control-X (CNOT) gate acting on $i$-th and $j$-th qubit:

\begin{equation}
 {\rm CNOT}_{i,j} = e^{i\frac{\pi}{4}(\mathbb{1} - \hat{Z}_i)(\mathbb{1}-\hat{X}_j)} = \hat{H}_j \text{CZ}_{ij}\hat{H}_j
\end{equation}

\item SWAP gate between $i$-th and $j$-th qubit
\begin{equation}
{\rm SWAP}_{ij} = \frac{1}{2}(\mathbb{1}  + \hat{X}_i\hat{X}_j + \hat{Y}_i \hat{Y}_j + \hat{Z}_i\hat{Z}_j)
\end{equation}

${\rm SWAP}_{ij}$ gate can be represented as three $\text{CNOT}$ gates, i.e.
\begin{equation}
    {\rm SWAP}_{ij} = {\rm CNOT}_{ij}{\rm CNOT}_{ji}{\rm CNOT}_{ij}.
\end{equation}
Now, we can express $\text{SWAP}_{ij}$ gate into rotational gates and ZZ-interaction gates as the following composition of three unitaries $\hat{A}_{1,2,3}$
\begin{equation}
\begin{split}
\text{SWAP}_{ij} & = \hat{A}_1\hat{A}_2\hat{A}_3\\
\hat{A}_1 & = \hat{H}_j \hat{R}^z_i(\pi) \hat{R}^z_j(\pi) \hat{R}^{zz}_{ij}\left(-\frac{\pi}{2}\right) \hat{H}_j\\
\hat{A}_2 & = \hat{H}_i \hat{R}^z_j(\pi) \hat{R}^z_i(\pi) \hat{R}^{zz}_{ji}\left(-\frac{\pi}{2}\right) \hat{H}_i\\
\hat{A}_3 &= \hat{H}_j \hat{R}^z_i(\pi) \hat{R}^z_j(\pi) \hat{R}^{zz}_{ij}\left(-\frac{\pi}{2}\right) \hat{H}_j.
\end{split}
\end{equation}

As we can see, in fact every single- and two-qubit quantum gate can be decomposed into single qubit rotations, and two-qubits interactions gates. Lastly, the most importantly, the two-qubit interactions gate are implemented in the evolution of a quantum system. This builds a bridge between \textit{analog} and \textit{quantum-gates} based quantum computers, where inherently present interactions between spins can be used to construct quantum gates. To see this analogy, let us consider the nearest-neighbor Ising Hamiltonian
\begin{equation}
    \hat{H}_{\text{Ising}} = -J\sum_{i=1}^{L}\hat{X}_i\hat{X}_j,
\end{equation}
where $J$ is coupling constant. The time evolution of the state is given by unitary transformation 
\begin{equation}
    \hat{U}(t) = e^{-i t \hat{H}_\text{Ising} } = e^{-i tJ \sum_{i}\hat{X}_i\hat{X}_{i+1}}.
\end{equation}
Now, if we tune the parameters, i.e. the time evolution and coupling constant, in such a way that $Jt = \frac{\theta}{2}$, then we implement $\hat{R}^{XX}_{i,i+1}(\theta)$ entangling gate acting on every neighboring pairs of spins.

\end{enumerate}

\section{Quantum Entanglement}

Quantum entanglement stands as a central and non-classical resource in quantum computing, enabling protocols and advantages over classical approaches \cite{jaeger2009entanglement}. Although bipartite entanglement—particularly for two qubits—is relatively well characterized, understanding and classifying multipartite entanglement, even at the level of three qubits, reveals a much richer structure \cite{bengtsson2016brief}. Multiple inequivalent entanglement classes emerge, and not all standard bipartite tools easily extend to these scenarios. For an excellent in-depth introduction to quantum entanglement, we suggest Ref.\cite{Bengtsson2006, Latorre2009, Horodecki2009, Srivastava2024,Horodecki2024}.

Here, we will start by briefly reviewing the resource theory of entanglement, introducing it from the foundations of quantum information (i.e. Hilbert space, states, measurements, and channels). This will allow us to review bipartite entanglement and later extend our brief discussion to multipartite systems. We will focus on three qubits as a minimal example of new complexities that arise, showing analytically the known hierarchy of classes, including fully separable, bi-separable, and genuinely tripartite classes.  In this way, we can start to understand how the structure of quantum entanglement can be learned or inferred by posing the question of whether (and the degree to which) a system is entangled or not in terms of basis dimensionality reduction. In some sense,  dimensionality reduction is the task of seeking a low-dimensional subspace on which our states (or data) live, in line with the \textit{manifold hypothesis} from Chapter~\ref{CH:UNSUPERVISED}. We will then be able to implement basis dimensionality reduction with techniques of unsupervised learning (PCA, t-SNE) to see how it can distinguish between different entanglement classes solely on density matrices. 


\subsection{Hilbert Space of Quantum States}

Quantum states of $d$ dimensions live in complex Hilbert spaces, $\mathcal{H}_d \cong \mathbb{C}^d$, and a pure state is represented by a normalized vector $|\psi\rangle \in \mathcal{H}_d$. Physical states are invariant under global phases, and all measurable physics emerges from projectors $|\psi\rangle\langle\psi|$.
However, not all states are pure. Mixed states, represented by density matrix $\hat{\rho}$, arise from statistical mixtures or open-system interactions. You can think of this as a mechanism for introducing classical uncertainty into quantum information, or representing some incomplete knowledge of a global, pure system.
Density operators are positive semidefinite and have trace $1$, with pure states corresponding to rank-1 projectors. When we want to extract information from a quantum system, we measure it. A projective measurement is defined by a set of orthogonal projectors $\{\hat{P}_a\}$ summing to the identity. By Born’s rule, the probability of outcome $a$ in state $|\psi\rangle$ is $p_a=\bra{\psi}\hat{P}_a\ket{\psi}$. Observables are Hermitian operators whose eigenvalues correspond to possible outcomes. Crucially, noncommuting observables reflect the intrinsic nonclassical uncertainty of quantum theory. The generalization of projective measurements is given by Positive-Operator Valued Measures (POVM).

\subsection*{Bipartite Entanglement Measures}

For composite systems $A|B$, the joint space is $\mathcal{H}_A\otimes\mathcal{H}_B$. Reduced states arise from partial traces, e.g. $\hat{\rho}_A=\mathrm{Tr}_B\hat{\rho}_{AB}$. Let us consider a bipartite pure state $|\psi_{AB}\rangle$. It is separable if it factorizes into $|\alpha\rangle_A\otimes|\beta\rangle_B$. If no such factorization exists, $|\psi_{AB}\rangle$ is entangled. For mixed states, separability requires that $\hat{\rho}_{AB}$ can be written as a convex combination of product states:
$$ \hat{\rho}_{AB}=\sum_i p_i(\hat{\rho}_i^{(A)}\otimes \hat{\rho}_i^{(B)}). $$
If no such decomposition is possible, we have entanglement. However, for a given quantum state, it is hard to check if such a decomposition exist, thus different criteria for entangled states have been constructed. 

One widely used test is the Peres-Horodecki criterion (also known as Positive-Partial Transpose, PPT), based on the partial transposition $T_B$, which acts only on subsystem $B$. If $\hat{\rho}^{T_B}\not\succeq 0$, then $\hat{\rho}_{AB}$ is entangled. For $2\times 2$ systems, PPT is both necessary and sufficient, providing a simple and powerful criterion. For higher dimensions, PPT remains necessary but not always sufficient, giving rise to PPT entangled states. Nonetheless, PPT is a foundational and easily implemented test, fundamental in low-dimensional entanglement detection. To go beyond PPT, we use entanglement witnesses. An entanglement witness $\hat{W}$ is a Hermitian operator that satisfies:
$$ \text{Tr}(\hat{W}\hat{\rho}_{\text{SEP}})\geq0 \text{ for all separable } \hat{\rho}_{\text{SEP}}, $$
but there exists at least one entangled $\hat{\sigma}$ such that
$$ \text{Tr}(\hat{W}\hat{\sigma})<0. $$

Because no single witness detects all entangled states, we often consider families of witnesses. They complement the PPT test and can sometimes detect states PPT fails to catch. For two qubits, a well-known entanglement witness is given by:
$$ \hat{W} = \frac{1}{2}\mathbb{I}\otimes \mathbb{I} - |\phi^+\rangle\langle\phi^+|, $$
where
$$ |\phi^+\rangle=\frac{|00\rangle+|11\rangle}{\sqrt{2}} $$
is a maximally entangled Bell state. This $\hat{W}$ detects all states that have “less overlap” with $|\phi^+\rangle$ than a certain threshold. In fact, for any separable state $\hat{\rho}_{\text{SEP}}$,
$$\text{Tr}(\hat{W}\hat{\rho}_{\text{SEP}})\ge0,$$
but for states close to $|\phi^+\rangle$, $\text{Tr}(\hat{W}\hat{\sigma})$ can become negative, thereby signaling entanglement. This witness is often considered a canonical example since it distinguishes entangled states close to a given Bell state.

More generally, one can construct witnesses tailored to detect entanglement along particular directions in state space. For instance, if we suspect that a state is close to $|\psi_{AB}\rangle$ which is known to be entangled, we can define:
$$ \hat{W}_{\psi}=\alpha\mathbb{I}\otimes\mathbb{I}-|\psi\rangle\langle\psi|, $$
where $\alpha$ is chosen so that all separable states yield nonnegative expectation values. By tuning parameters, we can detect different classes of entangled states.

\subsection{Multipartite Entanglement}

Multipartite entanglement is a natural generalization of bipartite entanglement, however it is  much more complex, because an 
$N$-partite system can be divided into subsystems in exponentially many ways, with each partition leading to its own set of locality constraints. As a result, no single theory or framework can fully capture all aspects of multipartite entanglement. Instead, its study depends on the specific goals and resources at hand.
In the following, we will define the structure of multipartite states and understand their unique features, and discuss key entanglement measures and invariants that help classify different types of multipartite entanglement.

Starting with pure states, an unentangled state vector in a multipartite Hilbert space $\mathcal{H} = \mathcal{H}_1 \otimes \mathcal{H}_2 \otimes ... \otimes \mathcal{H}_N$ of $N$-distinguishable constituents, is given by the product state of the form
\begin{equation}\label{Eq:pure_fullseparable}
    \ket{\Psi} = \ket{\psi_1} \otimes \ket{\psi_2} \otimes ... \otimes \ket{\psi_N}.
\end{equation}
Any state vector which is \emph{not} of this form is said to be entangled.

Unlike the case of bipartite systems, there are many ways in which one can create partitions in the multipartite system over the different parties, which gives rise to the notion of partial separability of a state based on these partitions.

\noindent\textbf{Partial separability in pure states}.-- The Hilbert space of $N$-parties can be coarse-grained by creating partitions $\{I_1, I_2,..., I_k\}$ such that $\cup_i^k I_i = I$ where $\{I_l\}$ are the disjoint subsets of indices, with $I = \{1,2,3,...,N\}$. Due to this, a state which may have been previously entangled will now be separable under such partitions. This is called partial separability and a state which is partially separable under such partitions is expressed as
\begin{equation}
    \ket{\Psi} = \ket{\psi_{I_1}}\otimes \ket{\psi_{I_2}} \otimes ... \otimes \ket{\psi_{I_k}},
\end{equation}
which looks similar to the state from Eq.~\eqref{Eq:pure_fullseparable}, with the only difference that $\ket{\psi_{I_i}}$ is a collective pure state of the parties belonging to the subset $I_i$.\\

Entanglement, or more generally speaking \textit{quantum correlations}, can be considered as a resource for different tasks, such as communication or computations. In resource theory, there are resource free operations which are physical (or permissible) operations, and resource free states which can be obtained from the action of resource free operations. When considering entanglement as a resource we define:
\begin{itemize}
    \item \emph{Resource free operations}: entanglement non-increasing operations or physically realizable operations(specifically, local operations with classical communication - LOCC)
    \item \emph{Resource free states}: separable states which can be generated from the resource free operations. An entangled state is automatically a resourceful state.
\end{itemize}

The resource theory of entanglement provides a framework for better characterization and quantification of entanglement, and study the manipulation and transformation of quantum states, and their utilization for specific quantum information tasks. Since entanglement is a purely non-local quantum phenomenon, any local operation \emph{cannot} increase the entanglement of the system. In this sense, entanglement is a resource; it is something we are given at the beginning of some process in the form of entangled states, and resource theory establishes what we can construct from it.

We start with the simplest class of resource-free operations which are the local unitary operations acting on each of the subsystems.\\

\noindent\textbf{Local unitary (LU) operations}.-- 
Any two state vectors $\ket{\Psi}, \ket{\Phi}$ are considered to be equivalently LU-entangled if they differ by a local unitary basis change:
\begin{equation}
    \ket{\Psi}\sim_{LU}\ket{\Phi} \equiv \ket{\Psi} = (U_1 \otimes U_2 \otimes ... \otimes U_N)\ket{\Phi},
\label{eq:LU}
\end{equation}
where $U_i$ are the local unitary operations acting on the respective parties. Having defined the LU-equivalence relation, the next step is to classify the set of entangled states into LU-inequivalent classes.\\

For the bipartite case, this classification into entangled and non-entangled states is much simpler due to the notion of so called Schmidt coefficients. A bipartite pure state can be expressed in the form
    \begin{equation}
        \ket{\Psi} = \sum_{i=1}\sqrt{p_i} \ket{\psi_i}\otimes\ket{\phi_i},
    \end{equation}
and the action of local unitaries $U_1$ and $U_2$ on the respective subsystems results in the basis change of the subsystems shown as
    \begin{equation}
        (U_1 \otimes U_2)\ket{\Psi} = \sum_{i=1}\sqrt{p_i}(U_1\ket{\psi_i})\otimes (U_2\ket{\phi_i}).
    \end{equation}
The Schmidt coefficients  $\{p_i\}$ are invariant under the action of local unitaries. The LU-inequivalent classes are completely described by these Schmidt coefficients in the bipartite case.

However, for the multipartite case, the characterization becomes more challenging. This is evident from counting the number of parameters necessary to describe a vector of $N$-qubits in the quotient space with respect the given inequivalent relation. LU-parameter counting in $N$-qubits requires taking $2^{N+1} - 2$ real parameters to specify a normalized quantum state in $\mathcal{H} = (\mathcal{C}^2)^{\otimes N}$, whereas the group of local unitary transformations $\text{SU}(2) \times ... \times \text{SU}(2)$ has $3N$ real parameters. This means that even in the case of $N-$qubits, one needs at least $2^{N+1}-3N-2$ real numbers to parameterize the sets of LU-inequivalent pure quantum states. Unlike the Schmidt coefficients from the bipartite case, most of these parameters do not have a physical interpretation.

In contrast to local unitary operations, \emph{global} unitary ones are able to create entanglement. 
Such features were quantified as \emph{entangling power}, which measures the (average or maximal) entanglement created on separable pure states, studied for both bipartite systems, as well as in the multipartite case \cite{Linowski2020}.

Now we shall extend the realm of unitary operations.  
Apart from LU, there exists a wider class of operations that cannot generate entanglement from separable states. 
It will give rise to a coarser notion of ``equivalent entanglement".\\

\noindent\textbf{Local operations and classical communication (LOCC)}.-- 
A more general form of the LU operations which can still be completely defined classically are the local operations with classical communication (LOCC). LOCC allows for exchange of classical information of the local measurement outcomes of the respective parties, hence, the name. Thus, the choice of the local operations of individual parties can be affected by the information of measurement outcomes by any other party. The best way to imagine this is with the help of ``distant laboratories model". There are $N$-particles each in their own laboratory. The particles may or may not be entangled in the first place. Each laboratory is capable of performing local measurements on its own particle, and convey the information regarding the outcome without exchanging the quantum systems among themselves. The LOCC map $\Lambda$ acts as
\begin{equation}
    \Lambda (\hat{\rho}) = \sum_k \left(\otimes_{i=1}^N K_{k,i}^{I_i}\right) \hat{\rho}\left(\otimes_{i=1}^N K_{k,i}^{I_i}\right)^{\dagger},
\end{equation}
where $K_{k,i}^{I_i}$ are the concatenated Kraus operators of each LOCC round, acting on the Hilbert space of party $I_i$ respectively. This is a direct generalisation of LU operations, where we considered the unitary matrices of Eq.~\eqref{eq:LU}, which are composed of only one Kraus operator acting on party $I_i$, given by $U_{I_i} = K_{k,i}^{I_i}$.
LOCC are considered the resource-free operations since any separable state $\ket{\Psi} \in \text{SEP}$ can be obtained from any other state in the Hilbert space by the action of LOCC. On the other hand, it is impossible to generate any entangled state from a separable state through LOCC alone.

\noindent\textbf{\emph{Stochastic}-LOCC (SLOCC)}.--
Any two states are SLOCC equivalent if they can be converted into each other by LOCC with some finite probability. Similar to LOCC, the SLOCC protocol consists of several rounds with each party performing operations based on measurement outcomes by other parties. The whole process can be imagined as a tree, and every measurement results in a new branch of the tree. If at least any one of the branch leads to the target state $\ket{\Phi}$ starting from $\ket{\Psi}$, the two states are SLOCC-equivalent. Such a conversion from $\ket{\Psi}\xrightarrow{\text{SLOCC}}\ket{\Phi}$ is possible under SLOCC if there exists a set of operators $\{A_i\}$ (compared to Eq.~\eqref{eq:LU}) such that,
\begin{equation}
\ket{\Psi}\sim_{\text{SLOCC}}\ket{\Phi} \equiv \ket{\Psi} = (A_1 \otimes A_2 \otimes ... \otimes A_N)\ket{\Phi}.
    \label{eq:SLOCC}
\end{equation}
In particular, the two states $\ket{\Psi},\ket{\Phi}$ are SLOCC-equivalent iff the matrices $\{A_i\}$ are invertible (or determinant $\text{det}\ A_i \neq 0$). The matrices $\{A_i\}$ for which $\text{det}\ A_i = 1$ which satisfy this property form the \emph{special linear} group SL. Similar to the case of $N-$qubits with LU-equivalence, one obtains a lower bound on the number of parameters required to define the SLOCC-equivalence classes given by $2^{N+1} - 6N - 2$ by substituting the $\text{SU}(2)$ group with $\text{SL}(\mathbb{C}^{2})$ group.\\

Now that we have defined the operations and transformations that are allowed within the resource theory framework of entanglement, we consider the example of three qubits to study the classification of states, and the corresponding invariants for each of the class.\\

\begin{figure}[t!]
    \centering
    \begin{mybox}[\hypertarget{box:3QEnt}{Box 5: Three-Qubit Entanglement Classes}]
    The tensor rank gives rise to the following different entanglement classes for \emph{pure states} of three qubits:
    \begin{itemize}
        \item \textbf{Separable states (S)} The first class of states that exist are the set of separable states (or free states) written as product states of the form $\ket{\Psi}  = \ket{\psi_1}\otimes\ket{\psi_2}\otimes\ket{\psi_3}$. In this particular case of separable states, $r(\rho_1) = r(\rho_2) = r(\rho_3) = 1$, where $r(\rho_i)$ is the rank of the reduced density matrix of subsystem-$i$. The rank is invariant under the action of invertible SLOCC.\\
    
        \item \textbf{Biseparable states (B)} Next, there are three classes of bipartite states based on the bipartitions $1|23, 2|13,$ and $3|12$. The set of pure product states in $i|jk$ form an SLOCC class, and are of the form $\ket{\psi_i}\otimes\ket{\Phi_{j,k}}$, where $\ket{\Phi_{j,k}}$ is a non-factoring entangled state of subsystems-$j,k$, where $r(\rho_i) = 1$, and $r(\rho_{j}) =r(\rho_{k}) = 2$. Thus, giving rise to three separate SLOCC-inequivalent classes of entanglement.\\
    
        \item \textbf{Genuinely entangled states} Furthermore, we have the set of states, for which the $r(\rho_i) = 2$ where $\rho_i$ is the reduced density matrices of party $i\in\{1,2,3\}$. It has been shown that there are two inequivalent classes that exist which are genuinely 3-qubit entangled. These two classes of states are the W and the GHZ states, 
        \begin{equation}
        \ket{\text{GHZ}} = \frac{1}{\sqrt{2}}\left( \ket{000} + \ket{111}\right), \ \ \ \  
        \ket{\text{W}} = \frac{1}{\sqrt{3}}\left( \ket{001}+\ket{010}+\ket{100}\right).  
    \end{equation}   
    Based on the definition of tensor rank, we can see that it is not possible to express $\ket{\text{W}}$ state with using only two product terms unlike $\ket{\text{GHZ}}$ state, i.e., $r_{\text{min}}(\ket{\text{GHZ}}) = 2$ and $r_{\text{min}}(\ket{\text{W}}) = 3$, hence they cannot be interconverted using SLOCC alone.

    \end{itemize}
    These four inequivalent classes, form the entanglement hierarchy in the convex sets \textit{S}$\subset$\textit{B}$\subset$\textit{W}$\subset$\textit{GHZ}, where \textit{S} is the set of all fully separable states, \textit{B} is the set of biseparable states, \textit{W} is the set of all states formed with LOCC from biseparable and W-class of states, and \textit{GHZ} is the set of all mixed states. 
    \end{mybox}
\end{figure}

The simplest model to extend our discussion on multipartite entanglement is a tripartite system of three qubits, i.e., with local dimension $d=2$. Based on the equivalence relations shown above with respect to different types of operations, quantum states of 3-qubits can be classified into six inequivalent entanglement classes.
Before advancing the discussion further towards the classification of 3-qubit entangled states, let us define the notion of class invariants.

SLOCC-invariants are referred to as the properties of the states that do not change under the action of SLOCC. These invariants play a crucial role in characterizing and distinguishing between different types of entangled states within the SLOCC framework. Tensor rank is the most known and studied SLOCC-invariants which we will  use to classify the genuinely tripartite entangled states. The \textit{rank of a  tensor} $T$ of order $d$ defined on the Hilbert space $\mathcal{H}_1 \otimes \mathcal{H}_2 \otimes \cdots \otimes \mathcal{H}_d$, is defined via:
$
r(T) = \min \left\{ r \,\middle|\, T = \sum_{i=1}^r \bigotimes_{j=1}^d v^{(j)}_i, \, v^{(j)}_i \in \mathcal{H}_j \right\},
$
where:
$r$ is the minimum number of separable product states $\bigotimes_{j=1}^d v^{(j)}_i$ in the decomposition, and
$v^{(j)}_i \in \mathcal{H}_j$ are vectors in the individual Hilbert spaces.  For a \emph{fully separable state}, $\mathrm{r}(T) = 1$, as the state can be written as a single product of local states.
A \emph{maximally entangled state} or a highly entangled many-body state will have $\mathrm{rank}(T) > 1$. Based on the tensor rank, we can are now in a position to write down the distinct entanglement classes of 3-qubit pure states. These are described in Box~\hyperlink{box:3QEnt}{4.2}.

\section{Quantum Measurements}

To have access to information about a considered quantum system, described by its density matrix $\hat{\rho}$ we have to \textit{measure the system}. Mathematically measurements are modelled via the quantum \textit{measurements} operators.

\subsubsection{Projective measurements}

A \textit{projective measurement} is a specific type of quantum measurement described by a set of Hermitian, orthogonal projection operators \(\{\hat{P}_i\}\) that satisfy the following conditions:
\begin{enumerate}
 \item  Hermiticity:
\begin{equation}
\hat{P}_i = \hat{P}_i^\dagger
\end{equation}
ensures the measurement outcomes are real and observable.

\item  Orthogonality
\begin{equation}
\hat{P}_i \hat{P}_j = \delta_{ij} \hat{P}_i
\end{equation}
indicates that different outcomes are mutually exclusive.

\item  Completeness
\begin{equation}
\sum_i \hat{P}_i = \mathbb{1},
\end{equation}
ensuring that one of the outcomes always occurs.

In this framework, when a projective measurement is applied to a quantum state $\hat{\rho}$, the probability of obtaining outcome $m_i$ is $p(i) = \mathrm{Tr}[\hat{P}_i \hat\rho]$,
and the post-measurement state becomes:
\begin{equation}
\hat{\rho} \leftarrow \frac{\hat{P}_i \hat{\rho} \hat{P}_i}{\mathrm{Tr}[\hat{P}_i \hat{\rho}]}.
\end{equation}
\end{enumerate}

\subsubsection[Positive Operator-Valued Measure]{Positive Operator-Valued Measure}

Projective measurements correspond to idealized measurements and are a subset of all possible quantum measurements. A generalization of projective measurements is given by \textit{Positive Operator-Valued Measure (POVM)} framework. POVM is defined by a set of positive semi-definite operators \(\{\hat{\pi}_i\}\), known as POVM elements, that satisfy:
\begin{enumerate}

\item  Positivity
\begin{equation}
    \hat{\pi}_i \geq 0
\end{equation}
ensuring that probabilities are non-negative.

\item  Completeness
\begin{equation}
\sum_i \hat{\pi}_i = \mathbb{1}.
\end{equation}
\end{enumerate}
Unlike projective measurements, the operators $\hat{\pi}_i$ do not need to be orthogonal or idempotent $(\hat{\pi}_i^2 \neq \hat{\pi}_i$).  This freedom allows POVMs to model measurements that are not tied to a single observable, e.g., joint or unsharp measurements, weak monitoring, or any scheme obtained by coupling the system to an ancilla and discarding part of the total state.

For an initial state \(\hat{\rho}\) the probability of outcome \(i\) is given by the Born rule
$p(i)=\operatorname{Tr}\!\bigl[\hat{\pi}_i\,\hat{\rho}\bigr]$
At this level the POVM fixes statistics only; it says nothing about how the state transforms on a given run.  
To describe the post-measurement state one must specify a \textit{measurement instrument}—a set of completely positive (CP) maps \(\{\mathcal E_i\}\) or, equivalently, Kraus operators \(\{\hat{M}_{i,\alpha}\}\) such that
\begin{equation}
\hat{\pi}_i=\sum_\alpha \hat{M}_{i,\alpha}^{\dagger}\hat{M}_{i,\alpha},
\qquad
\sum_{i,\alpha}\hat{M}_{i,\alpha}^{\dagger}\hat{M}_{i,\alpha}=\mathbb{1}.
\end{equation}
Conditional on outcome \(i\) the state updates according to
\begin{equation}
\hat{\rho}^{(i)}
   \;=\;
   \frac{1}{\displaystyle p(i)}\displaystyle
          \sum_\alpha M_{i,\alpha}\,\hat{\rho}\,M_{i,\alpha}^{\dagger},
\qquad
p(i)=\operatorname{Tr}\Bigl[\hat{\pi}_i\hat{\rho}\Bigr].
\end{equation}
Many distinct choices of \(\{\hat{M}_{i,\alpha}\}\) realize the same POVM and yield the same outcome probabilities but can leave the system in different conditional states; thus, the transformation rule is \textit{not unique} unless the physical implementation is specified.  
Sometimes, it is enough to consider \emph{square-root instrument}, where each result \(i\) is assigned just a \emph{single} Kraus operator,
\begin{equation}
  \hat{M}_i = \sqrt{\hat\pi_i} 
   \qquad
   \hat{M}'_i \equiv \hat{V}_i \hat{M}_i \qquad
   \sum_i \hat{M}_i^{\dagger}
   \hat{M}_i = \sum_i 
   \hat{M}_i^{'\dagger}
   \hat{M}'_i = \sum_i 
   \hat\pi_i = \mathbb{1},
\end{equation}
where $\hat{V}_i$ is an arbitrary unitary matrix. 
Because \(\hat{M}_i^\dagger \hat{M}_i = \hat\pi_i\), this choice reproduces the correct probabilities while yielding the conditional postmeasurement state.
\begin{equation}
   \hat\rho^{(i)}
   = \frac{\hat{M}_i\,\hat\rho\,\hat{M}_i^\dagger}{\operatorname{Tr}[\hat\pi_i\hat{\rho}]}
   = \frac{\sqrt{\hat\pi_i}\,\hat\rho\,\sqrt{\hat\pi_i}}{\operatorname{Tr}[\hat\pi_i\hat\rho]} .
\end{equation}

POVMs generalize projective measurements, enabling the description of more complex measurement scenarios, including measurements with noise or errors, weak measurements, and measurements in open quantum systems. Projective measurements (von Neumann) are a special case of POVMs where measurement operators are orthogonal projectors, that is, $\hat\pi_i = \hat{P}_i = \hat{M}_i$.

\subsubsection{Informationally Complete operators}

In quantum state tomography, an Informationally Complete (IC) set of operators is a collection of $d^2$ linearly independent operators $\{\hat{\pi}_i\}_{i=1}^{d^2}$ that fully determine the quantum state $\hat{\rho}$ in a $d$-dimensional Hilbert space. These operators enable reconstruction of $\hat{\rho}$ using the measured probabilities $p_i = \mathrm{Tr}[\hat{\rho} \hat{\pi}_i]$.
Such a collection of operator has to fulfill the following conditions:
\begin{enumerate}
    \item  Completeness: The operators span the space of Hermitian operators, allowing any $\hat{\rho}$ to be expressed in terms of $\{\hat{\pi}_i\}$.
    \item  Linearity: The set $\{\hat{\pi}_i\}$ must be linearly independent for unique reconstruction.
    \item  Reconstruction: The state $\hat{\rho}$ can be obtained as:
    $ \hat{\rho} = \sum_{i=1}^{d^2} c_i \hat{\pi}_i,
    $
    where $c_i$ are coefficients derived from $p_i$.
\end{enumerate}

As an examples of IC operators, we can consider:
\begin{enumerate}

\item Square-root POVM. -- The   set $\boldsymbol{\pi}$ of measurement operators consists of POVM generated by the square root measurements, defined as
\begin{equation}
    \hat{\pi}_i = \hat{H}^{-1/2}\ket{\phi_i}\bra{\phi_i}\hat{H}^{-1/2} \ \quad \hat{H} = \sum_{i\in [d^2]} \ket{\phi_i}\bra{\phi_i},
    \label{eq: root square povm}
\end{equation}
with $\{\ket{\phi_i}\in \mathcal{H}\}_{i\in [d^2]}$ are randomly generated Haar distributed pure states \cite{Mele2024}. \\

\item  Symmetric Informationally Complete POVM (SIC-POVM).--   The basis consists of the tensor product of local SIC-POVM, constructed by using the local vectors:
\begin{equation}\label{eq:POVMtetrahedral}
    \begin{split}
  \mathbf{s}_1 &= (0,0,1) \\
  \mathbf{s}_2 &=(\tfrac{2\sqrt{2}}{3}, 0, -\tfrac{1}{3} )\\
  \mathbf{s}_3 &=( -\tfrac{\sqrt{2}}{3},\sqrt{\tfrac{2}{3}},-\tfrac{1}{3} )\\
  \mathbf{s}_4 &=  (
 -\tfrac{\sqrt{2}}{3},-\sqrt{\tfrac{2}{3}},-\tfrac{1}{3} ) \ .
    \end{split}
\end{equation}
The space of Hermitian operators acting in the global Hilbert space $\mathcal{H} = [\mathbb{C}^2]^{\otimes L}$ can then be spanned by
\begin{equation}
    \hat{\boldsymbol{\pi}} = \left\{\hat{\pi}_{(a,b)\cong i}=\bigotimes_{b\in [L]}\frac{\hat{\sigma}_0 + \mathbf{s}_{a_b}\cdot \hat{\boldsymbol{\sigma}}}{4}  \right\}_{\{a_b \}_{b\in [L]}\in [4]^L} \ ,
\end{equation}
where $\hat{\boldsymbol{\sigma}}=(\hat{\sigma}_x, \hat{\sigma}_y, \hat{\sigma}_z)$ is the Pauli vector and $\hat{\sigma}_0 = \mathbb{1}_2$ is the identity acting in the local space $\mathbb{C}^2$.

Note that for any properly normalized state $\hat{\tau}$, $\mathbf{p}:=\mathrm{Tr}(\hat{\boldsymbol{\pi}}\hat{\tau})$ constitutes a valid probability distribution
\begin{equation}
    \forall i \in \left[2^{2L}\right],\ \  p_i\geq 0 \mbox{ and } \sum_{i\in \left[2^{2L}\right]}p_i = 1
\end{equation}
This observation is also true for the previous basis (square-root POVM) that we reviewed. \\

\item  Pauli basis.-- The Pauli basis constructed as:
\begin{equation}
    \hat{\boldsymbol{\pi}} = \left\{\bigotimes_{b\in [L]} \hat{\sigma}_{a_b}  \right\}_{\{a_b \}_{b\in [L]}\in \{0,x,y,z \}^L} \ .
\end{equation}
With respect to such a basis, expectation values can be evaluated experimentally by rotation of each qubit individually. This is also true for the SIC-POVM if evaluated with multiple settings. Such expectation values $\mathbf{p}$ no longer form a probability distribution (note that, in particular, such mean values can be negative). The reason why it will not lead to a probability distribution is that the basis does not form a POVM (i.e., its elements are not positive semi-definite and do not sum to $\mathbb{1}$). However, it spans equally well the whole space of Hermitian matrices supported in $[\mathbb{C}^{2}]^{\otimes L}= \mathcal{H}$.

\end{enumerate}

\section{Quantum State Tomography for Multi-Qubit Systems}

Modern quantum technologies rely on resources such as coherence, quantum entanglement, and Bell nonlocality. To evaluate the advantages these resources offer, it is essential to certify their presence. The resource content of a quantum system is inferred from the statistical properties (e.g., correlations) observed in the outcomes it produces. In the quantum framework, this information is encapsulated in the density matrix.

In practical scenarios, the density matrix can be reconstructed using limited data obtained from experimentally accessible measurements—a process referred to as quantum state tomography. However, conventional QST protocols are inherently affected by various sources of noise, including measurement calibration errors, dark counts, losses, and technical disturbances, among others. These noise factors are highly challenging to model accurately and can ultimately lead to decoherence in the system, diminishing or erasing the quantum resources present.

In recent years, deep learning has made significant strides in quantum technologies, providing innovative approaches to quantum state tomography (QST). Neural networks can learn noise directly from experimental data without requiring knowledge of its sources or underlying models. This enables them to mitigate not only shot noise, which is intrinsic to QST, but also other disturbances. Requiring only minimal assumptions about the system, neural networks are particularly well-suited for certification tasks. They effectively act as denoising filters for standard QST protocols, enhancing the fidelity of the reconstructed density matrix.

In the following sections, we present an introduction to QST in many-body quantum system, covering the foundational aspects of the measurement problem. We will present the most known QST protocols, and, finally, we will demonstrate how neural networks can enhance QST for quantum circuits. We will start with mathematical formulation of quantum system measurement protocol.

\subsection*{Density matrix reconstruction}

We consider the $d$-dimensional Hilbert space. A set of informationally complete (IC) measurement operators $\hat{\boldsymbol{\pi}}=\{ \hat{\pi}_i \}$, $i = 1, \dots, d^2$, in principle, allows unequivocally reconstructing the underlying target quantum state $\hat{\tau} \in \mathbb{C}^{d \times d}$
within the limit of an infinite number of ideal measurements.
After infinitely many measurements, one can infer the mean values
\begin{equation}
\label{eq:p_Born}
p_{i}  = \mathrm{Tr}[\hat{\tau} \hat{\pi}_{i}],
\end{equation}
and construct a valid vector of probabilities $\mathbf{p} = \{p_i \}$
for any proper state $\hat{\tau}\in \mathcal{S}$, where by $\mathcal{S}$ we denote the set of $d$-dimensional quantum states, i.e., containing all unit-trace, positive semi-definite (PSD) $d\times d$ Hermitian matrices.
Alternatively, $\hat{\boldsymbol{\pi}}$ can form a set of operators that spans the space of Hermitian matrices. In such a case, $\mathbf{p}$ can be evaluated from multiple measurement settings (e.g., Pauli basis) and is generally no longer a probability distribution.  In any case, there exists a one-to-one mapping $Q$ from the mean values $\mathbf{p}$ to the target density matrix $\hat{\tau}$:
\begin{align}
\label{eq:Q}
  Q\!:\,&\mathcal{F}_{\mathcal{S}}\longrightarrow \mathcal{S} \\
\nonumber
  &\mathbf{p}\longmapsto Q[\mathbf{p}]=\hat{\tau},
\end{align}
where $\mathcal{F}_{\mathcal{S}}$ is the space of accessible probability vectors. In particular, by inverting the Born's rule, Eq.~\eqref{eq:p_Born}, elementary linear algebra allows us to describe the map $Q$ as
\begin{equation}\label{eq:Q_p}
Q[\mathbf{p}] = \mathbf{p}^T \hat{G}^{-1}\hat{\boldsymbol{\pi}},
\end{equation}
where $\hat{G}$ is the Gram matrix of the measurements settings, with components $G_{ij} = \mathrm{Tr}(\hat{\pi}_i\hat{\pi}_j)$. The inference of the mean values $\mathbf{p}$ is only perfect in the limit of an infinite number of measurement shots, $N\rightarrow \infty$.

In a realistic scenario, with a finite number of experimental runs $N$, we have access to frequencies of relative occurrence $\mathbf{f} = \{f_i := n_i/N\}$, where $n_i$ is the number of times the outcome $i$ is observed.
Such counts allow us to estimate $\mathbf{p}$ within an unavoidable error dictated by the shot noise, whose amplitude typically scales as $1/\sqrt{N}$.
With only frequencies $\mathbf{f}$ available, we can use mapping $Q$ for estimation $\hat{\rho}$ of the target density matrix $\hat{\tau}$, i.e.,
\begin{equation}\label{eq:QST_f}
 \hat{\rho} = Q[\mathbf{f}].
\end{equation}
In the infinite number of trials $N\to \infty$, $f_i = p_i$ and $\hat{\rho} = \hat{\tau}$. Yet, in the finite statistics regime, as considered in this work, the application of the mapping as defined in Eq.~\eqref{eq:Q_p} to the frequency vector $\mathbf{f}$ will generally lead to nonphysical results (i.e.\ $\hat{\rho}$ not PSD).
In such case, as an example of proper mapping $Q$ we can consider different methods for standard tomography tasks, such as linear inversion (LI), or maximum likelihood estimation (MLE). As operators $\boldsymbol{\hat{\pi}}$, we can also consider positive operator-valued measures (POVMs) and a more experimentally appealing Pauli basis.\\

Let us review the most well-known approaches to quantum state reconstruction. \\

\noindent\textbf{Linear inversion (LI)}.-- By inverting Born's rule Eq.~\eqref{eq:p_Born}  we can express the state dependence on the mean values $\mathbf{p} = \{p_i \}$.
\begin{equation}
    \label{eq:LI}
    \hat{\tau} = \mathbf{p}^T \mathrm{Tr}[\hat{\boldsymbol{\pi}}\hat{\boldsymbol{\pi}}^T]^{-1} \hat{\boldsymbol{\pi}} \ .
\end{equation}
Note that the inverse of the Gram matrix $\mathrm{Tr}(\hat{\boldsymbol{\pi}}\hat{\boldsymbol{\pi}}^T)$ exists as the basis is informationally complete (IC). If it is~(informationally-) overcomplete, one needs to replace the inverse with the pseudoinverse. Finally, if it is under-complete (only partial information is available), it will determine the state up to a linear subspace.

The LI method infers $\hat{\rho}$ by replacing in Eq.~\eqref{eq:LI} the ideal expectation values $\mathbf{p}$ with the vector $\mathbf{f}$ of experimental frequencies (counts). This naive substitution generally leads to a negative matrix, $\hat{\rho}\nsucceq 0$. An \textit{optimal} way to tame its negative eigenvalues is to find the nearest physical state to $\hat{\rho}$ under the 2-norm. The drawback of LI is the fact that it can be afflicted by any type of noise.\\

\noindent\textbf{Least-squares estimation (LSE)}.-- Here, the reconstructed state is chosen to minimize the mean square error between the experimental frequencies $\mathbf{f}$ and the state probability distribution $\mathrm{Tr}(\hat{\boldsymbol{\pi}}\hat{\rho})$. The resulting problem can be expressed as,
\begin{align}
\label{eq:LSE_problem}
  \hat{\rho}_{\mathrm{LSE}} =  \arg\min_{\hat{\rho}\succeq 0} |\mathbf{f}- \mathrm{Tr}(\hat{\boldsymbol{\pi}}\hat{\rho})|^2 .
\end{align}

\noindent\textbf{Maximum likelihood estimation (MLE)}.-- In this case, the reconstructed state maximizes the likelihood of having produced the observed experimental outcomes,
\begin{equation}
\hat{\rho}_{\mathrm{MLE}}=\arg\max_{\hat{\rho}\succeq 0} \log P(\hat{\rho}|\mathbf{f}).
\end{equation}
Our counting experiment is modelled as a multinomial. Consequently, the log-likelihood is $\log P(\hat{\rho}|\mathbf{f}) = \mathbf{f}\cdot \log(\mathrm{Tr}[\hat{\boldsymbol{\pi}}\hat{\rho})])$, which is a concave function of the state, but the resulting task is not a DCP.
Therefore, solving it can be expensive, especially for a large Hilbert dimension space $d$.
The MLE is a robust estimator against noise; however, it is computationally demanding, suffering from the exponential scaling of the inputs.   \\

\noindent\textbf{Classical shadows (CS)}.-- This quantum state reconstruction approach differs from previously described, because is not based on IC set of operators, but rather on randomized measurements. For a system containing $L$ spins-$1/2$ described by a density matrix  $\hat{\tau}$ decomposed in the computational basis
$\{\ket{\vec{s}}\} = \{ |s_1, s_2,s_3,\dots,s_N\rangle\} = \{ \bigotimes_{j\in [L]}\ket{s_j} \}$, with $s_j = \pm 1$, the classical shadows tomography \cite{Huang2020,Struchalin2021,Huang2022} aims to reconstruct the target quantum state via $M$ randomized measurements. In $m$-th measurement, we apply a random unitary
$U_{m} = \bigotimes_{j \in [L]}u_j^{(m)}$
to the target state  $\hat{\tau}_{m} = U_{m}\hat{\tau}U_m^\dagger$, with $u^{(m)}_j$ being random operators from some ensemble ${\cal U}$.
After the projective measurement in a computational basis the bit-string $\{s_1^{(m)},\dots,s_N^{(m)}\}$ is used to construct the $m$-th classical shadow of the initial state as
\begin{equation}
    \hat{\rho}_{m}  = {\cal M}^{-1}\bigg[\bigotimes_{j\in [L]} u_j^{(m)\dagger} \ket{s_j^{(m)}}\bra{s_j^{(m)}}\hat{u}_j^{(m)}\bigg],
\end{equation}
with the inverse map ${\cal M}^{-1}$ determined by ${\cal U}$. For a Pauli measurements group, the inverse map factorizes,
${\cal M}^{-1} = \bigotimes_{j\in [L]}{\cal M}_1^{-1}$, where ${\cal M}^{-1}_n[\cdot] = (2^n+1)[\cdot] - \mathbb{1}_{2^n}{\rm Tr}([\cdot])$, and the $m$-th classical shadow reads
\begin{equation}
    \hat{\rho}_{m} = \bigotimes_{j\in [L]} \left[3 u_j^{(m)\dagger} \ket{s_j^{(m)}}\bra{s_j^{(m)}}\hat{u}_j^{(m)} - \mathbb{1}_{2}\right].
\end{equation}
The reconstructed density matrix $\hat{\rho}_{\rm cs}$ is average over $M$ instances of classical shadows $\hat{\rho}_{\rm m}$
\begin{equation}
     \hat{\rho}_{\rm CS} = \frac{1}{M}\sum_{m\in [M]}\hat{\rho}_{m}.
     \end{equation}
Details of the numerical implementation of the Classical Shadows protocol for QST, with a Pauli group, can be find in Ref.\cite{PlodzienClassicalShadowsImplementation_2023}.

\section{Variational Principle}

Consider trying to solve for the ground state energy, $E_{gs}$, for a quantum system described by its Hamiltonian $H$. If we are unable to solve the time-dependent Schr\"odinger equation\footnote{This is, frustratingly, often the case in many-body quantum systems!}, then the variational principle is a technique which yields an upper bound for $E_{gs}$. This is often all we need, and with appropriate care and design, we can tighten this upper bound to be close to the true value. This is because for \textit{any} normalised wavefunction $\ket{\psi}$,
\begin{equation}
    E_{gs} \leq \braket{\psi|\hat{H}|\psi}.
\end{equation}
That is, all states except the ground state $\ket{\psi_{gs}}$ of a system described by $H$ will be an overestimate of that system's ground state energy. For a short, neat proof of this fact, check out \textit{D.J. Griffith's Introduction to Quantum Mechanics 3rd Ed., Ch.8} \cite{griffiths2018introduction}. How then, can we create a tight upper bound for a system's ground state energy? 

Let's consider the simplest case of a one-qubit system described by $\ket{\psi} \in \mathcal{H}_2$, governed by the Hamiltonian
\begin{equation}
    \hat{H} = \hat{\sigma}_z.
\end{equation}
Clearly, we see that $\ket{\psi} = \ket{1}$ is the ground state of this system because it minimises
\begin{equation}
    \hat{H} \ket{\psi} =  \hat{\sigma}_z \ket{1} = -\ket{1},
\end{equation}
with energy $E = -1$ in natural units. However, let's imagine for a moment that we \textit{did not} know this. What if instead, we are given a generic, normalised state,
\begin{equation}
    \ket{\psi(\alpha)} = \alpha \ket{0} + (1 - \alpha^2)^{\frac{1}{2}} \ket{1},
\end{equation}
where $\alpha \in [0,1] \subset \mathbb{R}$. Then, applying the Hamiltonian to this state, we see
\begin{equation}
    \hat{H} \ket{\psi(\alpha)} =  \hat{\sigma}_z (\alpha \ket{0} + (1 - \alpha^2)^{\frac{1}{2}} \ket{1}) =   \alpha \ket{0} - (1 - \alpha^2)^{\frac{1}{2}} \ket{1}.
\end{equation}
Hence, we see that the expectation value $\braket{H}$ with respect to $\ket{\psi}$ reads
\begin{equation}
\begin{aligned}
     \braket{\hat{H}} = \braket{\psi(\alpha)|H|\psi(\alpha)} &= \left(
    \alpha \bra{0} + (1 - \alpha^2)^{\frac{1}{2}} \bra{1}
    \right)
    \left(
     \alpha \ket{0} - (1 - \alpha^2)^{\frac{1}{2}} \ket{1}
    \right) \\
    &= \alpha^2 - (1 - \alpha^2) = 2\alpha^2 - 1 
\end{aligned}
\end{equation}
As expected, $\ket{\psi(\alpha)}$ is an upper bound to $E_{gs} = 1$ for all $\alpha \in [0,1]$. To get the \textit{tightest} upper bound, notice that $\braket{H}$ is a continuous function of $\alpha$. All we therefore have to do is minimise it,
\begin{equation}
    \frac{d}{d \alpha} \braket{\hat{H}} = 2\alpha = 0 \implies \alpha = 0,
\end{equation}
giving us $\ket{\psi} = \ket{1}$ as expected. We can use the variational principle for \textit{any} trial wave function $\ket{\psi(\alpha)}$ which is differentiable in $\alpha$. This is because all we required to minimise was differentiability, and we know it forms an upper bound.

\begin{figure}[h!]
    \centering
    \begin{mybox}[\hypertarget{box:varPrincip}{Box 4: Variational Principle}]
    For any normalised wavefunction, $\ket{\psi}$, the variational principle states that
    \begin{equation}
        E_{gs} \leq \braket{\psi|\hat{H}|\psi}.
    \end{equation}
    That is, all states of a quantum system with Hamiltonian H overestimate the ground-state energy except for the ground state itself, $\ket{\psi_{gs}}$. Variational methods seek to find a tight upper bound to this. 
    \end{mybox}
\end{figure}

\section{Phase Space and Quantum Optics}
\label{sec:CVQM} 
The next topic we will briefly review is some basic properties of quantum states of light. These properties can be found in any standard quantum optics textbook, although we highlight \cite{schleich2015quantum} and \cite{klimov2009group} for a particularly insightful introduction in terms of phase space and group theory respectively. 

Recall that we can describe a quantum state of light using a \textit{Fock space}, which is the extension of a two level system to the basis of states of the quantum harmonic oscillator. Unlike spin systems, quantum operators in Fock space are continuous, with all continuous quantum variables (CQVs) coming in canonical conjugate pairs. Without loss of generality, let these be $\hat{x}, \hat{p}$ satisfying
\begin{equation*}
    [\hat{x},\hat{p}] = i.
\end{equation*}
It is well known in quantum mechanics that a wavefunction, $\ket{\Psi}$, can be described in different representations (a basis), each related by a unitary (reversible) transformation. For CQVs, this transformation is simply the Fourier Transform, i.e.
\begin{equation*}
    \braket{x|\Psi} = \frac{1}{\sqrt{2 \pi}} \int_{\mathbb{R}} \mathrm{d}p \;  e^{-ipx} \braket{p|\Psi},
\end{equation*}
\begin{equation*}
    \braket{p|\Psi} = \frac{1}{\sqrt{2 \pi}} \int_{\mathbb{R}} \mathrm{d}x \;  e^{ipx} \braket{x|\Psi},
\end{equation*}
where $\braket{x|\Psi} \equiv \psi(x)$ is the position representation of the wavefunction $\ket{\Psi}$ vice versa for $\braket{p|\Psi}$. Inner products and overlaps for CQVs are normalised using Dirac's Delta function. For example, in the position basis, one may write $\braket{x'|x}  = \delta(x' - x)$ for eigenstates $\ket{x},\; \ket{x'}$. Superposition and entanglement are handled analogously to discrete quantum mechanics, extending to continuity in the intuitive way. Superposition in some basis, $x \in \mathbb{R}$, is represented using a function $f:\mathbb{R} \rightarrow \mathbb{C}$, which describes the amplitude of the wavefunction $\ket{\Psi}$ at a given position $x$,
\begin{equation*}
    \ket{\Psi} = \int_{\mathbb{R}} dx \; f(x) \ket{x},
\end{equation*}
which has a form reminiscent of discrete superposition in some basis $\ket{e_i}$ s.t. $\ket{\Psi} = \sum_i c_i \ket{e_i}$ for complex amplitudes $c_i$. Similarly, entanglement is represented with a function $g:\mathbb{R} \rightarrow \mathbb{C}$ over two tensored Hilbert spaces. For example, a perfectly correlated wavefunction in the position basis can be written as
\begin{equation*}
    \ket{\Psi} = \int_{\mathbb{R}} dx \; g(x) \ket{x} \otimes \ket{x - c},
\end{equation*}
implying that a measurement on the first basis of $x = x_0$ would allow us to conclude with certainty that a measurement on the second basis would yield $x_0 - c$. However, measurements in CVQM are defined over \textit{intervals} rather than points. This is simply because probability amplitudes from projectors are of the form
\begin{equation}
    \hat{P}_{x} = \int_a^b dx \; \ket{x}\bra{x}
    \label{eq:projector}
\end{equation}
so the probability amplitude of measuring an exact position is 0. A simple calculation shows that the operator in Eq.(\ref{eq:projector}) is idempotent and normalised and is hence a valid representation of a rank 1 projective measurement.

A many-body system of $n$ bosons is then described by n-pairs of CQVs whose corresponding canonical commutation relations are 
\begin{equation}
    [\hat{x}_{k},\hat{p}_{l}] = i \hat{\delta}_{kl}.
    \label{eq:fieldQuadCommute}
\end{equation}
Note that this means bosons are non-interacting particles - their operators on different subspaces commute. We can also describe $n$ bosons using creation and annihilation operators, with commutation relations given by
\begin{equation}
    [\hat{a}_k,\hat{a}_l^{\dagger}] = \hat{\delta}_{kl}.
    \label{eq:CommuteRelation}
\end{equation}
As such, the Hilbert space structure of $n$ bosons, $\mathcal{H} = \bigotimes_{k = 1}^N \mathcal{F}_k$, is the tensor product of $N$ infinite-dimensional Fock spaces, $\mathcal{F}_k$, which is spanned by the (complete and orthonormal) number basis, $\{\ket{n}_k: n \in \mathbb{N}, k = 1,\ldots N\}$. Neglecting  zero point energy, the free (non-interacting) Hamiltonian of such a system is $\hat{H} = \sum_{k = 1}^n \hat{a}_k^{\dagger}\hat{a}_k$. Eigenstates of the free Hamiltonian are number states satisfying 
\begin{equation}
    \hat{a}_k^{\dagger}\hat{a}_k \ket{n}_k = \hat{n}_k \ket{n}_k = n_k \ket{n}_k,
    \label{eq:modeOpDef}
\end{equation}
where the $k$ indexes which subspace we are in, i.e. $\hat{a}^{\dagger}_1 \hat{a}_1 = \hat{a}^{\dagger} \hat{a} \otimes \mathbb{1}^{\otimes N - 1}$ and $\hat{a}^{\dagger}_2 \hat{a}_2 = \mathbb{1} \otimes \hat{a}^{\dagger} \hat{a} \otimes \mathbb{1}^{\otimes N - 2}$. We may also relate our two representations via
\begin{equation}
    \hat{x}_{k} = \frac{1}{2}(\hat{a}_k + \hat{a}_k^{\dagger}),\; \; \; \; \hat{p}_l = \frac{1}{2i}(\hat{a}_l - \hat{a}^{\dagger}_l).
    \label{eq:Qmodes}
\end{equation}
\subsection{Coherent States}
Coherent states, $\ket{\alpha} \in \mathcal{F}$, with 
\begin{equation}
    \ket{\alpha} = e^{-\frac{|\alpha|^2}{2}}
    \sum_{m \in \mathbb{N}} \frac{\alpha^n}{\sqrt{n!}} \ket{n},
\end{equation}
are minimal uncertainty states, and are eigenstates of the annihilation operator \cite{schleich2015quantum},
\begin{equation*}
    \hat{a}\ket{\alpha} = \alpha \ket{\alpha},\;\; \alpha \in \mathbb{C}.
    \label{eq:coherent_eigenequation}
\end{equation*}
They can also be written in terms of the unitary displacement operator $\hat{D}(\alpha) = e^{\alpha \hat{a}^{\dagger} - \alpha^* \hat{a}}$,
\begin{equation*}
    \ket{\alpha} = \hat{D}(\alpha)\ket{0}.
\end{equation*}
They are over-complete
\begin{equation*}
    \frac{1}{\pi}\int_{\mathbb{C}}d^2 \alpha\; \ket{\alpha}\bra{\alpha} = \hat{\mathbb{I}},
\end{equation*}
and their overlap is
\begin{equation}
    |\braket{\alpha|\beta}|^2 = e^{-|\alpha - \beta|^2}.
    \label{eq:overapFormulaYes}
\end{equation}

It is also possible to write the creation and annihilation operators in this basis, which read,
\begin{equation*}
    \hat{a}^{\dagger} \equiv \frac{\partial}{\partial \alpha} + \frac{\alpha^*}{2}, \; \; \; \; \; \hat{a} \equiv \frac{\partial}{\partial \alpha^*} + \frac{\alpha}{2}.
\end{equation*}
\subsection{Phase Space of Light}
Just how Spins can have different basis, so too can quantum states of light. For this course, we will be working the two different types of phase space functions called the Wigner function and Q-function. The Wigner function is the quantum phase space in the position-momentum basis,
\begin{equation}
        W_{\hat{f}}(x,p) = \int_{\mathbb{R}}da\; e^{-2iap}\braket{x + a|\hat{f}(\hat{x},\hat{p})|x-a},
    \label{eq:Wig}
\end{equation}
where $\hat{f}(\hat{x},\hat{p})$ is some operator function of $\hat{x}$ and $\hat{p}$, for example a Gaussian or Cat state. Given that the coherent states also form a (overcomplete) basis, we can write down a phase space representation in the coherent basis too. This is gives the Husimi Q-function,
\begin{equation}
    Q_{\hat{f}}(\alpha,\alpha^*) = \braket{\alpha|\hat{f}(\hat{a},\hat{a}^{\dagger})|\alpha},
\end{equation}
which we can interpret as expectation values in the coherent basis. For example, the Q-function of Fock states $\{\ket{n}_k: n \in \mathbb{N}\}$, reads
\begin{equation}
    Q_{\hat{n}}(\alpha) = \frac{1}{\pi}\braket{\alpha|n}\braket{n|\alpha} = \frac{1}{\pi}|\braket{\alpha|n}|^2 = \frac{1}{\pi}\frac{|\alpha|^{2n}}{n!}e^{-|\alpha|^2}.
\end{equation}
We can also perform an expansion of the operator directly in the coherent basis (rather than calculate an expectation value). This gives the P-function of an operator $\hat{f}(\hat{a},\hat{a}^{\dagger})$,
\begin{equation}
        \hat{f} = \frac{1}{\pi}\int_{\mathbb{C}}d\alpha d\alpha^*\;P(\alpha)_{\hat{f}}\ket{\alpha}\bra{\alpha}.
        \label{eq:pfuncdef}
    \end{equation}
Since phase space functions are representing the same physical operator, $\hat{f}(\hat{a},\hat{a}^{\dagger})$, we can imagine that they must be related in some way. In fact, $P$ and $Q$ functions are duals of each other, which means they are connected by
\begin{equation}
    \text{Tr}(\hat{f}\hat{g}) = \int_{\mathbb{C}}d^2\alpha \; P_f(\alpha)Q_g(\alpha) = \int_{\mathbb{C}}d^2\alpha\; Q_f(\alpha)P_g(\alpha).
    \label{eq:QPConnect},
\end{equation}
This allows us to take inner products in phase space, with the Wigner function being its own dual. This gives us a free choice for how to evaluate expectation values and probabilities for $\hat{f}$ with respect to some state $\hat{\rho}$. for example we could evaluate with
\begin{equation*}
    \braket{\hat{f}} = \text{Tr}(\hat{\rho} \hat{f}) = \int_{\mathbb{C}} d^2 \alpha \; P_{\hat{\rho}}(\alpha) Q_{\hat{f}}(\alpha) = \int_{\mathbb{C}}d^2 \alpha \; Q_{\hat{\rho}}(\alpha) P_{\hat{f}}(\alpha).
\end{equation*}

Notice that in order to evaluate this expectation value in phase space, we need to leverage the eigenvalue equation from Eq.~(\ref{eq:coherent_eigenequation}). This is what will allow us to convert a function of quantum operators, $\hat{f}(\hat{a},\hat{a}^{\dagger})$, into a function of complex-valued scalars. To do this scalar conversion without losing the non-commutative structure of quantum operators, we have to decide on a pre-specified ordering of the quantum operators that construct $\hat{f}(\hat{a},\hat{a}^{\dagger})$. This is in fact why there are many different phase space functions; each one corresponds to a pre-specified ordering of $\hat{f}(\hat{a},\hat{a}^{\dagger})$. For example, to calculate the Q-function of $\hat{f}(\hat{a},\hat{a}^{\dagger})$, we need to be able to write down the operator $\hat{f}(\hat{a},\hat{a}^{\dagger})$ in anti-normal ordering. In brief, anti-normal ordering is a way of assuring all the polynomial terms containing $\hat{a}$ and $\hat{a}^{\dagger}$ have the annihilation operators to the right. When this is not possible or too cumbersome to do, we can instead use a trick; since these phase space functions represent the same physical state, they must be related. This means we can usually choose an ordering which suits us, and then perform a transformation once we have \textit{just one} phase space function that represents the state. The mathematical details of ordering and transformations between phase spaces are not so important here, just know that we need to look out for this ordering, and that it is usually possible for operators and states of interest. 

Phase space functions over $N$-modes are defined in the intuitive way. For example, the $n$-boson $Q$-function is defined as
\begin{equation*}
    Q(\alpha_1,\ldots,\alpha_n) = \frac{1}{\pi^N}\braket{\alpha_1,\ldots,\alpha_N|\hat{f}(\hat{a_1},\hat{a}_1^{\dagger},\ldots,\hat{a}_N,\hat{a}_N^{\dagger})|\alpha_1,\ldots,\alpha_N},
\end{equation*}
Although this may look intimidating, it is a remarkably simple generalization from one mode to $N$-modes. Since bosons are non-interacting their multi-particle phase space functions are simply products of single mode phase space functions. This is analogous to constructing the joint distribution from its constituents. From here, we can compute moments and probabilities with the usual integration rules of continuous variable distributions. Although this formalism is enough to understand the phase-space material in Chapter~\ref{sec:DL_QT_intro}, it was recently proposed to do qubit-based quantum machine learning in phase-space too \cite{heightman2025quantum}. Such an approach recasts the curse of dimensionality from the size of the Hilbert space to the harmonic support of density functions on multi-qubit phase space manifolds.

%% file: chapters/chapter_5_deep_learning_quantum.tex
\chapter[Deep Learning Techniques in Quantum Science]{Deep Learning Techniques in Quantum  Science}
\label{sec:DL_QT_intro}
\markboth{Ch.~5 Deep Learning Techniques in Quantum Science}{Ch.~5 Deep Learning Techniques in Quantum Science}

\section{Introduction}
What makes deep learning well suited to quantum technology? The answer lies in \textit{data}. Consider a quantum system of $N$ qubits, whose wave function can be written as
\begin{equation}
    \ket{\psi} = \sum_{i_1,\ldots,i_N}
    c_{i_1,\ldots,i_N }\ket{i_1,\ldots,i_N},
\end{equation}
where $i\in{0,1}$ and $c_{i_1,\ldots,i_N }\in \mathbb{C}$ are coefficients in the computational basis. In general there are $2^N$ coefficients for $N$ qubit states. This is a \textit{huge} amount of data, and therefore quantum systems are great objects to study with deep learning, which is data intensive. Recall, for example, that in Chapter~\ref{CH:UNSUPERVISED}, we said the performance of a model is directly affected by the amount of training data made available to it. In this sense, quantum many-body systems are a great resource for deep learning; one has exponential amounts of data and the other needs it for good inference and training.

The structure of many-body Hilbert space means quantum many-body systems also have an exponentially large representation capacity. This is their premise for quantum advantage over classical systems, which would require exponential resources to simulate quantum dynamics of $N$-qubits. However, some of the biggest open problems to quantum advantage come from the lack of fine-grained control and interpretability of quantum computers. This comes from the fact that it is difficult to represent general quantum many-body states and operators in an efficient way. 

Therefore, we will focus on quantum states and operators \textit{of interest}, for example ground states, or sparse and local interactions. These systems occupy a low-dimensional subspace of their Hilbert space $\mathcal{H} = \mathcal{H}_2^{\otimes N}$. More formally, they span a \textit{manifold} which contains all their key features and characteristics. On the other hand, we have new tools from deep learning that are capable of capturing emergent behaviour from high-dimensional spaces. They can learn to represent high dimensional, complicated distributions, in line with the \textit{Manifold Hypothesis} from Chapter~\ref{CH:UNSUPERVISED}. This is the idea that the important features and characteristics of a high-dimensional dataset are contained in a smaller, lower-dimensional space. Therefore, to represent quantum many-body systems of interest, we can use our tools applied to datasets that come from quantum many-body systems.

The field of Quantum Machine Learning (QML) encompasses four concepts shown in Tab.~\ref{tab:qml_table}. They are divided by which techniques are used, and how data is gathered. For example, the \textit{quantum for classical} would use quantum algorithms techniques on classical datasets, like images, text, or audio. Whereas the bottom-left uses classical techniques on quantum datasets, often coming from measurements on a quantum system.

\begin{center}
\begin{tabular}{c|c|c}
    \label{tab:qml_table}
     & Classical & Quantum \\
    \hline
    Classical & ML and DL & Quantum for Classical \\
    \hline
    Quantum & Classical for Quantum & Quantum for Quantum
\end{tabular}
\end{center}

Having developed the top-left in the previous two chapters, the rest of our course will focus on the bottom left, that is the \textit{classical for quantum} branch of QML. We focus on this for a few reasons. First, current quantum hardware is noisy and prone to errors, meaning it is unlikely to outperform classical deep learning techniques for classical datasets anytime soon. Second, the astounding progress that classical deep learning has made in the last decade is thanks to \textit{emergent behaviour}. Deep learning systems tend to perform well when we give them a sufficiently large variational space. At its current scale, quantum hardware does not yet have sufficient capacity to reap the rewards of emergent behaviour in the same way classical deep learning has. As such, the current ``best fit'' for these two fields is to employ classical deep learning on quantum datasets. This is likely to change in future, with error correction and quantum hardware getting ever-closer to their goal of scalable quantum advantage. \\

In this chapter, we will see how to use deep learning techniques in a few different problems in quantum technology. These are :
\begin{enumerate}
    \item Unsupervised analysis of 3-qubit entanglement classes,
\item Phase diagram boundaries discovery with anomaly detection,
\item Variational Quantum Eigensolver (VQE)
\item Quantum Approximate Optimization Algorithm (QAOA),
\item Neural Quantum States (NQS)
\item Hamiltonian Learning (HL),
\item Quantum State Tomography (QST) with Normalizing Flows (NF) for quantum optics,
\item  QST for many-qubit systems.
\end{enumerate}

\section{Three-Qubit Entanglement Problem with Unsupervised Techniques}
\label{sec:PCA_t-SNE_entanglement_classification}

In this section we will use unsupervised ML techniques to tackle the problem of entanglement classes in pure three qubit system.

Let's consider the scenario where we are given a three-qubit density matrix, $\hat{\rho} \in \mathcal{H}_2^{\otimes 3}$, which has an eight-by-eight matrix representation $\hat{\rho} = \Re[\hat{\rho}] + i\Im[\hat{\rho}] \in \mathbb{C}^{8 \times 8}$, where
\begin{equation*}
\Re[\hat{\rho}] =
\begin{bmatrix}
0.03 & -0.04 &  0.01 & -0.02 & -0.09 & -0.03 & -0.03 & -0.01 \\
-0.04 &  0.39 & -0.01 &  0.16 & -0.11 &  0.09 & -0.07 &  0.03 \\
 0.01 & -0.01 &  0.01 & -0.01 & -0.04 & -0.01 & -0.02 & -0.00 \\
-0.02 &  0.16 & -0.01 &  0.07 & -0.02 &  0.04 & -0.02 &  0.02 \\
-0.09 & -0.11 & -0.04 & -0.02 &  0.39 &  0.04 &  0.16 &  0.02 \\
-0.03 &  0.09 & -0.01 &  0.04 &  0.04 &  0.03 &  0.01 &  0.01 \\
-0.03 & -0.07 & -0.02 & -0.02 &  0.16 &  0.01 &  0.07 &  0.01 \\
-0.01 &  0.03 & -0.00 &  0.02 &  0.02 &  0.01 &  0.01 &  0.01 \\
\end{bmatrix},
\end{equation*}
and
\begin{equation*}
\Im[\hat{\rho}] =
\begin{bmatrix}
 0.00 &  0.11 &  0.00 &  0.04 & -0.07 &  0.02 & -0.03 &  0.01 \\
-0.11 &  0.00 & -0.05 &  0.02 &  0.38 &  0.07 &  0.15 &  0.03 \\
-0.00 &  0.05 &  0.00 &  0.02 & -0.02 &  0.01 & -0.01 &  0.00 \\
-0.04 & -0.02 & -0.02 &  0.00 &  0.16 &  0.02 &  0.06 &  0.01 \\
 0.07 & -0.38 &  0.02 & -0.16 &  0.00 & -0.11 &  0.02 & -0.04 \\
-0.02 & -0.07 & -0.01 & -0.02 &  0.11 &  0.00 &  0.05 &  0.00 \\
 0.03 & -0.15 &  0.01 & -0.06 & -0.02 & -0.05 &  0.00 & -0.02 \\
-0.01 & -0.03 & -0.00 & -0.01 &  0.04 & -0.00 &  0.02 &  0.00 \\
\end{bmatrix}.
\end{equation*}

Can we say if the state given by $\hat{\rho}$ is an entangled state? In general, we can consider situation where we have large dataset pure 3 qubit density matrices for different entanglement classes, however we do not know the classes a-priori. The only assumption we make, is that each state is a maximal representative of the respective class (i.e. it is a maximally entangled state possible in each class). We may now ask the following question more concretely: how many distinct entanglement classes are present in the dataset\footnote{This is a pertinent question in device-independent entanglement certification, where we don't want to make any assumptions about a physical piece of hardware, and certify its entangling properties.}?

Note that we have posed this as an \textit{unsupervised} learning problem, as we are not trying to predict a given label. That is, the model is not evaluated by comparing its output to some ground truth, like Mean Square Error. Instead we are trying to use unsupervised learning to identify the existence of classes. The number of which is only known to us thanks to entanglement theory. However, to highlight the ability of different models, we will see if they can establish a correspondence between different classes of states from density matrices sampled from their class. Importantly, these models will do so with \textit{no knowledge of the fact that there are four classes of entangled states with three qubits}. In this sense, we are considering an unsupervised problem.

Given the complexity of analytic methods to solve this kind of problem, a natural approach is to use computational techniques to gain insights. By sampling large sets of random states and labeling them, we can attempt to visualize how states with different types of entanglement distribute themselves in the high-dimensional space of density operators.

Here, we consider a dataset containing $N=10^5$ that randomly generates pure state for each entanglement class of the $3$-qubit system, which being a maximal representative of each class (meaning that the states are maximally entangled within the class).

\begin{figure}[t!]
    \includegraphics[width=0.99\linewidth]{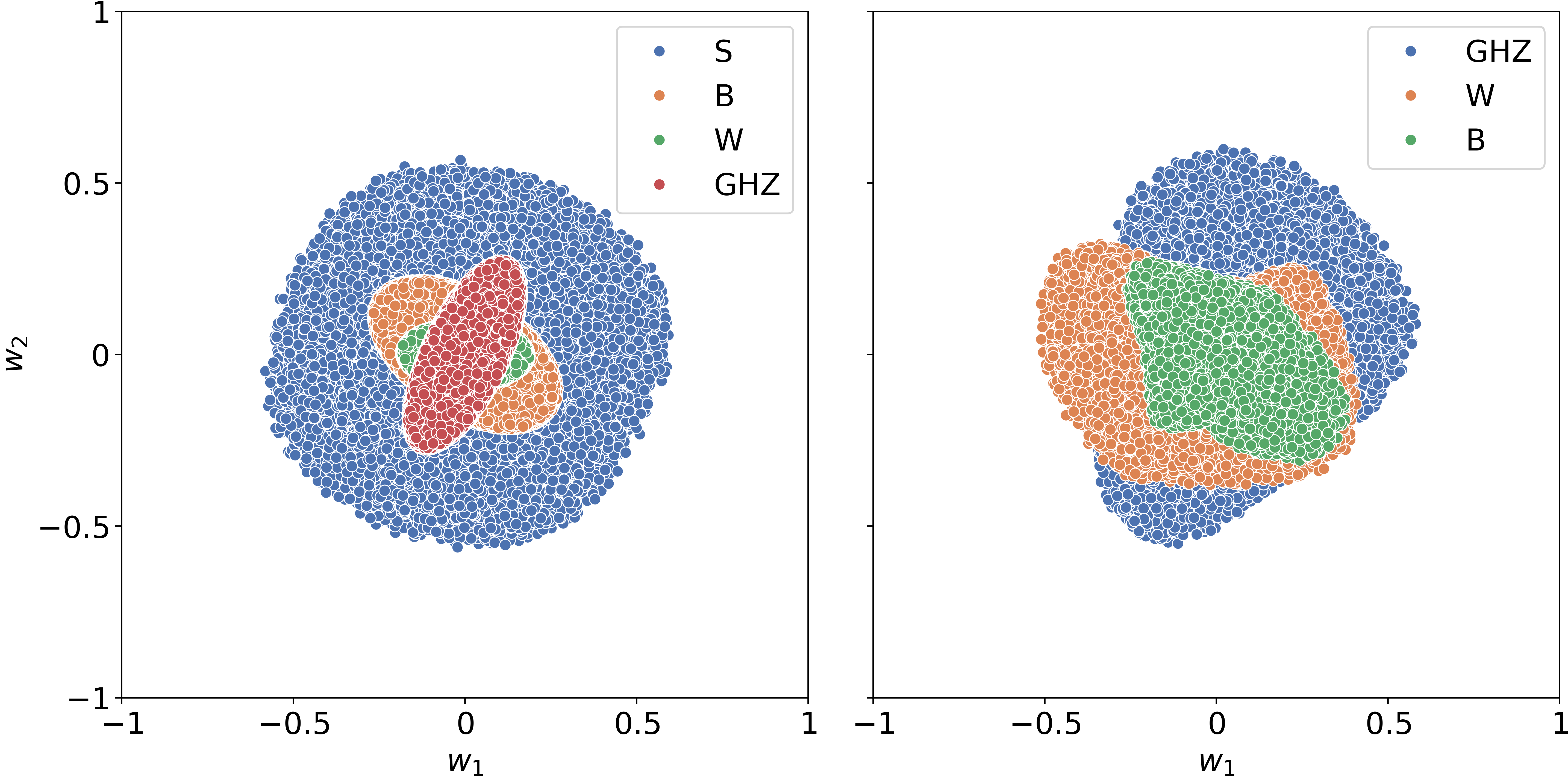}
    \caption{PCA analysis of maximal representative of 3-qubit entanglement classes, separable states (S), biseparable states (B), W-states (W), and GHZ-states (GHZ). PCA does not create separated clusters on two-dimensional plane, indicating that different classes separation of different classes cannot be simply done via linear transformation}
    \label{fig:fig_3_qubits_PCA_all}
\end{figure}

Let us start with PCA analysis. In Fig.\ref{fig:fig_3_qubits_PCA_all}, we present projection projection of dataset onto two dominant components of the correlation matrix from PCA.
As we can see, there is no visible cluster separation, indicating that indeed different entanglement classes do not have linearly separable features, and only thanks to label colouring can we see clusters. We label each class with colors here for us to see visually which of the three strategies works best for this task; a better strategy should separate each class spatially.

\begin{figure}[t!]
    \centering
    \includegraphics[width=0.8\linewidth]{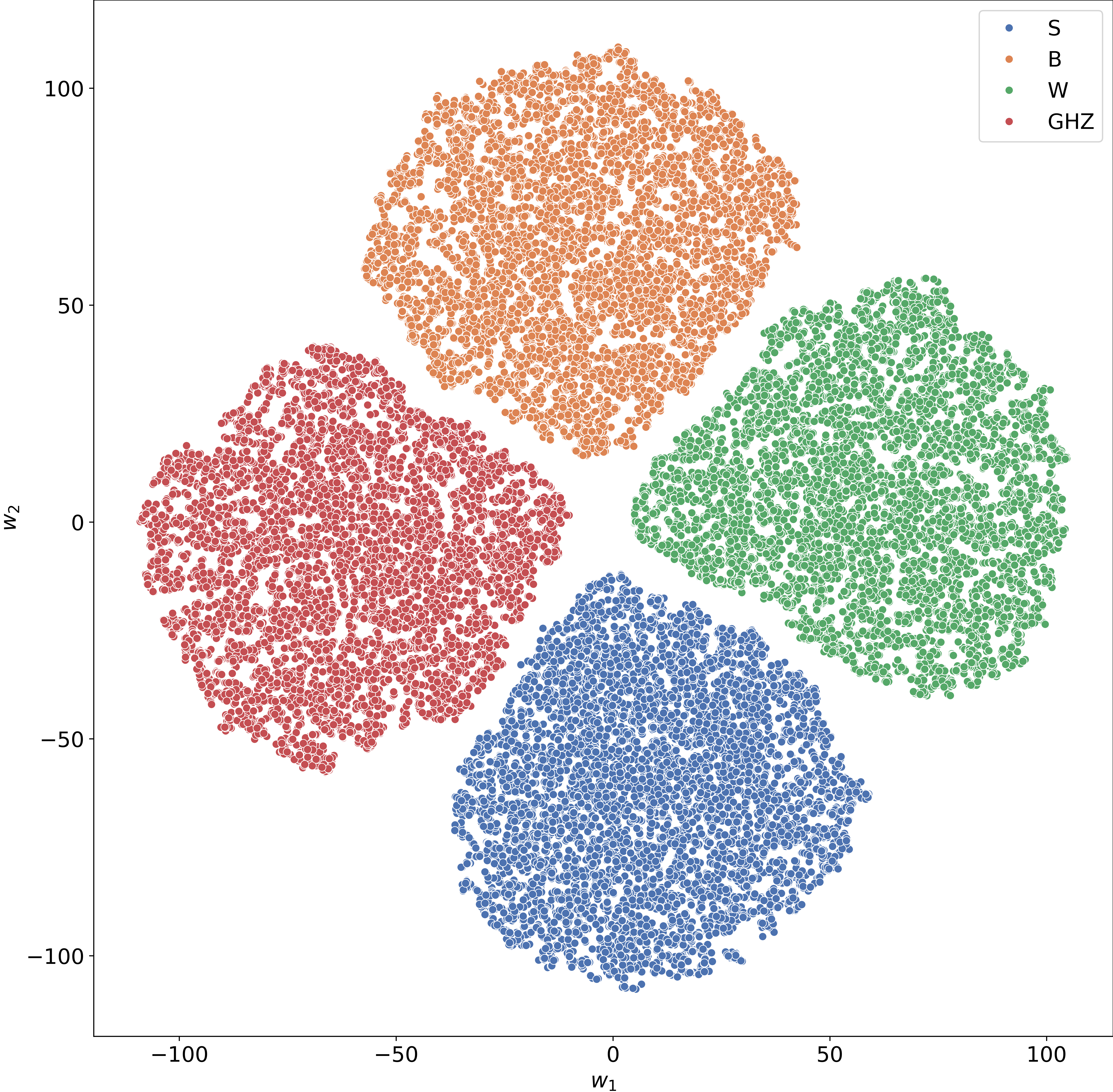}
    \caption{Non-linear dinmensionality reduction via tSNE show distint $4$ separated cluster indicating four entanglement classes in 3-qubit system.}
    \label{fig:fig_3_qubits_tSNE_all}
\end{figure}

As we have seen, PCA alone, as a linear transformation, cannot separate distinct entanglement classes into their own clusters on the two-dimensional plane. This invites us to consider \textit{non-linear} dimensionality reduction, seen in the t-SNE from Chapter~\ref{CH:UNSUPERVISED}. In Fig.\ref{fig:fig_3_qubits_tSNE_all}, we see that the t-SNE dimensionality reduction reveals four well-separated distinct clusters corresponding to four entanglement classes. 
Both PCA and tSNE are considered as Machine Learning algorithms for unsupervised problems. This is because they do not use neural networks. 
In fact, recently, the problem of 3-qubit state classification with respect to the entanglement class, both in the pure and mixed scenario, has been solved with the help of many-body Bell correlations and the notion on \textit{ non-k separability} \cite{Plodzien2024}.

However, what is important to note is that while tSNE could group data into well-separated clusters, it does not provide any insight into properties of distinct three-qubit entanglement classes.

\newpage

\section{Phase Diagram Discovery with Anomaly Detection and Auto Encoders}

Over the last years ML techniques have been used to study different problems in quantum many-body quantum systems. Using unsupervised techniques for studying quantum system properties is one of the most important aspect of modern machine learning and deep learning.
In this chapter, we will see how anomaly detection with autoencoder can be used to find phase boundaries of finite-size XXZ spin-1/2 chain.

The general one-dimensional spin-$1/2$ quantum Heisenberg Hamiltonian with transverse field (open boundary conditions) is given by:
\begin{equation}
H = - \sum_{i=1}^{L-1} \left( J_x\hat{\sigma}^x_i \hat{\sigma}^x_{i+1} + J_y\hat{\sigma}^y_i \hat{\sigma}^y_{i+1} + \Delta \hat{\sigma}^z_i \hat{\sigma}^z_{i+1} \right) 
    - h_z \sum_{i=1}^L \hat{\sigma}^z_i
\end{equation}
where $\hat{\sigma}^{x,y,z}_i$ are Pauli spin operators at site $i$,
 $\Delta$ is the anisotropy parameter,
$h_z$ is the transverse field in the $z$-direction. 
When $J_x = J_y = \Delta$ we deal with $XXX$ Hamiltonian, while when $J_y = \Delta = 0$, and $J_x\ne0$ we have quantum Ising model in a transvevrse field. The XXZ model (without transverse field) is exactly solvable by Bethe ansatz. These models a cornerstone in quantum many-body physics, providing insights into quantum entanglement and spin interactions (for introduction to spin systems see
\cite{Giamarchi2003,Lamers2015,Franchini2017}). Analytical solutions for general Heisenberg model, i.e. finding ground state, or partition function for finite temperatures, is possible only in few limiting cases, thus numerical tools are needed. 
To study numerically the model allowing extracting its thermodynamics limit (i.e. infinite system size) properties it is necessarry to consider large number of spins $L\gg 1$. The state-of-the-art numerical methods are based on tensor networks, Matrix Product States, and DMRG calculations. Here, we will work with small system sizes $L \le 10$, allowing us using Exact Diagonalization (ED).

\begin{figure}    
\includegraphics[width =0.98\linewidth]{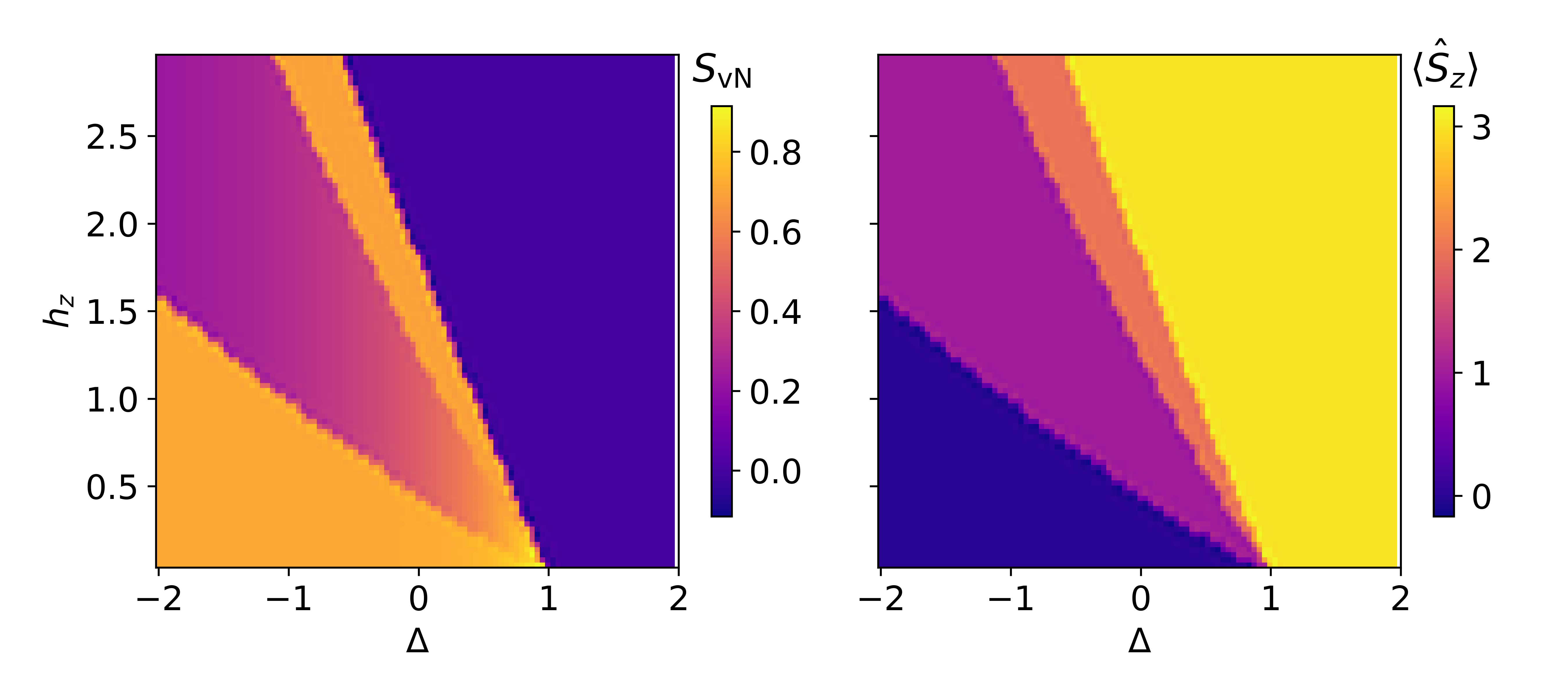}
    \caption{Density plot of bipartite entanglement von Neuman entropy $S_{\rm vN}$, $\langle \hat{S}_z\rangle$  in $\Delta-h_z$ plane for $L = 6$. We can distinguish  $4$ phases in the finite XXZ chain characterized by total magnetization.}
    \label{fig:fig_XXZ_1}
\end{figure}

Here, we focus on the XXZ model ($J_x = J_y = 1 \ne \Delta$) in a transverse field
$h_z \ge 0$:
\begin{equation}
H = - \sum_{i=1}^{L-1} \left( \hat{\sigma}^x_i \hat{\sigma}^x_{i+1} + \hat{\sigma}^y_i \hat{\sigma}^y_{i+1} + \Delta \hat{\sigma}^z_i \hat{\sigma}^z_{i+1} \right) 
    - h_z \sum_{i=1}^L \hat{\sigma}^z_i.
\end{equation}
We are interested in the following: can we distuinghish separate classes characterizing the ground states of XXZ for different sets $\{\Delta, h_z\}$ without knowing a priori the proper quantity/observable, allowing for such  classification? As starting point, let us have a look at properties of the system as a function of $\Delta$ and $h_z$. We will focus on
von Neuman entanglement entropy $S_{\rm vN} = -\hat{\rho}\log\hat{\rho}$, and
 expectation value of the total magnetization $\hat{S}_z = \frac{1}{2}\sum_{i=1}^L \hat{\sigma}^z_i$, see Fig.\ref{fig:fig_XXZ_1}. As we can see, the expectation value of the total magnetization $\langle \hat{S}_z\rangle$ clearly indicates four distinct regions on the density plot. The boundaries between phases are also visible on bipartite entanglement entropy $S_{\rm vN}$ phase diagram. However, in general, we do not know a priori which quantity we should analize to check distinct phases, on the other hand we know that all informations are encoded in the state vector $|\psi_{gs}\rangle$, thus we can directly studies state vector.

To study the phase boundaries between distinct phases of the finite size XXZ model we will start with unsupervised techniques based on dimensionality reduction, i.e. PCA and t-SNE. As a input data we consider the ground states of XXZ model $|\psi_{gs}\rangle$ for given $
\{L, \Delta, h_z\}$. However, instead of using the whole state vector $|\psi_{gs}\rangle$ we  will use Cholesky decomposition of its density matrix $\hat{\rho}_{gs}$. 
The Cholesky decomposition of Hermitian matrix $\hat{\rho}_{gs}$ reads $
\hat{\rho} = AA^\dagger$, 
where $A$ is lower-triangular matrix. Now, we can vectorize Cholesky matrix $A$, i.e. store its only lower-triangular elements as $A_{\rm vec} = [\Re[A], \Im[A]]$ --- having Cholesky vector we can restore original density matrix.

\begin{figure}
    \centering
    \begin{mybox}[\hypertarget{box:CholeskiDecomp}{Box 5.1: Cholesky Decomposition}]    

    Cholesky decomposition is a matrix factorization method used for symmetric and positive-definite matrices, i.e. for a given  Hermitian (or symmetric) positive-definite matrix $\rho$, we can find its representation in the form:
    \begin{equation}
    \rho = \frac{A A^\dagger}{Tr[A A^\dagger]},
    \end{equation}
    where $A$ is a lower triangular matrix with real and positive diagonal entries.
    The Cholesky decomposition of a matrix $\rho$ is computed element by element. For $\rho = A A^\dagger$, where $A$ is a lower triangular matrix, the entries of $A$ are calculated as:
    \begin{enumerate}
        \item 
        $A_{ii} = \sqrt{\rho_{ii} - \sum_{k=1}^{i-1} A_{ik}^2}$.
        \item 
        $A_{ij} = \frac{1}{A_{jj}} \left(\rho_{ij} - \sum_{k=1}^{j-1} A_{ik} A_{jk}\right)$, $i > j$.
    \end{enumerate}

    Cholesky decomposition requires that $\rho$ must be positive-definite, i.e. all eigenvalues of $\rho$ are greater than zero.
    If these conditions are not satisfied, the Cholesky decomposition does not exist.
    When $\rho$ is nearly positive-definite (e.g., due to numerical precision errors), Cholesky decomposition can fail. In such cases, regularization techniques are needed, by adding a small multiple of the identity matrix to $\rho$, i.e. $\rho' = (1-\epsilon)\rho + \frac{\epsilon}{dim(\rho)} \mathbb{1}$, where $\epsilon$ is a small positive number, and $\mathbb{1}$ is diagonal matrix. This adjustment ensures $\rho'$ is positive-definite and allows Cholesky decomposition to proceed.
    For $\epsilon \approx 10^{-6}$ the regularization only slightly alters $\rho$ but ensures numerical stability. Cholesky decomposition is widely used in analysis of  density matices in quantum mechanics.
     \end{mybox}
\end{figure}

\begin{figure}   
\includegraphics[width=0.99\linewidth]{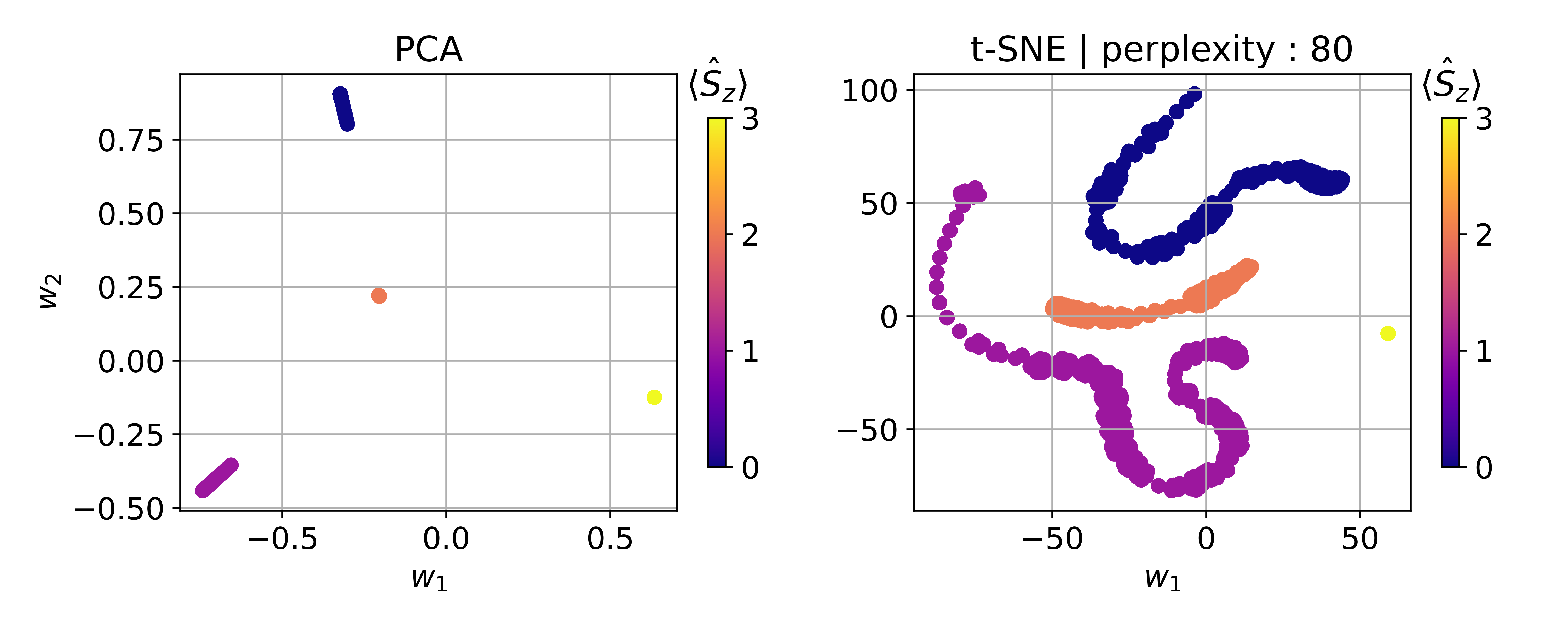}
    \caption{Dimensionality reduction of the ground state Cholesky vector $A_{\rm vec}$. Both PCA and tSNE (with perplexity parameter set to $80$) clearly creates $4$ distinct clusters colored with values of total magnetization $\langle \hat{S}^z\rangle$.}
    \label{fig:fig_XXZ_2}
\end{figure}

As a starting point, let us perform dimensionality reduction with PCA and t-SNE on Cholesky vectors $A_{\rm vec}$ for each ground state $|\psi_{gs}\rangle$, using $\langle\hat{S}_z\rangle$ value for coloring, see Fig.\ref{fig:fig_XXZ_2}. We can see that both linear PCA and non-linear t-SNE provide very clear clusterization of XXZ ground states - we see $4$ phases, each clearly defined by the total magnetization $\langle\hat{S}_z\rangle$. However, the dimensionality reduction techniques do not allow to investigate corresponding boundaries on the $\Delta-h_z$ phase diagram. In general, finding the boundary of distinct phases is based on brute force approach, via calculating ground state of the XXZ Hamiltonian for different set of parameters $\Delta$ an $h_z$, and calculating distinct expectation values - hoping that we choose proper observable which provides clear phase boundaries. 
This brute force approach is very costly, and one can think about smarter approach for phase boundaries discovery. One of the ML-based approaches is known as anomaly detection.

Anomaly detections relies on the analysis of the autoencoder reconstruction loss for considered parameter range of $\Delta$ and $h_z$. First, let us define the training dataset ${\cal D}_{\rm train} = (\{\Delta, h_z, A_{\rm vec}(\Delta, h_z)\}_i, i = 1\dots M)$, where $A_{\rm vec}(\Delta, h_z)$ is the Cholesky vector of the XXZ ground state vector $|\psi_{gs}\rangle$ for $M$ randomly chosen pairs  $\{\Delta, h_z\}$. Next, wetrain autoencoder \footnote{details of AE architecture are not curcial now} to reconstruct the Cholesky vectors from ${\cal D}_{\rm train}$. Finally, we 
evaluate autoencoder on test data ${\cal D}_{\rm train} \ne {\cal D}_{\rm train}$, and we plot reconstruction loss plot on $\Delta-h_z$ plane.

\begin{figure}    \includegraphics[width=0.99\linewidth]{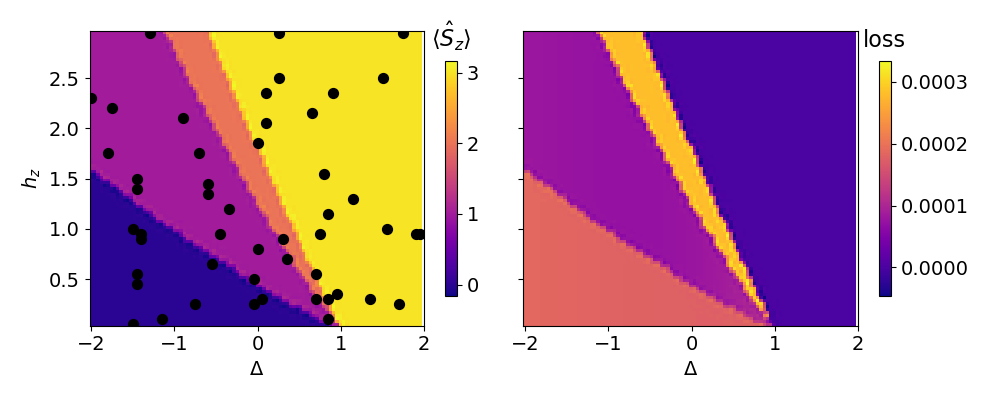}
    \caption{Phase boundaries detection with anomaly detection. We choose randomly $M=20$ points on which we train autoencoder to reconstruct Choleky vector $A_{\rm vec}$ of a ground state corresponding to tuple $\{\Delta, h_z\}$ (left panel). Next, we plot reconstruction loss for Cholesky vector ground state for all $\Delta$ and $h_z$ in the specific range. Density plot of the reconstruction loss reveals the phase boundaries of the XXZ model (right panel).}
    \label{fig:fig_XXZ_3}
\end{figure}

In a natural way, the boundaries between distinct phases appear on the reconstruction loss phase diagram: points on diagram comming from the same phase have smaller reconstruction loss, while for different phases - the reconstruction loss is larger, revealing phase boundaries of the model, see Fig.\ref{fig:fig_XXZ_3}.

\newpage
    
\section{Variational Quantum Eigensolver}
\label{sec:VQAs}
The key idea behind variational quantum algorithms (VQAs) is to realise the variational principle with gradient descent methods from Chapter~\ref{CH:FUNDAMENTALS} \cite{peruzzo2014variational}, and apply it to quantum circuits. Consider some quantum many-body Hamiltonian, $\hat{H} \in \mathcal{H}$, whose ground state we seek. We can use the variational principle by creating a state $\ket{\psi(\theta)}$ with variational parameters $\theta \in \mathbb{R}^n$. Using gradient descent with respect to $\theta$, we can solve
\begin{equation}
    \min_{\theta} \braket{\psi(\theta)|\hat{H}|\psi(\theta)}.
    \label{eq:VQE_problem}
\end{equation}
By the variational principle, this will always be an upper bound for $E_{gs}$, with the tightest bound being $\theta^*$, the solution to Eq.~(\ref{eq:VQE_problem}). This defines the \textit{variational quantum eigensolver} (VQE), a variational method for computing tight bounds on for the ground state of a quantum many-body Hamiltonian. This problem is of particular interest to quantum chemistry, where the ground state properties of a many-body system are highly sought after.

\begin{figure}[h!]
    \centering
    \begin{tikzpicture}
        \draw[thick] (0,0) -- (5,0);
        \draw[thick] (0,-1) -- (5,-1);
        \draw[thick] (0,-3) -- (5,-3);
        
        \node at (-0.5, 0) {\( \ket{0} \)};
        \node at (-0.5, -1) {\( \ket{0} \)};
        \node at (-0.5, -3) {\( \ket{0} \)};
        \node at (-0.5, -2) {\(\vdots\)};
        
        \draw[thick, fill = white] (1, 0.5) rectangle (4, -3.5);
        \node at (2.5, -1.5) {\( \hat{U}(\theta) \in SU(2^N) \)};
        
        \draw[thick] (4,0) -- (5,0);
        \draw[thick] (4,-1) -- (5,-1);
        \draw[thick] (4,-3) -- (5,-3);

        \node at (5.0,-2) {\(\vdots\)};
        \node at (6.5,-1.5) {\(\ket{\psi(\theta)} \in  \mathcal{H}_2^{\otimes N}\)};

    \end{tikzpicture}
    \caption{An $N$-qubit parametric quantum circuit realised by applying an $N$-qubit Unitary gate to $\ket{0}^{\otimes N}$.} \label{fig:PQC_unitary}
\end{figure}
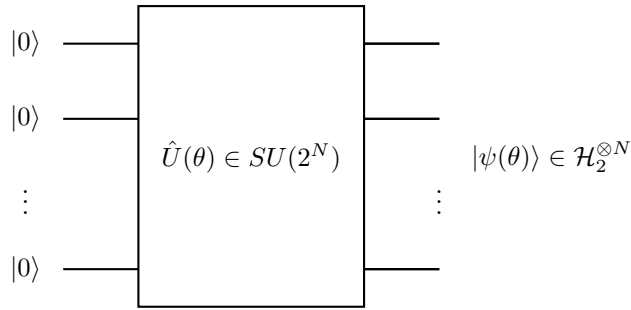

To run a VQE, we require two key ingredients. The first is a way of preparing $\ket{\psi(\theta)}$, and the second is a way of measuring $\braket{\psi(\theta)|\hat{H}|\psi(\theta)}$. First, we can prepare $\ket{\psi(\theta)}$ using a parametric quantum circuit,
\begin{equation}
    \ket{\psi(\theta)} = \hat{U}(\theta) \ket{0}^{\otimes N},
\end{equation}
which looks like Fig.~\ref{fig:PQC_unitary} in quantum circuit notation.

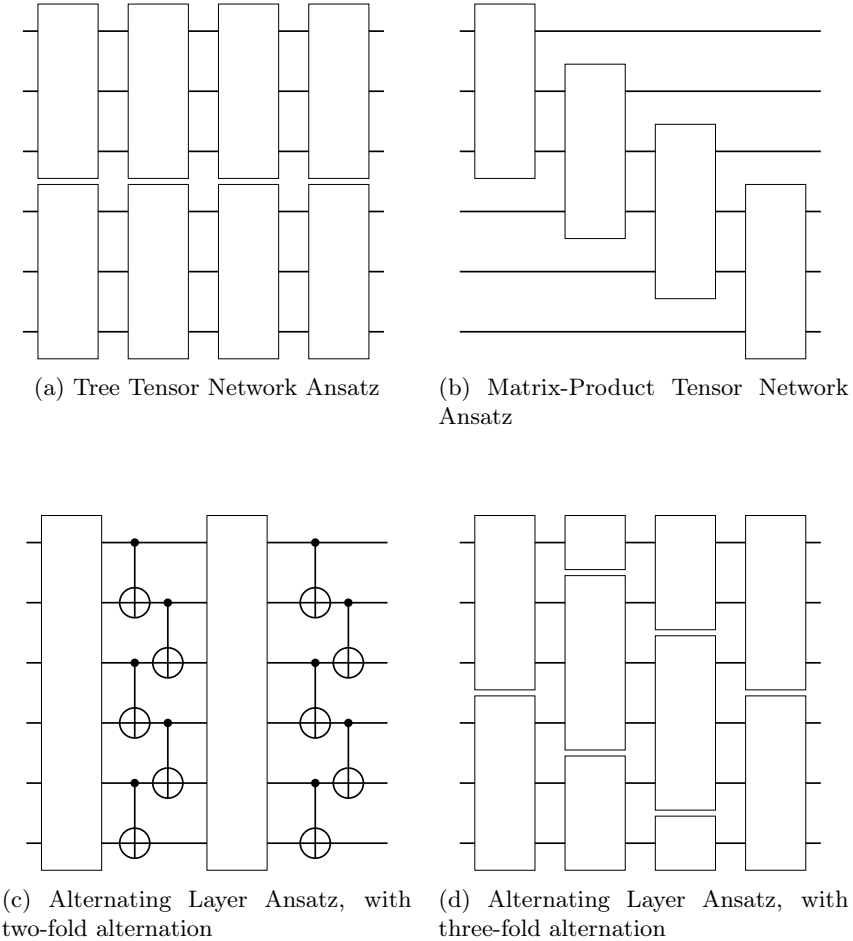
\begin{figure}
\centering
\begin{subfigure}[t]{0.45 \textwidth}

    \centering
   \scalebox{0.8}{\input{circuits/TEN_blocks.tikz}}
  \caption{Tree Tensor Network Ansatz}
  \label{fig:architectures_TEN}
\end{subfigure}
~
\begin{subfigure}[t]{0.45 \textwidth}
\centering
  \scalebox{0.8}{\input{circuits/MPS_blocks.tikz}}
  \caption{Matrix-Product Tensor Network Ansatz}
  \label{fig:architectures_AL}
\end{subfigure}

\vspace{3em}

\begin{subfigure}[t]{0.45 \textwidth}
\centering
  \scalebox{0.8}{\input{circuits/HAE_blocks.tikz}}
  \caption{Alternating Layer Ansatz, with two-fold alternation}
  \label{fig:architectures_HAE}
\end{subfigure}
~
\begin{subfigure}[t]{0.45 \textwidth}
\centering
  \scalebox{0.8}{\input{circuits/AL_blocks.tikz}}
  \caption{Alternating Layer Ansatz, with three-fold alternation}
  \label{fig:architectures_AE_blocks}
\end{subfigure}
\caption{Four example block structures for an $N$-qubit variational circuit with (a) and (b) based on Tensor Networks and (c) and (d) based on alternating layers. Each white box corresponds to a Unitary transform over the number of wires it connects. For example, a white box connecting three wires corresponds to a three-qubit unitary, $\hat{U} \in SU(8)$. }\label{fig:VQE_block_structures}
\end{figure}

\begin{figure}
\centering
\begin{subfigure}[t]{0.45 \textwidth}
\centering
  \scalebox{0.8}{\input{circuits/blocks_decomposition1.tikz}}
  \caption{Bi-partite decomposition}
  \label{fig:block_decomp(a)}
\end{subfigure}
~
\begin{subfigure}[t]{0.45\textwidth}
\centering
  \scalebox{0.8}{\input{circuits/blocks_decomposition2.tikz}}
  \caption{Bi-partite decomposition}
  \label{fig:block_decomp(b)}
\end{subfigure}

\vspace{3em}

\begin{subfigure}[t]{0.45 \textwidth}
\centering
  \scalebox{0.8}{\input{circuits/blocks_decomposition3.tikz}}
  \caption{Recursive decomposition}
  \label{fig:block_decomp}
\end{subfigure}
~
\begin{subfigure}[t]{0.45 \textwidth}
\centering
  \scalebox{0.8}{\input{circuits/blocks_decomposition4.tikz}}
  \caption{Recursive decomposition}
  \label{fig:block_decomp(d)}
\end{subfigure}

\caption{Three different block decompositions.  \ref{fig:block_decomp(a)} has a bi-partite layered structure which splits it into one- and two-local variational parameters on both sides. Whilst \ref{fig:block_decomp(b)} has fixed two-local nearest-neighbour entangling operations (right) of its bipartite structure, and single qubit variational gates (left) like 
\ref{fig:block_decomp(a)}. The second row shows a recursive block decomposition. On the left we see a three qubit gate decomposed into three two-qubit gates in sequence. We can decompose blocks in this way thanks to a simple recursion relation, with the four-qubit example shown in \ref{fig:block_decomp(d)} We see here why universality classes and our choice of quantum hardware are important. The universality of two-qubit gates means these are the smallest ``building blocks'' we need in order to construct larger unitary transforms. On the other hand, different quantum hardware have different universal gate sets, and usually dictate the structure of a block decomposition should we want it to be \textit{efficient} (we usually do).}
\label{fig:VQE_block_decomps}
\end{figure}
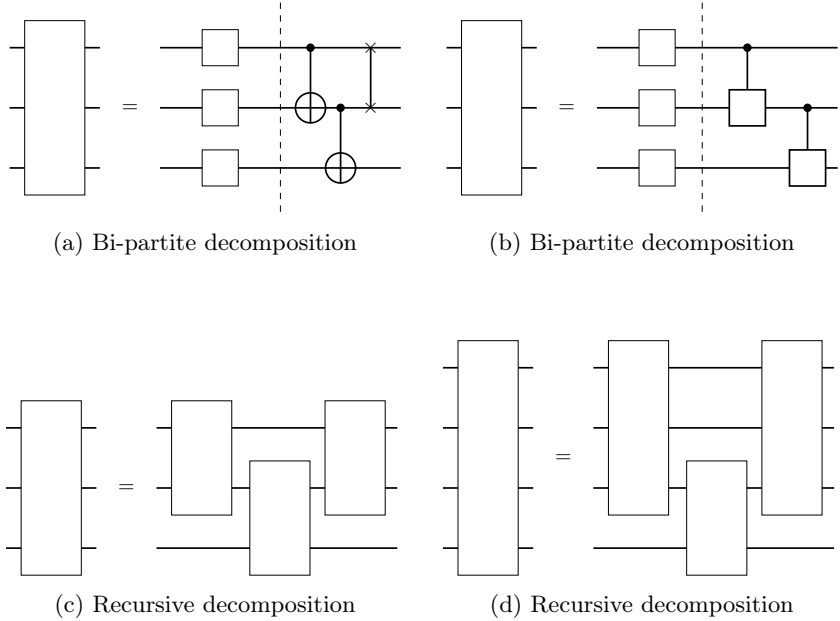

Our specific choice of decomposition for $U(\theta)$ constitutes a \textit{circuit ans\"atze}. There is no best ans\"atze in general, and a specific circuit choice is often informed by the structure of $H$ and the specifics of our hardware choice. Much research is currently being done so be able to properly tailor a circuit ans\"atze to a given problem. At the time of writing these notes, the current way of constructing ans\"atze is based on \textit{block structure} and \textit{block decomposition}. Block decomposition is a way of decomposing (sometimes called compiling) a given $N$-qubit Unitary in terms of lower-dimensional $M \leq N$-qubit unitaries. Whereas block structure is a way of \textit{combining} $N$-qubit unitaries into an overall circuit. For some examples of block structure and decomposition, see respectively Figs.~\ref{fig:VQE_block_structures} and \ref{fig:VQE_block_decomps}.

Onto our second ingredient - we need a way of evaluating
\begin{equation}
    \braket{\psi(\theta)|\hat{H}|\psi(\theta)} = \braket{0|\hat{U}(\theta) \hat{H} \hat{U}(\theta)|0}.
\end{equation}
This is nothing more than an expectation value, so we can estimate it by sampling from the distribution $\braket{\psi(\theta)|\hat{H}|\psi(\theta)}$. To do this, we can decompose $H$ as sum of measurable operators. Let's use the Pauli basis, as~we covered~this~in~Sec.~\ref{sec:DL_QT_intro}~Eq.~(\ref{eq:pauli_basis}). To that end, let
\begin{equation}
    \hat{H} = \sum_{j} c_j \hat{P}_j,
    \label{eq:hamiltonian_pauli_decomp}
\end{equation}
where $c_j \in \mathbb{R}$ are real coefficients, and $\hat{P}_j \in \mathcal{P}_N$ are Pauli strings (formally, members of the Pauli group $\mathcal{P}_N$, see \hyperlink{box:PauliGroup}{Box 5}). Then the expectation value $\braket{\psi(\theta)|H|\psi(\theta)}$ can be expressed as
\begin{equation}
    \braket{\psi(\theta)|\hat{H}|\psi(\theta)} = \sum_{j} c_j \braket{\psi(\theta)|\hat{P}_j|\psi(\theta)},
    \label{eq:hamiltonian_pauli_basis_VQE}
\end{equation}
where $\braket{\psi(\theta)|\hat{P}_j|\psi(\theta)}$ is the expectation value of the $j^{\text{th}}$ Pauli string in the decomposition of $H$ from Eq.~(\ref{eq:hamiltonian_pauli_decomp}). Since these are individually measurable on a quantum computer, we can estimate $\braket{\psi(\theta)|\hat{H}|\psi(\theta)}$ by repeatedly sampling $\braket{\psi(\theta)|\hat{P}_j|\psi(\theta)}$ for each $j$.

\begin{figure}[h!]
    \centering
    \begin{mybox}[\hypertarget{box:PauliGroup}{Box 5.2: Pauli Group}]
    An $N$-qubit Pauli operator is an operator $\hat{O}$ acting on Hilbert space, $\mathcal{H}_2^{\otimes N}$, of the form
    \begin{equation}
        \hat{O} = \hat{P}_1 \otimes \hat{P}_1 \otimes \ldots \otimes \hat{P}_N,
    \end{equation}
    where $\hat{P}_j \in \{\hat{\sigma}_x, \hat{\sigma}_y, \hat{\sigma}_z, \mathbb{I}\}$ is one of four Pauli matrices acting on each qubit. Notice that the tensor product structure combined with the Pauli \textit{algebra} $[\hat{\sigma}_i, \hat{\sigma}_j] = i \epsilon_{ijk} \hat{\sigma}_k$, makes the set, $\mathcal{P}_N$, of possible Pauli operators (sometimes called Pauli strings) forms a group, hereby referred to as the \textit{Pauli Group}. That is, we have a bi-linear (matrix) product 
    \begin{equation}
        \hat{O}_1 \hat{O}_2 = \hat{O}_3 \in \mathcal{P}_N,
    \end{equation}
    which is closed, has an identity $\mathbb{I}^{\otimes N} \in \mathcal{P}_N$, and a well defined inverse. In the QML literature, you might often see shorthand notation for Pauli strings, like $\hat{X} \leftrightarrow \hat{\sigma}_x$, or $\hat{X}_1 \hat{Y}_2 \hat{Z}_4 \leftrightarrow \sigma_x \otimes \sigma_y \otimes \mathbb{I} \otimes \sigma_z$. \\
    
    \end{mybox}
\end{figure}

With these two ingredients, we are ready to implement a gradient-based feedback loop to find the optimal value $\theta^*$ which offers the tightest upper bound to $E_{gs}$. Let's define the cost (loss) function as
\begin{equation}
    C(\theta) = \braket{\psi(\theta)|\hat{H}|\psi(\theta)}.
    \label{eq:VQE_cost_fn}
\end{equation}
Then, with a sampling-based estimate for $\braket{\psi(\theta)|\hat{H}|\psi(\theta)}$, we can apply
\begin{equation}
    \theta \gets \theta + \eta \nabla_{\theta}C,
\end{equation}
where $\eta$ is the learning rate, and $\nabla_{\theta}C$ is a vector containing the gradients of $C(\theta)$ with respect to each component of $\theta \in \mathbb{R}^n$. A summary of the VQE algorithm is detailed in Algorithm~\ref{alg:VQE_TrainStep_Gradient}.

\begin{algorithm}[h!]
\caption{VQE Train Step with Assumed Gradient Calculation}
\label{alg:VQE_TrainStep_Gradient}
\begin{algorithmic}[1]
\State \textbf{Input:} Current parameters $\boldsymbol{\theta}$,
Hamiltonian $\hat{H} = \sum_i c_i \hat{P}_i$, Ansatz circuit $U(\boldsymbol{\theta})$

\State Prepare quantum state $|\psi(\boldsymbol{\theta})\rangle = \hat{U}(\boldsymbol{\theta}) |0\rangle$
\State $C(\boldsymbol{\theta}) \gets 0$
\For{each term $c_i \hat{P}_i$ in $H$}
    \State Measure expectation value $\langle \hat{P}_i \rangle_{\boldsymbol{\theta}}$
    \State $C(\boldsymbol{\theta}) \gets C(\boldsymbol{\theta}) + c_i \langle \hat{P}_i \rangle_{\boldsymbol{\theta}}$
\EndFor
\State Compute gradient $\nabla_{\boldsymbol{\theta}} E(\boldsymbol{\theta})$ \Comment{See below for two different tricks!}
\State Update parameters $\boldsymbol{\theta}' \gets \boldsymbol{\theta} - \eta \nabla_{\boldsymbol{\theta}} E(\boldsymbol{\theta})$
\State \Return $\boldsymbol{\theta}'$
\end{algorithmic}
\end{algorithm}

As we can see, one of the key steps in the VQE is to compute the gradient of the energy cost function from Eq.~(\ref{eq:VQE_cost_fn}). There are several ways to do this, depending on how we prepare the quantum state $\ket{\psi(\theta)}$. If we are classically simulating our quantum circuit, then we can compute $\nabla_{\theta}C(\theta)$ via automatic differentiation from Chapter~\ref{CH:FUNDAMENTALS}. This is because we can construct $\ket{\psi} = U(\theta) \ket{0}^{\otimes N}$ such that it forms a differentiable computation graph. When running on quantum hardware, we can estimate the gradient $\nabla_{\theta}C(\theta)$ with a few different tricks which depend on our choice of circuit ans\"atze:
\begin{enumerate}
    \item If our Unitary circuit $U(\theta)$ is constructed from rotation-like gates, we can use the parameter shift rule. To see how this works, consider a one-qubit unitary gate of the form 
    \begin{equation}
        \hat{U}(\theta) = e^{i \theta \hat{\sigma}_j},
    \end{equation}
    where $\hat{\sigma}_j,\;j\in\{x,t,z\}$ is a Pauli operator with eigenvalues $\pm r$. Then this cost function satisfies,
    \begin{equation}
        \frac{\partial C(\theta)}{\partial \theta_j} = r \left[ C(\theta_j^+) - C(\theta_j^-) \right],
        \label{eq:param_shift_one_q}
    \end{equation}
    where $\theta^{\pm}_j = \theta_j \pm \frac{\pi}{4r}\mathbf{e}_j$, and $\mathbf{e}_j$ is the unit vector in the variational space of the component $\theta_j \in \mathbb{R}^N$. A one-parameter example, of this result follows from Eq.~(\ref{eq:euler_su(2)}). Try differentiating Eq.~(\ref{eq:euler_su(2)}) with respect to $\theta$, and rearranging the derivative to be expressed as the original function! With these types of unitary gates, we can therefore estimate the derivative of our cost function through sampling. We can do this because the RHS of Eq.~(\ref{eq:param_shift_one_q}) is entirely in terms of the original cost function. All we need to do is run our VQE circuit with the parameters set to $\theta^{\pm}$ and sample from the output! This is an example of trying to perform \textit{natural gradient descent}, the act of updating parameters by sampling in coordinates that are a natural description of the gradient operator $\nabla_{\theta}$. Whilst there is not time in this course to go deeper into this\footnote{Sadly, we would need to supplement the course with some basic differential geometry and more advanced Lie theory to cover this}, it is a fascinating and pertinent topic that is is at the forefront of contemporary research in QML. For the interested reader, we recommend \cite{hackl2020geometry} for a first introduction to these concepts, and \cite{stokes2020quantum} for an exposition on how this idea was first applied in VQAs. We refer readers unfamiliar with basic differential geometry to \cite{needham2021visual}, and to \cite{knapp1996lie} for more nuanced aspects of Lie theory that underpins this subfield of QML.
    
    \item More generally, we can apply finite different methods to do gradient estimation. For example, we could estimate the gradient with
    \begin{equation}
        \frac{\partial E(\theta)}{\partial \theta_j} \approx \frac{E(\theta + \delta \mathbf{e_j}) - E(\theta - \delta \mathbf{e_j})}{2\delta},
    \end{equation}
    where $\delta \ll 1 \in \mathbb{R}$ and $\mathbf{e}_j$ is a unit vector as above. We can estimate the gradient from this finite difference method by running our VQE circuit with parameters $\theta \pm \delta \mathbf{e}_j$. This allows for a sampling-based estimate of the gradient.

\end{enumerate}
When simulating a quantum circuit classically, we require that our simulation forms a \textit{differentiable computation graph}. As seen in Chapter~\ref{CH:FUNDAMENTALS}, we can do this with matrix functions, for example with neural networks. What matters is that we can calculate the gradient, $\nabla_{\theta}C(\theta)$, of our loss function so that we can perform \textit{meaningful} updates to our model. 

Notice here that the methods for calculating a VQE's loss function gradient are sampling based. That is, they involve repeatedly executing a quantum circuit and performing measurements. Often this is referred to as \textit{shots} of quantum circuit; more shots means more statistics, and therefore we would anticipate \textit{better} updates to a VQE model with more shots. We might therefore wonder whether we can learning else from these statistics. As it turns out, the gradient of the loss function vanishes exponentially in the size of our circuit. This is due to a phenomenon called \textit{Barren Plateaus}.

\subsubsection*{Barren Plateaus} 

In some sense, we can think of barren plateaus as the quantum analogue of vanishing gradients in Chapter~\ref{CH:FUNDAMENTALS}. Recall that this effect means our loss function gradient gets vanishingly small so that
\begin{equation}
    \theta \gets \theta + \alpha \nabla_{\theta} L(\theta)
\end{equation}
performs essentially no change to the values of our variational parameters $\theta$. We can identify barren plateaus in practise when we the loss function has the following exponential relation,
\begin{equation}
    L(\theta) \sim \mathcal{O}(b^{-N}),
\end{equation}
where $b > 1 \subset \mathbb{R}$, and $N$ is the number of bodies in our system (e.g. qubits for VQEs).

To understand why this can happen, let's consider the case where we are measuring the Hamiltonian in question via it's Pauli basis expansion\footnote{actually any basis works here for this derivation. But we have together seen the Pauli group, so let's use it!}, as was done in Eq.~(\ref{eq:hamiltonian_pauli_basis_VQE}). In this sense, our loss function,
\begin{equation}
    L(\ket{\psi(\theta)}, H) = \sum_j c_j \braket{\psi(\theta)|\hat{P}_j|\psi(\theta)}, 
\end{equation}
is constructed out of some combination of Pauli measurements. At the simplest level, we can therefore ask how meaningful a \textit{single} given Pauli measurement is by setting
\begin{equation}
    L(\ket{\psi(\theta)}, \hat{P}_j) = \braket{\psi(\theta)|\hat{P}_j|\psi(\theta)} = \braket{0|\hat{U}(\theta) \hat{P}_j \hat{U}(\theta)|0}.
\end{equation}
The loss function gradient $\nabla_{\theta} L$ will be vanishing whenever it is \textit{insensitive} to small changes $\theta \rightarrow \theta + \delta \theta$. This is because if the loss function $L(\ket{\psi(\theta + \delta \theta)}, \hat{P}_j) \approx L(\ket{\psi(\theta)}, \hat{P}_j)$, then the \textit{difference} (and therefore gradients) will be vanishingly small. If we understand when this can happen, then we can understand the more general settings where the loss function is just some combination of this simplest case. Furthermore, if the single Pauli measurement has vanishing sensitivity to small changes in $\theta$, an exponential number of shots would be required for us to make a meaningful update.

How then can we see that a single Pauli measurement is insensitive to small changes in $\theta$? To understand these origins, it is instructive to see how Pauli strings give rise to the Unitary group. Pauli strings come from combining different elements of the Pauli Group (see Box \hyperlink{box:PauliGroup}{5)}. 

Since the Pauli matrices form \textit{generators} of $SU(2)$, the set of Unitaries on $N$-qubits can be generated from exponentiating the $N$-qubit Pauli strings or any combination thereof (e.g. combining with addition). For example, all one-qubit gates $\hat{U} \in SU(2)$ can be written as $\hat{U} = e^{\theta_x \hat{X} + \theta_y \hat{Y} + \theta_z \hat{Z}}$, with $\theta_j \in \mathbb{C}$. Thanks to the tensor product structure, this result generalises to higher numbers of qubits. For example, the two-qubit Pauli group, $\mathcal{P}_2$ generates $SU(4)$.

In this sense, the \textit{size} of the Pauli group present in a quantum circuit or operator can be very helpful in understanding its properties. This is because our choice of variational architecture dictates the size of the \textit{irreducible} set of generators of a circuit. When this irreducible set increases in size (e.g. as we change our choice of architecture), there will necessarily be more coefficients to tune with gradient-based algorithms that we saw in CH1. When there are more coefficients to tune, each individual coefficient makes a \textit{smaller} contribution to the overall Unitary process. 

Given that we have access to the outcomes of this unitary process through measurements, it becomes immediately clear how just a \textit{single} measurement cannot be sensitive to these parameters in general. The only way to improve the situation would be to acquire more data, however this comes at the cost of running \textit{exponentially more shots}. In this sense, a loss function that tries to change these coefficients by making small changes according to their gradients is insensitive. This means that geometrically, BPs correspond to often featureless, flat loss landscapes.

We have deliberately skimmed over the mathematical details here, as this is just an introductory course. However, if this has piqued your interest you can read the recent seminal work on this topic \cite{ragone2023unified, diaz2023showcasing}. This is a collection of three papers which present a unified theory of barren plateaus using Lie theory; a great example of where good understanding of groups can provide a solid foundation to better understanding quantum processes and deep learning. There is also a great tutorial-style review of this topic \cite{larocca2024review}.

A natural question now arises; \textit{how do we design and train variational quantum algorithms in lieu of this?} That is, what can we do to avoid BPs systematically? This is currently an active and highly popular area of research in QML. 

\subsection*{Example}

\begin{figure}
   \includegraphics[width=0.99\linewidth]{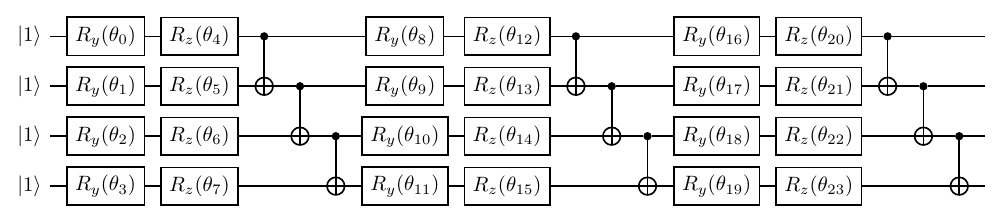}

    \caption{Parametrized Quantum Circuit used for Variational Quantum Eigensolver to find ground state of XXZ Hamiltonian, Eq.\eqref{eq:H_XXZ}.}
    \label{fig:PQC}
\end{figure}

Let us consider $L=4$ qubits. We are interested in finding ground state of the XXZ Hamiltonian
\begin{equation}\label{eq:H_XXZ}
    \hat{H} = -\frac{1}{4}\sum_{i=1}^{L-1}(\hat X_i\hat X_{i+1} + \hat Y_i\hat Y_{i+1}+\Delta\hat Z_i\hat Z_{i+1}).
\end{equation}
We consider ansatz for the ground state $|\psi(\boldsymbol\theta)\rangle
= \hat U^{(2)}\hat U^{(1)}\hat U^{(0)}|1\rangle^{\otimes 4}$,
where for $k=0,1,2$
\begin{equation}
\hat U^{(k)}
= \mathrm{CNOT}_{3,4}\,\mathrm{CNOT}_{2,3}\,\mathrm{CNOT}_{1,2}
\;\bigotimes_{i=1}^{4} \hat{R}^{z}_{i}\bigl(\theta_{8k+4+i-1}\bigr)
\;\bigotimes_{i=1}^{4} \hat{R}^{y}_{i}\bigl(\theta_{8k+i-1}\bigr),
\end{equation}
and $24$ trainable parameters are stored in vector $\vec{\theta} = \{\theta_0,\dots,\theta_{23}\}$, represented as a Parametrized Quantum Circuit (PQC), see Fig.~\ref{fig:PQC}
\begin{figure}[h!]

   \includegraphics[scale=0.35]{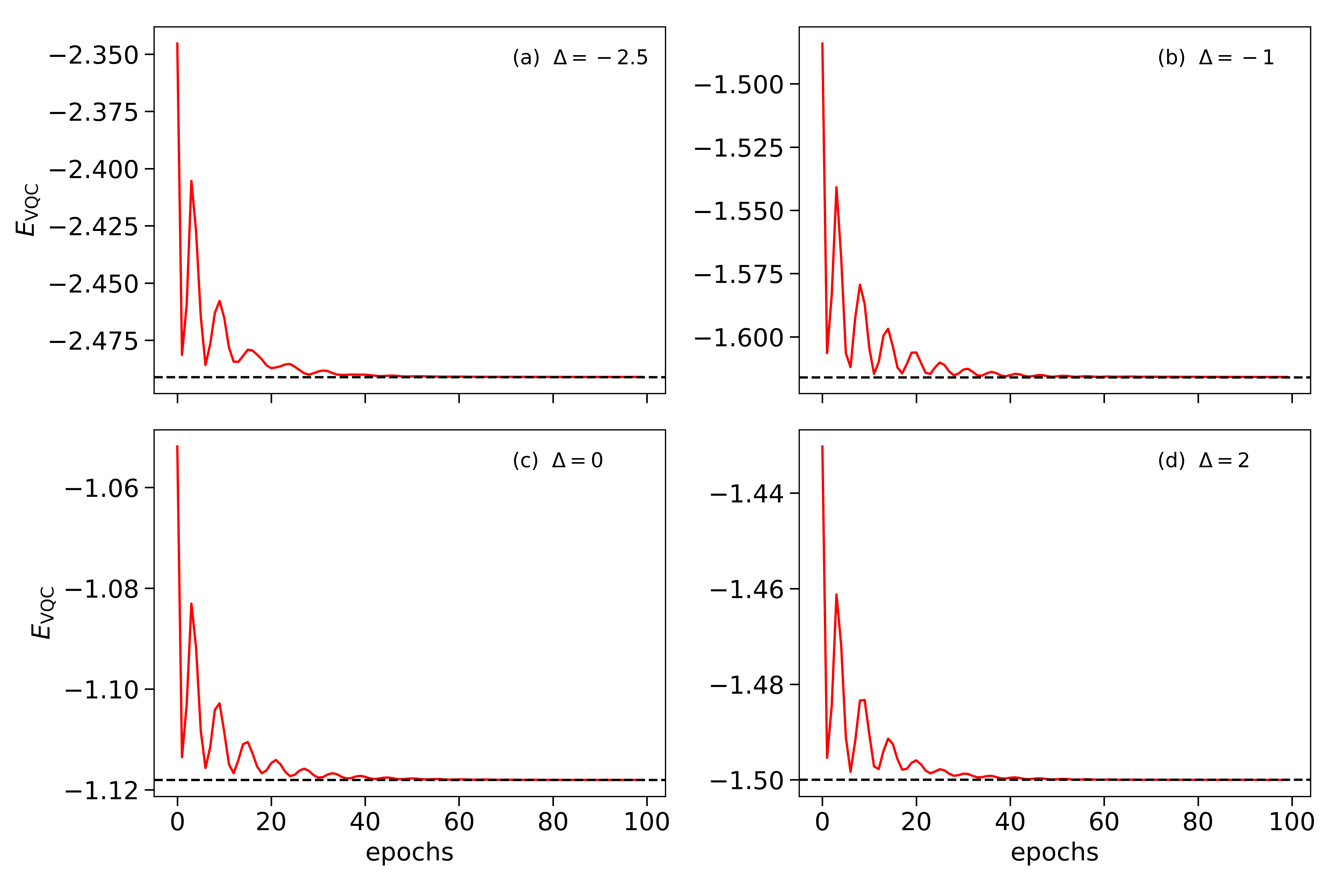}
   \includegraphics[scale=0.39]{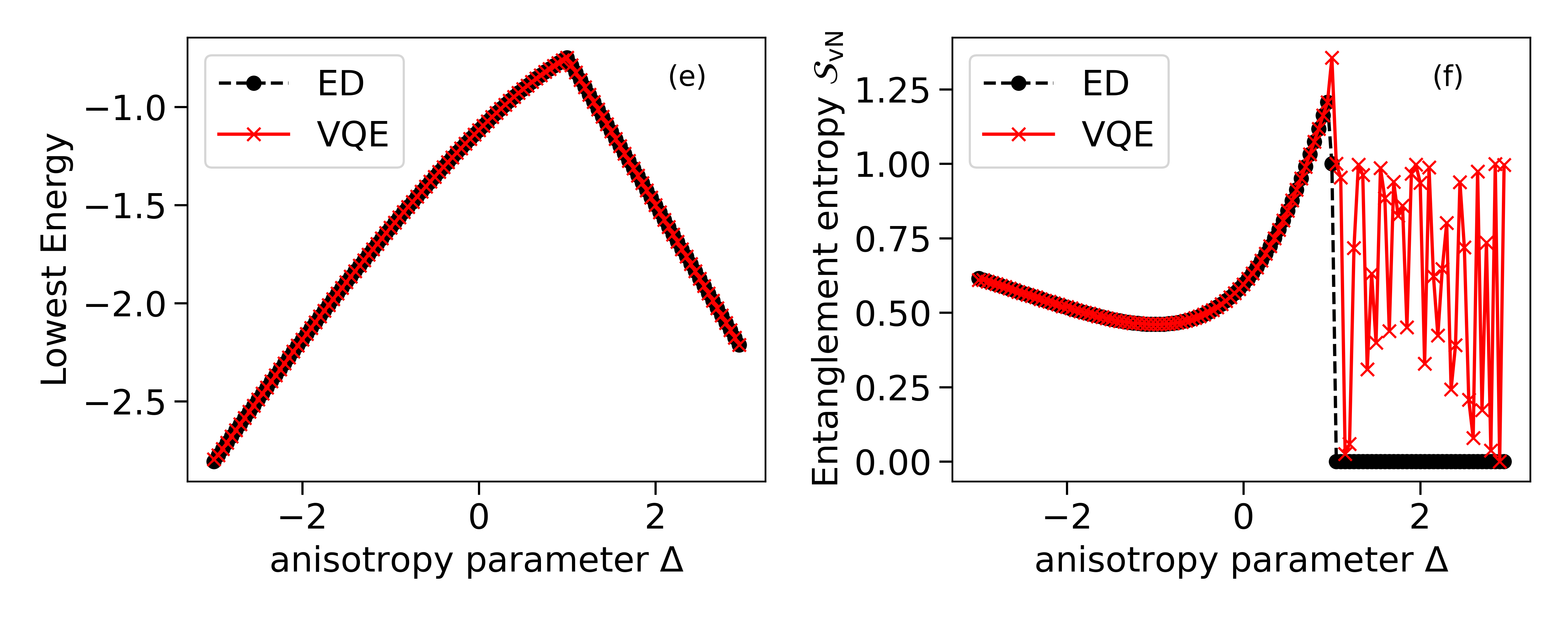}
        \centering
    \caption{Variational energy convergence vs epochs for $\Delta = -2, -1, 0, 2$ for ansatz $|\psi(\vec{\theta})\rangle$. Horizontal dashed lines correspond to Exact Diagonalization ground state energy $E_{\rm GS}$. 
    Energy, and bipartite entanglement entropy (left and right panels, respectively) vs anisotropy parameter $\Delta$. Black circles correspond to Exact Diagonalization results, while red stars to VQE solution. 
    }
    \label{fig:E_S_vN_vs_Delta}
\end{figure}

With Exact Diagonalization approach we calculate the true ground state $E_{\rm GS}$ of the system, and compare with variationally obtained $E_{\rm VQE}=\langle \psi(\vec{\theta}|\hat{H}|\psi(\vec{\theta})\rangle$ energy, as a function of the anisotropy parameter $\Delta$. Fig.\ref{fig:E_S_vN_vs_Delta} (e) shows the energy of the ansatz $E_{\rm VQE}$ during training. 
As we can see, accuracy of the optimized ground state, for fixed PQC, depends on the anisotropy parameter $\Delta$. For simple observable, such as ground state energy $E$, the ansatz perfectly captures ground state properties, (Fig.\ref{fig:E_S_vN_vs_Delta}(e)) in the whole range of anisotropy parameter $\Delta$. However, when considering bipartite entanglement entropy $S_{\text{vN}}$ situation complicates for $\Delta>1$ - the ground state is a product state, and the bipartite entanglement entropy vanishes, $S_{vN} = 0$, while the bipartite entropy in the ansatz is non-zero, indicating that ansatz does not capture properly properties of the ground state for this regime of anisotropy parameter. This is an important observation, indicating that a given ansatz might be good for capturing only some properties of the system (like ground state energy), while can fail when considering more sophisticated quantities, such as bipartite entanglement entropy.

\section{Quantum Approximate Optimization Algorithm  }

Having seen one instance of how to combine gradient descent with quantum circuits, we might wonder more generally if there are other problems we can solve by combining these techniques. This broadens the scope of our discussion to the more general class of \textit{variational quantum algorithms}. In this section, we will explore one other member of this class known as the Quantum Approximate Optimization Algorithm (QAOA). QAOA is a quantum algorithm introduced by Farhi, et al. in \cite{farhi2014quantum} which aims to probabilistically solve NP-hard {Quadratic Unconstrained Binary Optimization}  ({QUBO}) problems, which will be briefly introduced below.

The objective of QUBO problems is to optimize the  quadratic function $C$ subject to a set of constraints. QUBO problems are typically used to model combinatorial optimization problems, such as the travelling salesman problem, the maximum cut problem (Max-Cut), and the graph colouring problem. 

\begin{figure}[h]
    \centering
    \begin{subfigure}[b]{0.35\textwidth}
        \centering
        \begin{tikzpicture}
            \node [anchor=south west, inner sep=0] (image) at (2.0,0.0) {\includegraphics[width=\textwidth]{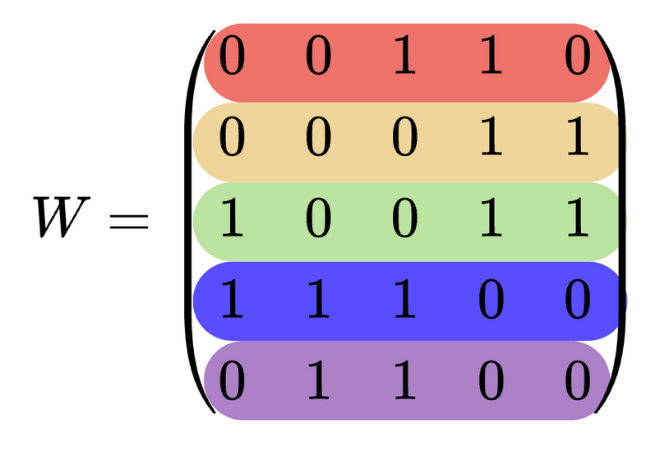}};
        \end{tikzpicture}
    \end{subfigure}

    \begin{subfigure}[b]{0.25\textwidth}
        \centering
        \begin{tikzpicture}
            \node [anchor=north west, inner sep=0] (image) at (-2,0) {\includegraphics[width=\textwidth]{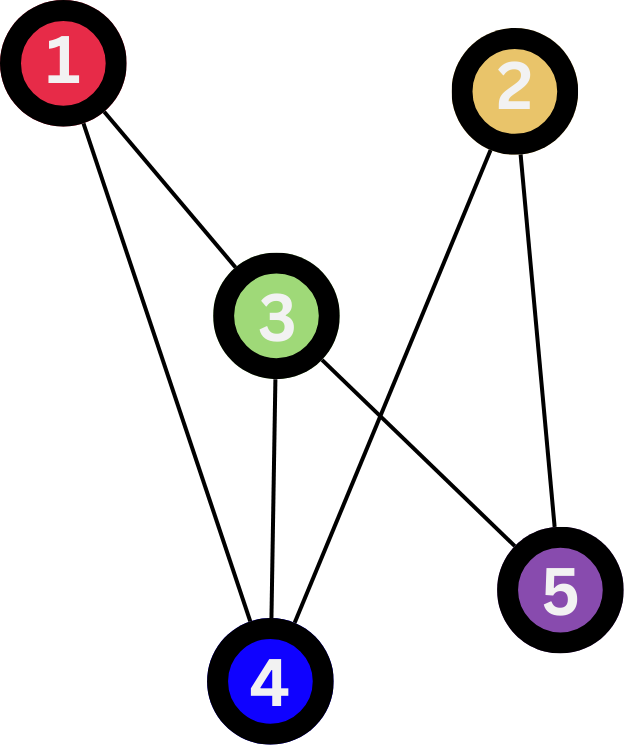}};

        \end{tikzpicture}
    \end{subfigure}
    \caption{Visualisation of QUBO problem for a small (five-component) system. Here, each row of the cost function's weight matrix, $W$, corresponds to a node in an undirected graph. In this example, the edges are unweighted, but it is possible to create a weighted QUBO instance by allowing the entries of $W$ to be positive reals instead of binary digits. Here, the colour coding shows how each row corresponds to a node in the undirected graph, and its binary-vector value indicates to which other nodes it is connected.}
    \label{fig:QUBO_basic}
\end{figure}

\begin{figure}[h]
    \centering
    \begin{tikzpicture}
        \node[anchor=south west,inner sep=0] (image) at (0,0) {\includegraphics[width=0.5\linewidth]{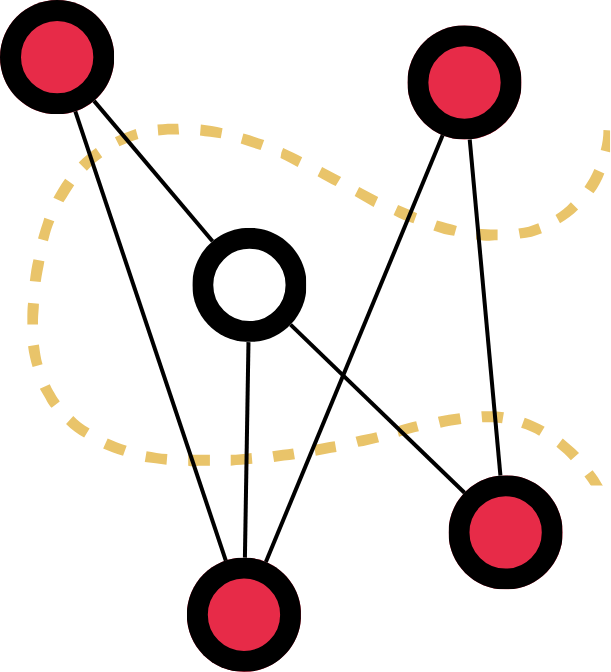}};
        \begin{scope}[x={(image.south east)},y={(image.north west)}]

            \node at (1.0,0.25) (scissors) {\Huge{\Leftscissors}};;
        \end{scope}
    \end{tikzpicture}
    \caption{Visualization of the MaxCut problem. The graph nodes are partitioned by a dashed line starting at the scissors and cutting every edge that the dashed line intersects. The aim of the max cut problem is to find a partition of graph into two sub-graphs that cuts a maximal number edges. In other words, the sub-graphs should be maximally connected, and thus a maximal number of edges is cut after the partition. In this figure, the partition is indicated by colouring the nodes red and white, making a binary assignment of $(0,0,1,0,0)$ with respect to the $s$-vector in Fig.\ref{fig:QUBO_basic}.}
    \label{fig:QUBO_maxcut}
\end{figure}

Mathematically QUBO problem is defined as follows: Let $\mathbb{S}$ be a set of binary vectors, $\mathbb{S} = \{ \vec{s} = (s_1,\dots,s_L) | L>0,  s_i \in (0,1) \}$.
In QUBO, we are given a real-valued upper-triangular matrix $W \in \mathbb{R}^{L\times L}$. We define a QUBO \textit{cost} function $C(\vec{s})$ as:
\begin{equation}
C(\vec{s}) = \sum_{i=1}^L\sum_{j=i}^{L} W_{ij}s_is_j.
\label{eq:QUBO_cost}
\end{equation}
The solution of a QUBO problem is a vector $\vec{s}_*$ that minimizes this cost function, i.e.:
\begin{equation}
\vec{s}_* = \arg \min_{\vec{s}\in\mathbb{S}} C(\vec{s}).
\end{equation}
As we can see, the search space for the classical optimization algorithm, i.e., size of the vector space $\mathbb{S}$, grows exponentially with   $L$, $\dim \mathbb{S} = 2^L$.  The reason QUBO problems are of great interest for graph problems is because we can associate the elements $W_{ij}$ from Eq.~\ref{eq:QUBO_cost} with a weighted edge for a graph. That is, given a graph with $L$ nodes, the QUBO cost function encodes the edges that join the nodes since it contains a weight matrix that joins pairs of indices $i,j \in \{1,\dots,L\}$ within  $\vec{s}$, see Fig.\ref{fig:QUBO_basic}. The solution to the QUBO problem is thus some binary assignment to each node, from which the vector $\vec{s}_*$ can be extracted. An example for the MaxCut problem can be seen in Fig.\ref{fig:QUBO_maxcut}.


What Fahri et. al \cite{farhi2014quantum} noticed in their seminal work is that the solution to the considered problem has to belong to a Fock space of the chain of $L$ spins-$1/2$. To see how this works, let us cast a QUBO problem onto the spin-$1/2$ system, where the classical cost function $C$ is represented as a 
\textit{cost} Hamiltonian $\hat{H}_C$ of the spin-$1/2$ system of size $L$. The optimal solution $s_*$ of the original problem can therefore be posed as a ground state search, $\ket{\psi_*}$, of the cost Hamiltonian $\hat{H}_C$. 

In practice, the solution $\vec{s}_*$ is given by the spin configuration (i.e. computational basis states) of a state vector $\vec{s}_k$ having the dominant contribution to the $\hat{H}_C$ ground state $\ket{\psi_*} = \sum_{k} \alpha_k |\vec{s}_k\rangle$:
\begin{equation}
 \begin{split}
   k_* &  = \arg \max_{k} |\braket{\vec{s}_k|\psi_*}|^2,\\
   \vec{s}_* &\equiv \vec{s}_{k_*}.
   \end{split}
\end{equation}
In other words, by repeatedly sampling from $\ket{\psi_*}$, we will find the dominant bitstring that makes a \textit{probabilistic} solution to the QUBO problem. As such, the objective is to find a many-body wave function with the lowest possible energy for the cost Hamiltonian $\hat{H}_C$. This means we can unpack many of the tools we developed for the VQE, and apply them here. 

Let's now consider a QAOA ansatz for the many-body wave function in the form a parametrized quantum circuit
\begin{equation}
\ket{\psi(\vec{\beta},\vec{\gamma})} = \prod_{j=1}^p \hat{U}_B(\beta_j)\hat{U}_C(\gamma_j)\prod_{i=1}^L \hat{H}_i \ket{0},
\end{equation}
where $\hat{H}_i$ is a Hadamard gate acting on a $i$-th qubit ($i = 1,\dots, L)$, and two parametrized unitary operators $\hat{U}_B$ and $\hat{U}_C$ are defined as
\begin{equation}
\begin{split}
\hat{U}_C(\gamma_j) &= e^{-i\gamma_j\hat{H}_C} \\
\hat{U}_B(\beta_j) &= e^{-i\beta_j\hat{H}_B},
\end{split}
\end{equation}
where $\hat{H}_C$ is a considered cost Hamiltonian, and $\hat{H}_B = \sum_i \hat{\sigma}^x_i$ is a so-called \textit{mixer} Hamiltonian. After the first layer of Hadamard gates, this constitutes an \textit{alternating layer ansatz} with twofold alternation, see Fig.~\ref{fig:VQE_block_structures}. Further, this means the variational parameters are $\theta = \{\vec{\beta},\vec{\gamma}\}$.

The objective of the algorithm is to find a set of parameters $\vec{\beta} = (\beta_1, \dots, \beta_p)$, $\vec{\gamma} = (\gamma_1,\dots,\gamma_p)$ which\footnote{Here $p\ge1$ is a hyperparameter of the algorithm describing how many alternating layers are in the circuit. Per our discussion on barren plateaus, a diligent choice should be made here, as well as a careful choice for the structure of the Unitary within a layer.} minimizes the cost function
\begin{equation}
F(\vec{\beta},\vec{\gamma}) = \bra{\psi(\vec{\beta},\vec{\gamma})}\hat{H}_C\ket{\psi(\vec{\beta},\vec{\gamma})}.
\end{equation}
Finding $\vec{\beta}_*$ and $\vec{\gamma}_*$ can be done with the classical optimization algorithms based on gradient descent, analogously to the VQE . In practice, we run QAOA many times  with random initialization of $\vec{\beta}$, and $\vec{\gamma}$ and we plot the histogram of the index $k_*$ of the optimal configuration given by the corresponding binary vector $\ket{\vec{s}_{k_*}}$. The solution is given by $\ket{\vec{s}_{k_*}}$ corresponding to a peak in the histogram of all $k_*$.

Having seen a general example of how to construct a QOAO solver, let us consider a concrete optimization problem: The Max-Cut problem on a graph. We consider a planar graph $G = (V,E)$, having $L$ vertices (or nodes) enumerated by index $i \in V$ ($i=1,\dots,L$), and edges being connected pairs of vertices, i.e. $E = \{ (i,j) | i,j\in V, i\ne j\}$. Each edge has a weigth $W_{ij}>0$. We consider an undirected graph, thus\footnote{In fact, this is in line with the fact that $W$ is upper-triangular in the original definition. There was no need to define the other direction for an undirected graph.} $W_{ij} = W_{ji}$. 
We define a $\textit{cut}$ of a graph $G$, as a division of graph nodes into two parts (subsets of the nodes) $S$, $T$ such that $V = S\cup T$. To each node belonging to subgraph $S$, we assign the value $z_i = +1$, while to each node in subgraph $T$, we assign the value of $z_i = -1$. To check if two nodes belong to the same subgraph, we can define a function
\begin{equation}
 c_{ij} = \frac{1-z_iz_j}{2},
\end{equation}
which takes value $0$ for $z_i = z_j$, i.e. if two nodes belong to the same subset, and $1$ otherwise. Next, we can define a $\textit{value of a cut}$ as
\begin{equation}
  P = \sum_{i,j} c_{ij}W_{ij}.
\end{equation}
In a Max-Cut problem,  we aim to  $\textit{cut}$ maximizing $P$. In other words, we aim to partition the graph into two subsets which are joined by a maximal number of edges. Rewriting $P$ as
\begin{equation}
 P = \sum_{i<j}W_{ij} - \sum_{i<j}W_{ij}z_iz_j = W - C(\vec{z}),
\end{equation} 
where $W \equiv \sum_{i<j}W_{ij}$ is a positive constant number, the Max-Cut problem reduces to minimize the cost function $C(\vec{z}) = \sum_{i<j}W_{ij}z_iz_j$, $\vec{z}=(z_1,\dots,z_L)$, meaning it is a type of QUBO problem.

From here, we can now express the Max-Cut problem as finding the ground state of a spin-1/2 system, as we saw in the above discussion. To the $i$-th node with the value $z_i$ we assign the state of the spin-1/2, i.e.:
\begin{equation}
 \begin{split}
    z_i = +1 \to & |0\rangle \equiv |\uparrow\rangle \\ 
    z_i = -1 \to & |1\rangle \equiv |\downarrow\rangle.
 \end{split}
\end{equation}
As such, minimizing the cost function $C(\vec{z})$, can be considered as finding the spin configuration, which minimizes the \textit{cost} Hamiltonian:
\begin{equation}
 \hat{H}_C = \sum_{ik} W_{ij} \hat{\sigma}^z_i\hat{\sigma}^z_j,
\end{equation}
where $\hat{\sigma}^z_i$ is a Pauli-Z operator acting on $i$-th qubit. While we consider a planar graph and enumerate spins with index $i$, the problem is equivalent to considering a one-dimensional chain of spins properly connected via edges $W_{ik}$.

The family of QUBO problems is a promising application for quantum computing because mapping the problem to a quantum system and using quantum mechanics allows for a potentially more efficient solution than classical methods.

\section[Neural Network Quantum States]{Neural Quantum States}
\label{sec:NQS}
Let's again consider the wavefunction of $N$ 2-level quantum systems, which we repeat here for convenience,
\begin{equation}
    \ket{\psi(t)} = \sum_{b_1,\ldots,b_N = 0}^1
    c_{b_1,\ldots,b_N }(t)\ket{b_1,\ldots,b_N}.
\end{equation}
There are an exponential number of coefficients to keep track of here because there are $2^N$ binary bitstrings of length $N$ that are summer over. We can see how solving for the dynamics or this state, or even representing it at a given moment in time quickly becomes infeasible as $N$ grows. We call this the \textit{quantum many-body problem}, and much attention has been given to finding clever ways to accurately approximate $\ket{\psi(t)}$ over the past 30 years. Here, we will discuss one trick to do this which is based on variational methods \cite{carleo2017solving}. 

Usually, when we think about the coefficients of a wave function, we just imagine them as complex numbers accompanying a given basis vector. However, we can also think of the coefficients, $c_{b_1,\ldots,b_N}$, as a \textit{function} $c:\{0,1\}^{N} \rightarrow \mathbb{C}$ from a binary vector (sometimes called a bitstring) to a complex number. We denote this function as $c(\mathbf{b})$ for a binary vector $\mathbf{b} \in \{0,1\}^{N}$. The key idea of neural quantum states (NQS) is to use a neural network to approximate this function. Recall from Chapter~\ref{CH:FUNDAMENTALS} that a sufficiently large neural network is capable of representing any continuous, differentiable function given enough data about it. A NQS $\psi_{\theta}(\mathbf{b})$, with variational parameters $\theta$ is therefore a map between binary bitstrings to a complex amplitude. So far, the neural architectures we have studied have only been used to output real numbers. How then, can we make a neural network output complex numbers?

\begin{figure}[h!]
    \begin{tikzpicture}[
    every edge/.style = {draw,->}
  ]
            \node[anchor=south west, inner sep=0] (image) at (0,0) {\includegraphics[width=0.5\textwidth]{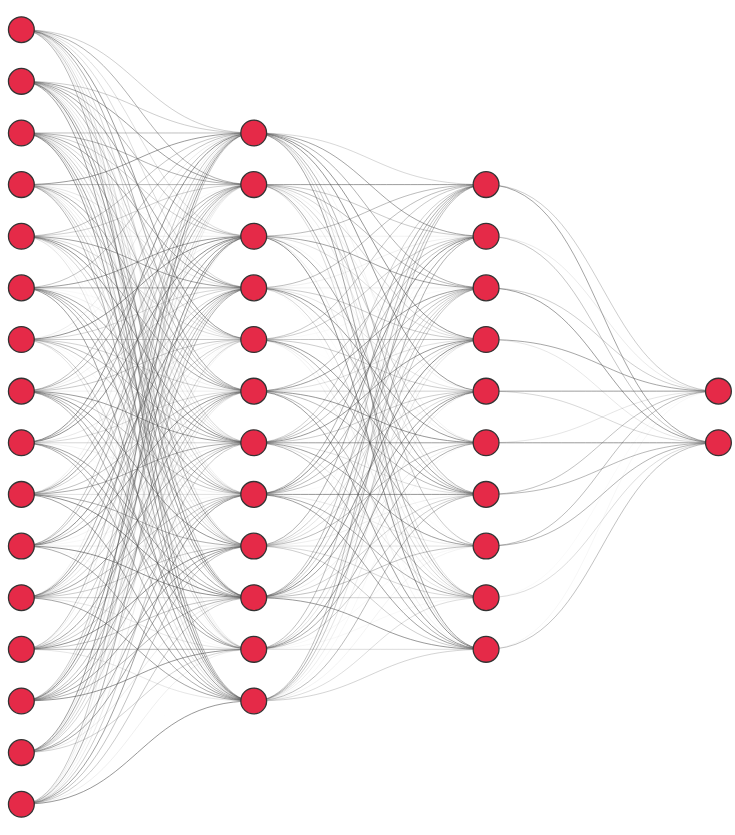}};
            \begin{scope}[x={(image.south east)}, y={(image.north west)}]
                \node (phi) at (1.06, 0.535) {$=\phi_{\theta}$};

                \node (rootp) at (1.08, 0.47) {$=\sqrt{p_{\theta}}$};

                \node (timesi) at (1.2, 0.535) {$\times$};

                \node (imagunit) at (1., 0.75) {$i \in \mathbb{C}$};

                \draw (phi) edge[out = 0, in = 180] (timesi);

                \draw (imagunit) edge[out = 270, in = 90] (timesi);

                \node (exp) at (1.375, 0.535) {$\text{exp}(\cdot)$};

                \draw (timesi) edge[out = 0, in = 180] (exp);

                \node (timesexp) at (1.55, 0.535) {$\times$};

                \draw (exp) edge[out = 0, in = 180] (timesexp);

                \draw (rootp) edge[out = 0, in = 270, looseness = 0.7] (timesexp);

                \node (cTheta) at (1.8, 0.535) {$= c_{\theta}(\mathbf{b}) = \braket{\mathbf{b}|\psi_{\theta}}$};
            \end{scope}
        \end{tikzpicture}
        \caption{An example NQS architecture based on a multi-layer perceptron, and post processing. Here, the MLP received a $16-$bit binary number $\mathbf{b} \in \{0,1\}^{16}$ and outputs a tuple of real numbers $(x, y)\in \mathbb{R}^2$ which we associate with $\phi_{\theta}$ and $\sqrt{p_{\theta}}$. Note we can enforce bounds or positivity by renormalising or by our choice of activation function in the final layer. The post-processing steps allow us to construct the amplitude $c_{\theta} = \sqrt{p_{\theta}} e^{i \phi_{\theta}} = \braket{\mathbf{b}|\psi_{\theta}}$. These are shown as a differentiable computation graph which as introduced in Chapter~\ref{CH:FUNDAMENTALS}. }
        \label{fig:NQS_MLP}
\end{figure}
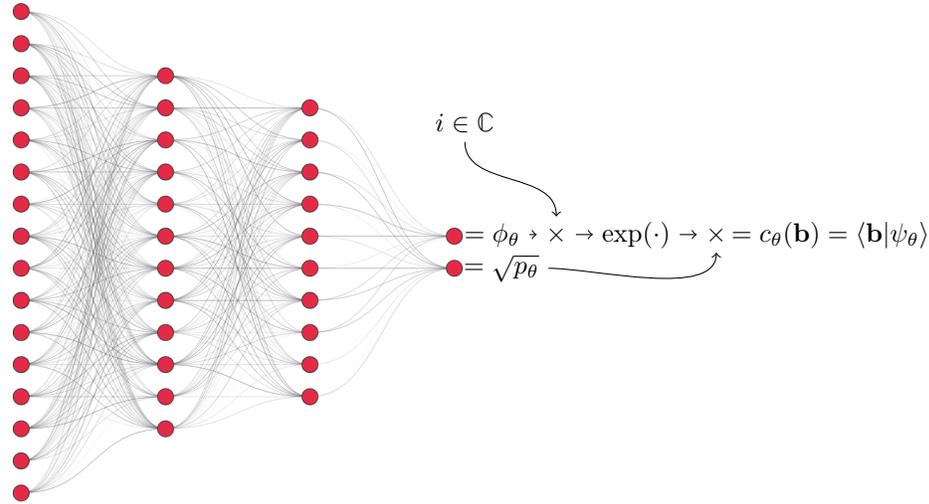

There are a few approaches. First, a neural networks can output complex numbers by having two output neurons, $\{p(\mathbf{b};\theta), \phi(\mathbf{b};\theta)\}$, to construct 
\begin{equation}
    \psi_{\theta}(\mathbf{b}) = \sqrt{p_{\theta}(\mathbf{b})} e^{i \phi_{\theta}(\mathbf{b})}.
\end{equation}
From a computational perspective, we can construct $\psi_{\theta}(\mathbf{b})$ in post-processing, as shown in Fig.~\ref{fig:NQS_MLP}. We could also have two separate neural networks for each of $\{p(\mathbf{b};\theta), \phi(\mathbf{b};\theta)\}$, both receiving the  bitstring $\mathbf{b}$ as input. Much recent success has also been found in allowing the neural network parameters $\theta$ to take complex values, $\theta \in \mathbb{C}^n$. The latter is a hot topic as of 2024, but comes with some interesting caveats surrounding back propagation as we shall see later.

Being an observant reader, you might wonder at this point whether this construction is at all useful. This is because in order to know the complete wavefunction, we would need to perform a forward pass of our neural network model for \textit{every} possible binary bitstring $\mathbf{b}$. This would mean there is exponential time complexity, making the whole exercise pointless; naively, we might think that NQS have simply displaced the space-complexity of holding $2^N$ coefficients into time-complexity of executing $2^N$ forward passes. 

However, let's think for a second about \textit{why} we want to write down a wavefunction in the first place. The postulates of quantum mechanics tell us that once we know a wavefunction, every observable quantity can be derived from it. Therefore, we could argue that what we are \textit{actually} interested in when using a wave function is computing statistics of observables. One way to do this is with exact wave-functions, inner products, and the Born rule. However, this is not the only way. Another way to yield predictions for observables and their statistics is the \textit{Variational Monte Carlo (VMC) method}.

VMC methods show that what we really need to emulate the statistics of quantum states are two things:
\begin{enumerate}
    \item The ability to efficiently (in polynomial time) generate amplitudes $\braket{\mathbf{b}|\psi}$  for any basis element $\ket{\mathbf{b}}$.
    \item The ability to efficiently sample from the distribution
    \begin{equation}
        p\big(\mathbf{b}|\psi_{\theta}\big) = \frac{|\!\braket{\mathbf{b}|\psi_{\theta}}\!|^2}{\braket{\psi_{\theta}|\psi_{\theta}}}.
        \label{eq:NQS_born_dist}
    \end{equation}
    This means our representation has a mechanism to execute the Born rule, that allows us to gather statistics efficiently.
\end{enumerate}

With these two properties, we can use sampling to estimate the expectation of arbitrary operators! Let's explore this in a little more depth. To begin with, recall that we can write the expectation value of some operator $\hat{O}$ with respect to $\ket{\psi_{\theta}}$ as
\begin{equation}
    \braket{\hat{O}} = \frac{\braket{\psi_{\theta} | \hat{O} | \psi_{\theta}}}{\braket{\psi_{\theta} | \psi_{\theta}}},
\end{equation}
where the denominator comes from renormalising in case $\braket{\psi_{\theta} | \psi_{\theta}} \neq 1$. Using one of two favourite tricks of mathematicians\footnote{In the wise words of my undergraduate mathematics lecturer: \textit{``The best thing you can do to make progress in maths problems is times by one, or add zero''}.}, we can resolve the identity $\mathbb{1} = \sum_{\mathbf{b}} \ket{\mathbf{b}}\!\bra{\mathbf{b}}$ and write,
\begin{equation}
    \braket{\hat{O}} = \frac{\sum_{\mathbf{b}, \mathbf{b'}} \braket{\psi_{\theta}|\mathbf{b}}\!\bra{\mathbf{b}} \hat{O} \ket{\mathbf{b'}} \! \braket{\mathbf{b'} | \psi_{\theta}}}{\sum_{\mathbf{b'}'} \braket{\psi_{\theta} | \mathbf{b''}} \braket{\mathbf{b''} | \psi_{\theta}}},
\end{equation}
where we resolved the identity three times. This might look like a painful thing to do, but notice that if we do it \textit{just one more time}, something cool happens. Let's write $1 = \braket{\mathbf{b}|\psi_{\theta}}/\braket{\mathbf{b}|\psi_{\theta}}$, to show
\begin{equation}
\begin{aligned}
    \sum_{\mathbf{b}, \mathbf{b'}}
    &\frac{
    \braket{\psi_{\theta}|\mathbf{b}}\!\bra{\mathbf{b}} \hat{O} \ket{\mathbf{b'}} \! \braket{\mathbf{b'} | \psi_{\theta}}
    }{
    \sum_{\mathbf{b'}'} \braket{\psi_{\theta} | \mathbf{b''}} \braket{\mathbf{b''} | \psi_{\theta}}
    }
    \cdot \frac{\braket{\mathbf{b}|\psi_{\theta}}}{\braket{\mathbf{b}|\psi_{\theta}}} \\
    &= \sum_{\mathbf{b}} \left(
    \frac{|\! \braket{\mathbf{b}|\psi_{\theta}}\!|^2}{\sum_{\mathbf{b''}}|\!\braket{\mathbf{b''}|\psi_{\theta}}\!|^2} \cdot \sum_{\mathbf{b'}} \frac{\braket{\mathbf{b}|\psi_{\theta}}}{\braket{\mathbf{b'}|\psi_{\theta}}} \braket{\mathbf{b}|\hat{O}|\mathbf{b'}}
    \right).
\end{aligned}  
\end{equation}
In this form, we can recognise the first term in the product as the distribution 
\begin{equation}
    p\big(\mathbf{b}|\psi_{\theta}\big) = \frac{|\!\braket{\mathbf{b}|\psi_{\theta}}\!|^2}{\braket{\psi_{\theta}|\psi_{\theta}}},
\end{equation}
which we assumed to admit efficient sampling. The second term in the product
\begin{equation}
    \sum_{\mathbf{b'}} \frac{\braket{\mathbf{b}|\psi_{\theta}}}{\braket{\mathbf{b'}|\psi_{\theta}}} \braket{\mathbf{b}|\hat{O}|\mathbf{b'}} := \braket{O^{\text{loc}}(\mathbf{b};\theta)}
    \label{eq:local_O}
\end{equation}
is simply the matrix elements of $\hat{O}$ in the computational basis, combined with two amplitudes, $\braket{\mathbf{b}|\psi_{\theta}}$ and $\braket{\mathbf{b'}|\psi_{\theta}}$. If $\hat{O}$ is given, then we can safely assume that its computational basis expansion is known. Provided that the elements $\braket{\mathbf{b}|\hat{O}|\mathbf{b'}} \neq 0$ for at most a polynomial number of bitstrings $\mathbf{b}$, then we can compute $\braket{O^{\text{loc}}(\mathbf{b};\theta)}$ efficiently. Although it does come with a caveat that the operator $\hat{O}$ is sparse. In other words, $\hat{O}$ contains terms acting on at most $k$ different qubits. For example, $k=2$ corresponds to quadratic operators, $k = 3$ are cubic, and so on. This constitutes a \textit{sparsity} assumption.

Together both terms constitute a \textit{useful} variational approximation, because amplitudes, $\braket{\mathbf{b}|\psi}$ can be generated efficiently, as can $\braket{O^{\text{loc}}(\mathbf{b};\theta)}$. Hence, we may write
\begin{equation}
    \braket{\hat{O}} = \sum_{\mathbf{b}} p(\mathbf{b}|\psi_{\theta}) \sum_{\mathbf{b'}} \frac{\braket{\mathbf{b}|\psi_{\theta}}}{\braket{\mathbf{b'}|\psi_{\theta}}} \braket{\mathbf{b}|\hat{O}|\mathbf{b'}} \approx \braket{O^{\text{loc}}(\mathbf{b};\theta)}_{\mathbf{b}},
\end{equation}
where the notation $\braket{\cdot}_{\mathbf{b}}$ denotes a Monte Carlo average. That is, $\braket{\cdot}_{\mathbf{b}}$ is a statistical average generated by sampling from the distribution in Eq.~(\ref{eq:NQS_born_dist}). Since we have limited ourselves to a sparsely interacting system, we can also assume that the interactions are $k$-local. That is, interactions between bodies limited to those which are space-like separated by at most $k$ times their average separation. 

This means we can express $\braket{\hat{O}}$ by decomposing it in terms of all the local observable outcomes
\begin{equation}
    \braket{\hat{O}} = \sum_{\mathbf{b}}p\left(\mathbf{b}|\psi_{\theta}\right)\braket{O^{\text{loc}}(\mathbf{b};\theta)}_{\mathbf{b}}.
\end{equation}
Therefore, sampling some $M$ bitstrings, $\mathbf{s}_1,\ldots,\mathbf{s}_M $,  from  $p(\mathbf{b}|\psi_{\theta})$, we can estimate $\hat{O}$ by the arithmetic mean
\begin{equation}
    \braket{\hat{O}} \approx \frac{1}{M}\sum_{j = 1}^M \braket{O^{\text{loc}}(\mathbf{s}_j;\theta)}.
\end{equation}
This is just another type of basis expansion, like Pauli or computational which we saw in the introduction. 

One nice thing about $\braket{O_\theta^{\text{loc}}(\mathbf{b};\theta)}_{\mathbf{b}}$ is that is has a controllable amount of error with respect to the true expectation value $\braket{\hat{O}}$. This is thanks so the fact that any statistical estimator will have an uncertainty of
\begin{equation}
    \epsilon = \sqrt{\sigma^2/M},
\end{equation}
where $\sigma^2$ is the variance from our samples $\mathbf{s}_1,\ldots,\mathbf{s}_M $ of $O^{\text{loc}}(\mathbf{s}_j;\theta)$. We can summarise computing local observables with Algorithm~\ref{alg:NQS_observables}.\\

\begin{algorithm}
\caption{Estimating Local Observables with a NQS}\label{alg:NQS_observables}
\begin{algorithmic}[1]
    \State \textbf{Input:} a NQS $\psi_{\theta}: \{0,1\}^N \rightarrow \mathbb{C}$ and a set of random Binary bitstrings $\mathbf{B} = \{\mathbf{b}_1, \mathbf{b}_2,\ldots,\mathbf{b}_M\}$
    \For{each $\mathbf{b}$ in $\mathbf{B}$ }
        \State Forward pass $\psi_{\theta}(\mathbf{b}) = \braket{\mathbf{b}|\psi_{\theta}}$ \Comment{Should be efficient}
        \State Compute $\braket{O^{\text{loc}}(\mathbf{b};\theta)}$ \Comment{See Eq.~(\ref{eq:local_O}), $\hat{O}$ inputted therefore efficient}
    \EndFor
    \State \textbf{Return} $\braket{\hat{O}} \approx \frac{1}{M}\sum_{\mathbf{b} \in \mathbf{B}}\braket{O^{\text{loc}}(\mathbf{b};\theta)}$
\end{algorithmic}
\end{algorithm}

So far, we have a basic idea of how the VMC method can be used to compute observable statistics of the wavefunction they are designed to represent. This assumed we are given some variational parameters $\theta$ to work with. However, the beauty of variational parameters is that they are tunable with gradient-based optimisers. We can now use VMC, combined with gradient-based optimisers for a number of different settings:
\begin{enumerate}
    \item To represent a given state $\ket{\psi}$ such that sampling from $\ket{\psi}$ and our neural representation are indistinguishable up to some error.
    \item To search for a new state satisfying some constraint. This could be a ground state search, like we saw in the VQE, finite temperature states, or open system density matrices, to name a few.
    \item To infer the time-trajectory of a given input state $\ket{\psi(t = 0)}$.
\end{enumerate}

As this is an introductory course, we will only go through NQS for the (arguably simplest) case of ground state searching. For a detailed (and great!) introduction and review of NQS and their applications, see \textit{From Architectures to Applications: A Review of Neural Quantum States} by Lange et. al \cite{lange2024architectures}.

In brief, the variational principle also applies to NQS. This is because using a neural network as a map from a binary bitstring (the index) to an amplitude will still be an upper bound for the ground state energy,
\begin{equation}
    E(\theta) = \frac{\braket{\psi_{\theta}|\hat{H}|\psi_{\theta}}}{\braket{\psi_{\theta}|\psi_{\theta}}} \geq E_{gs},
    \label{eq:NQS_variational_principle}
\end{equation}
where $H$ is a given system Hamiltonian. Recall that we can evaluate Eq.~(\ref{eq:NQS_variational_principle}) thanks to Monte-Carlo sampling, because $H$ admits a decomposition into a weighted sum of observables per our discussion on the VQE, see Eq.~(\ref{eq:hamiltonian_pauli_decomp}). Explicitly, we may write
\begin{equation}
    E(\theta) = \braket{\hat{H}} \approx \sum_{\mathbf{b}}p\left(\mathbf{b}|\psi_{\theta}\right)\braket{H^{\text{loc}}(\mathbf{b};\theta)}_{\mathbf{b}}.
    \label{eq:NQS_energy_loss}
\end{equation}
To create a training loop, all that remains is to show that our estimate of $\braket{\hat{H}}$ with respect to the NQS $\ket{\psi_{\theta}}$ is differentiable. This will allow us to use optimisers or vanilla gradient descent to train our NQS. Due to the fact that we are using neural networks to predict complex-valued amplitudes, an interesting and subtle point arises here due to complex-valued differentiation. \\

Let's consider the variational ground state search where we seek some optimal values, $\theta^{\text{opt}}$ which are the solution to
\begin{equation}
    \min_{\theta} \frac{\braket{\psi_{\theta}|\hat{H}|\psi_{\theta}}}{\braket{\psi_{\theta}|\psi_{\theta}}}.
\end{equation}
Like in the variational setting from SEC:VQE, we aim to solve this problem via gradient-based optimsiers on $\theta$. As before, we can define a loss-function as the average energy,
\begin{equation}
    E(\theta) = \frac{\braket{\psi_{\theta}|\hat{H}|\psi_{\theta}}}{\braket{\psi_{\theta}|\psi_{\theta}}},
\end{equation}
which is minimised when $\theta = \theta^{\text{opt}}$.
Whilst our situation may appear identical, recall that
in the VQE, everything in the differentiable feedback loop was real valued. This is because the entire state $\psi_{\theta}$ was constructed from a parametric quantum circuit, $U(\theta)$,
\begin{equation}
    \braket{\psi(\theta)|\hat{H}|\psi(\theta)} = \braket{0|\hat{U}(\theta)^{\dagger} \hat{H} \hat{U}(\theta)|0}.
\end{equation}
In NQS, our situation is different. We have a neural representation of a wavefunction, rather than access to it as the output of some unitary process. It receives a computational basis vector as input, and outputs a complex-valued amplitude. This meant that computing expectation values was based on Monte-Carlo sampling, rather than taking repeated shots like in the VQE. For each single sample, computing the energy-based loss means $E(\theta)$ is a map
\begin{equation}
    E(\psi_{\theta}): \mathbb{C} \rightarrow \mathbb{R},
\end{equation}
since the output of our neural representation is a complex number, and we have allowed\footnote{That is $\psi_{\theta}(\mathbf{b}): \{0,1\}^N \xrightarrow[]{\theta} \mathbb{C}$, where the $\theta$ themselves can be complex valued.} $\theta \in \mathbb{C}^{\text{dim}(\theta)}$. How then, do we back-propagate through a complex-to-real loss function? 

To answer this question, we need to recall a few basics from complex analysis. First, recall a holomorphic function of a complex variable $f(z,z^*)$ is one for which,
\begin{equation}
    \frac{\partial}{\partial z^*} f(z,z^*) = 0.
\end{equation}
In other words, holomorphic functions are functions only of a complex variable $z \in \mathbb{C}$, not its conjugate, $z^*$. Let's assume for simplicity that our NQS is Holomorphic, so
\begin{equation}
    \frac{\partial \psi_{\theta}}{\partial 
    \theta^*} = 0.
    \label{eq:holomorphic_defn}
\end{equation}
This is a reasonable assumption because the components of a complex neural network can easily be made holomorphic by ensuring the activation functions and sums over layers are taken over $\theta$ and not $\theta^*$. We also have the useful property that
\begin{equation}
    \frac{\partial E(\theta)}{\partial \theta} = \left( 
        \frac{\partial E(\theta)}{\partial \theta^*}
    \right)^*.
\end{equation} 
Hence, to compute $\frac{\partial E(\theta)}{\partial \theta}$ we can compute $\frac{\partial E(\theta)}{\partial \theta^*}$ and take the conjugate of the answer. To make the notation less cumbersome, let's let $\partial_{\theta}:= \frac{\partial}{\partial \theta}$. By the product and chain rules, we have
\begin{equation}
\begin{aligned}
    &\partial_{\theta^*} E(\theta) = \frac{\partial}{\partial \theta^*} 
    \left( \frac{\braket{\psi_{\theta}|\hat{H}|\psi_{\theta}}}{\braket{\psi_{\theta}|\psi_{\theta}}}
    \right)\\
    &= \frac{\big(\partial_{\theta^*}\!\bra{\psi_{\theta}}\big) \hat{H} \ket{\psi_{\theta}}}{\braket{\psi_{\theta}|\psi_{\theta}}} + \frac{\bra{\psi_{\theta}} \hat{H} \big(\partial_{\theta^*}\!\ket{\psi_{\theta}}\big)}{\braket{\psi_{\theta}|\psi_{\theta}}} - E(\theta) \frac{\big( \partial_{\theta^*} \!\braket{\psi_{\theta}|\psi_{\theta}} \big)}{\braket{\psi_{\theta}|\psi_{\theta}}}
\end{aligned}
\end{equation}
Immediately, the middle of these three terms is zero from the definition of a holomorphic function in Eq.~(\ref{eq:holomorphic_defn}). We also see $\partial_{\theta^*}\! \bra{\psi_{\theta}} = \bra{\partial_{\theta} \psi_\theta}$, so we have
\begin{equation}
    \partial_{\theta^*} E(\theta) = \frac{\bra{\partial_{\theta}\psi_{\theta}} \hat{H} \ket{\psi_{\theta}}}{\braket{\psi_{\theta}|\psi_{\theta}}} - E(\theta) \frac{ \!\braket{\partial_{\theta}\psi_{\theta}|\psi_{\theta}}}{\braket{\psi_{\theta}|\psi_{\theta}}}.
    \label{eq:deriv_half_way}
\end{equation}
We can now simplify this expression with the same tricks as before and resolve the identity $\mathbb{1} = \sum_x \ket{x}\!\bra{x}$. For clarity, we will work on each term separately, and add them at the end. For the second term in Eq.~(\ref{eq:deriv_half_way}), we see that
\begin{equation}
\begin{aligned}
    \frac{\braket{\partial_{\theta}\psi_{\theta}|\psi_{\theta}}}{\braket{\psi_{\theta}|\psi_{\theta}}}
    = \sum_x \braket{\partial_{\theta}\psi_{\theta}|x}\!\braket{x|\psi_{\theta}}\cdot\frac{1}{\braket{\psi_{\theta}|\psi_{\theta}}}\cdot \frac{\braket{\psi_{\theta}|x}}{\braket{\psi_{\theta}|x}}.
 \end{aligned}
\end{equation}
Recalling that
\begin{equation}
    p_{\theta}(x) = \frac{\braket{x|\psi_{\theta}} \braket{\psi_{\theta}|x}}{\braket{\psi_{\theta}|\psi_{\theta}}} = \frac{|\braket{x|\psi_{\theta}}|^2}{\braket{\psi_{\theta}|\psi_{\theta}}}
\end{equation}
 we can simplify this to
\begin{equation}
    \frac{\braket{\partial_{\theta}\psi_{\theta}|\psi_{\theta}}}{\braket{\psi_{\theta}|\psi_{\theta}}}
    = \sum_{x}p_{\theta}(x)  \frac{\braket{\partial_{\theta} \psi | x}}{\braket{\psi_{\theta}|x}} = \mathbb{E}_{x \sim p_{\theta}(x)}\bigg[\partial_{\theta^*} \log \psi_{\theta}^*(x)\bigg],
\end{equation}
where we have used $\braket{\psi_{\theta}|x} = \psi^{*}_{\theta}(x)$, and $\mathbb{E}_{x \sim p_{\theta}(x)}[\cdot]$ means the expectation value when x is sampled from the distribution $p_{\theta}(x)$. Using the same set of tricks on the first term, we find
\begin{equation}
    \begin{aligned}
        \frac{\bra{\partial_{\theta}\psi_{\theta}} \hat{H} \ket{\psi_{\theta}}}{\braket{\psi_{\theta}|\psi_{\theta}}} &= \sum_{x} \frac{\braket{\partial_{\theta} \psi_{\theta}|x} \! \braket{x|\hat{H}|\psi_{\theta}}}{\braket{\psi_{\theta}|\psi_{\theta}}} \cdot \frac{|\braket{x|\psi_{\theta}}|^2}{|\braket{x|\psi_{\theta}}|^2} \\
        &= \sum_{x} \frac{\braket{\partial_{\theta}\psi_{\theta}|x}}{\braket{\psi_{\theta}|x}} \cdot \frac{\braket{x|\hat{H}|\psi_{\theta}}}{\braket{x|\psi_{\theta}}} \cdot \frac{|\braket{x|\psi_{\theta}}|^2}{\braket{\psi_{\theta}|\psi_{\theta}}} \\
        &=\mathbb{E}_{x \sim p_{\theta}(x)} \bigg[ 
        \partial_{\theta} \log (\psi_{\theta}(x)) \hat{H}^{\text{loc}(x)}
        \bigg],
    \end{aligned}
\end{equation}
where $H^{\text{loc}}$ is the local observable decomposition of H, in line with Eq.~(\ref{eq:local_O}). Putting these together, we find 
\begin{equation}
    \partial_{\theta^*} E(\theta) = \mathbb{E}_{x \sim p_{\theta}(x)}\bigg[ 
    (\partial_{\theta} \log \psi_{\theta}(x))\left(\hat{H}^{\text{loc}} - \mathbb{E}_{x \sim p_{\theta}(x)}[\hat{H}^{\text{loc}}(x)]\right)
    \bigg].
\end{equation}

Since this gradient is based on expectation under distributions, we can approximate it with batches of samples $x \sim p_{\theta}(x)$ drawn from forward passes of a NQS. Explicitely, this means we would be estimating the gradient via
\begin{equation}
    \partial_{\theta^*} E(\theta) \approx \frac{1}{N} \sum_{i=1}^{M} \left( \partial_{\theta} \log \psi_{\theta}(x_i) \right) \left( \hat{H}^{\text{loc}}(x_i) - \frac{1}{N} \sum_{j=1}^{M} \hat{H}^{\text{loc}}(x_j) \right),
    \label{eq:NQS_grad_loss_approx}
\end{equation}
which is ``efficient'' in the sense that it is $\mathcal{O}(M^2)$ for a batch of size $M$. We now have all the ingredients we need to estimate observables with a NQS, and train its variational parameters. A summary of how to train a NQS for ground-state searching is provided in Algorithm~\ref{alg:NQS_train_step}. You should now be able to train your own NQS to perform ground-state searching, and sample expectation values from it.

\begin{algorithm}
\caption{Ground State Search with NQS Train Step}\label{alg:NQS_train_step}
\begin{algorithmic}[1]
    \State \textbf{Input:} NQS, $\psi_{\theta}$,  $\mathbf{B} = \{\mathbf{b}_1, \mathbf{b}_2, \ldots, \mathbf{b}_M\}$, $\hat{H}$ \Comment{A batch of bitstrings, $\mathbf{B}$, Hamiltonian $\hat{H}$ assumed to be $k$-local with known basis expansion.} 
    \State Set $E(\theta) = 0$ \Comment{Energy-based loss function from Eq.~(\ref{eq:NQS_energy_loss})}
    \For{$\mathbf{b}$ in $\mathbf{B}$}
        \State Forward pass $\psi_{\theta}(\mathbf{b}) = \braket{\mathbf{b}|\psi_{\theta}}$
        \State Evaluate $p(\mathbf{b} \big | \theta) = |\braket{\mathbf{b}|\psi_{\theta}}|^2$
        \State $E(\theta) \gets \left(\mathbf{b}|\psi_{\theta}\right)\braket{\hat{H}^{\text{loc}}(\mathbf{b};\theta)}_{\mathbf{b}}$ \Comment{Accumulate loss (can be parallelised!)}
    \EndFor
    \State Evaluate $\partial_{\theta^*} E(\theta)$ via Eq.~(\ref{eq:NQS_grad_loss_approx}).
    \State Evaluate $(\partial_{\theta^*}E(\theta))^*$ \Comment{Per our discussion above, $(\partial_{\theta^*}E(\theta))^* = \partial_{\theta} E(\theta)$}
    \State $\theta \gets \theta + \alpha \partial_{\theta}E(\theta)$ \Comment{Learning rate $\alpha$}
    \State \textbf{Output:} $\theta$.
\end{algorithmic}
\end{algorithm}

\newpage
\section{Hamiltonian Learning}

So far, our discussion on deep learning in quantum science has been focussed on inferring quantum states of interest. These usually come in the form of ground states of quantum many-body Hamiltonians. In the VQE, we saw how to use the variational principle on a parametric quantum circuit to find the smallest upper bound for the ground state of a given quantum many-body Hamiltonian. Whereas, in NQS, we found a way of representing a target wavefunction as a map from the computational basis to complex-valued amplitude. Tuning variational parameters and $\mathbb{C}\mathbb{R}$ calculus allowed us to also try and solve the ground-state problem of a given many-body Hamiltonian. 

We might therefore wonder whether deep learning might be able to help solve the \textit{inverse} problem. That is instead of finding quantum states of interest given fixed or tunable operators and (certified) knowledge of them, we can instead consider finding quantum operators of interest given (certified) knowledge of quantum states related to those operators.

Solving the inverse problem is often \textit{harder} than finding quantum states of interest. This is because there are lots of different ways that states and operators can be connected, for example
\begin{itemize}
    \item Stationary (eigenstates) states of operators
    \item Unitary time-evolution under the action of a Hamiltonian
    \item Steady states of an open quantum system 
    \item Transient states of open quantum systems
\end{itemize}

In this section, we will study the simplest dynamical case of this problem, which we refer to as \textit{Hamiltonian Learning}. Hamiltonian Learning is the task of inferring a \textit{unitary} quantum system's Hamiltonian given some dataset about that system. Even in this ``simplest'' case, this problem remains difficult for two reasons. 

First, there are an exponential number of different interactions a system can have. To see why this is the case, consider again an $N$-qubit wave-function repeated here for convenience:
\begin{equation}
    \ket{\psi(t)} = \sum_{b_1,\ldots,b_N}
    c_{b_1,\ldots,b_N }(t)\ket{b_1,\ldots,b_N}.
\end{equation}
This wavefunction already has an exponential number of coefficients\footnote{This was the motivating reason to define NQS in the first place!}. Any operator that acts on this wavefunction must therefore be \textit{at least as large}. Since quantum operators connect any element of our wave function's basis to any other, they are $d^2$-dimensional for a $d$-dimensional quantum state. In the case of qubits, this leaves us with an object with $\mathcal{O}(4^N)$ coefficients! 

\begin{figure}[htbp]
    \centering
    \begin{subfigure}[b]{0.45\textwidth}
         \includegraphics[width=\textwidth]{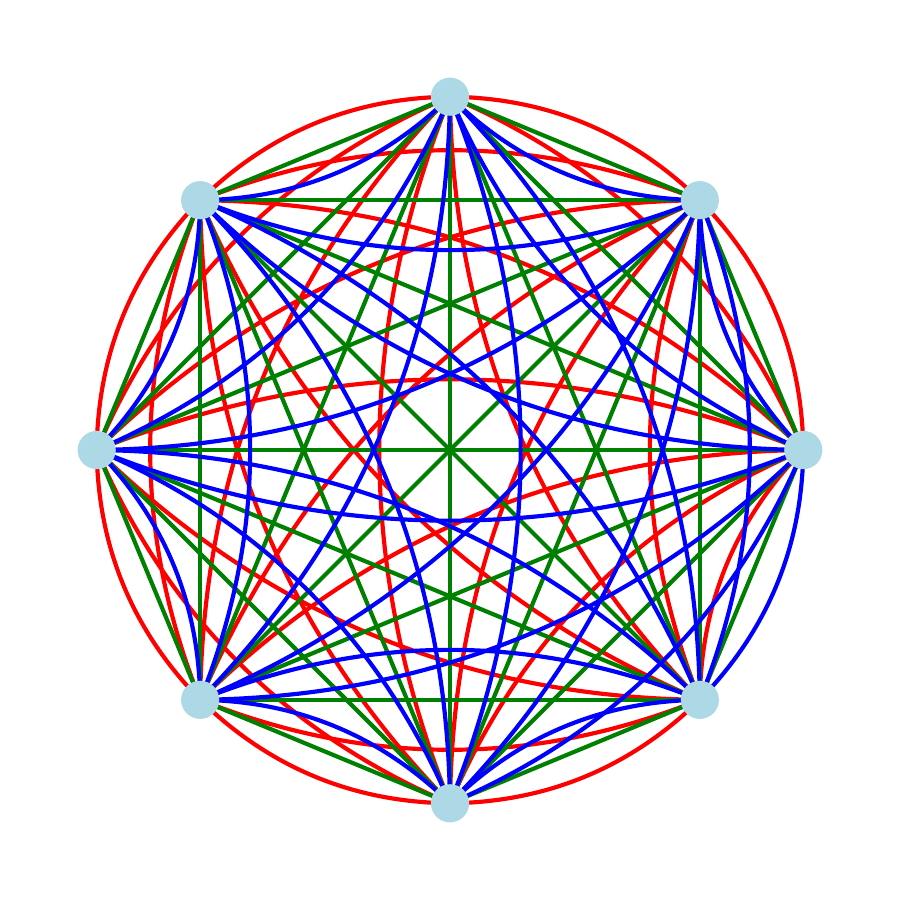}
        \caption{Fully Connected Pauli (Homogenous, dense, non-local)}
        \label{fig:MB_fully_connected}
    \end{subfigure}
    \hfill
    \begin{subfigure}[b]{0.45\textwidth}
        \includegraphics[width=\textwidth]{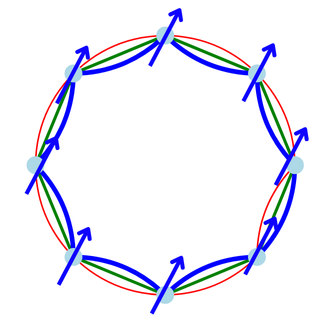} 
        \caption{Homogeneous, Local and Sparse \\(Heisenberg Model)}
        \label{fig:Heisenberg_model}
    \end{subfigure}

    \begin{subfigure}[b]{0.45\textwidth}
        \includegraphics[width=\textwidth]{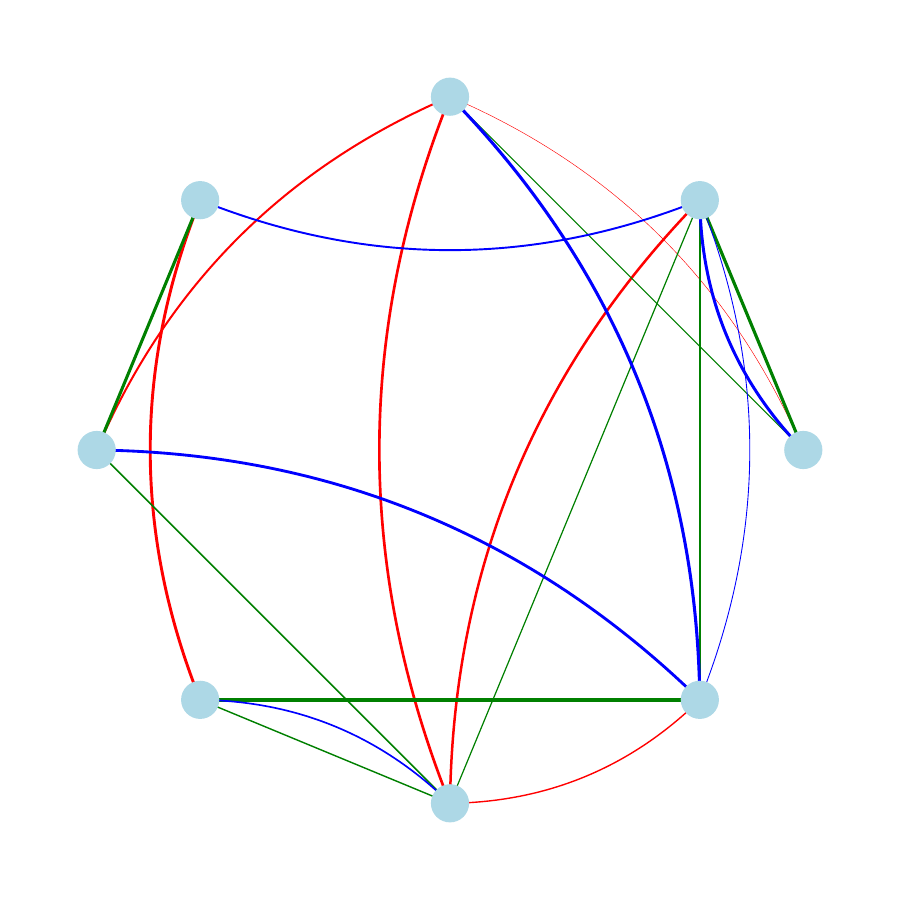}
        \caption{Sparse but Non-local
        and Inhomogeneous}
        \label{fig:MB_Sparse_Hetero}
    \end{subfigure}
    \hfill
    \begin{subfigure}[b]{0.45\textwidth}
        \includegraphics[width=\textwidth]{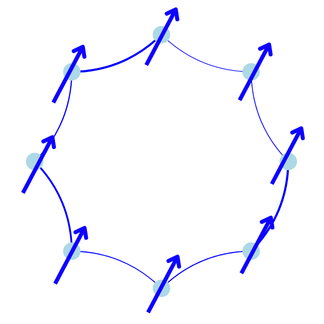} 
        \caption{Inhomogeneous, Local (and sparse) (Ising Model)}
        \label{fig:Ising_model}
    \end{subfigure}
    \caption{Visualisation of $N$ qubit quadratic Hamiltonians with periodic boundary conditions. Each node represents a local subspace $\mathcal{H}_2$, with edges showing a quadratic Pauli term in the Hamiltonian. Red edges correspond to $\hat{Y}$ operators, green to $\hat{Z}$ and blue to $\hat{X}$ respectively, and their thickness represents the weight of the coefficient of each coupling term. Homogeneous interacting terms have constant thickness, whilst inhomogeneous have varying thickness.}
    \label{fig:MB_Ising_Inhomogeneous}

\end{figure}

Second, there are sets of Hamiltonians that can be compatible with a given dataset acquired through measuring a quantum system. In this sense, our answer is not always unique. We can see this as a consequence of the exponential size of quantum operators. An generic $N$-qubit Hamiltonian has $\mathcal{O}(4^N)$ coefficients, so a dataset that could uniquely constrain it would need to be exponential in size. Even though deep learning is very data intensive, we want to avoid the situation where we demand exponential amounts of data to train our models. This is because it would take an exponential amount of time to acquire this data through sampling\footnote{Just try asking some experimental friends of yours how they would feel making $\mathcal{O}(4^N)$ measurements and you'll get a good idea of what you'd be up against. There is actually a deep connection here about replacing the concept of quantum oracles (from quantum algorithms) with sampling from their distributions. This is related to the idea of using sampling to do maximum likelihood estimation. It is in this sense that  exponential sized datasets are a hindrance to us when we only have access through sampling!}! Because of these difficulties, we play a physicists favourite game and make some assumptions that render our task considerably easier:
\begin{itemize}
    
    \item The Hamiltonian we are trying to learn is \textit{geometrically local}. This is a way of saying that the interactions between different parts of our quantum system decrease with distance, as might be seen in the Coulomb, or van der Waals interactions. See Figs.~\ref{fig:Heisenberg_model} and \ref{fig:Ising_model}.
    \item The Hamiltonian is \textit{sparse}. This means that of the potentially exponential number of coefficients to try to learn, we assume the vast majority are zero. Whilst this may sound like a bad assumption at face-value but is actually a pretty good one. Consider what the locality assumption above means in the case of nearest neighbours. This means in a 1D geometry, every qubit can connect to at most two others, meaning there are $2N^2$ interactions. Figs.~\ref{fig:MB_Sparse_Hetero},~\ref{fig:Heisenberg_model} and \ref{fig:Ising_model} have this property.
    \item The Hamiltonian is homogeneous between different sites. This means the interactions between different sites are the same as one another. Again, this assumption seems very sweeping, but again works well in practise. Each qubit in a circuit usually comes from the same physical process in terms of its hardware implementation. Usually this means a given qubit's environment is quite similar to another, hence our homogeneity assumption. Figs.~\ref{fig:MB_fully_connected} and \ref{fig:Heisenberg_model} have this property.
\end{itemize}

For the rest of this section, we will also simplify/specialize to the case of qubit quadratic Hamiltonians, as this is typically what we will encounter in practical scenarios for quantum hardware. We can also motivate these types of Hamiltonian by considering that the Ising model is contained within this definition; this model is NP-complete, so it is hard enough!

\begin{figure}
    \centering
    \begin{tikzpicture}
    \draw[thick] (2,-2.5) rectangle (5,-0.5);
    \node at (3.5,-1.5) {\( \hat{U}(t) = e^{-i \hat{H} t} \)};
    \draw[thick] (0.5,-2.15) -- (2,-2.15);
    \draw[thick] (0.5,-0.65) -- (2,-0.65);
    \draw[thick] (0.5, -1.0) -- (2, -1.0);
    \node at (1.0, -1.5) {$\vdots$};

    \draw[thick] (5,-1.5) -- (5.5,-1.5);
    
    \node[rotate = 90] at (-0.5,-1.5) {\(\ket{\psi(t = 0)}\)};
    
    \draw[thick] (5.5,-1.0) rectangle (6.5,-2.0);
    \draw[thick] (6.3,-1.7) arc[start angle=0,end angle=180,radius=0.3];
    \draw[->][thick] (6.0,-1.7) -- (6.4,-1.3);

    \node[right] at (7.0,-1.0) {\textbf{Promise}:};
    \node[right] at (7.0,-1.5) { \(\hat{U}(t) \hat{U}^{\dagger}(t) = \mathbb{I}\),};
    \node[right] at (7.0,-2.0) {\( \hat{H} = \sum_j c_j \hat{P}_j \),};
    \node[right] at (7.0,-2.5) {\(\hat{P_j}\) known, \( c_j \) unknown};
    
\end{tikzpicture}
    \caption{The simplest dynamical Hamiltonian Learning problem referred to as the white-box scenario. Here, we assume that we are given access to a box which contains the true Hamiltonian where dynamics of states being acted on by this Hamiltonian are 
    unitary. We also assume that we know \textit{which} Pauli strings have non-zero coefficients. This fixes the structure of the Hamiltonian, so that learning it from data means finding a good variational estimate for the coefficients $\theta = \{c_j\}$.}
    \label{fig:HL_whitebox}
\end{figure}
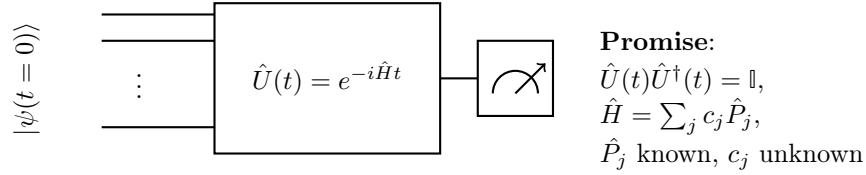

The techniques we will study in this case do however apply in more general settings like qu$d$its and continuous variable quantum systems. Since the Pauli basis is complete, the $N$-qubit Hamiltonian we seek to learn may always be written as,
\begin{equation}
    \hat{H} = \sum_{j = 0}^{4^N} c_j \hat{P}_j,
\end{equation}
where $hat{P_j}$ is one of $4^N$ possible \textit{Pauli strings} (see Box. \hyperlink{box:PauliGroup}{ 5} on the Pauli group). In general, there are $4^N$ of these coefficients in this Hamiltonian. This is because for each of $N$ qubits, there are four possible Pauli operators, $\{\mathbb{I}, \sigma_x, \sigma_y, \sigma_z\}$. With $N$ repeats of this choice, we get $4^N$ possible interactions in the Pauli basis. Let's now see how we can use our three assumptions to lower this number of free parameters as much as possible. First, our assumption of \textit{sparsity} means that a given qubit only connects to a few others (rather than \textit{every} other qubit). Second, out assumption of \textit{geometric locality} means that the a given qubit only connects to a few others which are space-like close to it. Third, our \textit{homogeneity} assumption means that the interactions between connected qubits are the same. Together, these three assumptions can completely fix the structure of the Hamiltonian, giving rise to the so-called white-box scenario of Hamiltonian Learning shown in Fig.~\ref{fig:HL_whitebox}.

Now that we have simplified the structure of our problem as much as possible, let's consider what kind of data we might receive to solve a Hamiltonian Learning problem. In the white-box scenario detailed in Fig.~\ref{fig:HL_whitebox}, we are able to prepare some initial states $\ket{\psi(t = 0)}$ and measure them after they have evolved. That is, we are able to perform measurements on
\begin{equation}
    \ket{\psi(t)} = e^{-i \hat{H} t} \ket{\psi(t = 0)},
\end{equation}
where $\hat{H}$ is the Hamiltonian to be learned. Intuitively, we can see this as using \textit{trajectories} of a few states (of interest) to try and learn what the Hamiltonian is. There are lots of other ways that we can create a dataset. We could use special time-independent states like eigenstates $\ket{\phi}$ satisfying $\hat{H} \ket{\phi} = \lambda \ket{\phi}$, or thermal states 
\begin{equation}
    \hat\rho = \frac{e^{-\beta \hat H}}{\text{Tr}[e^{-\beta\hat{H}}]}.
\end{equation}
However, the reason we will opt for trajectories is because we need not make extra assumptions about which states we can prepare. Instead, we can just evolve states which are ``easy'' to prepare. In this sense, we don't need any fine-grained control of the box; to generate data from measurements, we simply need to choose the time at which we measure\footnote{Again, ask your experimental friends which they will find easier: to evolve a few easy-to-prepare states up to some desired time, or to prepare eigenstates/thermal states reliably and control their evolution!}! Concretely, let us consider Pauli basis measurements on these time evolved states. This means our dataset looks like a collection of binary outcomes in the basis of each Pauli measurement. That is our dataset looks like
\begin{equation}
    \mathcal{D} = \{(\mathbf{b}, \hat{P}, t)\},
\end{equation}
where $\mathbf{b}$ is the binary bitstring outcome of length $N$, for a Pauli measurement $\hat{P}_j$ at time $t$. These outcomes will happen with probability,
\begin{equation}
    p\left(\mathbf{b}\bigg|\ket{\psi(t)}, \hat{P}_j\right) = \braket{\psi(t)|\hat{P}_j|\psi(t)} = \braket{\psi(0)|e^{i \hat{H} t} \hat{P}_j e^{-i \hat{H} t}|\psi(0)},
\end{equation}
where $\hat{H}$ is the Hamiltonian we wish to learn. We can summarise this dataset generation method by writing down a sub-routine for generating data of states' trajectories, see Alg.~\ref{alg:gen_HL_data}

\begin{algorithm}
\caption{Generating Hamiltonian Learning Trajectory Data}\label{alg:gen_HL_data}
    \begin{algorithmic}[1]
        \State \textbf{Input} $\{\ket{\psi_j(t = 0)}\}_{j = 0}^J$, $\{\hat{P}_{jk}\}_{k = 0}^K$, $\mathcal{D} = \empty $\Comment{$J$ initial states, $K$ Pauli measurements}
        \For{each j in $0, \ldots, J$:}
            \State Evolve $\ket{\psi_j(0)}$ to $\ket{\psi_j(t)} = e^{-i H t} \ket{\psi(0)}$ \Comment{Through white-box}
        \EndFor
        \For{Each k in $0, \ldots, K$:}
            \State Measure $\hat{P}_{jk}$ on each $\ket{\psi_j(t)}$ giving outcome $\mathbf{b}$ \Comment{Repeats as necessary, can be stored as separate tuples}
            \State $\mathcal{D} \gets (\mathbf{b}, k, t, j)$  \Comment{$j$ index which trajectory, and $k$ which measurement was applied}          
        \EndFor
        
        \State \textbf{Output} $\mathcal{D}$ \Comment{With $M$ repeats, $J$ initial states, and $K$ Pauli measurements, there are $MJK$ pieces of data.}
	\end{algorithmic} 
\end{algorithm}

Given a dataset generated by Alg.~\ref{alg:gen_HL_data}, we may now use the Maximum Likelihood Estimation technique (see \hyperlink{box:MLE}{1}) from Chapter~\ref{CH:UNSUPERVISED} to find a Hamiltonian which maximises the likelihood of observing the outcomes in our dataset, $\mathcal{D}$. In order to use this technique, we have to come up with some variational parameters that can be tuned with gradient-based methods per Chapter~\ref{CH:UNSUPERVISED}, and a loss function. Thanks to our assumptions in the white-box scenario which fixed the structure of the Hamiltonian to be learned, we can directly make our variational parameters the coefficients $c_j$. In other words, we can define an ansatz Hamiltonian $\hat{H}_A$ with the same structure as the true Hamiltonian,
\begin{equation}
    \hat{H}_A = \sum_j c_j \hat{P}_j
\end{equation}
where the variational parameters $\theta = \{c_j\}$. By simulating the evolution of initial states $\ket{\psi_j(t = 0)}$ under the action of our ansatz, we can create \textit{estimators} for the true trajectories,
\begin{equation}
    \ket{\tilde{\psi}_j(t)} = e^{-i \hat{H}_A(\theta) t} \ket{\psi_j(0)},
    \label{eq:estimator_state_HL}
\end{equation}
where the tilde on $ \ket{\tilde{\psi}_j(t)}$ is there as a helpful reminder that this is our estimator\footnote{i.e. to avoid confusion with true trajectories $\ket{\psi_j(t)}$.}. For a loss function, we can use negative log-likelihood (see Box.\href{box:MLE}{1}),
\begin{equation}
    L(\ket{\tilde{\psi}(t)}; \mathcal{D})= -\sum_{\mathbf{b} \in \mathcal{D'}} \log |\braket{\mathbf{b}|{\tilde{\psi}(t)}}|^2,
    \label{eq:log_likelihood_HL}
\end{equation}
where $\mathcal{D}'$ is a \textit{batch} from our training dataset, $\mathcal{D} \subset \mathcal{D}$. Intuitively this loss function tells us how likely it is that we would observe outcomes in our dataset given our estimator. That is, the overlap $|\braket{\mathbf{b}|{\tilde{\psi}(t)}}|^2$ is maximal for a best estimate, and the negative log-likelihood is hence minimised. We now have all the components to define the \textit{vanilla} Hamiltonian Learning algorithm for the white-box scenario, see Alg.~\ref{alg:vanilla_HL}.

\begin{algorithm}
\caption{Vanilla Hamiltonian Learning (White-Box) Train Step}\label{alg:vanilla_HL}
    \begin{algorithmic}[1]
        \State \textbf{Input} $\mathcal{D}$ and Ansatz Hamiltonian $H_A = \sum_j c_j \hat{P}_j$ \Comment{Variational parameters $\theta = \{c_j\}$}
        \For{Each initial state $\ket{\psi_j(0)}$ in $\mathcal{D}:$}
            \State Evolve $\ket{\tilde{\psi}_j(t)} = e^{-i \hat{H} t} \ket{\psi_j(0)}$
            \State Compute Loss $ L(\ket{\tilde{\psi}_j(t)}; \mathcal{D})$ from Eq.~(\ref{eq:log_likelihood_HL})
            \State $\theta \gets \theta + \nabla_{\theta} L(\ket{\tilde{\psi}_j(t)}; \mathcal{D})$ \Comment{We can also compute average loss and do one update. }
        \EndFor
    \end{algorithmic} 
\end{algorithm} 

The vanilla version (Alg.~\ref{alg:vanilla_HL}) has several equivalent variants, for example we could consider optimising with respect to one initial state at a time, or use more sophisticated optimisers like ADAM or ADAM-W to do the gradient descent step. We can also choose hyperparameters like batch size and averaging loss \`a la Mean Average Error from Chapter~\ref{CH:FUNDAMENTALS}. Moreover, there is the important question for how exactly we evolve initial states under our ansatz Hamiltonian. That is, how do we find $\ket{\tilde{\psi}_j(t)}$ from Eq.~(\ref{eq:estimator_state_HL})?

As this is a classical-for-quantum course,
we will consider simulating this evolution classically, although this comes with its own complexity-related problems, see \hyperlink{box:curse_of_dim}{Box. 5}. This is the premise of quantum advantage in quantum simulation. There are of course, other options like compiling the Unitary $\hat{U}(t) = e^{-i\hat{H}t}$ onto a quantum circuit, or even using a parametric quantum circuit \textit{as the ansatz}, like we saw with VQEs.

\begin{figure}[h!]
    \centering
    \begin{mybox}[\hypertarget{box:curse_of_dim}{Box 5.3: The Curse of Dimensionality}]
     The curse of dimensionality refers to the face that the size of quantum many-body Hilbert space scales exponentially in the number of bodies. For an ensemble of $N$-qubits, each with $2$-dimensional state space, $\mathcal{H}_2$, their collective wavefunction $\ket{\psi} \in \mathcal{H}_2^{\otimes N}$ has $\mathcal{O}(2^N)$ coefficients. This means representing general (unrestricted) quantum many-body states, and simulating their dynamics is NP-hard. Much attention and effort is focussed on finding special, restricted cases where representing states and simulating their dynamics becomes easier again. \\

     From a deep learning perspective, the curse of dimensionality means the amount of data we would need to generalise grows exponentially with the number of features (dimensions). There are many phenomena in high-dimensional spaces which seldom happen in low-dimensional settings. This makes our job \textit{harder} because we can't use toy-models, or simple settings to observe (and later mitigate for) these effects. Hence the name \textit{curse}...
    \end{mybox}
\end{figure}

Classical simulation comes with an added bonus that we should be able to directly establish differentiability, and therefore computational graphs for the time evolution of quantum states. To see how this works, let's write down the time-dependent Schr\"odinger equation of our estimator $\ket{\tilde{\psi}(t)}$ in integral form,
\begin{equation}
 \ket{\tilde{\psi}(t + \Delta t);\theta} = 
 \int_t^{t + \Delta t} \frac{d }{dt'} \ket{\tilde{\psi}(t')}dt' = 
 -i \int_t^{t + \Delta t} \hat{H}_{A}(\theta) \ket{\tilde{\psi}(t')}dt',
\end{equation}
where $\hat{H}_A$ is our ansatz Hamiltonian, and $\theta = \{c_j\}$ are the variational parameters in our computational graph. Note the explicit $\theta$ dependence shown here - we want to understand how to back-propagate through time evolution with respect to $\theta$. Lucky for us, lots of popular numerical integration techniques like Euler's method are differentiable with respect to $\theta$. As we are doing a deep learning course, we won't go into detail about numerical integration techniques here. In brief, Euler's method can approximate the time evolution of a quantum state \( \ket{\psi(t)} \) governed the time-dependent Schr\"odinger equation.
\begin{equation}
\frac{d}{dt} \ket{\psi(t)} = -i \hat{H}_A(\theta) \ket{\psi(t)},
\end{equation}
where \( \hat{H}_A(\theta) \) is the Hamiltonian. Given an initial state \( \ket{\psi(t_0)} \), Euler’s forward step is:
\begin{equation}
\ket{\psi(t_{n+1})} = \ket{\psi(t_n)} - \epsilon i \hat{H}_A(\theta) \ket{\psi(t_n)},
\end{equation}
where \( \epsilon \) is the time step. To ensure normalization, we renormalize \( \ket{\psi(t_{n+1})} \) after each step by setting 
\begin{equation}
\ket{\psi(t_{n+1})} \to \frac{\ket{\psi(t_{n+1})}}{\|\ket{\psi(t_{n+1})}\|},
\end{equation}
incurring a simulation error of $\mathcal{O}(\epsilon^2)$ on each step. Clearly this forward step is differentiable in $\theta$ provided $H_A(\theta)$ is differentiable. Since numerical methods execute this step recursively, the chain rule therefore allows us to differentiate numerical integration methods like Euler's method or the Runge-Kutta method with respect to $\theta$. 

Another option for differentiable quantum simulation would be to directly approximate the exponential $e^{-i H t}$ via its truncated Taylor expansion,
\begin{equation}
    e^{-i \hat{H} t} = \sum_{n = 0}^N \big(-i \hat{H}_A(\theta) t\big)^n.
\end{equation}
This allows us to approximate
\begin{equation}
    \ket{\tilde{\psi}(t)} \approx \sum_{n = 0}^N \big(-i \hat{H}_A(\theta) t\big)^n \ket{\tilde{\psi}(0)},
    \label{eq:matrix_exp_approx_time}
\end{equation}
at the cost of lots of matrix multiplication to enumerate the powers\footnote{this can be efficient when the powers of $\hat{H}_A(\theta)$ have cyclic structure. For example, Pauli matrices have $\sigma^2 = \mathbb{I}$, so $\sigma^{2n + 1} = \sigma$ and $\sigma^{2n} = \mathbb{I}$. Try this for yourself with pen and paper if you have not before, and you will have proven Eq.~\ref{eq:euler_su(2)}.} of $\hat{H}_A(\theta)^n$. Again, the chain rule ensures that Eq.~(\ref{eq:matrix_exp_approx_time}) is differentiable in $\theta$. 

At this point, we should now be satisfied that vanilla Hamiltonian Learning (Alg.~\ref{alg:vanilla_HL}) can be done with variational methods. However, so far the variational parameters we have chosen ($\theta = \{c_j\}$) are directly involved in the Hamiltonian's Pauli expansion. In that sense, vanilla Hamiltonian Learning is really a type of machine learning; we haven't leveraged neural networks, hidden layers, and much of the other machinery we saw in Chapter~\ref{CH:FUNDAMENTALS}. How (and why!) then, would we use neural networks to solve this problem?

To answer this question, let's consider again where our training data comes from. We are using the trajectories of states evolving under the true Hamiltonian from the white-box in Fig.~\ref{fig:HL_whitebox}. Our learned representation of the true Hamiltonian is created by evolving estimators for the true trajectories, which are tuned with gradient-based methods. In this sense, when our learned representation is correct, our estimators solve the time-dependent Schr\"odinger equation,
\begin{equation}
\frac{d}{dt} \ket{\tilde{\psi}(t)} = -i \hat{H} \ket{\tilde{\psi}(t)},
\end{equation}
where $H$ is the true Hamiltonian from the white-box in Fig.~\ref{fig:HL_whitebox}. We might therefore attempt to solve the TDSE for estimators' trajectories directly with deep learning methods. When our solution is correct, it means we must have a learned representation of the system's Hamiltonian, provided the dynamics we solve are unitary. 

\begin{figure}
\begin{subfigure}[t]{0.45 \textwidth}
  \centering
   \scalebox{0.9}{\input{figures/fig_Ch5_HL_vanilla.tikz}}
  \caption{}
  \label{fig:architectures_a}
\end{subfigure}
~
\begin{subfigure}[t]{0.45 \textwidth}
  \centering
   \scalebox{0.9}{\input{figures/fig_Ch5_HL_vanilla_ode.tikz}}
  \caption{}
  \label{fig:architectures_b}
\end{subfigure}

\vspace{10mm}

\begin{subfigure}{\textwidth}
  \centering
  \scalebox{0.8}{\input{figures/fig_Ch5_HL_state_vector_node.tikz}}
  \caption{}
  \label{fig:architectures_c}
\end{subfigure}
\caption{Architectures of the models used to simulate the time evolution of the parametrised Hamiltonian $\hat{H}(\theta)$. (a) The numerically exact time evolution under the Hamiltonian. (b) The approximate time evolution through integrating the Schr\"odinger equation, hereby referred to as the vanilla model. (c) The time evolution through integrating the Schr\"odinger equation with an added corrective term to the right-hand-side.}
\label{fig:HL_NODE_architectures}
\end{figure}
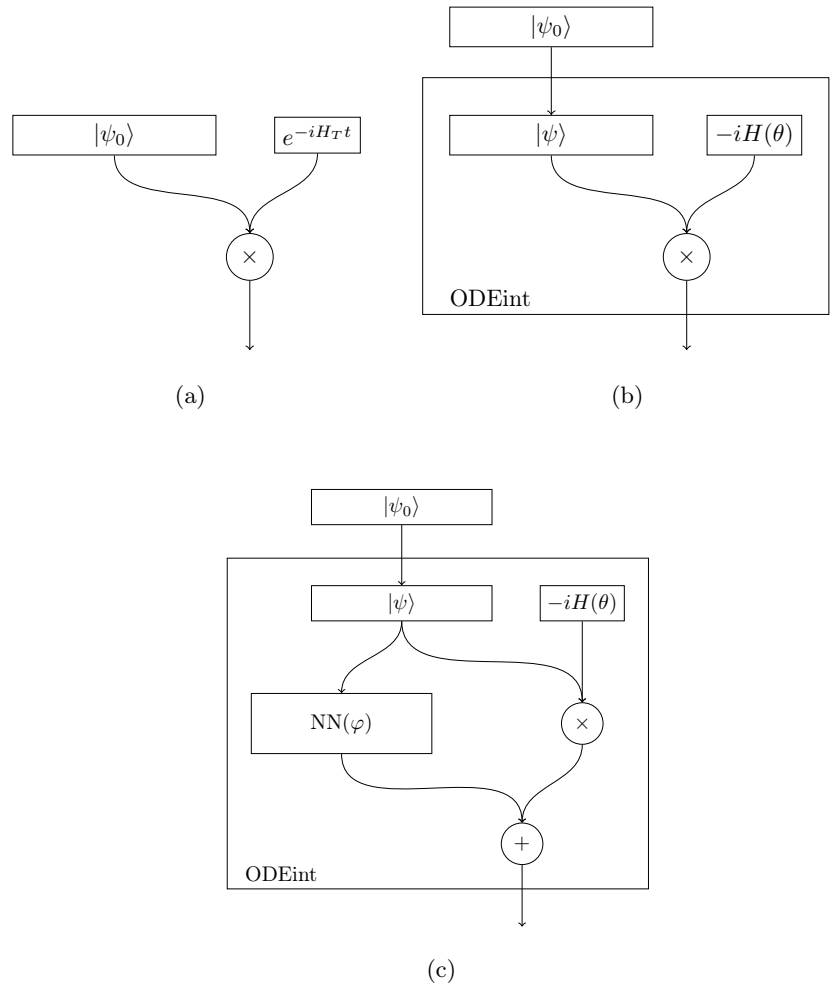

To that end, consider a representation of $H_T$ decomposed as the sum of an Ansatz Hamiltonian, $\hat{H}_A(\theta)$, and a NODE with a Neural Network (NN) architecture acting as a map $\text{NN}:\mathcal{H} \xrightarrow{\theta} \mathcal{H}$,
\begin{equation}
    \hat{H}_T = \hat{H}_A(\theta) + \text{NN}(\varphi),
\end{equation}
with independent, tunable parameters $\theta, \;\varphi$ for the Ansatz and NN respectively. Notice here that the NN represents an \textit{operator}, rather than a state as we saw in NQS. Under this decomposition, the Schrödinger equation reads,
\begin{equation}
    i \ket{\dot{\psi}(t)} = \hat{H}_{A}(\theta) \ket{\psi(t)} + \text{NN}(\ket{\psi(t)};\varphi),
    \label{eq:schrodinger_NODE}
\end{equation}
which can be expressed as,
\begin{equation}
\begin{split}
    \ket{\psi(t + \Delta t)} &= 
    \int_t^{t + \Delta t} \frac{d }{dt'} \ket{\psi(t')}dt' \\
    &=
    -i \int_{t}^{t + \Delta t}  \hat{H}_{A}(\theta) \ket{\psi(t')} dt' -i \int_{t}^{t + \Delta t} \text{NN}(\ket{\psi(t')};\varphi) \; dt'.
\end{split}
\label{eq:HL_NODE_integration}
\end{equation}
for numerical integration as we saw before. In integral form, we may represent the Schrödinger equation as a computational graph with a NODE component, as shown in Fig.~\ref{fig:HL_NODE_architectures}. Recall that this was because numerical integration techniques can be made differentiable with respect to variational parameters $\theta$ and $\phi$. This architecture sits in contrast with the Ansatz-based method alone.

We can understand Eq.~(\ref{eq:schrodinger_NODE}) as a type of \textit{Neural Differential Equation} (NDE). NDEs exploit the universal approximation theorem to approximate the \textit{rate of change} of a function, rather than the function directly. They are of great importance as a tool in deep learning on dynamical systems, and a brief summary of them is provided in Box~\hyperlink{box:NDE}{5.4}. As before, we can train our combined model (Ansatz and NN) with MLE based on samples from a dataset $\mathcal{D}$ from Alg.~\ref{alg:gen_HL_data}. A summary of Hamiltonian Learning is provided in Alg.~\ref{alg:HL_NODEs}, allowing you to build your very own HL solvers with neural networks.

\begin{figure}[h!]
    \centering
    \begin{mybox}[\hypertarget{box:NDE}{Box 5.4: Neural Differential Equations}]
     Neural Differential Equations (NDEs) are a way to use neural networks to \textit{specify} a given differential equation. Consider a first order, linear ODE
     \begin{equation}
         \frac{dx}{dt} = f_{\theta}(x,t),
     \end{equation}
     where $f_{\theta}(x,t)$ is a neural network function $f: \mathbb{R} \xrightarrow[]{\theta} \mathbb{R}$ with variational parameters $\theta$. Intuitively, we can see this as a way to use neural networks to predict the rate of change of a function. Note that this means the input and output shape of the neural network to be the same. This is because in higher dimensions (i.e with more features), we need to predict the rate of change of \textit{every} component should we ever want to \textit{solve} the differential equation! Computationally speaking, we usually solve NDEs with numerical integration techniques like the Euler Method or Runge-Kutta Integration. 

     To see why NDEs are a powerful tool, consider how they elegantly combine two key tools. On the one hand, we have differential equations; a battle-tested method in Physics which has revolutionised our understanding of the world. On the other, the Universal Approximation Theorem (UAT) from Chapter~\ref{CH:FUNDAMENTALS} says that given enough data and depth, a neural network can in principle approximate \textit{any} function. Since the rate of change (component wise) of a function has just as many elements as inputs, we can see how the UAT can be extended to differential equations. For a great proof, and pedagogical introduction to NDEs, see the thesis of Patrick Kidger \textit{On Neural Differential Equations} \cite{kidger2022neural}.
    \end{mybox}
\end{figure}

Finally, we turn our attention to \textit{why} we might want to use deep learning and neural networks to build HL solvers. The first reason is simple, by having a two-component architecture, room to get the Ansatz wrong and still have a working representation. However, our joint representation allows us to predict the properties of observables, as well as the time evolution of states. Often, this is all you want anyway out of a learned representation as we saw in NQS. 

At a deeper level, using a combined representation gives us direct control over the sensitivity of the log-likelihood loss function from Eq.~(\ref{eq:log_likelihood_HL}) to small changes in our variational parameters $\theta$. We saw in Chapter~\ref{CH:FUNDAMENTALS} that this is advantageous - we always want a loss function to be responsive to changes in our variational parameters so that we can perform meaningful updates. It can be shown that with an ansatz alone, this sensitivity is difficult to control, leaving loss landscapes which are much harder to navigate with gradient descent!

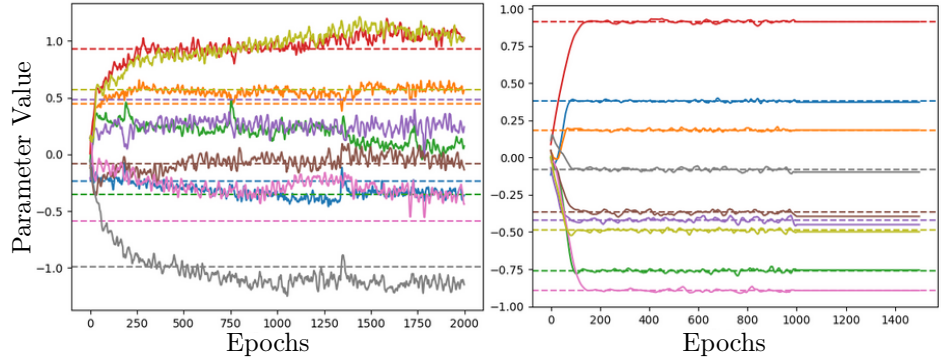
\begin{figure}[h]
    \centering
    \begin{tikzpicture}
        \node[anchor=south west,inner sep=0] (image) at (0,0) {\includegraphics[width=\linewidth]{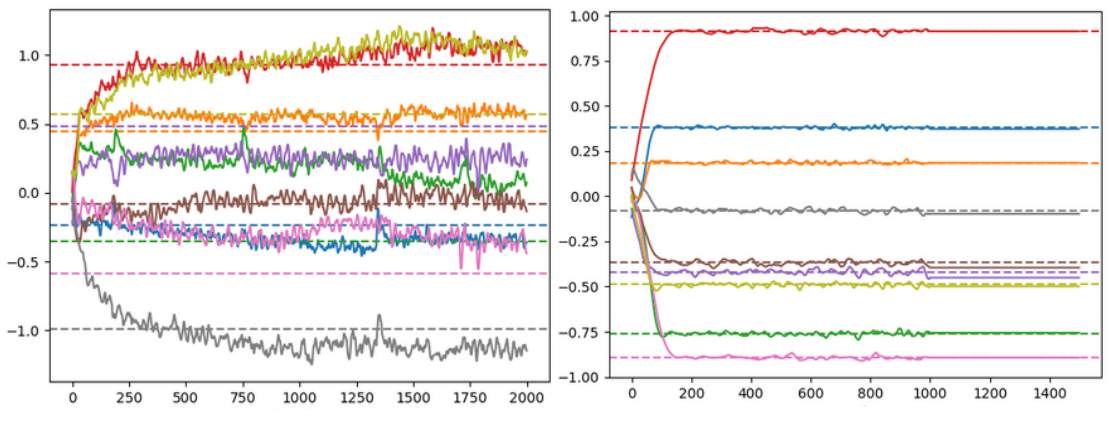}};
        \begin{scope}[x={(image.south east)},y={(image.north west)}]

             \node[rotate = 90] (ParameterValue) at (-0.01, 0.535) {Parameter Value};
             \node (Epochs) at (0.26, 0.0) {Epochs};\node (Epochs) at (0.76, 0.0) {Epochs};
        \end{scope}
    \end{tikzpicture}
    \caption{Comparison of convergence to ground truth parameters for a six-qubit Heisenberg XYZ model, with a vanilla architecture (see Fig.\ref{fig:architectures_b}) on the left, and a NODE architecture (see Fig.\ref{fig:architectures_c}) on the right. The parameters to fit were homogeneous couplings $\{J_x, J_y,J_z\}$, and in-homogeneous transverse fields $\{h^{(z)}_k\}_{k = 1}^{6}$ for a total of nine parameters. The ground truths for each are shown as dashed lines with their respective solid-line being the estimate for the true parameters. We see a strong difference in the convergence of the NODE model when compared against a vanilla one; parameters on the vanilla figure have not converged, and the updates on each epoch are still large, despite using exactly the same optimizers. Notice also that there are different ground truths here. The ground truth parameters are somewhat arbitrary, and to demonstrate robustness rigorously, one needs to test many instances of the problem and count the average convergence rate, as was done in \cite{heightman2024solving}. As such these are two plots selected at random from a test set, with arbitrary values for ground truth parameters samples uniformly on the interval $[-1,1]$. Training was done with Algorithms \ref{alg:HL_NODEs} and \ref{alg:vanilla_HL} respectively for NODE (left) and vanilla (right), with the dataset generated via Algorithm \ref{alg:gen_HL_data}.}
    \label{fig:HL_node_vs_vanilla}
\end{figure}

We can see this being the case via the following example. Let's consider a six-qubit Heisenberg XYZ-model,
\begin{equation}
    \hat{H} = \sum_{j=1}^6 \left(J^{x} \hat\sigma_j^{x}\hat\sigma_{j + 1}^x + J^{y} \hat\sigma_j^{y}\hat\sigma_{j + 1}^y + J^{z} \hat\sigma_j^{z}\hat\sigma_{j + 1}^z\right) + \sum_{j=1}^6 h^z_j \hat\sigma_{j}^z 
\end{equation}
which has three interaction terms, $\{J^{x}, J^{y}, J^{z}\}$, and six local field terms $\{h^z_j\}_{j = 1}^6$ for a total of nine learnable parameters. Fig.~\ref{fig:HL_node_vs_vanilla} shows how the parameter values change through a 2000-epoch training run using the vanilla architecture (Fig.\ref{fig:architectures_b}) and the NODE architecture (Fig.\ref{fig:architectures_c}). We can see that there is a clear difference in terms of convergence thanks to the expressibility of the MLP. Furthermore, by using deep
neural networks to explore the space of many-body Hamiltonians, we can relax our assumptions of sparsity, geometric locality and homogeneity (see Fig.5.12). This is because deep neural networks can (in their middle layers) correlate any part of the input wavefunction with any other. Considering the fact that they have non-linear activation functions, this also lets us explore strongly correlated systems. For readers interested in a more detailed study of these ideas, check out \cite{heightman2024solving}.

\begin{algorithm}[h!]
\caption{Hamiltonian Learning via NDEs (White-Box) Train Step}\label{alg:HL_NODEs}
    \begin{algorithmic}[1]
        \State \textbf{Input} $\mathcal{D}$, $\hat{H}_A = \sum_j c_j \hat{P}_j$, $NN_{\varphi}: \mathcal{H} \rightarrow \mathcal{H}$ \Comment{Variational parameters $\theta = \{c_j, \varphi\}$, $\mathcal{D}$ is batched data from Alg.~\ref{alg:gen_HL_data}}
        \For{Each initial state $\ket{\psi_j(0)}$ in $\mathcal{D}:$} 
            \State Numerically integrate 
            $\ket{\psi(0)}$ to $\ket{\tilde{\psi}_{\theta}(t)}$
            via Eq.~(\ref{eq:HL_NODE_integration}) \Comment{See Fig.~\ref{fig:HL_NODE_architectures}. }
            \State Compute Loss $ L(\ket{\tilde{\psi}_j(t)}; \mathcal{D})$ from Eq.~(\ref{eq:log_likelihood_HL})
            \State $\theta \gets \theta + \nabla_{\theta} L(\ket{\tilde{\psi}_j(t)}; \mathcal{D})$ \Comment{We can also compute average loss and do one update. }
        \EndFor
    \end{algorithmic} 
\end{algorithm} 

\newpage

\section{Photonic Quantum Tomography with Normalising Flows}

Having now seen a way to incorporate neural networks with quantum operators and dynamics on spin systems, we might wonder whether we can also leverage deep learning in bosonic settings. In this section, we will consider the simplest case of doing so, namely the task of inferring a quantum state from data generated by that state, referred to as quantum state tomography (QST). QST comes in many forms, but the common elements of all QST problems is a large collection of data, and the aim of trying to infer which state could have generated it. 

In the case of quantum states of a single boson, QST searches for a state $\ket{\psi} \in \mathcal{H}_{\mathcal{F}}$, living in a \textit{Fock space} $\mathcal{F} = \{\ket{n} \in \mathbb{Z}_+\}$. These were introduced in Sec.~\ref{sec:CVQM}, where we saw how these states have a phase space Q-function, 
\begin{equation}
    Q(\alpha,\alpha^*) = \frac{\braket{\alpha|\psi}\braket{\psi|\alpha}}{\braket{\psi|\psi}},
\end{equation}
where $\alpha,\alpha^* \in \mathbb{C}$ are the canonical conjugate phase space variables coming from eigenstates of $\hat{a}$ and $\hat{a}^{\dagger}$.

To infer a true state, $\hat\rho_T = \ket{\psi}\bra{\psi}$,
we will use a dataset $\mathcal{D}$ be constructed from 
measurements, for example from quadrature measurements arising from heterodyne detection. Since $Q(\alpha,\alpha^*) \geq 0$, 
we can treat it numerically as a probability distribution conditioned on \textit{both} $\alpha,\alpha^*$. This is simply because
\begin{equation}
    \frac{1}{\pi}\int_{\mathbb{C}} Q(\alpha,\alpha^*) d\alpha d\alpha^* = 1.
\end{equation}
The task of QST can be solved with a normalising flow model. This is because there is a one-to-one correspondence between Q-functions and states (these are phase space representations of them afterall). Therefore to infer the quantum state which maximises the likelihood of observing our dataset, it is sufficient to find its Q-function. Given a prior distribution, we can treat this as a normalising flow by performing density estimation of the optimal Q-function. Explicitly, normalising flows can perform non-linear transformations of phase space axes. Therefore with some (arbitrary, but sometimes informed) prior distribution, normalising flows can act generatively to go and search over the space of phase-space functions. For example, we could use the vacuum state $\rho = \ket{0}\bra{0}$ as a prior, and connect this distribution to any other with normalising flows thanks to the universal approximation theorem. In this sense, a sufficiently large neural network model can be \textit{complete} on the space of quantum distributions. 

To see how this works, let's first consider an analytical example. We will try to learn a coherent state, $\rho_T =  \ket{\beta} \bra{\beta}$ for some $\beta \in \mathbb{C}$, using the vacuum as a prior, and a coordinate transformation that is differentiable (a NF). To that end, let $\rho_b = \ket{0}\bra{0}$ be the base distribution, and $\tilde{\rho}(\theta)$ be the normalising flow \textit{estimator} of the true state $\rho_T$. The Q-function for $\rho_T$ reads
\begin{equation}
        Q_T(\alpha,\alpha^*) = \frac{1}{\pi}\braket{\alpha|\hat\rho|\alpha} = \frac{1}{\pi}\braket{\alpha|\beta}\braket{\beta|\alpha} = \frac{1}{\pi}e^{|\alpha - \beta|^2},
\end{equation}
which we will treat as the target (true) distribution for a given target value $\beta$ in the generative direction of our normalising flow. For the base distribution, let's use the vacuum state $\rho_b = \ket{0}\bra{0}$, whose Q-function reads
\begin{equation}
    Q_b(\alpha,\alpha^*) = \frac{1}{\pi}e^{- |\alpha|^2}.
\end{equation}

A pass in the generative direction makes the estimator's Q-function, $Q_{\theta}(\alpha',\alpha'^*)$. It has new phase space coordinates that are a result of the flow $\mathcal{C}^1(\mathbb{C}^2) \rightarrow \mathcal{C}^1(\mathbb{C}^2)$ of differentiable functions of two complex numbers, ${C}^1(\mathbb{C}^2)$.  This means if $\vec{\alpha} = (\alpha, \alpha^*)^T$, then 
\begin{equation}
    \vec{\alpha}' = f_{\theta}(\vec{\alpha}).
\end{equation}

\noindent For our analytical example, we will use a single NF layer which performs an affine transformation of the phase space axes,
\begin{equation}
    \vec{\alpha}' = f_{\vec{\theta}}(\vec{\alpha}) = \vec{\alpha} + \vec{\theta},
\end{equation}
where $\vec{\theta} = (\theta_1,\theta_2)^T \in \mathbb{C}^2$ are the variational parameters. For computational ease, we will split these into their real and imaginary parts, giving a total of four variational parameters for the model to tune. Notice that we did not act on the base distribution, but rather its coordinates. You can therefore think of this affine transformation as the change in coordinates that \textit{induces} the displacement operator in quantum optics.

The normalising direction, $g = f^{-1}$, then yields the base distribution, $\hat\rho_b = \ket{0}\bra{0}$. Hence, in terms of phase space axes, we have
\begin{equation}
    \vec{\alpha} = g_{\theta}(f_{\theta}(\vec{\alpha})),
\end{equation}
where
\begin{equation}
    g^{-1}(\vec{\alpha}') = \vec{\alpha}' - \vec{\theta}
\end{equation}
by inspection, per the invertible criterion from CH3. We now have all the ingredients to write our flow equation, which describes how the probabilities change under the NF. Namely to probability density transforms as
\begin{equation}
    p(\alpha'|Q_b(\alpha,\alpha^*)) = p(\alpha|Q_{\theta}( \alpha',\alpha'^*)) \left| \text{det}
    \left( 
    \frac{\partial \alpha }{\partial \alpha'} 
    \right) \right|^{-1}. 
\end{equation}

Next, we need an update rule. For data coming from quadrature measurements, we can compute the negative log-likelihood of seeing that data given our estimator distribution $Q_{\theta}(\alpha',\alpha'^*)$. Consider the dataset 
\begin{equation}
    D = \{(o_1, \hat{R}_1), \ldots (o_N, \hat{R}_N)\},
\end{equation} 
split into batches of size $B < N$, where $o_j$ is the $j^{\text{th}}$ outcome for the POVM $\hat{R}$ in the region, $R \subset \mathbb{C}$, of phase space,
\begin{equation}
    \hat{R} = \frac{1}{\pi}\int_R d\alpha d\alpha^* \ket{\alpha}\bra{\alpha}.
\end{equation}

Then the joint probability of observing a batch of data given the Q-function estimator $Q_{\theta}(\alpha',\alpha'^*)$ is given by
\begin{equation}
    p(B|Q_{\theta}(\alpha',\alpha'^*)) = \prod_{j = 1}^B p(o_j|Q_{\theta}(\alpha',\alpha'^*)),
\end{equation}
which we can write as
\begin{equation}
    \prod_{j = 1}^B p(o_j|Q_{\theta}(\alpha',\alpha'^*)) = \prod_{j = 1}^B p(o_j|Q_{b}(\vec{\alpha}'),\hat{R}_j)\left| \det \frac{\partial f_\theta^{-1}(\vec{\alpha}')}{\partial \vec{\alpha}},\right|,
\end{equation}
using the NF transformation. From here, we may compute a log-likelihood,
\begin{equation}
    \log p(B|Q_{\theta}) =
    \sum_{j = 1}^B \log p(o_j|Q_{b}(\vec{\alpha}'),\hat{R}_j) + \log \left| \det \frac{\partial f_\theta^{-1}(\vec{\alpha}')}{\partial \vec{\alpha}} \right|,
    \label{eq:log_likelihood_NFQST}
\end{equation}
which you may recognise as a Kullback-Leibler (KL) divergence. To compute the probability of outcomes $o_j$ on the base distribution, notice that 
\begin{equation}
    p(o_j| Q_{b}(\vec{\alpha}), \hat{R}_j) = \text{Tr}(
    \hat{\rho}_{b} \hat{R}_j) = \int_{\mathbb{C}} d\alpha \; Q_{b}(\vec{\alpha}) P_{\hat{R}_j}(\vec{\alpha}),
\end{equation}
where $P_{\hat{R}_j}(\vec{\alpha})$ is the P-function of the POVM $\hat{R}_j$. This is just phase space integration. Computationally this looks like something difficult to implement, however there are a wealth of different tools from theoretical physics to help us here. The mathematical details of executing phase space integrals do not matter for the purpose of this course, just know that it is possible to execute these integrals, and therefore we can compute log-likelihood in a way that is differentiable with respect to the variational parameters $\vec{\theta}$.

We can then back-propagate using negative log-likelihood as our loss function, as was explained in Ch3. We now have all the ingredients needed to write down the pseudocode to train a NF model for the task of QST. This is shown in Algorithm~\ref{alg:NF_density_Ch5}.

\begin{algorithm}
\caption{Boson Density Estimation with NFs (train step)}\label{alg:NF_density_Ch5}
\begin{algorithmic}[1]
\State Dataset (batched) $\mathcal{D} = \{(o_j, \hat{R}_j)\}$, number of iterations $N$, learning rate $\alpha$
\State Initialize parameters $\theta$ of the normalizing flow
\For{each batch, $B\subset\mathcal{D}$}
    \State $\vec{\alpha} = f_\theta^{-1}(\vec{\alpha'})$  
    \State Compute $\mathcal{L} = - \sum_{j \in B} \log p(o_j|Q_{\theta}(\vec{\alpha}'),\hat{R}_j)$  \Comment{Via Eq.~(\ref{eq:log_likelihood_NFQST})}
    \Comment{Negative log-likelihood}
    \State $\theta = \theta - \alpha \nabla_\theta \mathcal{L}$  \Comment{Update parameters}
\EndFor
\end{algorithmic}
\end{algorithm}

For the mathematically observant, you might be worried about encountering similar problems that we saw in NQS, namely that we have need a $\mathbb{C}\rightarrow\mathbb{R}$ calculus over two complex variables in order to do back-propagation. However, this is surprisingly not the case thanks to the structure of NF transformations\footnote{This subtle point is what makes NFs a great tool for probing quantum distributions in phase space!}.  To see why this is the case, it is useful to reconsider the fundamental properties of NFs.

Recall that a NF is simply a type of \textit{diffeomorphism}, which means it transforms the variables in a phase space distribution function, rather than the transforming the function itself. In our quantum optics setting, this means NFs act as a transformation over phase space axes, rather than a directly transforming the quasi-distribution on those phase space axes. Whilst these may sound similar, notice that \textit{one induces the other}. That is, transforming the axes of phase space can change the shape and behaviour of any function on that phase space, but not vice versa. Since normalising flows are acting on the axes themselves, the input-output size is \textit{linear} in the size of the Hilbert space. This means despite the fact that NFs would need to act on complex numbers, we are free to artificially introduce the complex unit outside of a differentiable computation graph, at a cost of doubling the (linear) input size. In the example of a single boson phase space NF, this means $\vec{\theta} \in \mathbb{R}^4$. There is a small connection here to Neural Quantum States and their motivation. In some sense, the Q-function of our state can be thought of as its amplitude distribution. This is because they are an equivalent representation of a state in Fock space, $\mathcal{F}$, which has an \textit{infinite} number of coefficients. However, neural quantum states are acting in the space of coefficients of a wavefunction in a fixed basis, whilst NFs act as a basis transformation, not on the coefficients. 

Beyond efficiency in terms of input-output size, NFs also come with another key advantage. By employing NFs, we are seeking to learn a map which can help us predict observable data. However, the output of the model with a given input phase space function gives another phase space function. In this sense, NFs can explore the space of quantum distributions, meaning they can be used for tomography in a way that NQS cannot. With NFs, the model outputs a phase space function which is \textit{interpretable}, we may compute a density matrix from it. Whereas other deep learning models like NQS have an output which is the neural network model itself. As we will explore in later chapters, this makes the problem of interpretability considerably more difficult; we can use the trained model as a map from the basis to an amplitude, but we cannot actually write down the wavefunction it corresponds to. Hence, NQS cannot solve the problem of recovering an interpretable representation of a wavefunction. However, as we discussed in Sec~\ref{sec:NQS}, often all we are interested in is the properties of observables, not the actual state. As QML practitioners, we should therefore take great care in curating which models may best fit a problem depending on how we need to use it.

Now that we understand how NFs act on phase space axes, we can understand a further reason why they are well suited to inferring quantum distributions; The transformations in a NF are coupling and mixing together different phase space axes. This coupling is what allows for correlation between the phase space axes, and therefore the ability to search over correlations between the constituents of a quantum many-body system. To see how this works, let's now consider trying to learn a three-body quantum state $\rho \in \mathcal{H}_{\mathcal{F}}^{\otimes 3}$ whose Q-function is Gaussian,
\begin{equation}
    Q_{\hat\rho}(\pmb{\alpha}) = \frac{1}{\sqrt{|\pi\hat\sigma_Q}|} e^{-\frac{1}{2}\pmb{\alpha}^T \sigma_Q^{-1} \pmb{\alpha}}.
\end{equation}
Here $\hat\sigma_Q =\hat\sigma + \mathbb{I}/2$, where $\sigma$ is the covariance matrix of the state $\rho$, and $|\cdot|$ indicates the determinant. Since there are three bodies, $\alpha \in \mathbb{C}^6$. Notice here that the scaling again is linear in the number of bodies, so NFs for the three-body states are flows on functions of six complex variables. This means we have NFs over 12 real parameters.

\begin{figure}[t!]
\centering
    \includegraphics[width=\textwidth]{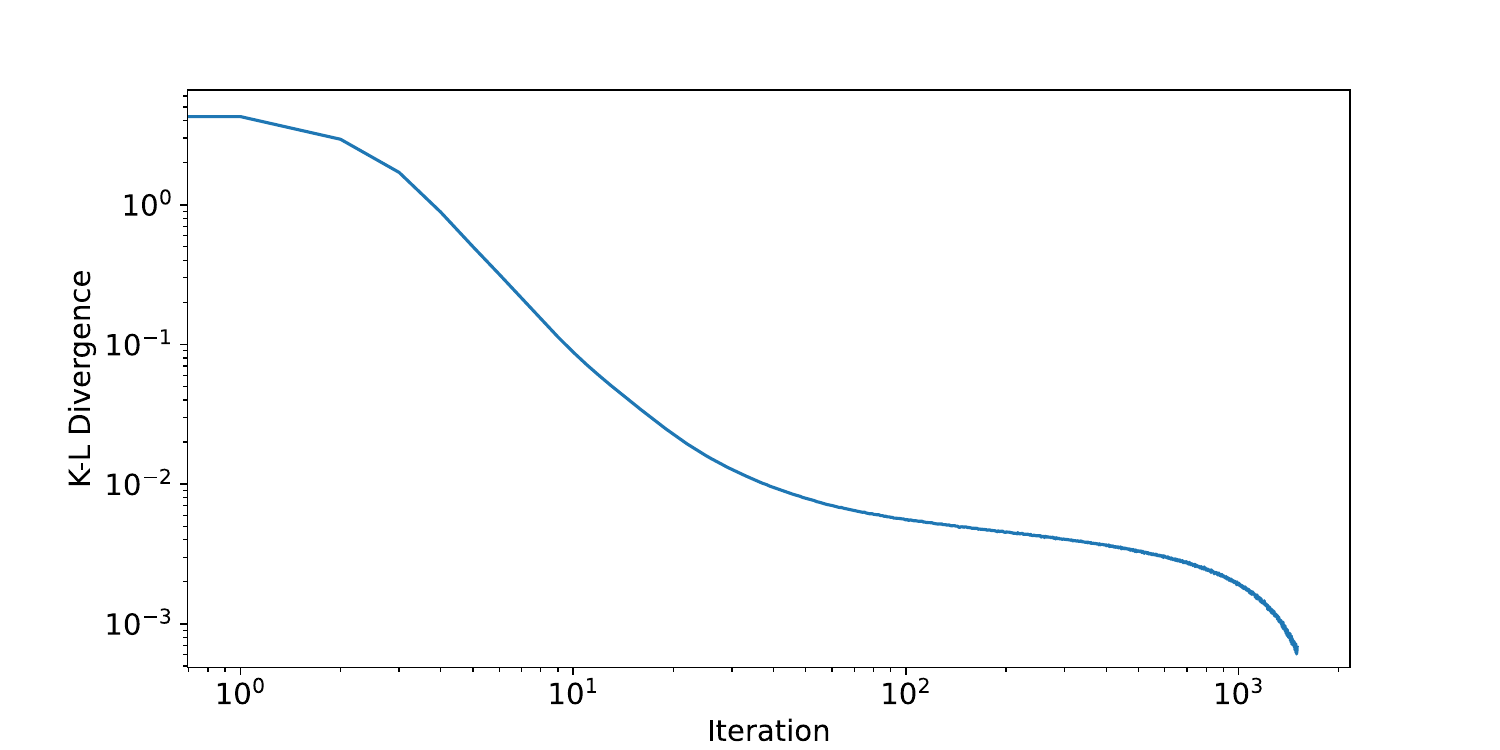}
    \caption{Loss function (K-L divergence) from Eq.~(\ref{eq:log_likelihood_NFQST}) plotted against epochs. Here training was terminated by an early stopping condition, although we can see the subminiature of grokking. This is where a deep learning architecture initially appears to converge to a final value (in this case around $7 \times 10 ^{-2}$) before continuing to decrease again around the $10^3$ epoch mark.}
    \label{fig:kl_loss_3_mode_eg}
\end{figure}

For our choice of covariance matrix, let's opt for a rotation of the squeezed three-mode vacuum state\footnote{For the more experimentally oriented, we are assuming the state is created by mixing three equally squeezed vacuum states on a tritter (3-mode symmetric beam splitter), then doing a phase shift on each mode.}. A squeezed vacuum (SV) has a covariance matrix with block structure,
\begin{equation}
    \sigma_{\text{SV}} = 
\begin{pmatrix}
A & C & C' \\
C & A & C \\
C' & C & A
\end{pmatrix}
\end{equation}
with \(A,C \in \mathbb{R}^{2 \times 2}\) given by 
\begin{equation}
    A = \frac{1}{6} \left(2 e^{2r} + e^{-4r} \right) \mathbb{I}_2, \quad C = \frac{1}{6} \left(e^{2r} - e^{-4r} \right) \hat\sigma_z,
\end{equation}
where $\sigma_z$ is the usual Pauli $z$-matrix and let's choose \( r = 1.5 \) as the squeezing parameter. Next, we apply a phase rotation $\hat{a}_j \rightarrow e^{i \theta_j} \hat{a}_j$ to this, which rotates each mode by the standard $2 \times 2$ rotation matrix,
\begin{equation}
    R(\theta) = \begin{pmatrix}
    \cos \theta & \sin \theta \\
    - \sin \theta & \cos \theta
\end{pmatrix}.
\end{equation}
This means the rotation of the global state reads,
\begin{equation}
    R = R(\theta_1) \oplus R(\theta_2) \oplus R(\theta_3)
\end{equation}
and our ground truth covariance is simply
\begin{equation}
    \sigma = R \cdot \sigma_{\text{SV}} \cdot R^T,
    \label{eq:3_mode_GT_cov}
\end{equation}
where $(\cdot)$ denotes standard matrix multiplication, and we will choose $\theta_j = e^{2 \pi i j/3}$ for $j = 1,2,3$. Putting this together, we have a ground truth Q-function, $Q: \mathbb{R}^6 \to \mathbb{R}$. Training for 1500 epochs with Alg.~\ref{alg:NF_density_Ch5} on a NF architecture with 5 layers and 32 hidden features per layer, we see the KL divergence between the true and learned distributions is well below $1\times10^{-3}$ in Fig.~\ref{fig:kl_loss_3_mode_eg}, at which point training was terminated.

\begin{figure}[t!]
\centering
        \includegraphics[width=\textwidth]{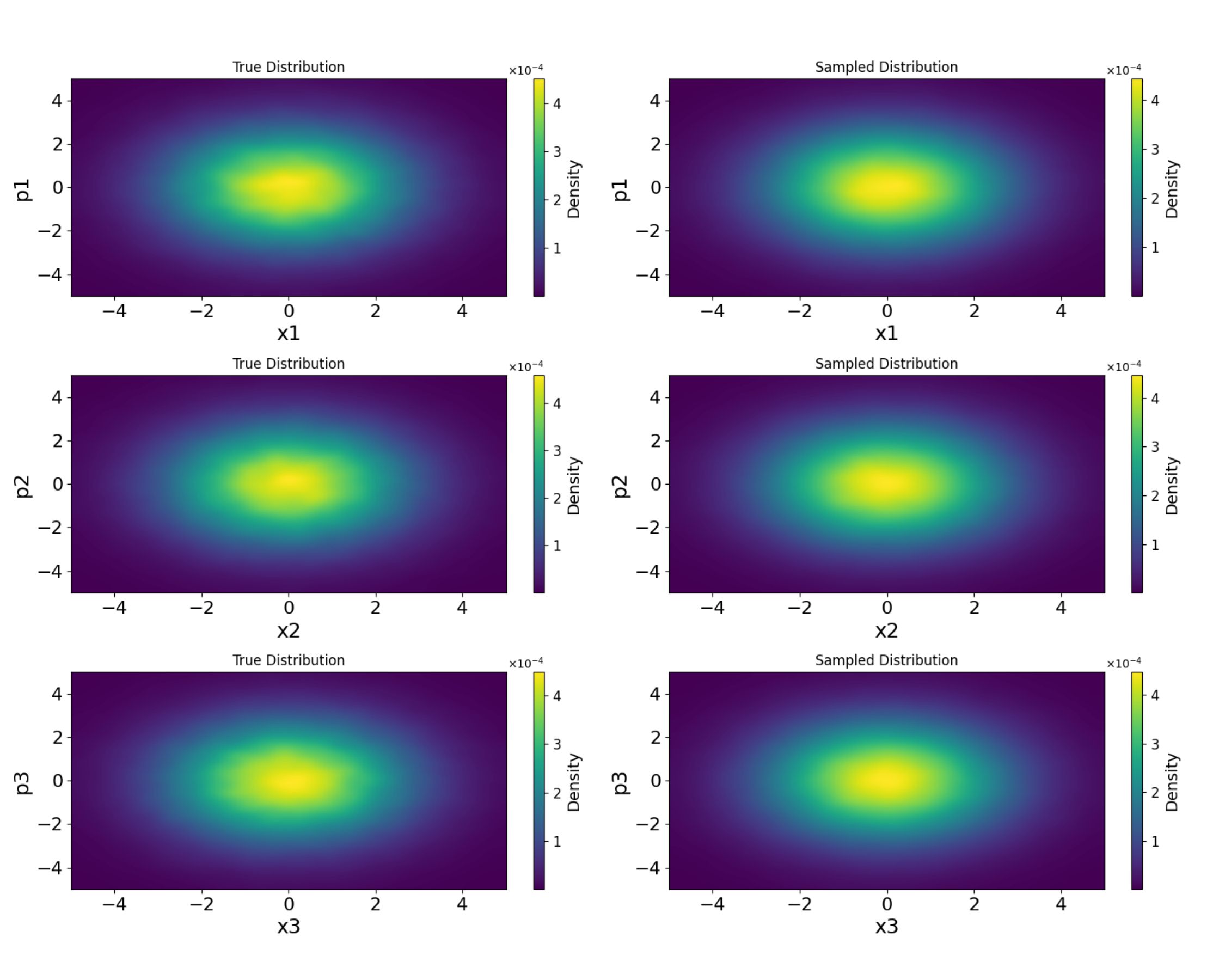}
    \caption{Marginal distribution of the ground-truth Q-function (left) and the trained NF model (right) after 20 epochs. }
    \label{fig:marginals_3_mode_eg}
\end{figure}

However, to better-visualise what the NF has learned, we can break down $Q:\mathbb{R}^6 \to \mathbb{R}$ into local distributions in two ways. First,  we can visualise the local \textit{marginals}. The local marginal distribution can be found by integrating out the remaining coordinates. For example
\begin{equation}
    Q_{\text{marg.}}(\alpha_1,\alpha_1^*) = \int_{\mathbb{R}^4} d \alpha_2 d\alpha_2^* d\alpha_3 d\alpha_3^*\;  Q_{\hat\rho}(\pmb{\alpha}),
\end{equation}
and so on for the other subsystems. This gives three heat maps for ground truth and the learned distribution (one per boson) shown in Fig.~\ref{fig:marginals_3_mode_eg}.

\begin{figure}[t!]
\centering
    \includegraphics[width=\textwidth]{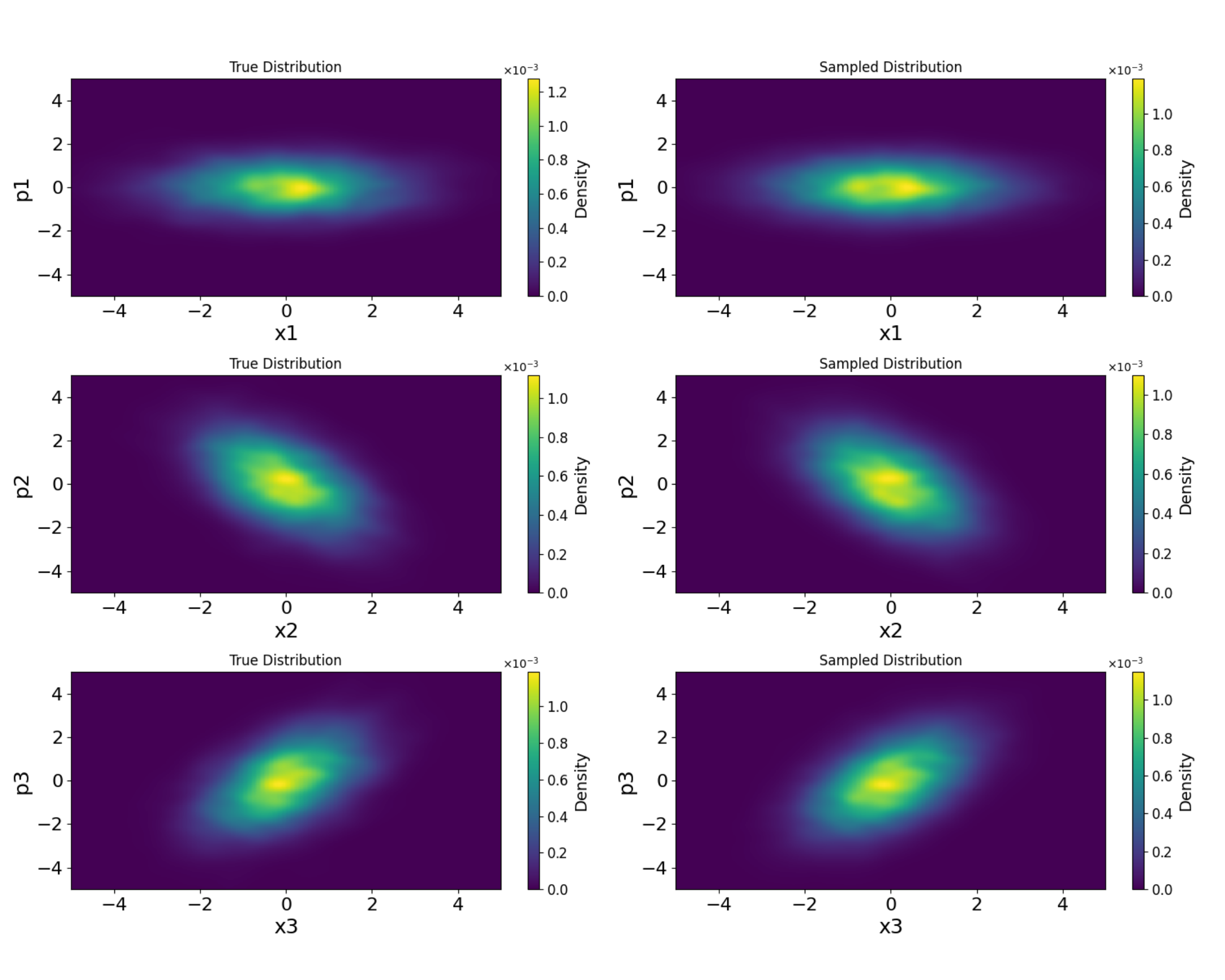}
    \caption{Samples from the conditional distribution of the ground-truth state Q-function in Eq.(\ref{eq:3_mode_GT_cov}) (left) and from the conditional distribution of the trained NF model (right) after 20 epochs. In all three plots, the marginal is conditioned on the other two at the origin. For example the left most plot's conditional marginal is $\rho(p_1,q_1)= \rho(p_j,q_j| 0,0,0,0)$. We can see more clearly the effects of the local rotation in the conditionals rather than the marginals. These are correctly captured by the NF model. Notice that we have plotted the heat-map for \textit{samples} of the ground truth. This is to highlight that the NF architecture isn't trained on the analytic function, rather it is trained on samples from it. You can see the effect of this most clearly in the third row, where the NF has over-fitted to samples of the global distribution, and this has filtered through the Gaussian KDE.}
    \label{fig:conditionals_3_mode_eg}
\end{figure}

Second, we can compute the local \textit{conditional} distributions. These are local functions conditioned on the others taking specific values. We will take
\begin{equation}
    Q_{\text{cond}}(\alpha_1,\alpha_1^*) = Q_{\hat\rho}(\boldsymbol{\alpha}, \boldsymbol{\alpha}| \alpha_2 = \alpha_3 = \alpha_2^* = \alpha_3^* = 0),
\end{equation}
which conditions the first boson's distribution on the other global coordinates at the origin. Likewise for the second and third boson, we will condition on the other coordinates being at the origin. Analytically, this produces samples with heat-maps (one per boson) shown in Fig.~\ref{fig:conditionals_3_mode_eg} (left).

Yielding a conditional distribution can be done with rejection sampling. However for a high-dimensional distribution in $\mathbb{R}^6$, this it becomes vanishingly unlikely that \textit{all four} out of six coordinates on two subspaces are within $\epsilon$ of the origin. Since the ground truth was Gaussian, we can resort to Kernel Density Estimation (KDE) as a special case to fit a Gaussian density function to the learned distribution, and then sample from the Gaussian fit's conditionals. Note that is a special case; a generic learned non-Gaussian  Q-function's conditionals could not be found by fitting a Gaussian to them
In our Gaussian case, we can then sample from the conditionals to produce three heat-maps (again one per boson) shown in Fig.~\ref{fig:conditionals_3_mode_eg}.

\subsection*{From inference to control}

As we have seen, neural networks are a highly versatile tool for inference problems for both stationary and dynamical settings in quantum processes. So far, we have learned how they can be used to retrospectively understand minimal energy states, entangled states, time evolution, and phase space. We might therefore wonder if we can employ neural networks to perform quantum control. That is, to actively steer a quantum system and drive it towards desired outcomes. After-all, quantum optimal control can be phrased as the problem of inferring a quantum operator that maximises likelihood of observing some target outcome. With quantum control, we can use training data in a different way. Instead of considering each forward pass independently, we can start to create deep learning architectures which run online with quantum processes. This has made deep learning an indispensable tool for quantum computing, with recent claims of supremacy relying on techniques like deep reinforcement learning for reliable error correction and control \cite{ai2024quantum}. As both fields progress, a deeper connection between them is flourishing that is currently state-of-the-art for control quantum dynamical processes. Whilst optimal control and deep reinforcement learning is beyond the scope of this course, we refer interested readers to \cite{niu2019universal} and \cite{sutton1998reinforcement}.

\section{Denoising Quantum State Tomography Reconstruction with Neural Networks}

In the previous section, we explored the application of Normalizing Flows for quantum state tomography in quantum optics. Here, we shift our focus to the QST task for many-body quantum systems with a local Hilbert space dimension 
$d$. Notably, when $d=2$, this corresponds to the case of many-qubit systems

Let us begin with the motivation behind QST: Modern quantum technologies rely on resources such as coherence, quantum entanglement, and Bell nonlocality. To evaluate the advantages these resources offer, it is essential to certify their presence. The resource content of a quantum system is inferred from the statistical properties (e.g., correlations) observed in the outcomes it produces. In the quantum framework, this information is encapsulated in the density matrix.

In practical scenarios, the density matrix can be reconstructed using limited data obtained from experimentally accessible measurements—a process referred to as quantum state tomography. However, conventional QST protocols are inherently affected by various sources of noise, including measurement calibration errors, dark counts, losses, and technical disturbances, among others. These noise factors are highly challenging to model accurately and can ultimately lead to decoherence in the system, diminishing or erasing the quantum resources present.

In recent years, deep learning has made significant strides in quantum technologies, providing innovative approaches to quantum state tomography (QST). Neural networks can learn noise directly from experimental data without requiring knowledge of its sources or underlying models. This enables them to mitigate not only shot noise, which is intrinsic to QST, but also other disturbances. Requiring only minimal assumptions about the system, neural networks are particularly well-suited for certification tasks. They effectively act as denoising filters for standard QST protocols, enhancing the fidelity of the reconstructed density matrix.

 In this example, we follow the original work proposing using neural network as a denoising filter for QST \cite{Palmieri2024}. 
Let us consider a quantum system on which we perform measurements, described by a density matrix $\hat{\tau}$ (target) which we want to estimate. With the help of any QST protocol we obtain its reconstruction, i.e. a density matrix $\hat{\rho}$. We consider the reconstructed density matrix $\hat{\rho}$ as a noisy version of the actual target density matrix $\hat{\tau}$. Our aim, is to train a neural network architecture in such a way, that after feed-forward the QST reconstructed density matrix $\hat{\rho}$, at the output we obtain its denoised version $\bar{\hat{\rho}} \simeq \hat{\tau}$. Because both $\hat{\tau}$, and $\hat{\rho}$ are Hermitian matrices, it is beneficial to perform its Cholesky decomposition of $\hat{\rho}$, i.e.
\begin{equation}
    \begin{split}
    \hat{\tau} & = \hat{C}_\tau \hat{C}^\dagger_\tau,\\
    \hat{\rho} & = \hat{C}_\rho \hat{C}^\dagger_\rho,
    \end{split}
\end{equation}
where $\hat{C}_{\tau,\rho}$ are lower triangular matrices. It allows us to reduce necessary input size of our denoising neural network architecture.  We treat the Cholesky matrix $C_\rho$ as a \textit{noisy} version of the target Cholesky matrix $C_\tau$, and we cast the training problem as a supervised learning. As an input to the neural network, we consider flattened Cholesky matrix $\hat{C}_\rho$ into the one-dimensional vector containing real, and imaginary part.

The neural network  $f_{\boldsymbol{\theta}}$ transforms vectorized Cholesky matrix $\hat{C}_{\rho}$ into an output $f_{\boldsymbol{\theta}}(\hat{C}_{\rho})  = C_{\bar{\rho}}$,
and denoised density matrix $\bar{\hat{\rho}}$ is defined as
\begin{equation}
    \bar{\hat{\rho}} = \frac{\hat{C}_{\bar{\rho}} \hat{C}_{\bar{\rho}}^\dagger}
    {\text{Tr}[\hat{C}_{\bar{\rho}} \hat{C}_{\bar{\rho}}^\dagger]},
\end{equation}
which is an improved approximation of the target density matrix, $\bar{\hat{\rho}} \simeq \hat{\tau}$.
To train our neural network, we have to define a proper cost function. The cost function for this task should be a quantum distance between two density matrices, i.e. target $\hat{\tau}$ and the reconstructed $\bar{\hat{\rho}}$ from the network output. On of measures of distance between density matrices is the Bures distance defined as
\begin{equation}
    D_B(\hat{\rho},\hat{\tau}) = 2 - 2\sqrt{F(\hat{\rho}, \hat{\tau})},
\end{equation}
where,
\begin{equation}
\sqrt{F(\hat{\rho}, \hat{\tau})} = \mathrm{Tr}\left[\sqrt{\sqrt{\hat{\rho}}\hat{\tau}\sqrt{\hat{\rho}}} \right], \end{equation}
is the square root of quantum fidelity between $\hat{\rho}$, and $\hat{\tau}$. The square root of the fidelity $ \sqrt{F(\hat{\rho},\hat{\tau})}$ can be expressed as (see Ref.\cite{Bengtsson2006}, Eq.~$9.30$)
\begin{equation}
   \label{eq:deforiginal}
    \sqrt{F(\hat{\rho},\hat{\tau})} = \max_{\substack{\hat{\rho} =  AA^\dagger \\
      \hat{\tau} = BB^\dagger}
    }\mathrm{Tr}[AB^\dagger + BA^\dagger]/2,
\end{equation}
where the maximization is over the \textit{complex amplitudes} $\{A,B\}$ which constitute a polar decomposition of $\{\hat{\rho},\hat{\tau} \}$ respectively. 
From Eq.~\eqref{eq:deforiginal} we see that for   the Cholesky decomposition, $\{\hat{\rho} = C_\rho C_\rho^\dagger, \hat{\tau} = C_\tau C_\tau^\dagger\}$ ), the following inequality always holds:
\begin{equation}
   \sqrt{ F(\hat{\rho},\hat{\tau})} \geq  \mathrm{Tr}[C_\rho C_\tau^\dagger +C_\tau C_\rho^\dagger]/2.
\end{equation}
Finally, rewriting $2$ as $1+1 = \mathrm{Tr}(\hat{\rho}+\hat{\tau}) = \mathrm{Tr}(C_\rho C_\rho^\dagger+C_\tau C_\tau^\dagger)$ we arrive at
\begin{equation}
\begin{split}
    D_{\mathrm{B}}(\hat{\rho},\hat{\tau})\leq & \mathrm{Tr}[C_\rho C_\rho^\dagger+C_\tau C_\tau^\dagger-C_\rho C_\tau^\dagger-C_\tau C_\rho^\dagger] \\
    =& \mathrm{Tr}[(C_\rho -C_\tau)(C_\rho-C_\tau)^\dagger] \\ =&  D_{\mathrm{HS}}^2(C_\rho,C_\tau),
\end{split}
\end{equation}
where $ D_{\mathrm{HS}}^2(C_\rho, C_\tau)$ is the Hilbert-Schmidt distance   defined as:
\begin{equation}
\label{eq:HSdistcomplex}
    D_{\mathrm{HS}}^2(C_\rho, C_\tau) = \mathrm{Tr}[(C_\rho- C_\tau)(C_\rho-C_\tau)^\dagger].
\end{equation}
As it was shown in Ref.\cite{Palmieri2024}, the Hilbert-Schmidt distance between two density matrices, is equivalent to a mean squared error (MSE)  of the matrix elements of their Cholesky decomposition. To see it, let us consider a $d\times d$ complex matrix $K$, $\{ K_{\alpha\beta}\}_{\alpha,\beta\in [d]\times [d]}$.
Next, let us introduce the vectorization $\vec{K}$ of its matrix elements as
\begin{equation}
    \vec{K} = \Re[{\mathbb{K}}]\oplus\Im[{\mathbb{K}}]\ ,
\end{equation}
where $\mathbb{K}$ is the flattening of the matrix, i.e., $\mathbb{K} = (K_{11},K_{12},..,K_{dd})$, and $\oplus$ the direct sum of vectors, $\vec{v}\oplus\vec{u} = (\vec{v},\vec{u})$. 
Let $K = C_\rho - C_\tau$, then the square HS distance, Eq.~\eqref{eq:HSdistcomplex}, reads
\begin{align}
D^2_{\rm HS}(C_\rho,C_\tau) &= \mathrm{Tr}(KK^\dagger) =\sum_{\alpha,\beta}|K_{\alpha\beta}|^2 \\
&= \sum_{\alpha,\beta}\Re{K_{\alpha\beta}}^2+\Im{K_{\alpha\beta}}^2 \\
&= (\Re{\mathbb{K}}\oplus\Im{\mathbb{K}})\cdot (\Re{\mathbb{K}}\oplus\Im{\mathbb{K}}) \\
&= ||\vec{C}_\rho - \vec{C}_\tau||^2  = \mathrm{MSE}(\vec{C}_\rho,\vec{C}_\tau).
\end{align}
As such, we see that MSE is the natural cost function of a standard feed-forward neural network, and as the loss function we tak
\begin{equation}
\label{eq:costf}
    {\cal L}(\boldsymbol{\theta})  = \| \vec{C}_{\tau}-\vec{C}_{\boldsymbol{\theta}}\|^2.
\end{equation}

To train the model, we consider a dataset containing $N_{\rm train}$ training samples $\{\hat{\rho}_l \}$. 
By minimizing Eq.~\eqref{eq:costf}, we obtain the set of optimal parameters $\bar{\boldsymbol{\theta}} $ for our model
$h_{\bar{\boldsymbol{\theta}}}$. The trained neural netwrok allows for the reconstruction of the target density matrix $\hat{\tau}$ via Cholesky matrix $C_{\bar{\rho}}$, i.e.,
\begin{equation}\label{eq:inference}
    \hat{\bar{\rho}} = \frac{C_{\bar{\rho}} C_{\bar{\rho}} ^\dagger}{\text{Tr}[C_{\bar{\rho}} C_{\bar{\rho}} ^\dagger]} \simeq \hat{\tau},
\end{equation}
where $C_{\bar{\rho}}$ is reshaped from $\vec{C}_{\bar{\rho}} =h_{\bar{\boldsymbol{\theta}}}[\vec{C}_\rho]$.
\newline

The choice of architecture should be motivated by capturing correlations between elements of the target density matrix $\hat{\tau}$. In particular, it can contain stack of one-dimensional convolutional layers $h_\mathrm{cnn}$  i.e.:
\begin{align}
\label{eq: transformer function}
     f_{\boldsymbol{\theta}}[\vec{C}_{\rho}] = \sigma(h_{\rm cnn}) 
      \circ\dots\circ \sigma(h_{\rm cnn}) \circ  \sigma(h_{\rm cnn})[\vec{C}_{\rho}] \ ,
\end{align}
 where $\sigma$ can be any non-linear function.

The practical aspect of the considered protocol rely on the fact that to train our model, we can use synthetically constructed random Haar density matrices \cite{Mele2024}, while infere its performance on a specific, experimentally-oriented, protocol. We do not have to construct experimentally training dataset, which would be very demanding, while we can adapt trained model directly to QST data obtained from experimental protocols. To construct our training dataset ${\cal D} = \{\vec{X}, \vec{y}\}$ we have to perform two steps.
In the first step, we  numerically generate random pure states, known a  \textit{Haar vectors} \cite{Mele2024}. 
Such ensemble is defined on the pure state space $\hat{\tau} = |\Psi\rangle\langle\Psi|$.
Haar vectors (or equivalently unitaries), can be drawn from i.i.d. random normal variables $\mathcal{N}(0,1)$ . The construction is as follows: Let $A$ be a complex square matrix sampled as before. Then, we apply the so-called QR decomposition $A = QR$, where $Q$ is unitary and $R$ is upper-triangular. Then $\hat{\tau} = Q\Lambda $, where $\Lambda = \mathrm{diag}(\{R_{\alpha\alpha}/|R_{\alpha\alpha}| \}_\alpha)$, is Haar random. Formally, it takes advantage of the link between states $\ket{\Psi}\in \mathcal{H}$, and the unitary group. The fundamental property of such measure is that it is invariant under unitary transformations, thus making any vector in $\mathcal{H}$ equiprobable. 
\footnote{In practice, we can use the python package \textit{QuTip} generate target density matrices $\hat{\tau}$ from Haar measure.}. Random, pure, Haar states form our labels $\vec{y}$.
In the second step, for each $\hat{\tau}$ we perform its reconstruction $\hat{\rho}$ via any QST method. Set of reconstructed density matrices forms our data $\vec{X}$.

%% file: circuits/TEN_blocks.tikz
\begin{tikzpicture}
        \tikzstyle{wire}=[thick]
        \tikzstyle{gate3}=[draw, fill=white, minimum width=1cm, minimum height=2.9cm]
    
        \foreach \i in {0,1,2,3,4,5} {
            \draw[wire] (0, -\i) -- (6, -\i);
        }
        
        \foreach \i in {0.25, 1.75, 3.25, 4.75} {
            \node[gate3] at (\i+0.5, -1.0) {};
            \node[gate3] at (\i+0.5, -4.0) {};
        }

    \node at (0.75, -1.0) {}; 
    \node at (2.25, -1.0) {}; 
    \node at (3.75, -1.0) {}; 
    \node at (5.25, -1.0) {}; 

    \node at (0.75, -4.0) {}; 
    \node at (2.25, -4.0) {}; 
    \node at (3.75, -4.0) {}; 
    \node at (5.25, -4.0) {}; 

    \end{tikzpicture}

%% file: circuits/MPS_blocks.tikz
\begin{tikzpicture}
    \tikzstyle{wire}=[thick]
    \tikzstyle{gate3}=[draw, fill=white, minimum width=1cm, minimum height=2.9cm] 
    \tikzstyle{gate1}=[draw, fill=white, minimum width=1cm, minimum height=0.9cm]
    \tikzstyle{gate2}=[draw, fill=white, minimum width=1cm, minimum height=1.9cm] 

    \foreach \i in {0,1,2,3,4,5} {
        \draw[wire] (0, -\i) -- (6, -\i);
    }

    \node[gate3] at (0.75, -1.0) {};

    \node[gate3] at (2.25, -2.0) {};

    \node[gate3] at (3.75, -3.0) {};

    \node[gate3] at (5.25, -4.0) {};

\end{tikzpicture}

%% file: circuits/HAE_blocks.tikz
\newcommand{\CNOT}[3]{
    \node[fill=black, circle, minimum size=4pt, inner sep=0pt] at (#1, #2) {};
    \draw[thick] (#1, #2) -- (#1, #3 - 0.25);
    \node[draw, thick, circle, minimum size=0.5cm, inner sep=0pt] at (#1, #3) {};
}
\begin{tikzpicture}
    \tikzstyle{wire}=[thick]
    \tikzstyle{gate3}=[draw, fill=white, minimum width=1cm, minimum height=2.9cm] 
    \tikzstyle{gate1}=[draw, fill=white, minimum width=1cm, minimum height=0.9cm]
    \tikzstyle{gate2}=[draw, fill=white, minimum width=1cm, minimum height=1.9cm] 
    \tikzstyle{gate6}=[draw, fill=white, minimum width=1cm, minimum height=5.9cm] 

    \foreach \i in {0,1,2,3,4,5} {
        \draw[wire] (0, -\i) -- (6, -\i);
        
    }

    \foreach \i in {0,2,4} {
        \CNOT{1.8}{-\i}{-\i - 1}
    }

    \foreach \i in {1,3} {
        \CNOT{2.35}{-\i}{-\i - 1}
    }

    \node[gate6] at (0.75, -2.5) {};

    \node[gate6] at (3.5, -2.5) {};
    
    \foreach \i in {0,2,4} {
        \CNOT{4.8}{-\i}{-\i - 1}
    }

    \foreach \i in {1,3} {
        \CNOT{5.35}{-\i}{-\i - 1}
    }

\end{tikzpicture}

%% file: circuits/AL_blocks.tikz
\begin{tikzpicture}
    \tikzstyle{wire}=[thick]
    \tikzstyle{gate3}=[draw, fill=white, minimum width=1cm, minimum height=2.9cm] 
    \tikzstyle{gate1}=[draw, fill=white, minimum width=1cm, minimum height=0.9cm]
    \tikzstyle{gate2}=[draw, fill=white, minimum width=1cm, minimum height=1.9cm] 

    \foreach \i in {0,1,2,3,4,5} {
        \draw[wire] (0, -\i) -- (6, -\i);
    }

    \node[gate3] at (0.75, -1.0) {};
    \node[gate3] at (0.75, -4.0) {};

    \node[gate3] at (2.25, -2.0) {};
    \node[gate2] at (2.25, -4.5) {}; 
    \node[gate1] at (2.25, 0.0) {};

    \node[gate3] at (3.75, -3.0) {};
    \node[gate1] at (3.75, -5.0) {}; 
    \node[gate2] at (3.75, -0.5) {};

    \node[gate3] at (5.25, -1.0) {};
    \node[gate3] at (5.25, -4.0) {};

\end{tikzpicture}

%% file: circuits/blocks_decomposition1.tikz
\newcommand{\CNOT}[3]{
    \node[fill=black, circle, minimum size=4pt, inner sep=0pt] at (#1, #2) {};
    \draw[thick] (#1, #2) -- (#1, #3 - 0.25);
    \node[draw, thick, circle, minimum size=0.5cm, inner sep=0pt] at (#1, #3) {};
}
\begin{tikzpicture}
    \tikzstyle{wire}=[thick]
    \tikzstyle{gate3}=[draw, fill=white, minimum width=1cm, minimum height=2.9cm] 
    \tikzstyle{gate1}=[draw, fill=white, minimum width=0.6cm, minimum height=0.6cm]
    \tikzstyle{gate2}=[draw, fill=white, minimum width=1cm, minimum height=1.9cm] 

    \foreach \i in {0,1,2} {
        \draw[wire] (0, -\i) -- (1.5, -\i);
    }

    \node[gate3] at (0.75, -1.0) {};

    \node at (2.0, -1.0) {$=$};

    \foreach \i in {0,1,2} {
        \draw[wire] (2.5, -\i) -- (6.5, -\i);
        \node[gate1] at (3.5,-\i) {};
    }

    \draw[dashed] (4.5,0.75) -- (4.5,-2.75);

    \CNOT{5.0}{0}{-1}
    \CNOT{5.5}{-1}{-2}
    \draw[thick] (6.0, 0) -- (6.0, -1); 
        \node at (6.0, 0) {$\times$};
        \node at (6.0, -1) {$\times$};

\end{tikzpicture}

%% file: circuits/blocks_decomposition2.tikz
\newcommand{\CU}[3]{
    \node[fill=black, circle, minimum size=4pt, inner sep=0pt] at (#1, #2) {};
    \draw[thick] (#1, #2) -- (#1, #3 + 0.3);
    \node[draw, thick,  minimum size=0.6cm, fill = white] at (#1, #3) {};
}
\begin{tikzpicture}
    \tikzstyle{wire}=[thick]
    \tikzstyle{gate3}=[draw, fill=white, minimum width=1cm, minimum height=2.9cm] 
    \tikzstyle{gate1}=[draw, fill=white, minimum width=0.6cm, minimum height=0.6cm]
    \tikzstyle{gate2}=[draw, fill=white, minimum width=1cm, minimum height=1.9cm] 

    \foreach \i in {0,1,2} {
        \draw[wire] (0, -\i) -- (1.5, -\i);
    }

    \node[gate3] at (0.75, -1.0) {};

    \node at (2.0, -1.0) {$=$};

    \foreach \i in {0,1,2} {
        \draw[wire] (2.5, -\i) -- (6.5, -\i);
        \node[gate1] at (3.5,-\i) {};
    }

    \draw[dashed] (4.25,0.75) -- (4.25,-2.75);

    \CU{5.0}{0}{-1}
    \CU{6.0}{-1}{-2}

\end{tikzpicture}

%% file: circuits/blocks_decomposition3.tikz
\begin{tikzpicture}
    \tikzstyle{wire}=[thick]
    \tikzstyle{gate3}=[draw, fill=white, minimum width=1cm, minimum height=2.9cm] 
    \tikzstyle{gate1}=[draw, fill=white, minimum width=0.6cm, minimum height=0.6cm]
    \tikzstyle{gate2}=[draw, fill=white, minimum width=1cm, minimum height=1.9cm] 

    \foreach \i in {0,1,2} {
        \draw[wire] (0, -\i) -- (1.5, -\i);
    }

    \node[gate3] at (0.75, -1.0) {};

    \node at (2.0, -1.0) {$=$};

    \foreach \i in {0,1,2} {
        \draw[wire] (2.5, -\i) -- (6.5, -\i);
    }

    \node[gate2] at (3.25, -0.5) {};
    \node[gate2] at (4.55, -1.5) {};
    \node[gate2] at (5.8, -0.5) {};

\end{tikzpicture}

%% file: circuits/blocks_decomposition4.tikz
\begin{tikzpicture}
    \tikzstyle{wire}=[thick]
    \tikzstyle{gate2}=[draw, fill=white, minimum width=1cm, minimum height=1.9cm] 
    \tikzstyle{gate3}=[draw, fill=white, minimum width=1cm, minimum height=2.9cm] 
    \tikzstyle{gate4}=[draw, fill=white, minimum width=1cm, minimum height=3.9cm] 

    \foreach \i in {0,1,2,3} {
        \draw[wire] (0, -\i) -- (1.5, -\i);
    }

    \node[gate4] at (0.75, -1.5) {};

    \node at (2.0, -1.5) {$=$};

    \foreach \i in {0,1,2,3} {
        \draw[wire] (2.5, -\i) -- (6.5, -\i);
    }

    \node[gate3] at (3.25, -1.0) {}; 
    \node[gate2] at (4.55, -2.5) {}; 
    \node[gate3] at (5.8, -1.0) {};  

\end{tikzpicture}

%% file: figures/fig_Ch5_HL_vanilla.tikz
\begin{tikzpicture}[
    every edge/.style = {draw,->}
  ]
    \node (state) at (0, 0) [draw, minimum width=3cm] {$|\psi_0\rangle$};
  \node (h) at (3, 0) [draw] {$e^{-iH_T t}$};
  \node (times) at (2, -1.8) [draw, circle] {$\times$};
  \node (end) at (2, -3.3) [] {};
  \draw (state) edge[out=270, in=90, looseness=1] (times);
  \draw (h) edge[out=270, in=90, looseness=1] (times);
  \draw (times) edge[] (end);
\end{tikzpicture}

%% file: figures/fig_Ch5_HL_vanilla_ode.tikz
\begin{tikzpicture}[
    every edge/.style = {draw,->}
  ]
  \node (start) at (0, 1.6) [draw, minimum width=3cm] {$|\psi_0\rangle$};
  \node (state) at (0, 0) [draw, minimum width=3cm] {$|\psi\rangle$};
  \node (h) at (3, 0) [draw] {$-iH(\theta)$};
  \node (times) at (2, -1.8) [draw, circle] {$\times$};
  \node (end) at (2, -3.3) [] {};
  \draw (state) edge[out=270, in=90, looseness=1] (times);
  \draw (h) edge[out=270, in=90, looseness=1] (times);
  \draw (times) edge[] (end);
  \draw (start) edge[] (state);
  \node (ode) at (1.1, -0.9) [draw, minimum width=6cm, minimum height=3.5cm, label={[shift={(-2,-3.5)}]ODEint}] {};
\end{tikzpicture}

%% file: figures/fig_Ch5_HL_state_vector_node.tikz
\begin{tikzpicture}[
    every edge/.style = {draw,->}
  ]
  \node (start) at (0, 1.6) [draw, minimum width=3cm] {$|\psi_0\rangle$};
  \node (state) at (0, 0) [draw, minimum width=3cm] {$|\psi\rangle$};
  \node (h) at (3, 0) [draw] {$-iH(\theta)$};
  \node (times) at (3, -2) [draw, circle] {$\times$};
  \node (mlp) at (-1, -2) [draw, minimum height=1cm, minimum width=3cm] {NN$(\varphi)$};
  \node (add) at (2, -4) [draw, circle] {$+$};
  \node (end) at (2, -5.5) [] {};
  \draw (state) edge[out=270, in=90, looseness=1] (times);
  \draw (h) edge[] (times);
  \draw (state) edge[out=270, in=90, looseness=1] (mlp);
  \draw (times) edge[out=270, in=90, looseness=1] (add);
  \draw (mlp) edge[out=270, in=90, looseness=1] (add);
  \draw (add) edge[] (end);
  \draw (start) edge[] (state);
  \node (ode) at (0.6, -2) [draw, minimum width=7cm, minimum height=5.5cm, label={[shift={(-2.6,-5.5)}]ODEint}] {};
\end{tikzpicture}

%% file: chapters/chapter_6_conclusion.tex
\chapter{Conclusion and Outlook}
In this chapter, we review the three main components on which deep learning relies. We then discuss their unifying structure, which invites a brief exposition on the concept of natural gradient descent. We then discuss two important broader aspects of deep learning and its position in the larger field of AI. These are interpretability, and the difference between pattern finding and knowledge acquisition. After placing natural gradients in this context, we invite readers to other seminal texts and references to further deepen their knowledge.

Deep Learning relies on three main ingredients. The first is artificial neural networks, i.e. variational non-linear functions optimized to solve a given task. The second one is automatic differentiation, allowing calculating gradients of neural network's variational parameters. The third one is a proper loss function which encodes the desired behaviour of an artificial neural network. It's minimization over variational parameters given the empirical data and results in a trained neural network providing solution for our task.

The unifying structure here is therefore that deep learning relies on being able to do gradient-based algorithms in some set of coordinates that we used to define Loss functions.  For quantum states, these were gradient based algorithms in Hilbert space and phase space coordinates. For dynamical processes, these were gradient-based algorithms to model operators through their group representation. All of these objects come naturally equipped with the notion of a derivative.

We might therefore wonder if there is a way to leverage these derivatives in order to update our parametrisations. Indeed, \textit{natural gradient descent} is a way of being able to do differential updates in a coordinate system that naturally reflects the symmetries of the problem at hand \cite{amari1998natural, martens2020new, stokes2020quantum}.  In fact, this is nothing more than an extension of Noether's theorem being used to help us perform updates based on symmetry properties again. What makes natural gradient descent special is that it works \textit{in the coordinates of the problem directly}. Therefore if we can perform natural gradient descent, our representation of the problem is in some sense \textit{minimal}. This is beneficial to us because it means
\begin{enumerate}
     \item We have fewer parameters to update, so small changes in the coordinates are more likely to make meaningful changes in the loss function. 
     \item We can start to talk about interpretability (at least semantically) because we have a way of interpreting how the differential updates behave in coordinates that are meaningful to us.
\end{enumerate}

Now let us consider the following: If we can encode a problem at hand in such a way that gradient-based updates happen in its coordinates, then one could argue that indeed these models \textit{do} have understanding. After-all, they are working differentially in the parametrisations that physicists have come to know and love. However, our models are themselves \textit{models}. Using mathematics as a language to investigate how nature behaves as we do in physics is nothing more than trying to make this so-called connection ourselves.  This is why deep learning can be so powerful.  With the right design, deep learning has the potential to unlock major gains in the fields of quantum information, many-body physics, and their dynamical processes. This idea leads us to the discussion interpretability and knowledge acquisition, two broader aspects of DL that place it in the context of the larger field of AI, and address our current consideration.

First, there is a notion of semantic intepretability, which involves investigating neuron activation patterns in middle-layers of (usually very-) deep neural networks. This is a similar idea to the $\beta$-VAE, where we tried to \textit{disentangle} features from one another, except this time we are not changing the parameters. Rather, we are trying to understand under what prior conditions they activate. If that activation is quasi-unique for a given feature or class across all data that demonstrate that feature, then we could postulate that this pattern \textit{corresponds} to this feature or class.  This is an active area of researchers of large language models \cite{templeton2024scaling, besta2024graph, tang2023large}.
As tempting as it might be to believe that this is evidence of consciousness or understanding, we would like to take this moment to recall the Chinese Room thought experiment. Deep learning models are simply learned representation of the so-called ``rules'' or equations that connect input data to our desired behaviour. 
When we can encode problems of interest into the language of deep learning, our models are merely connecting these in the parametric space we have given them. Interpreting the answers comes from \textit{our} understanding of what those parameters mean. There is a big difference between the result of a universal approximation, and true understanding of the process it is modelling. Researchers should always therefore be sceptical of the outputs of their models. Deep learning results should never be taken at the level of an ``oracle''. 

Second, on knowledge acquisition, we refer to \textit{World of Patterns. A global history of knowledge} \cite{Bod2022}, where Rens Bod argues that knowledge begins with spotting patterns. That is, regularities extracted from experience. However, knowledge matures only when we elevate those regularities into principles that explain them. Deep-learning systems sit squarely in the first half of this story; they mine vast data sets to yield phenomenological laws, yet they cannot \textit{yet} perform the second, asymmetric step in which a principled theory deduces the very patterns it was built to explain. In the full empirical cycle a theory produces new predictions, which, after empirical testing, subsequently have repercussions on the theory itself. Only then does genuine scientific understanding emerge. Patterns alone, however many, do not force the leap to underlying mechanism. Kepler’s planetary curves awaited Newton’s gravitation, classical Ising model in two-dimensions awaited Onsager's solution to understand phase transition and its critical temperature, just as today’s ML-discovered regularities await a deeper dynamical account in classical or quantum physics. Hence DL presently offers powerful phenomenological description, while the construction and refinement of the bottom-level theory, the engine that can generate and revise those descriptions, remains beyond the machine’s reach and firmly within the scientist’s (for now!).

Finally, we conclude by remarking that the purpose of these notes was to equip you with the basic tools to begin exploring the ML research field on your own. The field is developing rapidly year-on-year making it hard for entry-level readers to get to grips with it. 
It is our hope that these notes offer somewhat a ``timeless introduction'' to the field that you can refer back to, in that the basic concepts and high-level ideas will persist through such rapid progress. 
If you are interested in starting a career in quantum machine learning we can make the following literature recommendations. For those interested in gaining a better understanding of deep learning, we recommend the book of Bronstein ``\textit{Geometric deep learning: Grids, groups, graphs, geodesics, and gauges}'' \cite{bronstein2021geometric}, as well as its recent mathematical prerequisite review \cite{borde2025mathematical}. 
For readers interested in the interdisciplinary aspects of this field, we recommend the book by M. Schuld and F. Pettrucione ``\textit{Machine Learning with Quantum Computers, $2^{nd}$ Edition} \cite{schuld2021machine}'' and the book of A. Dawid et. al ``\textit{Modern applications of machine learning in quantum sciences}'' \cite{dawid2022modern}. 
It goes without saying, of course, that we recommend M. Nielsen and I. Chuang's ``\textit{Quantum Computation and Quantum Information}'' \cite{nielsen2010quantum} for a more in-depth introduction to quantum computing and information. Finally, for those interested in the intersection of AI and quantum computing, we refer to \cite{klusch2024, Acampora2025}.